# Modelling the molecular composition and nuclear-spin chemistry of collapsing prestellar sources*

P. Hily-Blant$^{1,2}$†, A. Faure$^2$, C. Rist$^2$, G. Pineau des Forêts$^{3,4}$, D. R. Flower$^5$
$^1$*Institut Universitaire de France*
$^2$*Université Grenoble Alpes, CNRS, IPAG, F-38000 Grenoble, France*
$^3$*IAS (UMR 8617 du CNRS), Bâtiment 121, Université de Paris Sud, F-91405 Orsay, France*
$^4$*LERMA (UMR 8112 du CNRS), Observatoire de Paris, 61 Avenue de l'Observatoire, F-75014 Paris, France*
$^5$*Physics Department, The University, Durham DH1 3LE, UK*

**ABSTRACT**

We study the gravitational collapse of prestellar sources and the associated evolution of their chemical composition. We use the *University of Grenoble Alpes Astrochemical Network* (UGAN), which includes reactions involving the different nuclear–spin states of $H_2$, $H_3^+$, and of the hydrides of carbon, nitrogen, oxygen, and sulfur, for reactions involving up to seven protons. In addition, species-to-species rate coefficients are provided for the ortho/para interconversion of the $H_3^+$ + $H_2$ system and isotopic variants. The composition of the medium is followed from an initial steady state through the early phase of isothermal gravitational collapse. Both the freeze–out of the molecules on to grains and the coagulation of the grains were incorporated in the model. The predicted abundances and column densities of the spin isomers of ammonia and its deuterated forms are compared with those measured recently towards the prestellar cores H-MM1, L16293E, and Barnard B1. We find that gas–phase processes alone account satisfactorily for the observations, without recourse to grain-surface reactions. In particular, our model reproduces both the isotopologue abundance ratios and the ortho:para ratios of $NH_2D$ and $NHD_2$ within observational uncertainties. More accurate observations are necessary to distinguish between full scrambling processes—as assumed in our gas-phase network—and direct nucleus- or atom-exchange reactions.

**Key words:**
Stars: formation—Molecular processes and data—Astrochemistry—ISM: abundances, molecules—ISM: individual objects H-MM1, L16293E, Barnard B1

## 1 INTRODUCTION

The gravitational collapse of gas and dust, which can lead ultimately to the formation of a star and planets, can be observed, in its early stages, through the molecular lines that are emitted by the contracting object. Many such observations have been performed in recent years, by means of ground-based and satellite observatories, and the ALMA array is now providing unprecedented angular resolution. Furthermore, there are converging lines of evidence that cometary ices carry signatures of prestellar core chemistry. Most recently, unexpectedly large abundances of $O_2$ and volatile $S_2$ in the coma of comet 67P/C-G were reported (Bieler et al. 2015; Calmonte et al. 2016), while observations with ground-based interferometers and with the sub-millimeter ESA/Herschel observatory indicate that more than about 80% of water from the prestellar core actually remains in solid form during the collapse of the protostellar envelope towards the protoplanetary disk (van Dishoeck et al. 2014).

As the object collapses, its density increases, and changes in the rates of physical and chemical processes cause the composition of the medium to evolve. These changes can be followed by means of models that incorporate both the dynamical processes associated with the collapse and the physical chemistry of the gas. Given that the timescale for gravitational collapse and those of many of the chemical processes are comparable, it is desirable that the numerical model should be able to solve simultaneously the dynamical and the chemical rate equations.

Our previous models of prestellar objects have either neglected the dynamics and assumed that the chemical com-position attains steady–state at a given density (Le Gal et al. 2014), or followed the contraction of the radius, $R$, of a core of mass $M$ by means of the equation of gravitational free–fall,

$$\frac{1}{R}\frac{dR}{dt} \equiv \frac{1}{x}\frac{dx}{dt} = -\frac{\pi}{2\tau_{\rm ff}x}\left(\frac{1}{x}-1\right)^{1/2} \quad (1)$$

---

* This paper is dedicated to the memory of Charles Malcolm Walmsley, outstanding astrophysicist and friend.
† E-mail: pierre.hily-blant@univ-grenoble-alpes.fr

(Flower et al. 2005), where $t$ is the time and $x \equiv R/R_0 \leq 1$. The initial radius, $R_0$, of the condensation is given by

$$R_0 = \left(\frac{3M}{4\pi\rho_0}\right)^{1/3} \quad (2)$$

with $\rho_0$ the initial mass density. The timescale, $\tau_{\rm ff}$, for free–fall collapse is

$$\tau_{\rm ff} = \left(\frac{3\pi}{32G\rho_0}\right)^{1/2} \quad (3)$$

(Spitzer 1978). By integrating Equ. (1) in parallel with the chemical rate equations, the chemical composition of the medium may be derived, as a function of the (time–dependent) density of the cloud.

Seminal studies of the gravitational collapse of condensations of material in interstellar clouds (Larson 1969, 1972; Penston 1969) have shown that, in the initial isothermal phase, the collapsing core becomes surrounded by an envelope whose density varies as $r^{-2}$, where $r$ is the radial distance to a point in the condensation. The central object subsequently acquires mass via an outwards–propagating expansion shock wave (Shu 1977). This model was adopted by Visser et al. (2009, 2011) in order to simulate the chemical evolution of a molecular medium that is undergoing gravitational collapse. Their (two–dimensional) simulation assumes the condensation to be rotating, and consequently some of the infalling material accumulates in a disk. More recently, Keto et al. (2015) have applied a one–dimensional ("Larson-Penston") dynamical model in their study of the chemistry of collapsing prestellar cores.

In our approach, we distinguish between a central object ("core"), whose initial mass is equal to the Jeans mass and which is undergoing free–fall collapse, and the surrounding object ("envelope"), which forms through mass–loss from the core; the total mass of the core plus the envelope remains constant and equal to the Jeans mass. As the core evolves, its density increases but its mass decreases. The increasing density leads to an enhanced rate of adsorption of the molecular gas on to the grains, whilst an increasing fraction of the molecular column density is contributed by the surrounding envelope. We simulate the structure and chemical composition of the envelope by means of a time–dependent chemical model, assuming a $r^{-2}$ density profile in the envelope and a constant temperature throughout. The column densities of atomic and molecular species are computed in the core and the envelope, and comparisons are made with the corresponding values observed in prestellar sources.

Our treatment of the chemistry of the medium distinguishes itself from other studies of protostellar collapse (Lesaffre et al. 2005; Visser et al. 2011; Aikawa et al. 2012; Keto et al. 2015) by including explicitly the various nuclear–spin states of the carbon, nitrogen, oxygen, and sulfur hydrides and their abundant deuterated forms. It has been known for over a decade that the (observed) deuterium enrichment of molecules such as ammonia is mediated by reactions with the deuterated forms of $H_3^+$ and is intimately linked to the evolution of ortho:para abundance ratios. Our model combines a chemistry of prestellar objects that incorporates isotopic and nuclear–spin modifications of the relevant species with a dynamical simulation of the initial phase of gravitational collapse that results in the formation of the prestellar object. The dynamical and chemical rate equations are solved numerically, in parallel, thereby allowing, in a self-consistent manner, for departures of the physical and chemical states of the medium from steady–state.

In Section 2, we summarize the characteristics of the model; Section 3 contains our results and comparisons with observations of prestellar objects, and Section 4 our concluding remarks.

## 2  THE MODEL

### 2.1  A new dynamical model

We consider a spherically–symmetric, isothermal, non-rotating medium, consisting of neutral, positively– and negatively–charged gas and dust. The mass contained in the collapsing sphere is

$$M_0 = \frac{4\pi}{3}R_0^3\rho_0$$

where $R_0$ is the initial radius and $\rho_0$ is the initial, uniform, mass density. Both the Jeans mass, $M_{\rm J}$, and the critical mass for stability of a Bonnor–Ebert sphere, $M_{\rm crit}$, are of the same order, given by

$$M_{\rm J} \approx M_{\rm crit} \approx \left(\frac{\pi k_{\rm B} T}{\mu G}\right)^{3/2}\left(\frac{1}{\rho_0}\right)^{1/2},$$

where $\mu = 2.33 m_{\rm H} = 3.9 \times 10^{-24}$ g is the mean molecular mass and $T = 10$ K is the kinetic temperature of the gas. Taking $\rho_0 \equiv n_0\mu = 1.4 n_{\rm H} m_{\rm H}$, where $n_0 = 6.0 \times 10^3$ cm$^{-3}$ is the initial number density of the gas (corresponding to molecular gas with $n_{\rm H} = n({\rm H}) + 2n({\rm H}_2) = 10^4$ cm$^{-3}$ and $n({\rm He})/n_{\rm H} = 0.10$), yields $M_{\rm J} \approx 7$ M$_\odot$; the corresponding free–fall time is $\tau_{\rm ff} = 4.4 \times 10^5$ yr.

We compute the chemical evolution of a fluid particle that flows inwards, at speed, $v(R)$, where $R$ is the radius of the free–falling core. Following the early study of Larson (1969), the free–falling core, of radius $R$, has a uniform density $\rho$ and is progressively surrounded by an envelope whose density $\rho_{\rm env}$ evolves as

$$\rho_{\rm env}(R_{\rm env}) = \rho_0\left(\frac{R_0}{R_{\rm env}}\right)^2 = \rho\left(\frac{R}{R_{\rm env}}\right)^2, \quad (4)$$

which implies that the mass of the core is

$$M(R) = M_0\frac{R}{R_0}. \quad (5)$$

The equation of motion under conditions of free–fall gravitational collapse is

$$\frac{d^2R}{dt^2} = -\frac{GM(R)}{R^2} \quad (6)$$

where $G$ is the gravitational constant. Using Equ. (5), Equ. (6) may be integrated, yielding

$$v(R) = \frac{dR}{dt} = -\left[\frac{2GM_0}{R_0}\ln\left(\frac{R_0}{R}\right)\right]^{1/2}. \quad (7)$$

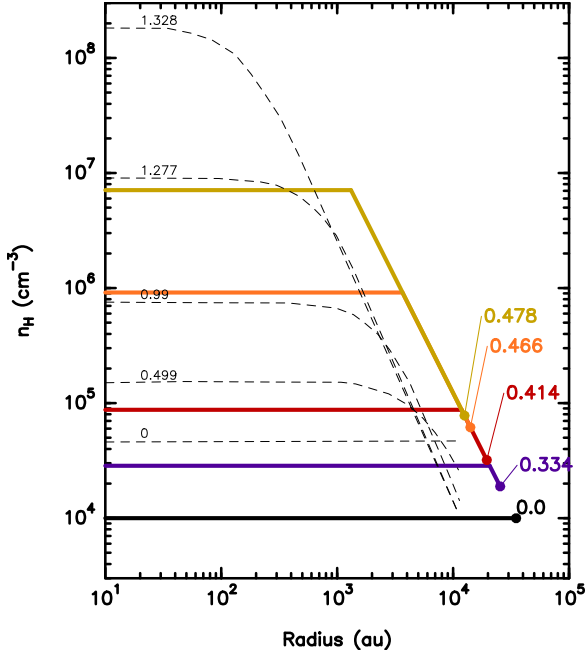

**Figure 1.** The evolution, in our "Larson-Penston" model, of the density profile of the gas (full curves), $n_H$, as a function of the radius for a mass $M_0 = 7 M_\odot$ and an initial density of $n_H = 10^4 \text{cm}^{-3}$. The time elapsed from the initial configuration (in Myr) is indicated for each curve. The filled circles indicate the location of the outer edge of the envelope, which varies with the core radius, in order that the total mass is conserved. The results of Larson (1969), for a $1 M_\odot$ contracting cloud of initial density and radius $n_H = 4.7 \times 10^4 \text{cm}^{-3}$ and $1.1 \times 10^4$ au respectively, are also shown for comparison (broken curves, labeled with the times in units of $10^{13}$ s).

with the initial condition that $v(R) = 0$ when $R = R_0$.

We note that Equs. (4) and (5) have the form of the density and mass distributions of the "singular isothermal sphere" considered by Shu (1977) (his Equ. (2)). When transformed to his dimensionless variables, the infall velocity that we compute exhibits a profile that is similar to the isothermal collapse solutions plotted in his Fig. 3.

Our calculation traces the free–fall collapse of a particle at the edge of the core but is not truly Lagrangian. The flux of matter escaping outwards, across the boundary, $R$, of the collapsing condensation, is given by

$$4\pi R^2 \rho u = -\frac{dM}{dt} = -M_0 v(R)/R_0 \qquad (8)$$

which can be written in the form

$$4\pi R_0^2 \rho_0 u = -\frac{dM}{dt} = -\frac{4\pi}{3} R_0^2 \rho_0 \frac{dR}{dt}. \qquad (9)$$

It follows that the speed at which matter crosses the surface of the sphere is

$$u = -v(R)/3 \qquad (10)$$

and hence the flow speed of this matter in an inertial frame is $2/3 v(R)$. The outer radius $R_{out}$ of the envelope contracts at this same speed (see Fig. 1), thereby ensuring conservation of the total mass of the core and the envelope.

In the following, comparisons to observations will be based on both abundances and column densities. To com-

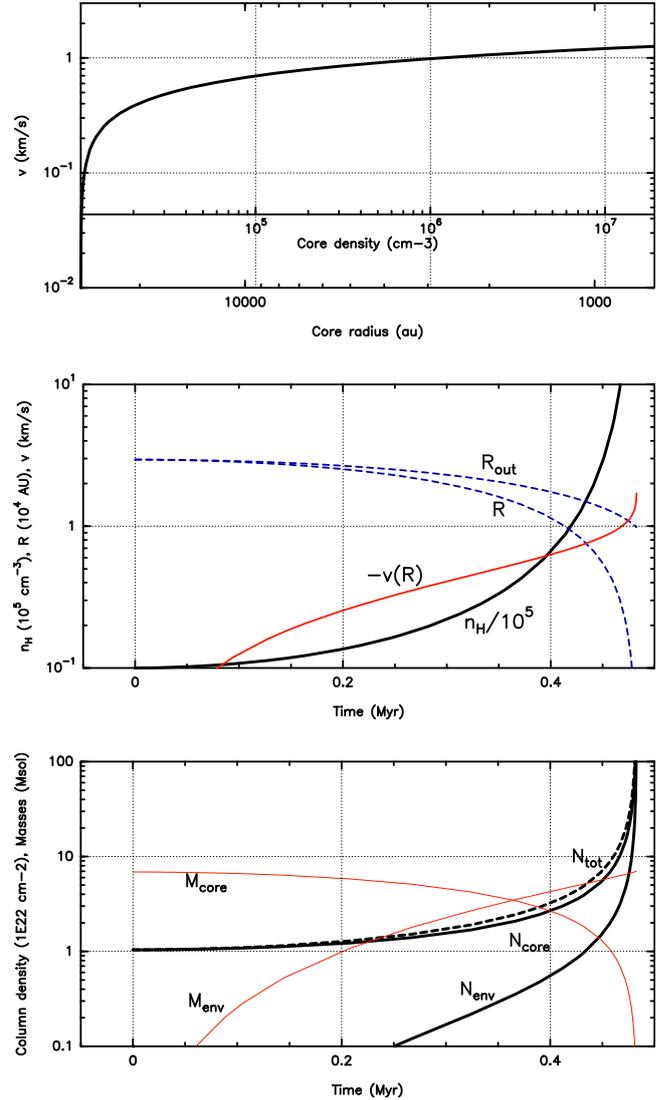

**Figure 2.** Our "Larson-Penston" model. *Top panel:* the velocity profile of the inner collapsing core as a function of its radius (in au) and its central density (in $\text{cm}^{-3}$). *Middle panel:* the temporal profiles of the total H nuclei density $n_H$ (in units of $10^5 \text{cm}^{-3}$), of the radii (in astronomical units) of the core ($R$) and the outer edge of the cloud ($R_{out}$), and of the infall velocity ($-v(R)$, Equ. 7). The parameters are those of Table 1. *Bottom panel:* contributions of the core and the envelope to the mass (in $M_\odot$), $M_{core}$ and $M_{env}$, and to the total column density (in units of $10^{22} \text{cm}^{-2}$), $N_{core}$ and $N_{env}$. In all panels, the time is expressed in Myr.

pute the column densities, we note that during a time interval $dt$, the core contracts by a distance $-dR$ and the column density of any species in the envelope increases by

$$dN_{env} = -\frac{1}{3} n_H(R) X(R) dR, \qquad (11)$$

where $n_H(R)$ and $X(R)$ represent the density and the abundance, with respect to total H nuclei, at radius $R$, respectively. Integration of that quantity thus gives the contribution of the envelope to the total column density. On the other hand, the column density of the uniform collapsing core is

Table 1. Initial parameters of the Flower et al. (2006b) (F06) and of the reference model used in our "Larson-Penston" collapse.

| Parameter | Notation | Unit | Value F06 | Reference |
|---|---|---|---|---|
| Mass | $M_0$ | $M_\odot$ | – | 7 |
| Initial cloud radius | $R_0$ | au§ | – | $3.5 \times 10^4$ |
| Initial Density | $n_H$ | cm$^{-3}$ | $10^4$ | $10^4$ |
| Kinetic temperature | $T$ | K | 10 | 10 |
| Cosmic ray ionization rate† | $\zeta$ | s$^{-1}$ | $1 \times 10^{-17}$ | $3 \times 10^{-17}$ |
| Initial radius of the refractory grain core | $a_g$ | $\mu$m | 0.05 | 0.10 |
| Initial free-fall time | $\tau_{ff}$ | Myr | 0.43 | 0.43 |

§ The abbreviation "au" is used here for the astronomical unit of distance ($1.5 \times 10^{13}$ cm).

† $\zeta$ is the rate of cosmic-ray ionization of $H_2$, namely the sum of the ionization and dissociative ionization of $H_2$.

simply

$$N_{core}(R) = n_H(R)X(R)R. \qquad (12)$$

Note that in our approach, the chemical composition of the envelope is not computed explicitly but is determined by that of the free-falling inner core (see Section 3.3.3). In the following, the contributions to the column densities from both the front and the rear of the cloud are included, assuming spherical symmetry. Although the mass of the envelope becomes greater than that of the contracting core early in the collapse process—when the density, $n_H$, of the core is about $2 \times 10^4$ cm$^{-3}$—the total column density remains dominated by the contracting core. As will be seen below, this is not necessarily the case for the chemical species, owing to freeze-out.

Our approach simulates well the density profile and, qualitatively, the velocity profile calculated by Larson (1969): the evolution of the density profile of the envelope and core is shown in Fig. 1. Initially, when $R = R_0$, the gas is at rest in an inertial frame. Subsequently, the infall velocity increases as $R$ decreases, as may be seen from Equ. (7) and the results in Fig. 2, in which the contributions of the collapsing core and of the envelope to the total column density and mass are also shown. On the other hand, the infall velocity predicted by the "Larson-Penston" solution of Keto et al. (2015) increases with $R$. Furthermore, the gas is not initially at rest, which is inconsistent with one of the conditions specified by Larson (1969). Thus, whilst the profile of the density of the gas that we calculate is analogous to that of Keto et al., the velocity profile is qualitatively different.

The reason for the discrepancy in the velocity profiles is that the "Larson-Penston" profile of Keto et al. derives from the similarity solution considered in Appendix C of Larson (1969) and not from his numerical results. The similarity solution for the infall speed tends, at large radius, to a value of 3.28 times the isothermal sound speed; but Larson notes that the computed velocity profile approaches the similarity solution only slowly, and the collapse must be followed over at least 12 orders of magnitude in density to approach the limiting value. In practice, we are concerned with the early stage of the collapse process, during which the density increases by only a few orders of magnitude from its initial value and the gas at large radius remains approximately at rest.

In our model, the differential equations are solved where $R$ is the only variable. The elapsed time is then computed as $t = \int_{R_0}^{R} dx/v(x)$. The equation of mass conservation applied to each of the neutral, positively- and negatively-charged fluids in the free-falling core takes the generic form

$$\frac{1}{\rho}\frac{d\rho}{dR} = \frac{S}{\rho \dot{R}} - \frac{2}{R} \qquad (13)$$

where $\rho(R)$ is the mass density, $S(R)$ is the rate of production of mass per unit volume, for the appropriate fluid, and $\dot{R} = v(R)$. In the case of a free-falling core with no envelope, such as in F06, the second term on the right hand side would be $3/R$. In our case, the core loses mass, hence building up the envelope (see Equ. 5), such that $dM/M = dR/R$ which yields $d\rho/\rho = -2dR/R$ instead of $-3dR/R$.

Chemical processes result in mass exchange amongst the three fluids (e.g. in the production of a neutral by recombination of a positive and a negative species), but overall mass conservation is maintained, i.e. $S_n + S_+ + S_- \equiv 0$. The three fluids are assumed to have a common flow speed, $v$, and kinetic temperature, $T$.

As the core contracts, from $R_0$ to $R$, its composition evolves. This chemical evolution is followed by means of

$$\frac{1}{n}\frac{dn}{dR} = \frac{N}{n\dot{R}} - \frac{2}{R} \qquad (14)$$

where $n(R)$ is the number density of the species and $N(R)$ is its rate of (chemical) formation per unit volume. Over 200 species, comprising gas- and solid-phase atomic and molecular species and grains, participate in a set of over 3000 chemical reactions, which change the composition and degree of ionization of the medium, notably by the freeze-out of neutral species (including, where relevant, nuclear-spin variants and deuterated forms) on to the surfaces of grains.

The first-order differential equations representing the dynamical and physico-chemical processes were integrated by means of the DVODE algorithm for solving, in particular, "stiff" differential equations (Hindmarsh 1983). We allowed for the coagulation of grains during the collapse, following the description of this process by Flower et al. (2005) and using the measurements of Poppe & Blum (1997). Coagulation reduces the total surface area of the grains and hence the rate of adsorption of the neutral species on to the grains.

The physical parameters used for the two series of models—labeled "F06" and "reference"—of our study are summarized in Table 1. We adopted the initial distribution of the elements amongst the gas and solid phases summarized in Tables 2 and 3 (?)see also[]flower2005; the solid phase consists of grains with a refractory core and an ice mantle that increases in size during the collapse process, owing to freeze-out.

**Table 2.** Fractional elemental abundances, with respect to H nuclei, and their partitioning into volatiles (gas and grain mantles) and refractory (grain cores), and the corresponding levels of depletion ($\delta$, in %) from the gas phase. Numbers in parentheses are powers of 10.

| Element | Total | Volatile | | Refractory | $\delta$ |
| --- | --- | --- | --- | --- | --- |
| | | Gas | Mantles | | |
| H  | 1.0       | 1.0       |           |           | 0   |
| He | 1.00(-1)  | 1.00(-1)  |           |           | 0   |
| C  | 3.55(-4)  | 8.27(-5)  | 5.55(-5)  | 2.17(-4)  | 77  |
| N  | 7.94(-5)  | 6.39(-5)  | 1.55(-5)  |           | 20  |
| O  | 4.42(-4)  | 1.24(-4)  | 1.78(-4)  | 1.40(-4)  | 72  |
| S  | 1.86(-5)  | 6.00(-7)  | 1.82(-5)  |           | 98  |
| Mg | 3.70(-5)  |           |           | 3.70(-5)  | 100 |
| Si | 3.37(-5)  |           |           | 3.37(-5)  | 100 |
| Fe | 3.23(-5)  | 1.50(-9)  |           | 3.23(-5)  | 100 |

**Table 3.** Initial fractional abundances in the grain mantles, expressed relative to the total H nuclei density $n_{\rm H}=10^4\,{\rm cm}^{-3}$. Numbers in parentheses are powers of 10.

| Species | Fractional abundance |
| --- | --- |
| $H_2O$    | 1.03(-4) |
| $CO$      | 8.27(-6) |
| $CO_2$    | 1.34(-5) |
| $CH_4$    | 1.55(-6) |
| $CH_3OH$  | 1.86(-5) |
| $H_2CO$   | 6.20(-6) |
| $HCOOH$   | 7.24(-6) |
| $NH_3$    | 1.55(-5) |
| $H_2S$    | 1.80(-5) |
| $OCS$     | 2.07(-7) |

### 2.2 A new chemical network

Since the Flower et al. (2006b) (hereafter F06) paper, there have been significant changes to the rate coefficients for chemical reactions, notably those determining ortho↔para inter-conversion, e.g. for $H_3^+ + H_2$ and isotopic variants (see below). In addition, new THz facilities have provided the abundances of spin isomers of several hydride molecules, such as $H_2D^+$, which motivated a complete revision of the spin-separated astrochemical network of F06. In the process, the rates of cosmic–ray–induced photo-reactions, which are significant in establishing the steady–state abundances, have also been revised. We provide here a summary of the most salient features of the new University of Grenoble Alpes Astrochemical Network (UGAN), while full details, and the entire network, can be found in the Appendices.

#### 2.2.1 New conversion rates for $H_3^+ + H_2$ and isotopic variants

An important addition to the reaction network is the exoergic proton–exchange reaction

$$oH_3^+ + oH_2 \longrightarrow pH_3^+ + oH_2 \quad (15)$$

which was excluded from the reaction set of F06 on the grounds that the change of proton spin orientation in the $H_3^+$ ion must be accompanied by $oH_3^+ + oH_2 \longrightarrow pH_3^+ + pH_2$, yielding para-$H_2$. However, both forms of $H_2$ can be represented by the combinations of the "spin-up" ($\alpha$) and "spin-down" ($\beta$) single–particle nuclear spin functions, namely $2^{-\frac{1}{2}}(\alpha_1\beta_2 + \alpha_2\beta_1)$ and $2^{-\frac{1}{2}}(\alpha_1\beta_2 - \alpha_2\beta_1)$ for the ortho and para $H_2$ respectively, where the subscripts 1 and 2 label the protons. It follows that a quantum-statistical calculation, such as that undertaken by Hugo et al. (2009), can give rise to non-zero rate coefficients for processes that would be excluded in a more simplistic approach. In fact, Hugo et al. obtained a rate coefficient for reaction (15) that is 4 times larger than for the reaction leading to $pH_2$. As a consequence, the ortho:para-$H_3^+$ ratio calculated using their data may be expected to be lower than when reaction (15) is neglected, as in F06. This expectation is confirmed by the results of the steady–state calculations, as may be seen in Table 5: even though the ortho:para-$H_2$ ratio is lower, the ortho:para-$H_3^+$ ratio is also lower than was calculated by F06[1].

For the important ortho↔para inter-conversion reactions in the $H_3^+ + H_2$ system and isotopic variants, we replaced the ground-state–to–species kinetic rates, published by Hugo et al. (2009) and used extensively in the astrophysical community, by *species-to-species* rates. To do so, the state-to-state rate coefficients computed by Hugo et al. (2009) were averaged, adopting LTE level populations for temperatures up to 50 K. For each reaction, the (logarithm of the) resulting temperature-dependent rates were fitted by a modified Arrhenius law. While the modifications are small for the lightest species—$H_2$, $H_3^+$, and $H_2D^+$—they become more important for the heavier, doubly- or triply-deuterated species. We note that our rates agree to within a factor of two with the species-to-species rates calculated by Sipilä et al. (2017) using a similar procedure. We also checked that our new rates correctly recover thermalized ortho-to-para ratios for $H_3^+$ and its deuterated analogs, at temperatures larger than $\approx 16$ K (Le Bourlot 1991; Flower et al. 2006a). These new rates are provided in Tables D7–D10.

#### 2.2.2 State-of-the-art, fully separated hydride chemistry

Faure et al. (2013); Rist et al. (2013) and Le Gal et al. (2014) have revisited and revised the chemistry of nitrogen–bearing species, which now includes specifically the various nuclear-spin states of the carbon, nitrogen, oxygen, and sulfur hydrides and their abundant deuterated forms. The nuclear-spin state separation was performed using the permutation symmetry approach of Quack (1977). During this process, the rate coefficients of many bi-molecular reactions were updated, following a literature survey. Full details of the assumptions behind this new chemical network are provided in Appendices B and C. In Appendix D we specify the rate coefficients for these reactions that have been adopted in the present calculations, including the steady–state computations reported in columns 3 and 4 of Table 4.

#### 2.2.3 Grain-surface processes

It should be noted that no explicit grain–surface reactions are included in the chemical network (except the formation of $H_2$ and isotopologues). The rates of adsorption of neutral species were computed with allowance for the contribution of the grain mantle to the grain cross section, as described in appendix B of Walmsley et al. (2004). Our treatment of the thermal desorption of molecules, following cosmic ray impact, follows Section 5 of Flower et al. (1995). Ortho- and para-$\mathrm{H}_2$ are assumed to form on grains in the statistical 3:1 ratio of their corresponding nuclear spin states, $I = 1$ and $I = 0$. On the other hand, an ortho-to-para equal to unity is assumed for all other species upon desorption. The desorption of molecules by the cosmic ray–induced ultraviolet radiation field (see Appendix A) is also included.

---

[1] Available at 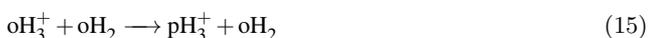 http://massey.dur.ac.uk/drf/protostellar/chemistry_species

**Table 4.** Steady–state abundances, expressed relative to $n_\mathrm{H} = n(\mathrm{H}) + 2n(\mathrm{H}_2) = 10^4\,\mathrm{cm}^{-3}$, as calculated using the chemical reaction set of Flower et al. (2006b) (F06) and our present, updated chemistry. Two values of the cosmic-ray ionization rate, $\zeta$, and of the initial refractory grain core, $a_\mathrm{g}$, have been used, which are indicated. Where the nuclear–spin state is not specified (e.g. $\mathrm{NH}_2\mathrm{D}$), the fractional abundance is the sum of the abundances of the individual nuclear–spin states. Numbers in parentheses are powers of 10.

| Species | F06[§] | Present | |
|---|---|---|---|
| | $\zeta = 1\times10^{-17}\,\mathrm{s}^{-1}$ | $1\times10^{-17}$ | $3\times10^{-17}$ |
| | $a_\mathrm{g} = 0.05\,\mu\mathrm{m}$ | 0.05 | 0.10 |
| H | 2.2(-05) | 2.3(-05) | 6.9(-05) |
| $p\mathrm{H}_2$ | 5.0(-01) | 5.0(-01) | 5.0(-01) |
| $o\mathrm{H}_2$ | 1.4(-03) | 4.1(-04) | 5.4(-04) |
| N | 1.2(-05) | 5.3(-06) | 8.4(-06) |
| $p\mathrm{NH}_3$ | 4.1(-08) | 2.3(-09) | 5.0(-09) |
| $o\mathrm{NH}_3$ | 1.4(-08) | 1.2(-09) | 2.7(-09) |
| $\mathrm{NH}_2\mathrm{D}$ | 4.6(-09) | 2.6(-09) | 4.7(-09) |
| $\mathrm{NHD}_2$ | 2.1(-11) | 6.6(-11) | 1.0(-10) |
| $\mathrm{ND}_3$ | 2.0(-14) | 2.7(-13) | 4.1(-13) |
| CN | 9.2(-09) | 2.0(-09) | 7.4(-09) |
| HCN | 1.1(-07) | 1.6(-09) | 2.8(-09) |
| HNC | 1.4(-07) | 1.4(-09) | 2.9(-09) |
| $\mathrm{N}_2$ | 2.6(-05) | 2.9(-05) | 2.8(-05) |
| $p\mathrm{H}_3^+$ | 7.4(-10) | 1.5(-09) | 4.1(-09) |
| $o\mathrm{H}_3^+$ | 1.1(-09) | 5.5(-10) | 1.8(-09) |
| $o\mathrm{H}_2\mathrm{D}^+$ | 2.0(-11) | 9.6(-11) | 2.7(-10) |
| $p\mathrm{H}_2\mathrm{D}^+$ | 3.4(-12) | 5.1(-11) | 1.3(-10) |
| $o\mathrm{HD}_2^+$ | 1.4(-12) | 2.2(-11) | 5.9(-11) |
| $p\mathrm{HD}_2^+$ | 5.2(-13) | 7.2(-12) | 1.9(-11) |
| $o\mathrm{D}_3^+$ | 3.0(-14) | 8.8(-13) | 2.2(-12) |
| $m\mathrm{D}_3^+$ | 3.8(-14) | 8.2(-13) | 2.1(-12) |
| $p\mathrm{D}_3^+$ | | 4.3(-14) | 1.1(-13) |
| $\mathrm{N}_2\mathrm{H}^+$ | 5.0(-10) | 5.2(-10) | 9.3(-10) |
| $\mathrm{N}_2\mathrm{D}^+$ | 2.5(-12) | 1.8(-11) | 3.1(-11) |

[§]http://massey.dur.ac.uk/drf/protostellar/chemistry_species

**Table 5.** *Top:* steady–state ortho:para (and also m:p in the cases of $\mathrm{D}_3^+$ and $\mathrm{ND}_3$) abundance ratios for the species listed in Table 4. Also given are the corresponding thermalized ratios at 10 K and in the limit of high temperature. *Bottom:* abundances of the deuterated isotopologues of ammonia, relative to $\mathrm{NH}_3$. The abundances are sums over the contributions of the individual nuclear–spin states. Numbers in parentheses are powers of 10.

| Species | Present | | F06[§] | Thermalized | |
|---|---|---|---|---|---|
| | $1\times10^{-17}$ | $3\times10^{-17}$ | $1\times10^{-17}$ | 10 K | $\infty$ |
| $\mathrm{H}_2$ | 7.4(-4) | 1.1(-3) | 2.8(-3) | 3.5(-7) | 3 |
| $\mathrm{H}_3^+$ | 3.6(-1) | 4.3(-1) | 1.4 | 7.5(-2) | 1 |
| $\mathrm{H}_2\mathrm{D}^+$ | 1.8 | 2.1 | 5.8 | 1.9(-3) | 3 |
| $\mathrm{HD}_2^+$ | 3.0 | 3.1 | 2.7 | 9.8(1) | 2 |
| $\mathrm{D}_3^+$ (o:p) | 20.9 | 19.5 | | 4.1(2) | 16 |
| $\mathrm{D}_3^+$ (m:p) | 19.2 | 19.3 | | 1.8(3) | 10 |
| $\mathrm{NH}_3$ | 5.9(-1) | 5.4(-1) | 3.4(-1) | 3.7 | 1 |
| $\mathrm{NH}_2\mathrm{D}$ | 1.7 | 1.7 | | 2.9 | 3 |
| $\mathrm{NHD}_2$ | 2.1 | 2.1 | | | 2 |
| $\mathrm{ND}_3$ (o:p) | 16.6 | 16.5 | | | 16 |
| $\mathrm{ND}_3$ (m:p) | 10.8 | 10.7 | | | 10 |
| $\mathrm{NH}_3$ | 1.0 | 1.0 | 1.0 | | |
| $\mathrm{NH}_2\mathrm{D}$ | 3.3(-1) | 2.2(-1) | 8.3(-2) | | |
| $\mathrm{NHD}_2$ | 8.2(-3) | 4.6(-3) | 3.8(-4) | | |
| $\mathrm{ND}_3$ | 4.3(-5) | 1.9(-5) | 3.6(-7) | | |

[§]http://massey.dur.ac.uk/drf/protostellar/chemistry_species

### 2.2.4 $H_2$-driven non-LTE ortho:para ratios

The thermal equilibrium (LTE) ratio of the population densities of the lowest ortho and para states of $\mathrm{H}_3^+$ is

$$\frac{n(J_K = 1_0)}{n(J_K = 1_1)} = \frac{4}{2} \exp\left(-\frac{32.9}{T}\right)$$

where 4 and 2 are the statistical weights of the ortho and para states, respectively, and 32.9 K is the energy of the lowest ortho state above that of the lowest para state. At a kinetic temperature $T = 10$ K, the LTE ortho:para-$\mathrm{H}_3^+$ ratio is 0.075; Tables 4 and 5 show that their populations are indeed strongly inverted (i.e. the population ratio exceeds its value in LTE). This inversion is a consequence of the dominance of ortho-$\mathrm{H}_2$ in establishing the relative populations of the ortho- and para-$\mathrm{H}_3^+$ states and the inversion of the populations of the $I = 1$ and $I = 0$ states of ortho- and para-$\mathrm{H}_2$, respectively, through the grain–formation process. Similarly strong deviations from LTE (thermalized) ratios are found for most species (Faure et al. 2013).

## 3 RESULTS AND DISCUSSION

The initial distribution of the elements is specified in Tables 2 and 3.

### 3.1 Steady–state composition

It is instructive to compare the initial (steady–state) abundances of species incorporated in the model with the values obtained by F06 for a cloud having a uniform density ($n_\mathrm{H} = 10^4\,\mathrm{cm}^{-3}$) and kinetic temperature ($T = 10$ K). In these models, the cosmic-ray ionization rate of $\mathrm{H}_2$ is $\zeta = 10^{-17}\,\mathrm{s}^{-1}$ and the initial radius of the refractory grain core is $a_\mathrm{g} = 0.05\,\mu\mathrm{m}$ (see Table 1).

As may be seen from a comparison of columns 2 and 3 of Table 4, the recent revisions to the chemistry have repercussions on the composition of the gas. The corresponding ortho:para ratios are listed in the upper part of Table 5. Faure et al. (2013) calculated, in steady state at $T = 10$ K, an ortho:para-$\mathrm{H}_2$ ratio of $9\times10^{-4}$, for a model that is similar to that considered in Tables 4 and 5; their ortho:para-$\mathrm{NH}_3$ ratio was 0.7. As might be expected, these values are closer to the present results than those of F06. In the lower part of Table 5 are the degrees of deuteration of ammonia, which are seen to be greater in the present calculations than in those of F06. This deuteration process is mediated by the deuterated forms of $\mathrm{H}_3^+$, which are produced in reactions of $\mathrm{H}_3^+$ with HD; these reactions were also studied by Hugo et al. (2009). It is the revisions to these rate coefficients that lead to higher abundances of the deuterated isotopologues of ammonia in the present study.

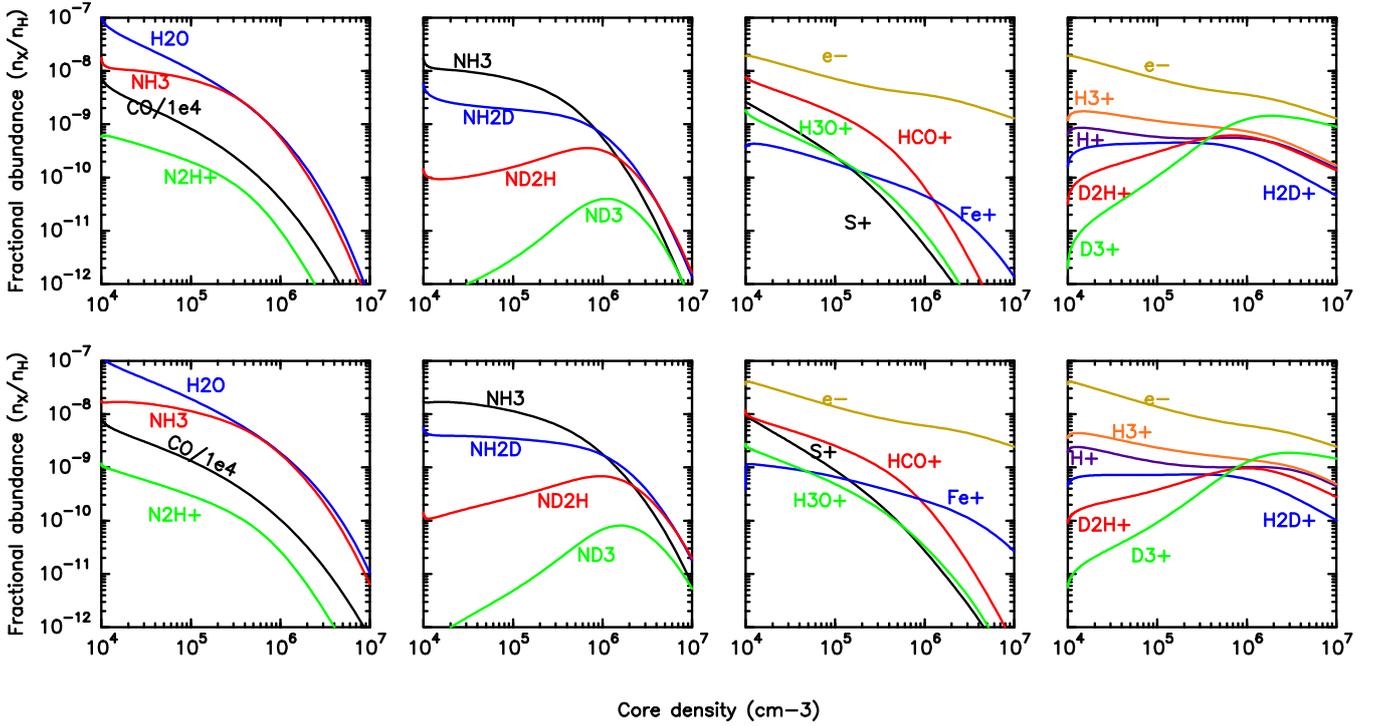

**Figure 3.** Evolution of the fractional abundances of selected chemical species—expressed relative to $n_{\rm H}$—in a cloud that is undergoing contraction in a free–fall collapse (four upper panels), and from our "Larson-Penston" model (four lower panels). In each case, the two left (right) panels focus on neutral (charged) species. The chemical network is described in Section 3.1 and Appendix D. *Upper panels:* models with $\zeta = 1 \times 10^{-17}$ s$^{-1}$, and an initial radius of the refractory grain core $a_{\rm g} = 0.05$ $\mu$m. *Lower panels:* initial parameters of the *reference* model of Table 1.

The significance of the steady–state fractional abundance of any given species depends on the associated equilibrium timescale, which can attain values in excess of 1 Myr under the conditions considered here. Although condensations may remain close to a quasi-static equilibrium for timescales as long as 500 kyr (Brünken et al. 2014), it is unlikely that the chemical composition of the cloud will evolve unperturbed over such long periods of time. Indeed, the timescale for free–fall collapse of the cloud is less than 1 Myr. It follows that the steady–state composition of the gas has limited relevance to observations of prestellar cores. Instead, the composition must be computed under conditions appropriate to collapse, as is described in the following Sections. Both our free–fall and Larson-Penston simulations assume an initial steady-state composition for the physical conditions of our reference model (see Table 1), except where otherwise specified.

### 3.2 Free–fall simulation

It is instructive to consider first the case of free–fall collapse, which provides a point of reference for subsequent models. In Fig. 3 are shown the variations with the collapsing cloud density, $n_{\rm H}$, of the fractional abundances of selected species, notably some of those containing nitrogen.

The variations of the fractional abundances with the density of the medium that are seen in Fig. 3 are qualitatively similar to the results of Flower et al. (2005), who also considered a free–fall model. For the purpose of this comparison, we adopt the same values of the cosmic–ray ionization rate ($\zeta = 1 \times 10^{-17}$ s$^{-1}$) and the initial radius of the refractory grain core ($a_{\rm g} = 0.05$ $\mu$m) as Flower et al. (2005). On the other hand, there are quantitative differences, notably for H$_3^+$ and its isotopologues, which relate to the revisions of the chemistry, as discussed in Section 3.1.

### 3.3 Larson-Penston simulation

At the onset of gravitational collapse, an envelope begins to form around a core that is initially the principal contributor to the total column densities of the molecular species. As the collapse proceeds, the core contracts according to Equ. (6) and its density increases from its initial value; the density profile of the envelope follows from Equ. (4). The envelope becomes increasingly important and, as we shall see, its contribution ultimately dominates the total column densities— in part because the total mass of material advected into the envelope from the core during free–fall is greater than the residual mass of the core.

#### 3.3.1 Fractional abundances

In Fig. 3 are shown the variations in the fractional abundances of selected neutral and charged species as functions of the current density, $n_{\rm H}$, of the core (and hence at the current interface between the core and the envelope). As the density increases, atoms and molecules are adsorbed by the grains, whose radius (of the core and the ice mantle) increases from its initial value of 0.13 μm as more ice is deposited. In these calculations, we assumed the same value, $S = 1$, of the "sticking coefficient"[2] for all adsorbing species. As is evident from Fig. 3, the depletion timescale is slightly longer in the L-P than in free–fall model, as a consequence of the following factors: in the L-P model, the contraction timescale is $4.8 \times 10^5$ yr, 10% longer than the free-fall time, and the value of $\zeta$ (and hence the cosmic–ray desorption rate) is three times larger; but, most significant, the initial radius of the grain core is smaller by a factor of 2 (and the total grain surface area correspondingly larger) in the free–fall than in the L-P model.

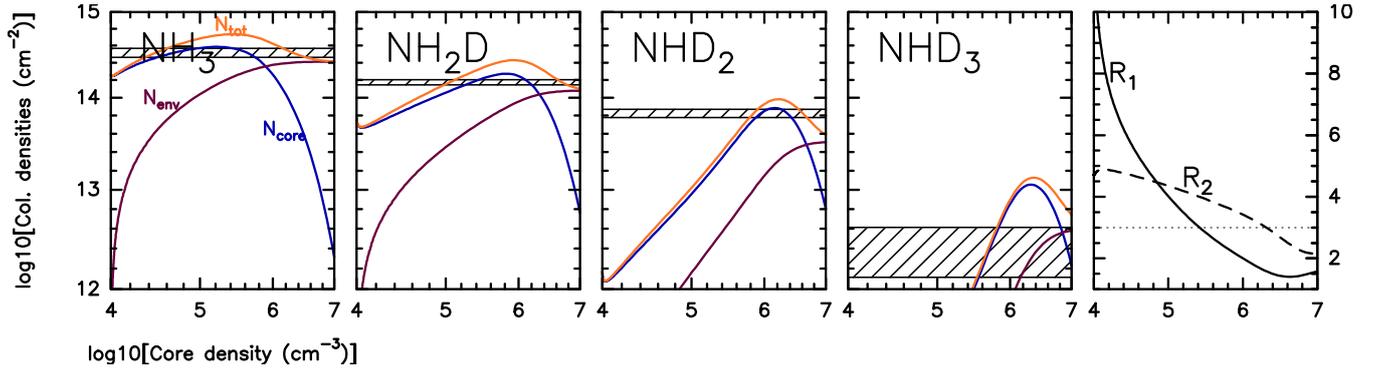

**Figure 4.** Contributions, in our reference "Larson-Penston" collapse model (Table 1), of the core, $N_{\rm core}$, and the envelope, $N_{\rm env}$, to the total column density $N_{\rm tot} = N_{\rm core} + N_{\rm env}$ of the ammonia molecule and its deuterated forms, as functions of the density of the core. The column densities are evaluated as described in Section 2. The LTE column densities in the starless core Ophiuchus/H-MM1 (Harju et al. 2017) are shown for comparison (see also Table 6.) *Far right panel:* the column density double-ratios, $R_1$ and $R_2$, defined in Equs. (16) and (17).

### 3.3.2 Total column densities

In order to illustrate the transition from the preponderance of the core to the preponderance of the envelope, we compare, in Fig. 4, the contributions of the core, $N_{\rm core}$, and the envelope, $N_{\rm env}$, to the column density of the ammonia molecule, $NH_3$, and its deuterated forms, $NH_2D$, $NHD_2$, $ND_3$. Column densities in the core and in the envelope are evaluated as described in Section 2. The total column densities, $N_{\rm tot} = N_{\rm core} + N_{\rm env}$, are also plotted, as are the column densities deduced by Harju et al. (2017) from their observations of the starless core Ophiuchus/H-MM1 by assuming a single excitation temperature and statistical ortho:para ratios (see Table 6).

It is clear from Fig. 4 that the total column density comes to be dominated by the contribution of the envelope, as the core contracts and its density increases but its mass decreases. The composition of the envelope is not uniform, as may be seen from Fig. 3: in the vicinity of the core, adsorption on to the grains depletes the neutral species, which has consequences for the degree of ionization of the gas. Agreement, to better than a factor of 2, with the observed column densities of the ammonia molecule and all of its deuterated forms is found at a density, $n_H$, in the range $5 - 10 \times 10^5$ cm$^{-3}$. Furthermore, since deriving mass, age, or size is not the chief objective of this work, no attempt was made to actually find the best fit to the data. This suggests that gas–phase processes can account for the deuteration of $NH_3$, without the necessity of invoking additional grain–surface reactions, as concluded by Le Gal et al. (2014). In any case, our gas–grain chemistry is dominated by the adsorption of neutral species in the course of gravitational collapse; desorption processes occur too slowly to affect significantly the gas–phase molecular abundances. On the other hand, if molecules were returned to the gas phase by a mechanism that is intrinsically faster than the cosmic ray–induced desorption processes considered here, the molecular column densities might become dependent on chemical reactions on the surfaces of grains. Furthermore, surface processes, and especially those possibly involved in spin conversion, are still poorly constrained (Turgeon et al. 2017), and the impact of such processes on the observed gas-phase ortho:para ratios is difficult to anticipate. Finally, we note that the agreement with the observations is obtained at densities for which the main contribution to the column density of ammonia and deuterated analogs comes from the free–falling core, before freeze-out begins to be dominant.

### 3.3.3 Column densities of spin isomers

Column densities of specific isomers of ammonia and its deuterated forms have been measured towards starless and prestellar cores (Daniel et al. 2016; Harju et al. 2017). Figure 5 shows the comparison with the column densities predicted by our reference isothermal collapse model (see Table 1). We note first that agreement to better than a factor 3 is obtained for all sources and species. In the case of H-MM1, the agreement is obtained at core densities in the range $0.3 - 1 \times 10^6$ cm$^{-3}$, as compared with the value of $1.2 \times 10^6$ cm$^{-3}$ derived from the dust emission (Harju et al. 2017). Similarly, in the case of Barnard B1, our model suggests core densities above approximately $10^6$ cm$^{-3}$, consistent with the value of $3 \times 10^6$ cm$^{-3}$ determined by Daniel et al. (2013). The observation that the core density is higher in B1 than in H-MM1 is correctly reproduced. However, Barnard B1 is likely to be at a more advanced stage of evolution than is described by our model (Gerin et al. 2015), and the agreement that we obtain suggests that ammonia is not tracing the innermost and densest regions. In addition, our calculations slightly under-produce p–$NH_3$ and o–$NH_2D$, which may indicate that gas–depletion through freeze–out is proceeding too rapidly in the model. Finally, in the case of the I16293E starless core, our model intercepts the observed ranges of column densities for o–$NHD_2$ and m–$ND_3$ at two densities, between approximately $10^6$ and $10^7 \, cm^{-3}$. Although the latter density may seem too high for this type of object, we note that it is close to the value of $1.4 \times 10^7 \, cm^{-3}$ obtained from the dust-emission profiles (Bacmann et al. 2016). Overall, the level of agreement with the observations reported in Fig. 5 (and in Fig. 4) lends credence to the validity of the gas–phase chemical network.

---

[2] The sticking coefficient $S$ is not to be confused with the mass production rate in Equ. (13).

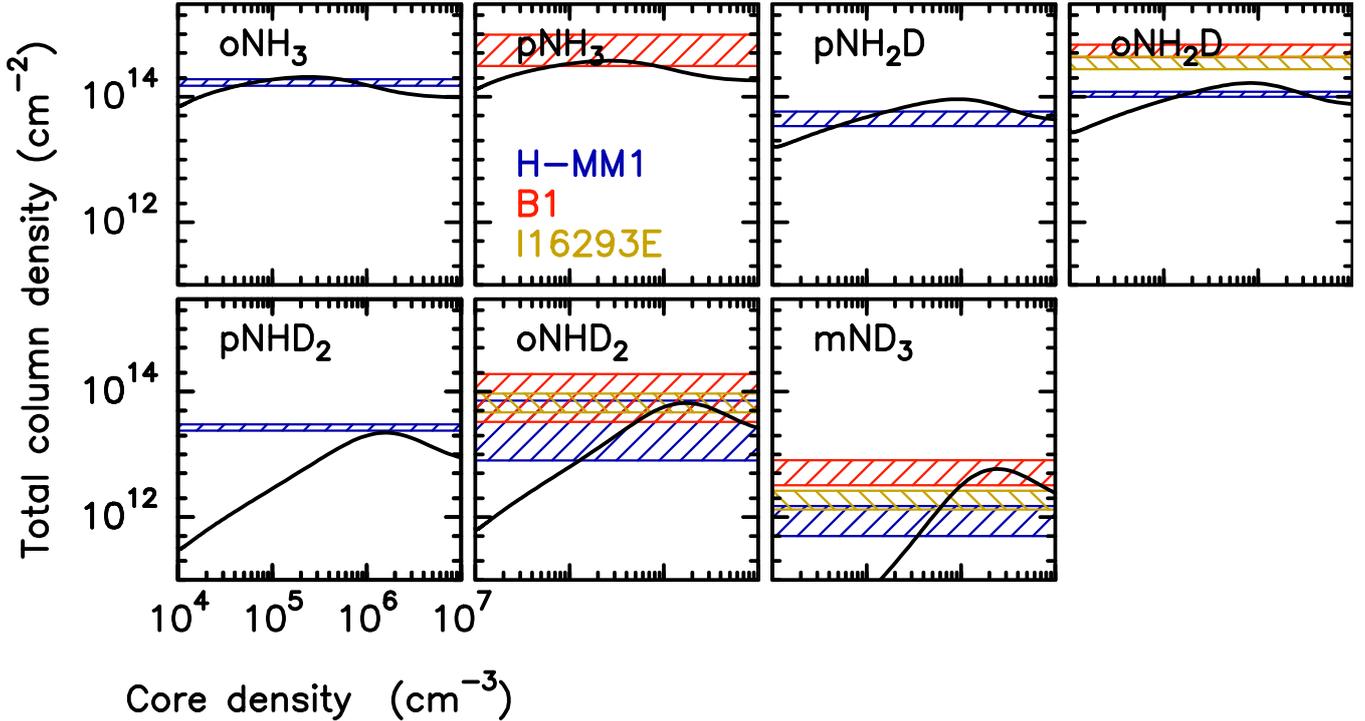

**Figure 5.** Column densities of ammonia and its deuterated forms, as predicted from our "Larson-Penston" collapse model (full lines) and from observations (hashed regions) of the I16293E, Barnard-B1, and H-MM1 prestellar cores (Daniel et al. 2016; Harju et al. 2017) (see Table 6). The initial parameters are those of Table 1.

Further support for our gas–phase network comes from the deuteration of ammonia, from $NH_3$ through to $ND_3$. In the far right panel of Fig. 4, we have plotted the ratios

$$R_1 = \frac{NH_2D/NH_3}{NHD_2/NH_2D} \qquad (16)$$

and

$$R_2 = \frac{NHD_2/NH_2D}{ND_3/NHD_2} \qquad (17)$$

Observationally, both ratios are close to 3, within their uncertainties (Daniel et al. 2016; Harju et al. 2017). If hydrogenation and deuteration were determined by statistics only—as is presumably the case in ices—both these ratios would be exactly equal to 3. However, as Fig. 4 shows, ratios close to 3 can be produced by gas–phase chemistry at the core densities relevant to prestellar cores. It follows that ratios of 3 are not reliable indicators of deuteration taking place in ices, rather than in the gas–phase (Harju et al. 2017). We note also that these ratios evolve significantly from their steady–state values as the density increases.

Whilst our simulation of the Larson-Penston collapse enables us to include the contribution of the envelope to the column densities of the chemical species, the temporal evolution of the chemical composition of the envelope is not calculated explicitly; this approximation could lead to over-estimating the contribution to the column densities of the outer regions of the envelope, owing to an underestimation of freeze-out. However, as freeze-out occurs least rapidly at the densities (lower than $10^5 \, cm^{-3}$) prevailing in the outer envelope, and because the envelope contributes only marginally to the total column density until massive freeze-out in the core takes place, this approximation was verified to have negligible impact on our results.

### 3.3.4 Ortho-to-para ratios

In Fig. 6, we plot the spin-isomer abundance ratios of $NH_3$ and $H_3^+$ and of their deuterated analogues. The calculated $H_3^+$ ortho:para ratio tends to approximately 0.7 as the core density increases. In the case of $NH_3$, our assumption that the deuteration reactions proceed via long–lived intermediate complexes, in which complete scrambling of the nuclei occurs, leads to values of the ortho:para ratio of $NH_2D$ and $NHD_2$ of approximately 1.8 and $2.0 - 2.8$, respectively, over the density range $10^5$-$10^6 \, \mathrm{cm}^{-3}$. These values are consistent, within observational uncertainties, with the ratios measured in H-MM1 and listed in Table 6. The chemical model of Harju et al. (2017) predicts a similar ortho:para ratio for $NH_2D$, but a significantly larger one (in the range $3-5$) for $NHD_2$. On the other hand, the observed ratios are also consistent with the statistical values of 3 and 2, respectively. Ortho:para ratios close to their statistical values were observed by Daniel et al. (2016) in Barnard B1b also.

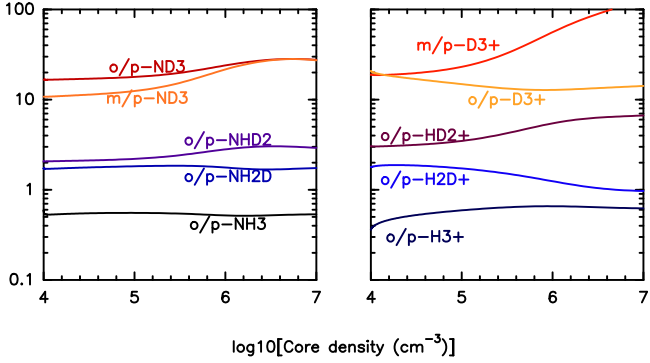

**Figure 6.** Variation with the core density of the ratios of the total column densities of the various nuclear spin states of $NH_3$ (left) and $H_3^+$ (right), and of their respective deuterated analogs. For reference, the corresponding thermalized ratios at 10 K are given in Table 5.

Statistical ratios could indicate that direct gas-phase proton/deuteron hops or hydrogen/deuterium abstraction—instead of full scrambling of nuclei in an intermediate reaction complex—plays an important role in the deuteration process (Harju et al. 2017). Deuteration was indeed found to inhibit the permutation of protons in the $H_5^+$ ion (Lin & McCoy 2015) thus preventing the full scrambling scheme adopted in the Harju et al. and our networks. To investigate the origin of the different predictions for $NHD_2$ from our model and from Harju et al. (2017), we have tested the influence of the reaction $NH_3 + D_3^+ \longrightarrow NH_2D_2^+ + HD$ which eventually leads to $NHD_2$ upon dissociative recombination. This, and similar, reactions are not included in our network because we assume that single particle (H, $H^+$, D or $D^+$) hop is the dominant outcome of the complex forming reaction. This particular reaction results is an increase of the ortho:para ratio of $NHD_2$ by less than 10%. Although other similar reactions have not been included in our network, it appears unlikely that the factor of 1.5-2 difference between our results and those of Harju et al. (2017) arises only from these reactions. We also checked that the difference is not due to the new species-to-species inter-conversion rates (see Section 2.2.1). Alternatively, it could be related to differences in the raw, unfractionated and unseparated, chemical networks. Furthermore, we emphasize that the details of the nuclear spin selection rules in deuteration reactions such as $NH_3 + D_3^+$ are currently not known, and the observational error bars are too large to allow a distinction to be made between complete scrambling and particle hop.

**Table 6.** Column densities (or their $\log_{10}$) of ammonia and its deuterated forms towards three prestellar cores.

| Source | Species | $N$ cm$^{-2}$ | o/p |
|---|---|---|---|
| H-MM1[†] | $oNH_3$ | $1.7\pm0.2\times10^{14}$ | |
| | $oNH_2D$ | $1.1\pm0.1\times10^{14}$ | $2.4\pm0.7$ |
| | $pNH_2D$ | $4.6\pm1.2\times10^{13}$ | |
| | $oNHD_2$ | $4.0\pm3.2\times10^{13}$ | $1.5\pm1.2$ |
| | $pNHD_2$ | $2.7\times10^{13}$ | |
| | $mND_3$ | $1.0\pm0.5\times10^{12}$ | |
| B1[§] | $pNH_3$ | $14.74\pm0.25$ | |
| | $oNH_2D$ | $14.73\pm0.10$ | $\sim 3$ |
| | $oNHD_2$ | $13.90\pm0.38$ | |
| | $mND_3$ | $12.71\pm0.20$ | |
| I16293E[§] | $oNH_2D$ | $14.54\pm0.10$ | |
| | $oNHD_2$ | $13.82\pm0.15$ | |
| | $mND_3$ | $12.27\pm0.15$ | |

[†] LTE column densities from Harju et al. (2017).
[§] $\log_{10}$ values from Daniel et al. (2016).

*3.3.5 Depletion and collapse timescales*

There remains an underlying issue with the dynamical model: the flow velocity, $u$ (Equ. 10), increases too rapidly as a function of the density of the gas, attaining approximately 1 km s$^{-1}$ when $n_H = 10^6 \, \mathrm{cm}^{-3}$; such velocities exceed the limiting values, of the order of 0.1 km s$^{-1}$, that are deduced from the observed line profiles (?)cf.[]bergin2007 and are predicted by hydrodynamical calculations (Foster & Chevalier 1993; Lesaffre et al. 2005; Keto et al. 2015). We remark that the singular isothermal–sphere model of Shu (1977) also produces excessive values of the collapse velocity over a significant fraction of the cloud. As pointed out by Keto et al. (2015), it is the velocity towards the outer edge of the cloud that is responsible for the differences in the shape of the emergent spectral lines. Although the velocity at large radii in our model is probably too high, radiative–transfer calculations are needed to confirm this conclusion.

A consequence of reducing the collapse velocity, and hence extending the duration of the collapse, is to deplete molecules on to grains more rapidly, as a function of $n_H$, resulting in molecular column densities that are lower than observed. This issue is unlikely to be resolved by slowing the freeze–out process, which would require improbably low sticking coefficients or high rates of grain coagulation. It seems more likely that molecules are returned to the gas phase at a rate which exceeds that of desorption by secondary photons (Appendix A) or as a consequence of heating by cosmic-ray impact (Leger et al. 1985), which are the only desorption processes presently incorporated in the model.

## 4 CONCLUDING REMARKS

We have studied the chemical composition of gas under the physical conditions that are believed to be appropriate to collapsing prestellar cores. A set of chemical reactions has been assembled that not only comprises the most recent determinations of the rates of gas–phase reactions but also discriminates between the various nuclear–spin states of $H_2$, $H_3^+$, and the abundant carbon, nitrogen, oxygen, and sulfur hydrides and their deuterated forms. The composition of the medium has been investigated under both steady–state conditions and during the initial stage of gravitational collapse, which ultimately leads to the formation of a low–mass star. The freeze–out of the gas on to grains and the coagulation of the grains themselves are taken into account.

We compared the computed values of the column densities of key molecular species, notably of $NH_3$ and its deuterated forms, with the values observed in starless cores and found that the gas–phase chemistry alone yields satisfactory agreement between the model and the observations, without recourse to grain–surface reactions. Regarding the relative column densities of different nuclear-spin states of $NH_3$ and its deuterated forms, we find that our full scrambling assumption for proton and deuteron exchange yields good agreement with the observations, in contrast with the proposition of Harju et al. (2017) that direct nucleus– or atom–exchange reactions dominate the deuteration process. However, accurate observational measurements are needed to disentangle between the two processes.

The dynamical model that we used derives from the studies of gravitational collapse by Larson (1969, 1972) and Penston (1969). The gravitationally collapsing core loses mass to the surrounding envelope at a rate that ensures that the density profile in envelope $\rho_{env}(R_{env}) \propto R_{env}^{-2}$, where $R_{env}$ is the radius; the total mass of the core and envelope is conserved. This simple model not only simulates the Larson–Penston density profile but also reproduces qualitatively their velocity profile. In order to test more quantitatively the kinematical aspects of the model, we are currently extending our study by computing the line profiles of key species such as CN and HCN and their isotopologues, including their hyperfine structure.

**ACKNOWLEDGMENTS**

The authors thank the anonymous referee for a careful reading and for useful and detailed comments which improved the quality of the paper. PH-B thanks the Institut Universitaire de France for its support. DRF acknowledges support from STFC (ST/L00075X/1), including provision of local computing resources. This work has been supported by the Agence Nationale de la Recherche (ANR-HYDRIDES), contract ANR-12-BS05-0011-01.

## APPENDIX A: COSMIC–RAY–INDUCED PHOTO-REACTIONS

As was discussed by Gredel et al. (1989), secondary electrons produced by cosmic–ray ionization of hydrogen can collisionally excite $H_2$ molecules in the gas, giving rise to ultraviolet photons in the subsequent radiative decay. Most of these photons are absorbed by dust, but some ionize atoms and dissociate or ionize molecules in the medium.

Equ. (5) of Gredel et al. (1989)—for the rate $R_M$ per unit volume of ionization/dissociation of species M by the secondary photons—should read

$$R_M = \zeta n_{H_2} \frac{X(M) \int \sigma_M(\nu) P(\nu) d\nu}{(1-\omega) X(g) \sigma_g + \sum_M X(M) \int \sigma_M(\nu) P(\nu) d\nu}$$

where $\zeta n_{H_2}$ is the rate per unit volume of production of the secondary electrons, and the ratio is the relative probability that a photon is absorbed by an atomic or molecular species, M, rather than the grains, g; $\omega$ is the grain albedo, and $n_H$ is the total number density of hydrogen nuclei. This equation may be rewritten

$$R_M = \zeta \frac{X(H_2) n(M) \int \sigma_M(\nu) P(\nu) d\nu}{(1-\omega) X(g) \sigma_g + \sum_M X(M) \int \sigma_M(\nu) P(\nu) d\nu} \quad (A1)$$

which becomes

$$R_M \approx \zeta X(H_2) n(M) \frac{p_M}{(1-\omega)}$$

in the limit in which absorption of the secondary photons by the dust dominates; $p_M$ is a photo-ionization or -dissociation probability. The factor $X(H_2)$ in this expression was omitted by Gredel et al. (1989), as was recognized by Woodall et al. (2007; section 3.6) and Flower et al. (2007; Appendix A); $X(H_2) = 0.5$ in mainly molecular gas. We have used Equ. A1, together with the more recent determinations of the photo-ionization/dissociation probabilities, $p_M$, of Heays et al. (2017) (Table 20), and adopted $\omega = 0.5$.

During the collapse, $X(g)\sigma_g$, where $\sigma_g = \pi a_g^2$ is the geometrical cross section of the grain, will tend to decrease, given that $X(g) \propto a_g^{-3}$ and that the grain radius, $a_g$, increases owing to coagulation. However, the fractional abundances $X(M)$ also decrease, owing to freeze–out on to the grains.

The cosmic–ray–induced photons also desorb molecules $M^*$ in the ice mantles, returning them to the gas phase. The rate of this process is given by

$$R_M^* \approx \zeta X(H_2) \frac{n(M^*)}{\sum_{M^*} n(M^*)} n_H \frac{\int Y(\nu) P(\nu) d\nu}{(1-\omega)} \quad (A2)$$

where $Y(\nu)$ is the desorption yield (probability of any molecule in the mantle being desorbed, per photon incident on the grain). Taking $\zeta = 1 \times 10^{-17}\,\text{s}^{-1}$, $n_H = 10^4\,\text{cm}^{-3}$, $Y(\nu) = 10^{-3}$, independent of $\nu$ (Hollenbach et al. 2009), and estimating $\int P(\nu) d\nu \approx 0.4$ from fig. 1 of Gredel et al. (1989), yields a timescale (in yr) for this process

$$\frac{n(M^*)}{R_M^*} \approx 10^9$$

which is too long to be significant in the context of star formation. We note that Hollenbach et al. (2009) overestimate the rate of this process by about 2 orders of magnitude: they calculate the flux of the far–ultraviolet secondary photons, $F_{FUV}$, to be of the order of $10^5\,\text{cm}^{-2}\,\text{s}^{-1}$, whereas

$$F_{FUV} \approx \zeta \frac{\int P(\nu) d\nu}{\langle n_g \sigma_g \rangle / n_H}$$

and, taking $\zeta = 1 \times 10^{-17}\,\text{s}^{-1}$ and $\langle n_g \sigma_g \rangle / n_H = 2 \times 10^{-21}\,\text{cm}^2$ from Gredel et al. (1989), as did Hollenbach et al., we obtain $F_{FUV} \approx 2 \times 10^3\,\text{cm}^{-2}\,\text{s}^{-1}$. We note that this value of $F_{FUV}$ agrees with that derived from Fig. 4 of Shen et al. (2004; $E_0 = 400$ MeV, corresponding to $\zeta = 3.1 \times 10^{-17}\,\text{s}^{-1}$), when scaled to the same value of $\zeta$. Hollenbach et al. (2009) appear to have been unaware of the work of Shen et al. (2004).

## APPENDIX B: THE UGAN CHEMICAL NETWORK

The *University of Grenoble Alpes Astrochemical Network* (UGAN) is a significantly upgraded version of the gas-phase network of F06, which included reactions involving species containing H, D, He, C, N, O and S. F06 distinguished between the different nuclear–spin states of $H_2$, $H_2^+$, $H_3^+$ and their deuterated isotopologues, and between those of nitrogen hydrides. Nuclear-spin branching ratios were derived from simple statistical considerations (and conservation of the total nuclear spin) but without recourse to symmetry conservation. A first update of the F06 network consisted in a revision of nitrogen hydrides chemistry (excluding deuterated species) (Le Gal et al. 2014), with special attention to the most recent experimental results—in particular, for the conversion of N to $N_2$ through radical–radical reactions and for the dissociative recombination of the $NH_n^+$ ($n=2$-4) ions. The nuclear–spin selection rules were derived by the method of Oka (2004), as described in Rist et al. (2013) and Faure et al. (2013). These rules result from the conservation of the total nuclear–spin of identical nuclei and the conservation of their permutation symmetry (Quack 1977) or, alternatively, of their rotational symmetry (Oka 2004). Although conceptually different, the two methods are closely related and predict the same spin statistics when applied to multiple H nuclei (fermions). However, in the case of deuterium nuclei (bosons), the one–to–one correspondence between the nuclear–spin angular momentum and the permutation symmetry breaks down. A new method unifying the rotational and permutation symmetries was proposed recently (Schmiedt et al. 2016).

The present network extends the work of Le Gal et al. (2014) to the entire F06 chemical network in a systematic fashion for all hydrides containing C, N, O, and S atoms, and their deuterated forms. Nuclear–spin selection rules were derived from the permutation symmetry approach of Quack (1977), as explained below. Furthermore, many reaction rate coefficients were updated from a literature survey, including recommendations from the KIDA (KInetic Database for Astrochemistry[3]) data sheets.

After spin isomer separation on H and D, the complete new network contains 207 species—including grains (neutral and charged) and adsorbed species—and 3266 reactions (when including adsorption and desorption). The

---
[3] http://kida.obs.u-bordeaux1.fr/

unseparated—or condensed—network contains 151 species and 1136 reactions. The list of chemical species and the corresponding chemistry files are provided in Appendix D.

## B1 Nuclear-spin-state separation

In principle, all molecules with identical nuclei of non-zero spin in equivalent positions should be treated as distinct nuclear–spin isomers. However, distinguishing between nuclear–spin states leads to a large increase in the number of chemical reactions that have to be considered, especially when deuterated species are included. Therefore, we have restricted the spin–state separation to hydrides and their deuterated isotopologues, i.e. to molecules with a single heavy atom and two or more hydrogen or deuterium nuclei; this includes molecules like $NH_2D_2^+$ but excludes species like $C_3H_2$.

### B1.1 Exothermic reactions

For strongly exothermic reactions, we assume (i) conservation of the total nuclear spin and (ii) full scrambling of hydrogen and/or deuterium nuclei in the intermediate or activated complex. In such reactions, we assume that the intermediate complex is highly energetic and decays statistically to the many rotational states of the products. To obtain the nuclear–spin branching ratios, the probabilities of forming an intermediate complex in a given nuclear–spin state are multiplied by the probabilities for this complex to decay towards the nuclear–spin states of the products. In the process, the conservation of the total nuclear spin and of the permutation symmetry of identical nuclei is taken into account. The results of these calculations are then summed. This procedure also requires the nuclear-spin–symmetry statistical weights for reactions involving many (here up to seven) identical particles (?)see Appendix C and ][]sipila2015b. In practice, for a reaction involving identical hydrogen or deuterium nuclei,

$$R_i + R_j \rightarrow C_n \rightarrow P_k + P_l, \quad (B1)$$

where $n = i+j = k+l$ is the total number of identical particles, one needs to determine the possible permutation–symmetry species for the reactants ($\Gamma_i \otimes \Gamma_j$, where $\otimes$ denotes a direct product), of the intermediate complex ($\Gamma_n$) and of the products ($\Gamma_k \otimes \Gamma_l$) in their respective permutation–symmetry groups ($S_i \otimes S_j$, $S_n$ and $S_k \otimes S_l$). One also needs the correlation tables between the symmetry group of the complex and the symmetry subgroups representing the reactants and the products. These tables, or matrices, are obtained using group theory and provide the required induction and subduction statistical weights $W_{(\Gamma_i \otimes \Gamma_j \uparrow \Gamma_n)}$. The statistical weights are the number of independent states of symmetry $\Gamma_n$ induced from the $\Gamma_i$ and $\Gamma_j$ symmetries of the reactants. We note that the induction and subduction statistical weights are equal according to the Frobenius reciprocity theorem (Quack 1977). The correlation tables are given in Appendix C (?)see also][]hugo2009, sipila2015b. From these tables, simplified expressions for the nuclear-spin branching ratios can be derived which involve matrix products only:

$$P_{(\Gamma_i \otimes \Gamma_j \rightarrow \Gamma_k \otimes \Gamma_l)} = \frac{\sum_{\Gamma_n} W_{(\Gamma_i \otimes \Gamma_j \uparrow \Gamma_n)} \times W_{(\Gamma_k \otimes \Gamma_l \uparrow \Gamma_n)}}{(\sum_{\Gamma'_n} W_{(\Gamma_i \otimes \Gamma_j \uparrow \Gamma'_n)}) \times (\sum_{\Gamma'_k \Gamma'_l} W_{(\Gamma'_k \otimes \Gamma'_l \uparrow \Gamma_n)})}, \quad (B2)$$

with $\sum_{\Gamma_k \Gamma_l} P_{(\Gamma_i \otimes \Gamma_j \rightarrow \Gamma_k \otimes \Gamma_l)} = 1$[4]. Based on this formalism, we developed an automated FORTRAN routine, spinstate.f90, to calculate the branching ratios for reactions involving up to seven hydrogen (e.g. $CH_4 + H_3^+$) and six deuterium nuclei (e.g. $D_3^+ + ND_3$).

It should be noted that because hydrogen and deuterium nuclei are distinguishable, reactions involving both multiple hydrogen and multiple deuterium nuclei were divided into two parts, each with separate branching ratios that were combined (multiplied) afterwards. In addition, for exothermic reactions involving a hydride and a non-hydride—but "separable"—species whose nuclear–spin isomers are ignored (e.g. $C_3H_2$), we adopted the above procedure to compute the branching ratios. Where the non-hydride species is a reactant, we assume equal abundances of the nuclear–spin isomers of the non-hydride species, while, if it is a product, the associated nuclear–spin branching ratios are summed. Finally, for charge–exchange reactions, we assumed that the nuclear spin of each reactant isomer is conserved.

### B1.2 Inter-conversion reactions

For inter-conversion reactions involving $H_2$ or $D_2$ (e.g. $H^+$ + para-$H_2$ ↔ $H^+$ + ortho-$H_2$), rate coefficients were taken from either specific calculations or measurements in the literature, or estimated on a case–by–case basis. Indeed, for such "nearly thermo-neutral" reactions, the previous approach no longer applies because the full scrambling hypothesis is not guaranteed as the intermediate complex has small excess energy, and only a small amount of this energy is available for the products (Rist et al. 2013). In this case, a state–to–state (rotationally resolved) analysis is required to derive the nuclear–spin branching ratios. We have adopted a number of recent theoretical results from the literature for the inter-conversion of $H_2$ and $D_2$. We note that inter-conversion processes were neglected in reactions involving highly–exothermic channels (e.g. $NH_3^+ + H_2$) because the reactive channels (in this example, leading to $NH_4^+ + H$) proceed much faster than the inter-conversion (?)see ][and references therein]faure2013. Specifically, for the reactions between all isotopic variants of $H_3^+$ and $H_2$, we have employed the state–to–state rate coefficients calculated by Hugo et al. (2009). These rate coefficients were used to derive *species-to-species* rate coefficients by assuming thermal population of the rotational levels within each nuclear–spin species; this supposes local thermal equilibrium (LTE), which should apply at densities above approximately $10^5$ cm$^{-3}$. An alternative would be to use *ground–state–to–species* rate coefficients, assuming that all isotopologues are in their ground

---

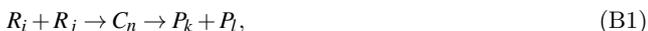

[4] We note that the two sums in the denominator of Eq. (B2) can be expressed from $\dim(\Gamma)$ and $f(\Gamma)$, the dimension and frequency of the $\Gamma$ permutation symmetry representation, as $f(\Gamma_i \otimes \Gamma_j) \times \dim(\Gamma_i) \times \dim(\Gamma_j)$ and $f(\Gamma_k \otimes \Gamma_l) \times \dim(\Gamma_k) \times \dim(\Gamma_l)$, respectively.

rotational states. Intermediate situations are also possible, as discussed in the recent work of Sipilä et al. (2017). However, only the LTE situation allows us to track anomalies in nuclear–spin isomer ratios, i.e. departures from thermal equilibrium ratios.

For the inter-conversion reaction between $H^+$ and $H_2$, we have employed state–to–state rate coefficients (Honvault et al. 2011, 2012), assuming that only the ground states of para-$H_2$ ($j=0$) and ortho-$H_2$ ($j=1$) are populated at low temperature. For the deuterated variant, $D^+ + D_2$, a statistical method was applied (Rist et al. 2013). This method—inspired by the work of Gerlich (1990)—was found to yield satisfactory agreement (typically better than a factor of 2) with time–independent quantum calculations (Honvault et al. 2011). The same statistical method was applied to the inter-conversion reactions between $HCO^+$ and $H_2$ and $DCO^+$ and $D_2$.

*B1.3 Endothermic reactions and exothermic reactions with an activation barrier*

For endothermic reactions or exothermic reactions with an activation barrier, our strategy was to include only those reactions whose activation energy is less than 1000 K. The spin–state separation of the original network of F06 was restricted to reactions that are significant at low temperatures (in practice, below 50 K). These reactions are generally of secondary importance; but we note the exception of $N^+ + H_2$, which is at the root of the ammonia chemistry. For this reaction, we adopted the rate coefficients of Dislaire et al. (2012), as did Le Gal et al. (2014). For the deuterated variant $N^+ + HD$, we used the experimental rate coefficient of Marquette et al. (1988).

### B2 Condensed network

An unseparated or "condensed" network, where the distinction between nuclear–spin states is removed, is also provided for applications where ortho:para ratios are not required. In doing so, some prescriptions are necessary for a number of reactions whose rate coefficients depend on the ortho:para ratio(s) of the reactant(s). A good example is $N^+ + H_2$ (**?**)see[]dislaire2012. In such case, the KIDA value—when validated by experts—is generally adopted by default. For the deuterated variants of $H_3^+ + H_2$, we adopt representative values of the rate coefficients at low temperature ($\sim 10$ K), based on a close inspection of the individual nuclear-spin reactions.

## APPENDIX C: NUCLEAR SPIN SYMMETRY INDUCTION AND SUBDUCTION MATRICES

We recall the pure nuclear spin symmetry induction (and subduction) statistical weights matrices for several hydrogen systems and for several deuterium systems used in the chemistry network. These matrices give the total number of nuclear spin states allowed from permutation symmetry conservation in reactions involving $n > 1$ identical nuclei. We denote the induction weight $W_{\Gamma_i \otimes \Gamma_j \uparrow \Gamma_{n=i+j}}$, the number of "complex" spin states in a symmetry species $\Gamma_n$ formed from two reactants respectively in the symmetry species $\Gamma_i$ and $\Gamma_j$. The corresponding subduction weight $W_{\Gamma_{n=i+j} \downarrow \Gamma_i \otimes \Gamma_j}$ gives the number of allowed products spin states respectively in the symmetry species $\Gamma_i$ and $\Gamma_j$ from the decay of a complex in the symmetry species $\Gamma_n$. The reciprocity Froebenius theorem shows that $W_{\Gamma_{n=i+j} \downarrow \Gamma_i \otimes \Gamma_j} = W_{\Gamma_i \otimes \Gamma_j \uparrow \Gamma_{n=i+j}}$.

**Table C1.** Pure permutation symmetry induction/subduction statistical weights $W_{\Gamma_n \otimes \Gamma_1 \uparrow \Gamma_{n+1}} = W_{\Gamma_{n+1} \downarrow \Gamma_n \otimes \Gamma_1}$ for $H_n + H \longrightarrow H_{n+1}$.

| H + H | → | $H_2$ (Hugo et al. 2009) | | |
|---|---|---|---|---|
| | | A | B | Total |
| A | A | 3 | 1 | 4 |
| Total | | 3 | 1 | $4 = 2^2$ |

| $H_2$ + H | → | $H_3$ (Hugo et al. 2009) | | |
|---|---|---|---|---|
| | | $A_1$ | E | Total |
| A | A | 4 | 2 | 6 |
| B | A | 0 | 2 | 2 |
| Total | | 4 | 4 | $8 = 2^3$ |

| $H_3$ + H | → | $H_4$ (Hugo et al. 2009) | | | |
|---|---|---|---|---|---|
| | | $A_1$ | E | $F_1$ | Total |
| $A_1$ | A | 5 | 0 | 3 | 8 |
| E | A | 0 | 2 | 6 | 8 |
| Total | | 5 | 2 | 9 | $16 = 2^4$ |

| $H_4$ + H | → | $H_5$ (this work) | | | |
|---|---|---|---|---|---|
| | | $A_1$ | $G_1$ | $H_1$ | Total |
| $A_1$ | A | 6 | 4 | 0 | 10 |
| E | A | 0 | 0 | 4 | 4 |
| $F_1$ | A | 0 | 12 | 6 | 18 |
| Total | | 6 | 16 | 10 | $32 = 2^5$ |

| $H_5$ + H | → | $H_6$ (this work) | | | | |
|---|---|---|---|---|---|---|
| | | $A_1$ | $H_1$ | $H_4$ | $L_1$ | Total |
| $A_1$ | A | 7 | 5 | 0 | 0 | 12 |
| $G_1$ | A | 0 | 20 | 0 | 12 | 32 |
| $H_1$ | A | 0 | 0 | 5 | 15 | 20 |
| Total | | 7 | 25 | 5 | 27 | $64 = 2^6$ |

| $H_6$ + H | → | $H_7$ (this work) | | | | |
|---|---|---|---|---|---|---|
| | | $A_1$ | I | X | Y | Total |
| $A_1$ | A | 8 | 6 | 0 | 0 | 14 |
| $H_1$ | A | 0 | 30 | 0 | 20 | 50 |
| $H_4$ | A | 0 | 0 | 10 | 0 | 10 |
| $L_1$ | A | 0 | 0 | 18 | 36 | 54 |
| Total | | 8 | 36 | 28 | 56 | $128 = 2^7$ |

### C1 Multi Hydrogen systems

See Tables C1 to C3.

### C2 Multi Deuterium systems

See Tables C4 and C5.

**Table C2.** Pure permutation symmetry induction/subduction statistical weights $W_{\Gamma_n \otimes \Gamma_2 \uparrow \Gamma_{n+2}} = W_{\Gamma_{n+2} \downarrow \Gamma_n \otimes \Gamma_2}$ for $H_n + H_2 \longrightarrow H_{n+2}$.

| $H_2 + H_2$ | $\rightarrow$ | $H_4$ (Hugo et al. 2009) | | | |
|---|---|---|---|---|---|
| | | $A_1$ | $E$ | $F_1$ | Total |
| A | A | 5 | 1 | 3 | 9 |
| A | B | 0 | 0 | 3 | 3 |
| B | A | 0 | 0 | 3 | 3 |
| B | B | 0 | 1 | 0 | 1 |
| Total | | 5 | 2 | 9 | $16 = 2^4$ |

| $H_3 + H_2$ | $\rightarrow$ | $H_5$ (Hugo et al. 2009) | | | |
|---|---|---|---|---|---|
| | | $A_1$ | $G_1$ | $H_1$ | Total |
| $A_1$ | A | 6 | 4 | 2 | 12 |
| $A_1$ | B | 0 | 4 | 0 | 4 |
| E | A | 0 | 8 | 4 | 12 |
| E | B | 0 | 0 | 4 | 4 |
| Total | | 6 | 16 | 10 | $32 = 2^5$ |

| $H_4 + H_2$ | $\rightarrow$ | $H_6$ (Sipilä et al. 2015) | | | | |
|---|---|---|---|---|---|---|
| | | $A_1$ | $H_1$ | $H_4$ | $L_4$ | Total |
| $A_1$ | A | 7 | 5 | 0 | 3 | 15 |
| $A_1$ | B | 0 | 5 | 0 | 0 | 5 |
| E | A | 0 | 0 | 0 | 6 | 6 |
| E | B | 0 | 0 | 2 | 0 | 2 |
| $F_1$ | A | 0 | 15 | 3 | 9 | 27 |
| $F_1$ | B | 0 | 0 | 0 | 9 | 9 |
| Total | | 7 | 25 | 5 | 27 | $64 = 2^4$ |

| $H_5 + H_2$ | $\rightarrow$ | $H_7$ (this work) | | | | |
|---|---|---|---|---|---|---|
| | | $A_1$ | I | X | Y | Total |
| $A_1$ | A | 8 | 6 | 0 | 4 | 18 |
| $A_1$ | B | 0 | 6 | 0 | 0 | 6 |
| $G_1$ | A | 0 | 24 | 8 | 16 | 48 |
| $G_1$ | B | 0 | 0 | 0 | 16 | 16 |
| $H_1$ | A | 0 | 0 | 10 | 20 | 30 |
| $H_1$ | B | 0 | 0 | 10 | 0 | 10 |
| Total | | 8 | 36 | 28 | 56 | $128 = 2^7$ |

**Table C3.** Pure permutation symmetry induction/subduction statistical weights $W_{\Gamma_n \otimes \Gamma_3 \uparrow \Gamma_{n+3}} = W_{\Gamma_{n+3} \downarrow \Gamma_n \otimes \Gamma_3}$ for $H_n + H_3 \longrightarrow H_{n+3}$.

| $H_3 + H_3$ | $\rightarrow$ | $H_6$ (Sipilä et al. 2015) | | | | |
|---|---|---|---|---|---|---|
| | | $A_1$ | $H_1$ | $H_4$ | $L_1$ | Total |
| $A_1$ | $A_1$ | 7 | 5 | 1 | 3 | 16 |
| $A_1$ | E | 0 | 10 | 0 | 6 | 16 |
| E | $A_1$ | 0 | 10 | 0 | 6 | 16 |
| E | E | 0 | 0 | 4 | 12 | 16 |
| Total | | 7 | 25 | 5 | 27 | $64 = 2^6$ |

| $H_4 + H_3$ | $\rightarrow$ | $H_7$ (Sipilä et al. 2015) | | | | |
|---|---|---|---|---|---|---|
| | | $A_1$ | $H_1$ | $H_4$ | $L_1$ | Total |
| $A_1$ | $A_1$ | 7 | 5 | 0 | 3 | 15 |
| $A_1$ | E | 0 | 5 | 0 | 0 | 5 |
| E | $A_1$ | 0 | 0 | 0 | 6 | 6 |
| E | E | 0 | 0 | 2 | 0 | 2 |
| $F_1$ | $A_1$ | 0 | 15 | 3 | 9 | 27 |
| $F_1$ | E | 0 | 0 | 0 | 9 | 9 |
| Total | | 7 | 25 | 5 | 27 | $64 = 2^6$ |

**Table C4.** Pure permutation symmetry induction/subduction statistical weights $W_{\Gamma_n \otimes \Gamma_1 \uparrow \Gamma_{n+1}} = W_{\Gamma_{n+1} \downarrow \Gamma_n \otimes \Gamma_1}$ for $D_n + D \longrightarrow D_{n+1}$.

| $D + D$ | $\rightarrow$ | $D_2$ (Hugo et al. 2009) | | |
|---|---|---|---|---|
| | | A | B | Total |
| A | A | 6 | 3 | 9 |
| Total | | 6 | 3 | $9 = 3^2$ |

| $D_2 + D$ | $\rightarrow$ | $D_3$ (Hugo et al. 2009) | | | |
|---|---|---|---|---|---|
| | | $A_1$ | $A_2$ | E | Total |
| A | A | 10 | 0 | 8 | 18 |
| B | A | 0 | 1 | 8 | 9 |
| Total | | 4 | 4 | | $27 = 3^3$ |

| $D_3 + D$ | $\rightarrow$ | $D_4$ (Hugo et al. 2009) | | | | |
|---|---|---|---|---|---|---|
| | | $A_1$ | E | $F_1$ | $F_2$ | Total |
| $A_1$ | A | 15 | 0 | 15 | 0 | 30 |
| $A_2$ | A | 0 | 0 | 0 | 3 | 3 |
| E | A | 0 | 12 | 30 | 6 | 48 |
| Total | | 15 | 12 | 45 | 9 | $81 = 3^4$ |

| $D_4 + D$ | $\rightarrow$ | $D_5$ (this work) | | | | | |
|---|---|---|---|---|---|---|---|
| | | $A_1$ | $G_1$ | $H_1$ | $H_2$ | I | Total |
| $A_1$ | A | 21 | 24 | 0 | 0 | 0 | 45 |
| E | A | 0 | 0 | 30 | 6 | 0 | 36 |
| $F_1$ | A | 0 | 72 | 45 | 0 | 18 | 135 |
| $F_2$ | A | 0 | 0 | 0 | 9 | 18 | 27 |
| Total | | 21 | 96 | 75 | 15 | 36 | $243 = 3^5$ |

# APPENDIX D: THE UGAN REACTION RATE COEFFICIENTS

In this section, we provide our separated *University of Grenoble Alpes Astrochemical Network* (UGAN). Our convention for the names of the isotopologues follows Maue's rule (Maue 1937), namely that ortho species—noted with prefix o—have the largest statistical weight, and para species—with prefix p—have the smallest statistical weight. Species with intermediate degeneracy have prefix m, l, etc... in increasing order. Furthermore, for species that are separable with respect to both hydrogen and deuterium—such as $NH_2D_2$—the isomers are indicated with one letter for each nucleus. In such instances, two letters are used with the previous convention applying equally to H and D, e.g. $ooNH_2D_2$.

In the following Tables, the rate coefficients are given in the traditional form of a modified Arrhenius law,

$$k(T) = \alpha (T/300)^\beta \exp(-\gamma/T) \quad \text{cm}^3\,\text{s}^{-1}.$$

Note however that the reactions of formation of $H_2$, HD, and $D_2$, which occur on dust surfaces, are computed as bimolecular reactions using a rate coefficient which is not given by the modified Arrhenius law, but which is computed internally from the density and the grain parameters. In the following Tables, the species noted $\gamma$ represents a photon (not to be confused with the exo/endothermicity), CRP are primary cosmic-ray particles (essentially protons), and $\gamma_2$ are secondary photons.

**Table C5.** Pure permutation symmetry induction/subduction statistical weights $W_{\Gamma_n \otimes \Gamma_2 \uparrow \Gamma_{n+2}} = W_{\Gamma_{n+2} \downarrow \Gamma_n \otimes \Gamma_2}$ for $D_n + D_2 \longrightarrow D_{n+2}$ and $W_{\Gamma_n \otimes \Gamma_3 \uparrow \Gamma_{n+3}} = W_{\Gamma_{n+3} \downarrow \Gamma_n \otimes \Gamma_3}$ for for $D_n + D_3 \longrightarrow D_{n+3}$.

| $D_2 + D_2$ | | $D_4$ (Hugo et al. 2009) | | | | Total |
|---|---|---|---|---|---|---|
| | | $A_1$ | E | $F_1$ | $F_2$ | |
| A | A | 15 | 6 | 15 | 0 | 18 |
| A | B | 0 | 0 | 15 | 3 | 18 |
| B | A | 0 | 0 | 15 | 3 | 9 |
| B | B | 0 | 6 | 0 | 3 | 9 |
| Total | | 15 | 12 | 45 | 9 | $81 = 3^4$ |

| $D_3 + D_2$ | | $D_5$ (Hugo et al. 2009) | | | | | Total |
|---|---|---|---|---|---|---|---|
| | | $A_1$ | $G_1$ | $H_1$ | $H_2$ | I | |
| $A_1$ | A | 21 | 24 | 15 | 0 | 0 | 60 |
| $A_1$ | A | 0 | 24 | 0 | 0 | 6 | 30 |
| $A_2$ | A | 0 | 0 | 0 | 0 | 6 | 6 |
| $A_2$ | B | 0 | 0 | 0 | 3 | 0 | 3 |
| E | A | 0 | 48 | 30 | 6 | 12 | 96 |
| E | A | 0 | 0 | 30 | 6 | 12 | 48 |
| Total | | 21 | 96 | 75 | 15 | 36 | $243 = 3^5$ |

| $D_4 + D_2$ | | $D_6$ (Sipilä et al. 2015) | | | | | | | Total |
|---|---|---|---|---|---|---|---|---|---|
| | | $A_1$ | $H_1$ | $H_3$ | $H_4$ | $L_1$ | $M_1$ | S | |
| $A_1$ | A | 28 | 35 | 0 | 0 | 27 | 0 | 0 | 90 |
| $A_1$ | A | 0 | 35 | 0 | 0 | 0 | 10 | 0 | 45 |
| E | A | 0 | 0 | 2 | 0 | 54 | 0 | 16 | 72 |
| E | A | 0 | 0 | 0 | 20 | 0 | 0 | 16 | 36 |
| $F_1$ | A | 0 | 105 | 0 | 30 | 81 | 30 | 24 | 270 |
| $F_1$ | B | 0 | 0 | 0 | 0 | 81 | 30 | 24 | 135 |
| $F_2$ | A | 0 | 0 | 0 | 0 | 0 | 30 | 24 | 54 |
| $F_2$ | B | 0 | 0 | 3 | 0 | 0 | 0 | 24 | 27 |
| Total | | 28 | 175 | 5 | 50 | 243 | 100 | 128 | $729 = 3^6$ |

| $D_3 + D_3$ | | $D_6$ (Sipilä et al. 2015) | | | | | | | Total |
|---|---|---|---|---|---|---|---|---|---|
| | | $A_1$ | $H_1$ | $H_3$ | $H_4$ | $L_1$ | $M_1$ | S | |
| $A_1$ | $A_1$ | 28 | 35 | 0 | 10 | 27 | 0 | 0 | 100 |
| $A_1$ | $A_1$ | 0 | 0 | 0 | 0 | 0 | 10 | 0 | 10 |
| $A_1$ | E | 0 | 70 | 0 | 0 | 54 | 20 | 16 | 160 |
| $A_2$ | $A_1$ | 0 | 0 | 0 | 0 | 0 | 10 | 0 | 10 |
| $A_2$ | $A_1$ | 0 | 0 | 1 | 0 | 0 | 0 | 0 | 1 |
| $A_2$ | E | 0 | 0 | 0 | 0 | 0 | 0 | 0 | 16 |
| E | $A_1$ | 0 | 70 | 0 | 0 | 54 | 20 | 0 | 160 |
| E | $A_1$ | 0 | 0 | 0 | 0 | 0 | 0 | 0 | 16 |
| E | E | 0 | 0 | 4 | 40 | 108 | 40 | 0 | 256 |
| Total | | 28 | 175 | 5 | 50 | 243 | 100 | 128 | $729 = 3^6$ |

**Table D1.** Species of our non-separated network (total: 151).

|   |   |   |   |   |   |   |   |   |
|---|---|---|---|---|---|---|---|---|
| 1  | H        | $H_2$       | D          | HD         | $D_2$      | He        | O         | OH       |
| 2  | $O_2$    | OD          | $H_2O$     | HDO        | $D_2O$     | C         | CH        | $CH_2$   |
| 3  | $CH_3$   | $CH_4$      | $C_2$      | $C_2H$     | $C_2H_2$   | $C_3$     | $C_3H$    | $C_3H_2$ |
| 4  | N        | $N_2$       | NH         | $NH_2$     | $NH_3$     | ND        | NHD       | $ND_2$   |
| 5  | $NH_2D$  | $NHD_2$     | $ND_3$     | S          | SH         | $H_2S$    | CO        | NO       |
| 6  | CN       | SO          | CS         | HNC        | HCN        | $CO_2$    | $SO_2$    | OCS      |
| 7  | $H_2CO$  | $CH_3OH$    | $HCO_2H$   | $H^+$      | $D^+$      | $He^+$    | $H_2^+$   | $HD^+$   |
| 8  | $D_2^+$  | $H_3^+$     | $H_2D^+$   | $HD_2^+$   | $D_3^+$    | $O^+$     | $OH^+$    | $H_2O^+$ |
| 9  | $OD^+$   | $O_2^+$     | $HDO^+$    | $D_2O^+$   | $H_3O^+$   | $H_2DO^+$ | $HD_2O^+$ | $D_3O^+$ |
| 10 | $C^+$    | $CH^+$      | $CH_2^+$   | $CH_3^+$   | $CH_4^+$   | $CH_5^+$  | $C_2^+$   | $C_2H^+$ |
| 11 | $C_2H_2^+$ | $C_2H_3^+$ | $C_3^+$   | $C_3H^+$   | $C_3H_2^+$ | $C_3H_3^+$ | $N^+$   | $NH^+$   |
| 12 | $ND^+$   | $N_2^+$     | $N_2H^+$   | $NH_2^+$   | $NH_4^+$   | $N_2D^+$  | $NH_3^+$  | $NHD^+$  |
| 13 | $ND_2^+$ | $NH_3D^+$   | $NH_2D^+$  | $NHD_2^+$  | $ND_3^+$   | $NH_2D_2^+$ | $NHD_3^+$ | $ND_4^+$ |
| 14 | $S^+$    | $SH^+$      | $SD^+$     | $H_2S^+$   | $H_3S^+$   | $HCO^+$   | $HCS^+$   | $HCO_2^+$ |
| 15 | $SO^+$   | $CO^+$      | $CS^+$     | $DCO^+$    | $NO^+$     | $HCN^+$   | $C_2N^+$  | $HCNH^+$ |
| 16 | $HNO^+$  | $CN^+$      | $HSO^+$    | $HOCS^+$   | $DCO_2^+$  | $HSO_2^+$ | $H_2NC^+$ | Fe       |
| 17 | $Fe^+$   | Gr          | $Gr^-$     | $Gr^+$     | $CH_4^*$   | $H_2O^*$  | $O_2^*$   | $CO^*$   |
| 18 | $CO_2^*$ | $N^*$       | $NH_3^*$   | $N_2^*$    | $H_2S^*$   | $OCS^*$   | $Fe^*$    | $HDO^*$  |
| 19 | $D_2O^*$ | $NH_2D^*$   | $NHD_2^*$  | $ND_3^*$   | $CH_3OH^*$ | $H_2CO^*$ | $HCO_2H^*$ |         |

**Notes:**
Species labeled with '*' are species adsorbed in grain mantles.

**Table D2.** Separated species (total: 207).

|   |   |   |   |   |   |   |   |   |
|---|---|---|---|---|---|---|---|---|
| 1  | H          | $oH_2$       | $pH_2$       | D            | HD           | $oD_2$       | $pD_2$       | He           |
| 2  | O          | OH           | $O_2$        | OD           | $oH_2O$      | $pH_2O$      | HDO          | $oD_2O$      |
| 3  | $pD_2O$    | C            | CH           | $oCH_2$      | $pCH_2$      | $oCH_3$      | $pCH_3$      | $mCH_4$      |
| 4  | $pCH_4$    | $oCH_4$      | $C_2$        | $C_2H$       | $C_2H_2$     | $C_3$        | $C_3H$       | $C_3H_2$     |
| 5  | N          | $N_2$        | NH           | $oNH_2$      | $pNH_2$      | $oNH_3$      | $pNH_3$      | ND           |
| 6  | NHD        | $oND_2$      | $pND_2$      | $oNH_2D$     | $pNH_2D$     | $oNHD_2$     | $pNHD_2$     | $mND_3$      |
| 7  | $pND_3$    | $oND_3$      | S            | SH           | $oH_2S$      | $pH_2S$      | CO           | NO           |
| 8  | SO         | CN           | CS           | HNC          | HCN          | $CO_2$       | $SO_2$       | OCS          |
| 9  | $H_2CO$    | $CH_3OH$     | $HCO_2H$     | Fe           | $H^+$        | $D^+$        | $He^+$       | $oH_2^+$     |
| 10 | $pH_2^+$   | $HD^+$       | $oD_2^+$     | $pD_2^+$     | $oH_2D^+$    | $pH_2D^+$    | $oH_3^+$     | $pH_3^+$     |
| 11 | $oHD_2^+$  | $pHD_2^+$    | $mD_3^+$     | $oD_3^+$     | $pD_3^+$     | $O^+$        | $O_2^+$      | $OH^+$       |
| 12 | $oH_2O^+$  | $pH_2O^+$    | $OD^+$       | $HDO^+$      | $oD_2O^+$    | $pD_2O^+$    | $oH_3O^+$    | $pH_3O^+$    |
| 13 | $oH_2DO^+$ | $pH_2DO^+$   | $oHD_2O^+$   | $pHD_2O^+$   | $mD_3O^+$    | $oD_3O^+$    | $pD_3O^+$    | $C^+$        |
| 14 | $CH^+$     | $oCH_2^+$    | $pCH_2^+$    | $C_2^+$      | $C_2H^+$     | $C_2H_3^+$   | $C_3^+$      | $C_3H^+$     |
| 15 | $oCH_3^+$  | $pCH_3^+$    | $pCH_5^+$    | $oCH_5^+$    | $mCH_5^+$    | $mCH_4^+$    | $pCH_4^+$    | $oCH_4^+$    |
| 16 | $C_2H_2^+$ | $C_3H_3^+$   | $C_3H_2^+$   | $N^+$        | $N_2^+$      | $ND^+$       | $NH^+$       | $N_2H^+$     |
| 17 | $N_2D^+$   | $oNH_2^+$    | $pNH_2^+$    | $mNH_4^+$    | $oNH_4^+$    | $pNH_4^+$    | $oNH_3^+$    | $pNH_3^+$    |
| 18 | $NHD^+$    | $oND_2^+$    | $pND_2^+$    | $oNH_3D^+$   | $pNH_3D^+$   | $oNH_2D^+$   | $pNH_2D^+$   | $oNHD_2^+$   |
| 19 | $pNHD_2^+$ | $mND_3^+$    | $oND_3^+$    | $pND_3^+$    | $ooNH_2D_2^+$ | $poNH_2D_2^+$ | $opNH_2D_2^+$ | $ppNH_2D_2^+$ |
| 20 | $mNHD_3^+$ | $oNHD_3^+$   | $pNHD_3^+$   | $lND_4^+$    | $oND_4^+$    | $pND_4^+$    | $mND_4^+$    | $S^+$        |
| 21 | $SH^+$     | $SD^+$       | $oH_2S^+$    | $pH_2S^+$    | $oH_3S^+$    | $pH_3S^+$    | $HCO^+$      | $HCO_2^+$    |
| 22 | $SO^+$     | $CO^+$       | $CS^+$       | $NO^+$       | $CN^+$       | $HCS^+$      | $DCO^+$      | $HCN^+$      |
| 23 | $C_2N^+$   | $HCNH^+$     | $HNO^+$      | $HSO^+$      | $HOCS^+$     | $DCO_2^+$    | $HSO_2^+$    | $H_2NC^+$    |
| 24 | $Fe^+$     | Gr           | $Gr^-$       | $Gr^+$       | $CH_4^*$     | $H_2O^*$     | $O_2^*$      | $CO^*$       |
| 25 | $CO_2^*$   | $N^*$        | $NH_3^*$     | $N_2^*$      | $H_2S^*$     | $OCS^*$      | $Fe^*$       | $HDO^*$      |
| 26 | $D_2O^*$   | $NH_2D^*$    | $NHD_2^*$    | $ND_3^*$     | $CH_3OH^*$   | $H_2CO^*$    | $HCO_2H^*$   |              |

**Notes:**

o: ortho; m:meta; p:para; The statistical weights are such that $p < m < l < o$, i.e. $lND_4^+$ has a statistical weight larger than that of $mND_4^+$ and smaller than $oND_4^+$;

$poNH_2D_2$: the first letter, p, refers to D nuclei symmetry, and the second, o, refers to symmetry of the H nuclei wavefunction.

Species labeled with '*' are species adsorbed in grain mantles.

**Table D3.** Sample of the "condensed" version of the UGAN chemical network (total: 1136 reactions).

| # | Reactants | | Products | | | $\alpha$<br>cm$^3$ s$^{-1}$ | $\beta$ | $\gamma$<br>K |
|---|---|---|---|---|---|---|---|---|
| 1  | H    | H   | H$_2$  |       |        | 1.0e+00 |      |         |
| 2  | H    | D   | HD     |       |        | 1.0e+00 |      |         |
| 3  | D    | D   | D$_2$  |       |        | 9.9e-01 |      |         |
| 4  | H    | CRP | H$^+$  | e$^-$ |        | 4.6e-01 | 0.00 | 0.0     |
| 5  | D    | CRP | D$^+$  | e$^-$ |        | 4.6e-01 | 0.00 | 0.0     |
| 6  | He   | CRP | He$^+$ | e$^-$ |        | 5.0e-01 | 0.00 | 0.0     |
| 7  | H$_2$| CRP | H      | H     |        | 1.0e-01 | 0.00 | 0.0     |
| 8  | H$_2$| CRP | H$^+$  | H     | e$^-$  | 2.3e-02 | 0.00 | 0.0     |
| 9  | H$_2$| CRP | H$_2^+$| e$^-$ |        | 9.8e-01 | 0.00 | 0.0     |
| 10 | HD   | CRP | H$^+$  | D     | e$^-$  | 2.0e-02 | 0.00 | 0.0     |
| 11 | HD   | CRP | D$^+$  | H     | e$^-$  | 2.0e-02 | 0.00 | 0.0     |
| 12 | HD   | CRP | H      | D     |        | 1.5e+00 | 0.00 | 0.0     |
| 13 | HD   | CRP | HD$^+$ | e$^-$ |        | 9.6e-01 | 0.00 | 0.0     |
| 14 | D$_2$| CRP | D$^+$  | D     | e$^-$  | 4.0e-02 | 0.00 | 0.0     |
| 15 | D$_2$| CRP | D      | D     |        | 1.5e+00 | 0.00 | 0.0     |
| 16 | D$_2$| CRP | D$_2^+$| e$^-$ |        | 9.6e-01 | 0.00 | 0.0     |
| 17 | H    | H$_2^+$ | H$_2$ | H$^+$ |        | 6.4e-10 | 0.00 | 0.0     |
| 18 | H    | H$_3^+$ | H$_2^+$ | H$_2$ |      | 2.1e-09 | 0.00 | 20000.0 |
| 19 | H    | HD$^+$  | H$^+$ | HD    |        | 6.4e-10 | 0.00 | 0.0     |
| 20 | H    | HD$^+$  | H$_2^+$ | D   |        | 1.0e-09 | 0.00 | 154.0   |

**Table D4.** Sample of the "separated" version of the UGAN chemical network (total: 3266 reactions split into 43 tables).

| # | Reactants | | Products | | | $\alpha$<br>cm$^3$ s$^{-1}$ | $\beta$ | $\gamma$<br>K |
|---|---|---|---|---|---|---|---|---|
| 1  | H      | H   | oH$_2$    |       |       | 7.5e-01 |      |     |
| 2  | H      | H   | pH$_2$    |       |       | 2.5e-01 |      |     |
| 3  | H      | D   | HD        |       |       | 1.0e+00 |      |     |
| 4  | D      | D   | oD$_2$    |       |       | 6.6e-01 |      |     |
| 5  | D      | D   | pD$_2$    |       |       | 3.3e-01 | 0.50 | 0.0 |
| 6  | H      | CRP | H$^+$     | e$^-$ |       | 4.6e-01 | 0.00 | 0.0 |
| 7  | D      | CRP | D$^+$     | e$^-$ |       | 4.6e-01 | 0.00 | 0.0 |
| 8  | He     | CRP | He$^+$    | e$^-$ |       | 5.0e-01 | 0.00 | 0.0 |
| 9  | oH$_2$ | CRP | H         | H     |       | 1.0e-01 | 0.00 | 0.0 |
| 10 | pH$_2$ | CRP | H         | H     |       | 1.0e-01 | 0.00 | 0.0 |
| 11 | oH$_2$ | CRP | H$^+$     | H     | e$^-$ | 2.3e-02 | 0.00 | 0.0 |
| 12 | pH$_2$ | CRP | H$^+$     | H     | e$^-$ | 2.3e-02 | 0.00 | 0.0 |
| 13 | oH$_2$ | CRP | oH$_2^+$  | e$^-$ |       | 9.8e-01 | 0.00 | 0.0 |
| 14 | pH$_2$ | CRP | pH$_2^+$  | e$^-$ |       | 9.8e-01 | 0.00 | 0.0 |
| 15 | HD     | CRP | H$^+$     | D     | e$^-$ | 2.0e-02 | 0.00 | 0.0 |
| 16 | HD     | CRP | D$^+$     | H     | e$^-$ | 2.0e-02 | 0.00 | 0.0 |
| 17 | HD     | CRP | H         | D     |       | 1.5e+00 | 0.00 | 0.0 |
| 18 | HD     | CRP | HD$^+$    | e$^-$ |       | 9.6e-01 | 0.00 | 0.0 |
| 19 | oD$_2$ | CRP | D$^+$     | D     | e$^-$ | 4.0e-02 | 0.00 | 0.0 |
| 20 | pD$_2$ | CRP | D$^+$     | D     | e$^-$ | 4.0e-02 | 0.00 | 0.0 |

**Table D5.** The "condensed" version of the UGAN chemical network (total: 1136 reactions split into 15 Tables).

| # | Reactants | | Products | | | $\alpha$ | $\beta$ | $\gamma$ |
|---|---|---|---|---|---|---|---|---|
| 1 | H | H | $H_2$ | | | 1.0e+00 | | |
| 2 | H | D | HD | | | 1.0e+00 | | |
| 3 | D | D | $D_2$ | | | 9.9e-01 | | |
| 4 | H | CRP | $H^+$ | $e^-$ | | 4.6e-01 | 0.00 | 0.0 |
| 5 | D | CRP | $D^+$ | $e^-$ | | 4.6e-01 | 0.00 | 0.0 |
| 6 | He | CRP | $He^+$ | $e^-$ | | 5.0e-01 | 0.00 | 0.0 |
| 7 | $H_2$ | CRP | H | H | | 1.0e-01 | 0.00 | 0.0 |
| 8 | $H_2$ | CRP | $H^+$ | H | $e^-$ | 2.3e-02 | 0.00 | 0.0 |
| 9 | $H_2$ | CRP | $H_2^+$ | $e^-$ | | 9.8e-01 | 0.00 | 0.0 |
| 10 | HD | CRP | $H^+$ | D | $e^-$ | 2.0e-02 | 0.00 | 0.0 |
| 11 | HD | CRP | $D^+$ | H | $e^-$ | 2.0e-02 | 0.00 | 0.0 |
| 12 | HD | CRP | H | D | | 1.5e+00 | 0.00 | 0.0 |
| 13 | HD | CRP | $HD^+$ | $e^-$ | | 9.6e-01 | 0.00 | 0.0 |
| 14 | $D_2$ | CRP | $D^+$ | D | $e^-$ | 4.0e-02 | 0.00 | 0.0 |
| 15 | $D_2$ | CRP | D | D | | 1.5e+00 | 0.00 | 0.0 |
| 16 | $D_2$ | CRP | $D_2^+$ | $e^-$ | | 9.6e-01 | 0.00 | 0.0 |
| 17 | H | $H_2^+$ | $H_2$ | $H^+$ | | 6.4e-10 | 0.00 | 0.0 |
| 18 | H | $H_3^+$ | $H_2^+$ | $H_2$ | | 2.1e-09 | 0.00 | 20000.0 |
| 19 | H | $HD^+$ | $H^+$ | HD | | 6.4e-10 | 0.00 | 0.0 |
| 20 | H | $HD^+$ | $H_2^+$ | D | | 1.0e-09 | 0.00 | 154.0 |
| 21 | H | $HD^+$ | $H_2D^+$ | $\gamma$ | | 1.2e-17 | 1.80 | -20.0 |
| 22 | H | $D^+$ | $H^+$ | D | | 1.0e-09 | 0.00 | 0.0 |
| 23 | H | $D^+$ | $HD^+$ | $\gamma$ | | 3.9e-19 | 1.80 | -20.0 |
| 24 | H | $H_2D^+$ | $H_3^+$ | D | | 1.0e-09 | 0.00 | 597.8 |
| 25 | H | $HD_2^+$ | $H_2D^+$ | D | | 1.0e-09 | 0.00 | 549.8 |
| 26 | H | $D_2^+$ | $HD^+$ | D | | 1.0e-09 | 0.00 | 430.0 |
| 27 | H | $D_2^+$ | $D_2$ | $H^+$ | | 6.4e-10 | 0.00 | 0.0 |
| 28 | H | $D_3^+$ | $HD_2^+$ | D | | 1.0e-09 | 0.00 | 642.3 |
| 29 | $H_2$ | $He^+$ | $H^+$ | H | He | 3.3e-15 | 0.00 | 0.0 |
| 30 | $H_2$ | $He^+$ | $H_2^+$ | He | | 9.6e-15 | 0.00 | 0.0 |
| 31 | $H_2$ | $H^+$ | H | $H_2^+$ | | 6.4e-10 | 0.00 | 21300.0 |
| 32 | $H_2$ | $H_2^+$ | $H_3^+$ | H | | 2.1e-09 | 0.00 | 0.0 |
| 33 | $H_2$ | $HD^+$ | $H_2D^+$ | H | | 1.1e-09 | 0.00 | 0.0 |
| 34 | $H_2$ | $HD^+$ | $H_3^+$ | D | | 1.1e-09 | 0.00 | 0.0 |
| 35 | $H_2$ | $D^+$ | $H^+$ | HD | | 2.1e-09 | 0.00 | 0.0 |
| 36 | $H_2$ | $D_2^+$ | $H_2D^+$ | D | | 1.1e-09 | 0.00 | 0.0 |
| 37 | $H_2$ | $D_2^+$ | $HD_2^+$ | H | | 1.1e-09 | 0.00 | 0.0 |
| 38 | HD | $He^+$ | $H^+$ | D | He | 5.5e-14 | -0.24 | 0.0 |
| 39 | HD | $He^+$ | $D^+$ | H | He | 5.5e-14 | -0.24 | 0.0 |
| 40 | HD | $H^+$ | $D^+$ | $H_2$ | | 1.0e-09 | 0.00 | 464.0 |
| 41 | HD | $H_2^+$ | $H_2D^+$ | H | | 1.1e-09 | 0.00 | 0.0 |
| 42 | HD | $H_2^+$ | $H_3^+$ | D | | 1.1e-09 | 0.00 | 0.0 |
| 43 | HD | $HD^+$ | $H_2D^+$ | D | | 1.1e-09 | 0.00 | 0.0 |
| 44 | HD | $HD^+$ | $HD_2^+$ | H | | 1.1e-09 | 0.00 | 0.0 |
| 45 | HD | $D^+$ | $H^+$ | $D_2$ | | 1.0e-09 | 0.00 | 0.0 |
| 46 | HD | $D_2^+$ | $HD_2^+$ | D | | 1.1e-09 | 0.00 | 0.0 |
| 47 | HD | $D_2^+$ | $D_3^+$ | H | | 1.1e-09 | 0.00 | 0.0 |
| 48 | D | $H^+$ | $D^+$ | H | | 1.0e-09 | 0.00 | 41.0 |
| 49 | D | $H^+$ | $HD^+$ | $\gamma$ | | 3.9e-19 | 1.80 | -20.0 |
| 50 | D | $H_2^+$ | $H_2$ | $D^+$ | | 6.4e-10 | 0.00 | 0.0 |
| 51 | D | $H_2^+$ | $H_2D^+$ | $\gamma$ | | 7.0e-18 | 1.80 | -20.0 |
| 52 | D | $H_2^+$ | $HD^+$ | H | | 1.0e-09 | 0.00 | 0.0 |
| 53 | D | $H_3^+$ | $H_2D^+$ | H | | 1.0e-09 | 0.00 | 0.0 |
| 54 | D | $HD^+$ | $D_2^+$ | H | | 1.0e-09 | 0.00 | 0.0 |
| 55 | D | $HD^+$ | $D^+$ | HD | | 6.4e-10 | 0.00 | 0.0 |
| 56 | D | $H_2D^+$ | $HD_2^+$ | H | | 1.0e-09 | 0.00 | 0.0 |
| 57 | D | $HD_2^+$ | $D_3^+$ | H | | 1.0e-09 | 0.00 | 0.0 |
| 58 | D | $D_2^+$ | $D_2$ | $D^+$ | | 6.4e-10 | 0.00 | 0.0 |
| 59 | $D_2$ | $He^+$ | $D^+$ | D | He | 1.1e-13 | -0.24 | 0.0 |
| 60 | $D_2$ | $He^+$ | $D_2^+$ | He | | 2.5e-14 | 0.00 | 0.0 |
| 61 | $D_2$ | $H^+$ | $D^+$ | HD | | 2.1e-09 | 0.00 | 405.0 |
| 62 | $D_2$ | $H_2^+$ | $H_2D^+$ | D | | 1.1e-09 | 0.00 | 0.0 |
| 63 | $D_2$ | $H_2^+$ | $HD_2^+$ | H | | 1.1e-09 | 0.00 | 0.0 |
| 64 | $D_2$ | $HD^+$ | $HD_2^+$ | D | | 1.1e-09 | 0.00 | 0.0 |
| 65 | $D_2$ | $HD^+$ | $D_3^+$ | H | | 1.1e-09 | 0.00 | 0.0 |
| 66 | $D_2$ | $D_2^+$ | $D_3^+$ | D | | 2.1e-09 | 0.00 | 0.0 |
| 67 | Fe | $H^+$ | $Fe^+$ | H | | 7.4e-09 | 0.00 | 0.0 |
| 68 | Fe | $H_3^+$ | $Fe^+$ | $H_2$ | H | 4.9e-09 | 0.00 | 0.0 |
| 69 | Fe | $H_2D^+$ | $Fe^+$ | $H_2$ | D | 4.9e-09 | 0.00 | 0.0 |
| 70 | Fe | $HD_2^+$ | $Fe^+$ | $D_2$ | H | 1.6e-09 | 0.00 | 0.0 |
| 71 | Fe | $HD_2^+$ | $Fe^+$ | HD | D | 3.3e-09 | 0.00 | 0.0 |
| 72 | Fe | $D_3^+$ | $Fe^+$ | $D_2$ | D | 4.9e-09 | 0.00 | 0.0 |
| 73 | $H^+$ | $e^-$ | H | $\gamma$ | | 3.6e-12 | -0.75 | 0.0 |
| 74 | $H_2^+$ | $e^-$ | H | H | | 2.5e-07 | -0.50 | 0.0 |

**Table D5** – *continued* (part 2)

| | Reactants | | Products | | | | $\alpha$ | $\beta$ | $\gamma$ |
|---|---|---|---|---|---|---|---|---|---|
| 75 | $H_2^+$ | $e^-$ | $H_2$ | $\gamma$ | | | 2.2e-07 | -0.40 | 0.0 |
| 76 | $HD^+$ | $e^-$ | H | D | | | 9.0e-09 | -0.50 | 0.0 |
| 77 | $D^+$ | $e^-$ | D | $\gamma$ | | | 3.6e-12 | -0.75 | 0.0 |
| 78 | $He^+$ | $e^-$ | He | $\gamma$ | | | 4.5e-12 | -0.67 | 0.0 |
| 79 | $H_2^+$ | $\gamma_2$ | $H^+$ | H | | | 6.1e+02 | 0.00 | 0.0 |
| 80 | $H^+$ | $Gr^-$ | Gr | H | | | 1.6e-06 | 0.50 | 0.0 |
| 81 | $H_3^+$ | $Gr^-$ | Gr | $H_2$ | H | | 4.6e-07 | 0.50 | 0.0 |
| 82 | $H_3^+$ | $Gr^-$ | Gr | H | H | H | 4.6e-07 | 0.50 | 0.0 |
| 83 | $He^+$ | $Gr^-$ | Gr | He | | | 8.0e-07 | 0.50 | 0.0 |
| 84 | $H^+$ | Gr | $Gr^+$ | H | | | 1.6e-06 | 0.50 | 0.0 |
| 85 | $H_3^+$ | Gr | $Gr^+$ | $H_2$ | H | | 4.6e-07 | 0.50 | 0.0 |
| 86 | $H_3^+$ | Gr | $Gr^+$ | H | H | H | 4.6e-07 | 0.50 | 0.0 |
| 87 | $He^+$ | Gr | $Gr^+$ | He | | | 8.0e-07 | 0.50 | 0.0 |
| 88 | $D^+$ | $Gr^-$ | Gr | D | | | 1.1e-06 | 0.50 | 0.0 |
| 89 | $H_2D^+$ | $Gr^-$ | Gr | $H_2$ | D | | 1.3e-07 | 0.50 | 0.0 |
| 90 | $H_2D^+$ | $Gr^-$ | Gr | HD | H | | 2.7e-07 | 0.50 | 0.0 |
| 91 | $H_2D^+$ | $Gr^-$ | Gr | D | H | H | 4.0e-07 | 0.50 | 0.0 |
| 92 | $D^+$ | Gr | $Gr^+$ | D | | | 1.1e-06 | 0.50 | 0.0 |
| 93 | $H_2D^+$ | Gr | $Gr^+$ | $H_2$ | D | | 1.3e-07 | 0.50 | 0.0 |
| 94 | $H_2D^+$ | Gr | $Gr^+$ | HD | H | | 2.7e-07 | 0.50 | 0.0 |
| 95 | $H_2D^+$ | Gr | $Gr^+$ | D | H | H | 4.0e-07 | 0.50 | 0.0 |
| 96 | $HD_2^+$ | $Gr^-$ | Gr | HD | D | | 2.4e-07 | 0.50 | 0.0 |
| 97 | $HD_2^+$ | $Gr^-$ | Gr | $D_2$ | H | | 1.2e-07 | 0.50 | 0.0 |
| 98 | $HD_2^+$ | $Gr^-$ | Gr | D | D | H | 3.6e-07 | 0.50 | 0.0 |
| 99 | $HD_2^+$ | Gr | $Gr^+$ | HD | D | | 2.4e-07 | 0.50 | 0.0 |
| 100 | $HD_2^+$ | Gr | $Gr^+$ | $D_2$ | H | | 1.2e-07 | 0.50 | 0.0 |
| 101 | $HD_2^+$ | Gr | $Gr^+$ | D | D | H | 3.6e-07 | 0.50 | 0.0 |
| 102 | $D_3^+$ | $Gr^-$ | Gr | $D_2$ | D | | 3.3e-07 | 0.50 | 0.0 |
| 103 | $D_3^+$ | $Gr^-$ | Gr | D | D | D | 3.3e-07 | 0.50 | 0.0 |
| 104 | $D_3^+$ | Gr | $Gr^+$ | $D_2$ | D | | 3.3e-07 | 0.50 | 0.0 |
| 105 | $D_3^+$ | Gr | $Gr^+$ | D | D | D | 3.3e-07 | 0.50 | 0.0 |
| 106 | O | CRP | $O^+$ | $e^-$ | | | 2.8e+00 | 0.00 | 0.0 |
| 107 | O | OH | $O_2$ | H | | | 4.0e-11 | 0.00 | 0.0 |
| 108 | $O^+$ | $H_2$ | $OH^+$ | H | | | 1.2e-09 | 0.00 | 0.0 |
| 109 | $O^+$ | H | $H^+$ | O | | | 6.0e-10 | 0.00 | 0.0 |
| 110 | O | $H^+$ | $O^+$ | H | | | 6.0e-10 | 0.00 | 227.0 |
| 111 | O | $H_2^+$ | $OH^+$ | H | | | 1.5e-09 | 0.00 | 0.0 |
| 112 | O | $H_3^+$ | $OH^+$ | $H_2$ | | | 8.0e-10 | -0.16 | 1.4 |
| 113 | O | $H_3^+$ | $H_2O^+$ | H | | | 3.4e-10 | -0.16 | 1.4 |
| 114 | O | $H_2D^+$ | $OH^+$ | HD | | | 5.3e-10 | 0.00 | 0.0 |
| 115 | O | $H_2D^+$ | $OD^+$ | $H_2$ | | | 2.7e-10 | 0.00 | 0.0 |
| 116 | O | $HD_2^+$ | $OH^+$ | $D_2$ | | | 2.7e-10 | 0.00 | 0.0 |
| 117 | O | $HD_2^+$ | $OD^+$ | HD | | | 5.3e-10 | 0.00 | 0.0 |
| 118 | O | $D_3^+$ | $OD^+$ | $D_2$ | | | 8.0e-10 | 0.00 | 0.0 |
| 119 | $O_2$ | $He^+$ | $O^+$ | O | He | | 1.0e-09 | 0.00 | 0.0 |
| 120 | $O_2$ | $H^+$ | $O_2^+$ | H | | | 1.2e-09 | 0.00 | 0.0 |
| 121 | OH | $He^+$ | $OH^+$ | He | | | 5.5e-10 | 0.00 | 0.0 |
| 122 | OH | $He^+$ | $O^+$ | H | He | | 5.5e-10 | 0.00 | 0.0 |
| 123 | OH | $H^+$ | $OH^+$ | H | | | 2.1e-09 | 0.00 | 0.0 |
| 124 | OH | $H_2^+$ | $H_2$ | $OH^+$ | | | 7.6e-10 | 0.00 | 0.0 |
| 125 | OH | $H_3^+$ | $H_2O^+$ | $H_2$ | | | 1.3e-09 | 0.00 | 0.0 |
| 126 | OH | $H_2D^+$ | $H_2O^+$ | HD | | | 8.7e-10 | 0.00 | 0.0 |
| 127 | OH | $H_2D^+$ | $HDO^+$ | $H_2$ | | | 4.3e-10 | 0.00 | 0.0 |
| 128 | OH | $HD_2^+$ | $H_2O^+$ | $D_2$ | | | 4.3e-10 | 0.00 | 0.0 |
| 129 | OH | $HD_2^+$ | $HDO^+$ | HD | | | 8.7e-10 | 0.00 | 0.0 |
| 130 | OH | $D_3^+$ | $HDO^+$ | $D_2$ | | | 1.3e-09 | 0.00 | 0.0 |
| 131 | OD | $He^+$ | $OD^+$ | He | | | 5.5e-10 | 0.00 | 0.0 |
| 132 | OD | $He^+$ | $O^+$ | D | He | | 5.5e-10 | 0.00 | 0.0 |
| 133 | OD | $H^+$ | $OD^+$ | H | | | 2.1e-09 | 0.00 | 0.0 |
| 134 | OD | $H_3^+$ | $HDO^+$ | $H_2$ | | | 1.3e-09 | 0.00 | 0.0 |
| 135 | OD | $H_2D^+$ | $HDO^+$ | HD | | | 8.7e-10 | 0.00 | 0.0 |
| 136 | OD | $H_2D^+$ | $D_2O^+$ | $H_2$ | | | 4.3e-10 | 0.00 | 0.0 |
| 137 | OD | $HD_2^+$ | $HDO^+$ | $D_2$ | | | 4.3e-10 | 0.00 | 0.0 |
| 138 | OD | $HD_2^+$ | $D_2O^+$ | HD | | | 8.7e-10 | 0.00 | 0.0 |
| 139 | OD | $D_3^+$ | $D_2O^+$ | $D_2$ | | | 1.3e-10 | 0.00 | 0.0 |
| 140 | $H_2O$ | $He^+$ | $OH^+$ | H | He | | 2.3e-10 | -0.94 | 0.0 |
| 141 | $H_2O$ | $He^+$ | $H_2O^+$ | He | | | 4.9e-11 | -0.94 | 0.0 |
| 142 | $H_2O$ | $He^+$ | $H^+$ | OH | He | | 1.6e-10 | -0.94 | 0.0 |
| 143 | $H_2O$ | $H^+$ | H | $H_2O^+$ | | | 8.2e-09 | 0.00 | 0.0 |
| 144 | $H_2O$ | $H_2^+$ | $H_2$ | $H_2O^+$ | | | 3.9e-09 | 0.00 | 0.0 |
| 145 | $H_2O$ | $H_2^+$ | $H_3O^+$ | H | | | 3.4e-09 | 0.00 | 0.0 |
| 146 | $H_2O$ | $H_3^+$ | $H_3O^+$ | $H_2$ | | | 4.3e-09 | 0.00 | 0.0 |
| 147 | $H_2O$ | $H_2D^+$ | $H_3O^+$ | HD | | | 2.9e-09 | 0.00 | 0.0 |
| 148 | $H_2O$ | $HD_2^+$ | $H_2DO^+$ | HD | | | 2.9e-09 | 0.00 | 0.0 |

**Table D5** – *continued* (part 3)

| | Reactants | | Products | | | $\alpha$ | $\beta$ | $\gamma$ |
|---|---|---|---|---|---|---|---|---|
| 149 | $H_2O$ | $HD_2^+$ | $H_3O^+$ | $D_2$ | | 1.4e-09 | 0.00 | 0.0 |
| 150 | $H_2O$ | $D_3^+$ | $H_2DO^+$ | $D_2$ | | 4.3e-09 | 0.00 | 0.0 |
| 151 | $H_2O$ | $H_2D^+$ | $H_2DO^+$ | $H_2$ | | 1.4e-09 | 0.00 | 0.0 |
| 152 | HDO | $He^+$ | $OD^+$ | H | He | 2.3e-10 | -0.94 | 0.0 |
| 153 | HDO | $He^+$ | $HDO^+$ | He | | 4.9e-11 | -0.94 | 0.0 |
| 154 | HDO | $He^+$ | $H^+$ | OD | He | 1.6e-10 | -0.94 | 0.0 |
| 155 | HDO | $H^+$ | $HDO^+$ | H | | 8.2e-09 | 0.00 | 0.0 |
| 156 | HDO | $H_3^+$ | $H_2DO^+$ | $H_2$ | | 4.3e-09 | 0.00 | 0.0 |
| 157 | HDO | $H_2D^+$ | $H_2DO^+$ | HD | | 2.9e-09 | 0.00 | 0.0 |
| 158 | HDO | $H_2D^+$ | $HD_2O^+$ | $H_2$ | | 1.4e-09 | 0.00 | 0.0 |
| 159 | HDO | $HD_2^+$ | $HD_2O^+$ | HD | | 2.9e-09 | 0.00 | 0.0 |
| 160 | HDO | $HD_2^+$ | $H_2DO^+$ | $D_2$ | | 1.4e-09 | 0.00 | 0.0 |
| 161 | HDO | $D_3^+$ | $HD_2O^+$ | $D_2$ | | 4.3e-09 | 0.00 | 0.0 |
| 162 | $D_2O$ | $He^+$ | $OD^+$ | D | He | 2.3e-10 | -0.94 | 0.0 |
| 163 | $D_2O$ | $He^+$ | $D_2O^+$ | He | | 4.9e-11 | -0.94 | 0.0 |
| 164 | $D_2O$ | $He^+$ | $D^+$ | OD | He | 1.6e-10 | -0.94 | 0.0 |
| 165 | $D_2O$ | $H^+$ | $D_2O^+$ | H | | 8.2e-09 | 0.00 | 0.0 |
| 166 | $D_2O$ | $H_3^+$ | $HD_2O^+$ | $H_2$ | | 4.3e-09 | 0.00 | 0.0 |
| 167 | $D_2O$ | $H_2D^+$ | $HD_2O^+$ | HD | | 2.9e-09 | 0.00 | 0.0 |
| 168 | $D_2O$ | $H_2D^+$ | $D_3O^+$ | $H_2$ | | 1.4e-09 | 0.00 | 0.0 |
| 169 | $D_2O$ | $D_3^+$ | $D_3O^+$ | $D_2$ | | 4.3e-09 | 0.00 | 0.0 |
| 170 | $D_2O$ | $HD_2^+$ | $D_3O^+$ | HD | | 2.9e-09 | 0.00 | 0.0 |
| 171 | $D_2O$ | $HD_2^+$ | $HD_2O^+$ | $D_2$ | | 1.4e-09 | 0.00 | 0.0 |
| 172 | $OH^+$ | $H_2$ | $H_2O^+$ | H | | 1.0e-09 | 0.00 | 0.0 |
| 173 | $OH^+$ | HD | $HDO^+$ | H | | 5.1e-10 | 0.00 | 0.0 |
| 174 | $OH^+$ | HD | $H_2O^+$ | D | | 5.1e-10 | 0.00 | 0.0 |
| 175 | $OD^+$ | $H_2$ | $HDO^+$ | H | | 1.0e-09 | 0.00 | 0.0 |
| 176 | $OD^+$ | HD | $D_2O^+$ | H | | 5.1e-10 | 0.00 | 0.0 |
| 177 | $OD^+$ | HD | $HDO^+$ | D | | 5.1e-10 | 0.00 | 0.0 |
| 178 | $OD^+$ | $D_2$ | $D_2O^+$ | D | | 1.1e-09 | 0.00 | 0.0 |
| 179 | $H_2O^+$ | $H_2$ | $H_3O^+$ | H | | 8.3e-10 | 0.00 | 0.0 |
| 180 | $H_2O^+$ | HD | $H_2DO^+$ | H | | 4.2e-10 | 0.00 | 0.0 |
| 181 | $H_2O^+$ | HD | $H_3O^+$ | D | | 4.2e-10 | 0.00 | 0.0 |
| 182 | $HDO^+$ | $H_2$ | $H_2DO^+$ | H | | 8.3e-10 | 0.00 | 0.0 |
| 183 | $HDO^+$ | HD | $HD_2O^+$ | H | | 4.2e-10 | 0.00 | 0.0 |
| 184 | $HDO^+$ | HD | $H_2DO^+$ | D | | 4.2e-10 | 0.00 | 0.0 |
| 185 | $HDO^+$ | $D_2$ | $HD_2O^+$ | D | | 8.3e-10 | 0.00 | 0.0 |
| 186 | $D_2O^+$ | $D_2$ | $D_3O^+$ | D | | 8.3e-10 | 0.00 | 0.0 |
| 187 | $D_2O^+$ | $H_2$ | $HD_2O^+$ | H | | 8.3e-10 | 0.00 | 0.0 |
| 188 | $D_2O^+$ | HD | $HD_2O^+$ | D | | 4.2e-10 | 0.00 | 0.0 |
| 189 | $D_2O^+$ | HD | $D_3O^+$ | H | | 4.2e-10 | 0.00 | 0.0 |
| 190 | $H_3O^+$ | Fe | $Fe^+$ | $H_2O$ | H | 3.1e-09 | 0.00 | 0.0 |
| 191 | $O_2^+$ | Fe | $Fe^+$ | $O_2$ | | 1.1e-09 | 0.00 | 0.0 |
| 192 | $O^+$ | $e^-$ | O | $\gamma$ | | 3.4e-12 | -0.64 | 0.0 |
| 193 | $O_2^+$ | $e^-$ | O | O | | 2.0e-07 | -0.70 | 0.0 |
| 194 | $OH^+$ | $e^-$ | O | H | | 3.7e-08 | -0.50 | 0.0 |
| 195 | $H_2O^+$ | $e^-$ | OH | H | | 7.8e-08 | -0.50 | 0.0 |
| 196 | $H_2O^+$ | $e^-$ | O | $H_2$ | | 3.4e-08 | -0.50 | 0.0 |
| 197 | $H_2O^+$ | $e^-$ | O | H | H | 1.5e-07 | -0.50 | 0.0 |
| 198 | $H_3O^+$ | $e^-$ | OH | H | H | 2.6e-07 | -0.50 | 0.0 |
| 199 | $H_3O^+$ | $e^-$ | $H_2O$ | H | | 1.1e-07 | -0.50 | 0.0 |
| 200 | $H_3O^+$ | $e^-$ | OH | $H_2$ | | 6.0e-08 | -0.50 | 0.0 |
| 201 | $H_3O^+$ | $e^-$ | $H_2$ | H | O | 5.6e-09 | -0.50 | 0.0 |
| 202 | $OD^+$ | $e^-$ | O | D | | 3.7e-08 | -0.50 | 0.0 |
| 203 | $HDO^+$ | $e^-$ | OD | H | | 3.1e-08 | -0.50 | 0.0 |
| 204 | $HDO^+$ | $e^-$ | OH | D | | 1.5e-08 | -0.50 | 0.0 |
| 205 | $HDO^+$ | $e^-$ | O | HD | | 1.5e-08 | -0.50 | 0.0 |
| 206 | $HDO^+$ | $e^-$ | O | H | D | 8.9e-08 | -0.50 | 0.0 |
| 207 | $D_2O^+$ | $e^-$ | OD | D | | 7.8e-08 | -0.50 | 0.0 |
| 208 | $D_2O^+$ | $e^-$ | O | $D_2$ | | 3.4e-08 | -0.50 | 0.0 |
| 209 | $D_2O^+$ | $e^-$ | O | D | D | 1.5e-07 | -0.50 | 0.0 |
| 210 | $H_2DO^+$ | $e^-$ | OD | H | H | 8.6e-08 | -0.50 | 0.0 |
| 211 | $H_2DO^+$ | $e^-$ | OH | D | H | 1.7e-07 | -0.50 | 0.0 |
| 212 | $H_2DO^+$ | $e^-$ | $H_2O$ | D | | 3.6e-08 | -0.50 | 0.0 |
| 213 | $H_2DO^+$ | $e^-$ | HDO | H | | 7.2e-08 | -0.50 | 0.0 |
| 214 | $H_2DO^+$ | $e^-$ | OD | $H_2$ | | 2.0e-08 | -0.50 | 0.0 |
| 215 | $H_2DO^+$ | $e^-$ | OH | HD | | 4.0e-08 | -0.50 | 0.0 |
| 216 | $H_2DO^+$ | $e^-$ | HD | H | O | 3.7e-09 | -0.50 | 0.0 |
| 217 | $H_2DO^+$ | $e^-$ | $H_2$ | D | O | 1.9e-09 | -0.50 | 0.0 |
| 218 | $HD_2O^+$ | $e^-$ | OH | D | D | 8.6e-08 | -0.50 | 0.0 |
| 219 | $HD_2O^+$ | $e^-$ | OD | D | H | 1.7e-07 | -0.50 | 0.0 |
| 220 | $HD_2O^+$ | $e^-$ | $D_2O$ | H | | 3.6e-08 | -0.50 | 0.0 |
| 221 | $HD_2O^+$ | $e^-$ | HDO | D | | 7.2e-08 | -0.50 | 0.0 |
| 222 | $HD_2O^+$ | $e^-$ | OH | $D_2$ | | 2.0e-08 | -0.50 | 0.0 |

**Table D5** – *continued* (part 4)

| | Reactants | | Products | | | | $\alpha$ | $\beta$ | $\gamma$ |
|---|---|---|---|---|---|---|---|---|---|
| 223 | $HD_2O^+$ | $e^-$ | OD | HD | | | 4.0e-08 | -0.50 | 0.0 |
| 224 | $HD_2O^+$ | $e^-$ | HD | D | O | | 3.7e-09 | -0.50 | 0.0 |
| 225 | $HD_2O^+$ | $e^-$ | $D_2$ | H | O | | 1.9e-09 | -0.50 | 0.0 |
| 226 | $D_3O^+$ | $e^-$ | OD | D | D | | 2.6e-07 | -0.50 | 0.0 |
| 227 | $D_3O^+$ | $e^-$ | $D_2O$ | D | | | 1.1e-07 | -0.50 | 0.0 |
| 228 | $D_3O^+$ | $e^-$ | OD | $D_2$ | | | 6.0e-08 | -0.50 | 0.0 |
| 229 | $D_3O^+$ | $e^-$ | $D_2$ | D | O | | 5.6e-09 | -0.50 | 0.0 |
| 230 | OH | $\gamma$ | O | H | | | 4.7e+02 | 0.00 | 0.0 |
| 231 | $OH^+$ | $\gamma$ | $O^+$ | H | | | 8.6e+00 | 0.00 | 0.0 |
| 232 | $H_2O$ | $\gamma$ | OH | H | | | 1.0e+03 | 0.00 | 0.0 |
| 233 | $H_2O$ | $\gamma$ | $H_2O^+$ | $e^-$ | | | 2.3e+01 | 0.00 | 0.0 |
| 234 | $O_2$ | $\gamma$ | O | O | | | 7.8e+02 | 0.00 | 0.0 |
| 235 | $O_2$ | $\gamma$ | $O_2^+$ | $e^-$ | | | 2.8e+01 | 0.00 | 0.0 |
| 236 | $O_2^+$ | $\gamma$ | $O^+$ | O | | | 7.0e+01 | 0.00 | 0.0 |
| 237 | $H_3O^+$ | Gr | $Gr^+$ | $H_2O$ | H | | 3.7e-07 | 0.50 | 0.0 |
| 238 | $H_3O^+$ | $Gr^-$ | Gr | $H_2O$ | H | | 3.7e-07 | 0.50 | 0.0 |
| 239 | O | $H_2$ | OH | H | | | 1.5e-13 | 2.80 | 2980.0 |
| 240 | $O_2$ | H | OH | O | | | 1.6e-09 | -0.90 | 8750.0 |
| 241 | OH | H | O | $H_2$ | | | 7.0e-14 | 2.80 | 1950.0 |
| 242 | OH | $H_2$ | $H_2O$ | H | | | 9.5e-13 | 2.00 | 1490.0 |
| 243 | $H_2O$ | H | OH | $H_2$ | | | 5.2e-12 | 1.90 | 9265.0 |
| 244 | $H_3O^+$ | H | $H_2O^+$ | $H_2$ | | | 6.1e-10 | 0.00 | 20500.0 |
| 245 | $C^+$ | H | $CH^+$ | $\gamma$ | | | 7.0e-17 | 0.00 | 0.0 |
| 246 | $C^+$ | $H_2$ | $CH_2^+$ | $\gamma$ | | | 2.0e-16 | -1.30 | 23.0 |
| 247 | $C^+$ | CH | $C_2^+$ | H | | | 3.8e-10 | 0.00 | 0.0 |
| 248 | $C^+$ | CH | $CH^+$ | C | | | 3.8e-10 | 0.00 | 0.0 |
| 249 | $C^+$ | $CH_2$ | $CH_2^+$ | C | | | 5.2e-10 | 0.00 | 0.0 |
| 250 | $C^+$ | $CH_2$ | $C_2H^+$ | H | | | 5.2e-10 | 0.00 | 0.0 |
| 251 | $C^+$ | $CH_4$ | $C_2H_2^+$ | $H_2$ | | | 3.3e-10 | 0.00 | 0.0 |
| 252 | $C^+$ | $CH_4$ | $C_2H_3^+$ | H | | | 9.8e-10 | 0.00 | 0.0 |
| 253 | $C^+$ | $C_2H$ | $C_3^+$ | H | | | 1.0e-09 | 0.00 | 0.0 |
| 254 | $C^+$ | $C_2H_2$ | $C_3H^+$ | H | | | 2.2e-09 | 0.00 | 0.0 |
| 255 | $C^+$ | Fe | $Fe^+$ | C | | | 2.6e-09 | 0.00 | 0.0 |
| 256 | $CH^+$ | H | $C^+$ | $H_2$ | | | 1.5e-10 | 0.00 | 0.0 |
| 257 | $CH^+$ | $H_2$ | $CH_2^+$ | H | | | 1.2e-09 | 0.00 | 0.0 |
| 258 | $CH_2^+$ | $H_2$ | $CH_3^+$ | H | | | 7.0e-10 | 0.00 | 0.0 |
| 259 | $CH_3^+$ | $H_2$ | $CH_5^+$ | $\gamma$ | | | 3.8e-16 | -2.30 | 21.5 |
| 260 | $CH_4^+$ | H | $CH_3^+$ | $H_2$ | | | 2.0e-10 | 0.00 | 0.0 |
| 261 | $CH_4^+$ | $H_2$ | $CH_5^+$ | H | | | 4.0e-11 | 0.00 | 0.0 |
| 262 | $C_2^+$ | $H_2$ | $C_2H^+$ | H | | | 1.4e-09 | 0.00 | 0.0 |
| 263 | $C_2H^+$ | $H_2$ | $C_2H_2^+$ | H | | | 1.7e-09 | 0.00 | 0.0 |
| 264 | $C_3^+$ | $H_2$ | $C_3H^+$ | H | | | 3.0e-10 | 0.00 | 0.0 |
| 265 | $C_3H^+$ | $H_2$ | $C_3H_3^+$ | $\gamma$ | | | 3.0e-13 | -1.00 | 0.0 |
| 266 | C | CRP | $C^+$ | $e^-$ | | | 1.8e+00 | 0.00 | 0.0 |
| 267 | C | H | CH | $\gamma$ | | | 1.0e-17 | 0.00 | 0.0 |
| 268 | C | $H_2^+$ | $CH^+$ | H | | | 2.4e-09 | 0.00 | 0.0 |
| 269 | C | $H_3^+$ | $CH^+$ | $H_2$ | | | 2.0e-09 | 0.00 | 0.0 |
| 270 | CH | H | C | $H_2$ | | | 1.2e-10 | 0.26 | 0.0 |
| 271 | CH | $He^+$ | $C^+$ | H | He | | 1.1e-09 | 0.00 | 0.0 |
| 272 | CH | $H^+$ | H | $CH^+$ | | | 1.9e-09 | 0.00 | 0.0 |
| 273 | CH | $H_2^+$ | $H_2$ | $CH^+$ | | | 7.1e-10 | 0.00 | 0.0 |
| 274 | CH | $H_2^+$ | $CH_2^+$ | H | | | 7.1e-10 | 0.00 | 0.0 |
| 275 | CH | $H_3^+$ | $CH_2^+$ | $H_2$ | | | 1.2e-09 | 0.00 | 0.0 |
| 276 | CH | $H_2D^+$ | $CH_2^+$ | HD | | | 1.2e-09 | 0.00 | 0.0 |
| 277 | CH | $HD_2^+$ | $CH_2^+$ | $D_2$ | | | 1.2e-09 | 0.00 | 0.0 |
| 278 | $CH_2$ | H | CH | $H_2$ | | | 2.2e-10 | 0.00 | 0.0 |
| 279 | $CH_2$ | $He^+$ | $C^+$ | $H_2$ | He | | 7.5e-10 | 0.00 | 0.0 |
| 280 | $CH_2$ | $He^+$ | $CH^+$ | H | He | | 7.5e-10 | 0.00 | 0.0 |
| 281 | $CH_2$ | $H^+$ | $CH^+$ | $H_2$ | | | 1.4e-09 | 0.00 | 0.0 |
| 282 | $CH_2$ | $H^+$ | H | $CH_2^+$ | | | 1.4e-09 | 0.00 | 0.0 |
| 283 | $CH_2$ | $H_2^+$ | $CH_3^+$ | H | | | 1.0e-09 | 0.00 | 0.0 |
| 284 | $CH_2$ | $H_2^+$ | $H_2$ | $CH_2^+$ | | | 1.0e-09 | 0.00 | 0.0 |
| 285 | $CH_2$ | $H_3^+$ | $CH_3^+$ | $H_2$ | | | 1.7e-09 | 0.00 | 0.0 |
| 286 | $CH_2$ | $H_2D^+$ | $CH_3^+$ | HD | | | 1.7e-09 | 0.00 | 0.0 |
| 287 | $CH_2$ | $HD_2^+$ | $CH_3^+$ | $D_2$ | | | 1.7e-09 | 0.00 | 0.0 |
| 288 | $CH_3$ | $He^+$ | $CH^+$ | $H_2$ | He | | 9.0e-10 | 0.00 | 0.0 |
| 289 | $CH_3$ | $He^+$ | $CH_2^+$ | H | He | | 9.0e-10 | 0.00 | 0.0 |
| 290 | $CH_3$ | $H^+$ | H | $CH_3^+$ | | | 3.4e-09 | 0.00 | 0.0 |
| 291 | $CH_3$ | $H_3^+$ | $CH_4^+$ | $H_2$ | | | 2.1e-09 | 0.00 | 0.0 |
| 292 | $CH_3$ | $H_2D^+$ | $CH_4^+$ | HD | | | 2.1e-09 | 0.00 | 0.0 |
| 293 | $CH_3$ | $HD_2^+$ | $CH_4^+$ | $D_2$ | | | 2.1e-09 | 0.00 | 0.0 |
| 294 | $CH_4$ | $He^+$ | $H^+$ | $CH_3$ | He | | 4.0e-10 | 0.00 | 0.0 |
| 295 | $CH_4$ | $He^+$ | $CH^+$ | $H_2$ | H | He | 2.6e-10 | 0.00 | 0.0 |
| 296 | $CH_4$ | $He^+$ | $CH_2^+$ | $H_2$ | He | | 8.5e-10 | 0.00 | 0.0 |

**Table D5** – *continued* (part 5)

| | Reactants | | Products | | | $\alpha$ | $\beta$ | $\gamma$ |
|---|---|---|---|---|---|---|---|---|
| 297 | $CH_4$ | $He^+$ | $CH_3^+$ | H | He | 8.0e-11 | 0.00 | 0.0 |
| 298 | $CH_4$ | $He^+$ | $CH_4^+$ | He | | 1.6e-11 | 0.00 | 0.0 |
| 299 | $CH_4$ | $H^+$ | $CH_3^+$ | $H_2$ | | 2.3e-09 | 0.00 | 0.0 |
| 300 | $CH_4$ | $H^+$ | H | $CH_4^+$ | | 1.5e-09 | 0.00 | 0.0 |
| 301 | $CH_4$ | $H_3^+$ | $CH_5^+$ | $H_2$ | | 1.9e-09 | 0.00 | 0.0 |
| 302 | $CH_4$ | $H_2D^+$ | $CH_5^+$ | HD | | 1.9e-09 | 0.00 | 0.0 |
| 303 | $CH_4$ | $HD_2^+$ | $CH_5^+$ | $D_2$ | | 1.9e-09 | 0.00 | 0.0 |
| 304 | C | $H_2D^+$ | $CH^+$ | HD | | 2.0e-09 | 0.00 | 0.0 |
| 305 | C | $HD_2^+$ | $CH^+$ | $D_2$ | | 2.0e-09 | 0.00 | 0.0 |
| 306 | $C_2$ | $H^+$ | $C_2^+$ | H | | 3.1e-09 | 0.00 | 0.0 |
| 307 | $C_2H$ | $He^+$ | $C^+$ | CH | He | 5.1e-10 | 0.00 | 0.0 |
| 308 | $C_2H$ | $He^+$ | $CH^+$ | C | He | 5.1e-10 | 0.00 | 0.0 |
| 309 | $C_2H$ | $He^+$ | $C_2^+$ | H | He | 5.1e-10 | 0.00 | 0.0 |
| 310 | $C_2H$ | $H^+$ | $C_2^+$ | $H_2$ | | 1.5e-09 | 0.00 | 0.0 |
| 311 | $C_2H$ | $H^+$ | $C_2H^+$ | H | | 1.5e-09 | 0.00 | 0.0 |
| 312 | $C_2H$ | $H_3^+$ | $C_2H_2^+$ | $H_2$ | | 1.7e-09 | 0.00 | 0.0 |
| 313 | $C_2H$ | $H_2D^+$ | $C_2H_2^+$ | HD | | 1.7e-09 | 0.00 | 0.0 |
| 314 | $C_2H$ | $HD_2^+$ | $C_2H_2^+$ | $D_2$ | | 1.7e-09 | 0.00 | 0.0 |
| 315 | $C_3H$ | $He^+$ | $C_3^+$ | H | He | 2.0e-09 | 0.00 | 0.0 |
| 316 | $C_3H$ | $H^+$ | $C_3^+$ | $H_2$ | | 2.0e-09 | 0.00 | 0.0 |
| 317 | $C_3H$ | $H^+$ | $C_3H^+$ | H | | 2.0e-09 | 0.00 | 0.0 |
| 318 | $C_3H$ | $H_3^+$ | $C_3H_2^+$ | $H_2$ | | 2.0e-09 | 0.00 | 0.0 |
| 319 | $C_3H$ | $H_2D^+$ | $C_3H_2^+$ | HD | | 2.0e-09 | 0.00 | 0.0 |
| 320 | $C_3H$ | $HD_2^+$ | $C_3H_2^+$ | $D_2$ | | 2.0e-09 | 0.00 | 0.0 |
| 321 | $C_2H_2$ | $He^+$ | $CH^+$ | CH | He | 7.7e-10 | 0.00 | 0.0 |
| 322 | $C_2H_2$ | $He^+$ | $C_2^+$ | $H_2$ | He | 1.6e-09 | 0.00 | 0.0 |
| 323 | $C_2H_2$ | $He^+$ | $C_2H^+$ | H | He | 8.7e-10 | 0.00 | 0.0 |
| 324 | $C_2H_2$ | $He^+$ | $C_2H_2^+$ | He | | 2.4e-10 | 0.00 | 0.0 |
| 325 | $C_2H_2$ | $H^+$ | $C_2H^+$ | $H_2$ | | 2.0e-09 | 0.00 | 0.0 |
| 326 | $C_2H_2$ | $H^+$ | $C_2H_2^+$ | H | | 2.0e-09 | 0.00 | 0.0 |
| 327 | $C_2H_2$ | $H_3^+$ | $C_2H_3^+$ | $H_2$ | | 2.9e-09 | 0.00 | 0.0 |
| 328 | $C_2H_2$ | $H_2D^+$ | $C_2H_3^+$ | HD | | 2.9e-09 | 0.00 | 0.0 |
| 329 | $C_2H_2$ | $HD_2^+$ | $C_2H_3^+$ | $D_2$ | | 2.9e-09 | 0.00 | 0.0 |
| 330 | $C_3H_2$ | $H^+$ | $C_3H^+$ | $H_2$ | | 2.0e-09 | 0.00 | 0.0 |
| 331 | $C_3H_2$ | $H^+$ | $C_3H_2^+$ | H | | 2.0e-09 | 0.00 | 0.0 |
| 332 | $C_3H_2$ | $He^+$ | $C_3H^+$ | H | He | 1.0e-09 | 0.00 | 0.0 |
| 333 | $C_3H_2$ | $He^+$ | $C_3^+$ | $H_2$ | He | 1.0e-09 | 0.00 | 0.0 |
| 334 | $C_3H_2$ | $H_3^+$ | $C_3H_3^+$ | $H_2$ | | 2.0e-09 | 0.00 | 0.0 |
| 335 | $C_3H_2$ | $H_2D^+$ | $C_3H_3^+$ | HD | | 2.0e-09 | 0.00 | 0.0 |
| 336 | $C_3H_2$ | $HD_2^+$ | $C_3H_3^+$ | $D_2$ | | 2.0e-09 | 0.00 | 0.0 |
| 337 | $C^+$ | $e^-$ | C | $\gamma$ | | 4.4e-12 | -0.61 | 0.0 |
| 338 | $CH^+$ | $e^-$ | C | H | | 1.5e-07 | -0.42 | 0.0 |
| 339 | $CH_2^+$ | $e^-$ | C | $H_2$ | | 1.2e-07 | -0.50 | 0.0 |
| 340 | $CH_2^+$ | $e^-$ | CH | H | | 1.2e-07 | -0.50 | 0.0 |
| 341 | $CH_3^+$ | $e^-$ | C | $H_2$ | H | 3.0e-07 | -0.30 | 0.0 |
| 342 | $CH_3^+$ | $e^-$ | CH | H | H | 1.6e-07 | -0.30 | 0.0 |
| 343 | $CH_3^+$ | $e^-$ | CH | $H_2$ | | 1.4e-07 | -0.30 | 0.0 |
| 344 | $CH_3^+$ | $e^-$ | $CH_2$ | H | | 4.0e-07 | -0.30 | 0.0 |
| 345 | $CH_4^+$ | $e^-$ | $CH_3$ | H | | 3.0e-07 | -0.50 | 0.0 |
| 346 | $CH_4^+$ | $e^-$ | $CH_2$ | H | H | 3.0e-07 | -0.50 | 0.0 |
| 347 | $CH_5^+$ | $e^-$ | CH | $H_2$ | $H_2$ | 8.7e-08 | -0.30 | 0.0 |
| 348 | $CH_5^+$ | $e^-$ | $CH_2$ | $H_2$ | H | 8.7e-08 | -0.30 | 0.0 |
| 349 | $CH_5^+$ | $e^-$ | $CH_3$ | $H_2$ | | 8.7e-08 | -0.30 | 0.0 |
| 350 | $CH_5^+$ | $e^-$ | $CH_4$ | H | | 8.7e-08 | -0.30 | 0.0 |
| 351 | $C_2^+$ | $e^-$ | C | C | | 3.0e-07 | -0.50 | 0.0 |
| 352 | $C_2H^+$ | $e^-$ | $C_2$ | H | | 1.4e-07 | -0.50 | 0.0 |
| 353 | $C_2H^+$ | $e^-$ | CH | C | | 1.4e-07 | -0.50 | 0.0 |
| 354 | $C_2H_2^+$ | $e^-$ | $C_2H$ | H | | 1.5e-07 | -0.50 | 0.0 |
| 355 | $C_2H_2^+$ | $e^-$ | CH | CH | | 1.5e-07 | -0.50 | 0.0 |
| 356 | $C_2H_3^+$ | $e^-$ | $C_2H$ | $H_2$ | | 1.4e-07 | -0.50 | 0.0 |
| 357 | $C_2H_3^+$ | $e^-$ | $CH_2$ | CH | | 1.4e-07 | -0.50 | 0.0 |
| 358 | $C_2H_3^+$ | $e^-$ | $C_2H_2$ | H | | 3.0e-08 | -0.50 | 0.0 |
| 359 | $C_3^+$ | $e^-$ | $C_2$ | C | | 3.0e-07 | -0.50 | 0.0 |
| 360 | $C_3H^+$ | $e^-$ | $C_2$ | CH | | 1.5e-07 | -0.50 | 0.0 |
| 361 | $C_3H^+$ | $e^-$ | $C_2H$ | C | | 1.5e-07 | -0.50 | 0.0 |
| 362 | $C_3H_2^+$ | $e^-$ | $C_3H$ | H | | 1.5e-07 | -0.50 | 0.0 |
| 363 | $C_3H_2^+$ | $e^-$ | $C_2H$ | CH | | 1.5e-07 | -0.50 | 0.0 |
| 364 | $C_3H_3^+$ | $e^-$ | $C_3H_2$ | H | | 1.5e-07 | -0.50 | 0.0 |
| 365 | $C_3H_3^+$ | $e^-$ | $C_2H_2$ | CH | | 1.5e-07 | -0.50 | 0.0 |
| 366 | C | $\gamma$ | $C^+$ | $e^-$ | | 2.6e+02 | 0.00 | 0.0 |
| 367 | CH | $\gamma$ | C | H | | 1.1e+03 | 0.00 | 0.0 |
| 368 | CH | $\gamma$ | $CH^+$ | $e^-$ | | 5.8e+02 | 0.00 | 0.0 |
| 369 | $CH_2$ | $\gamma$ | C | $H_2$ | | 2.9e+02 | 0.00 | 0.0 |
| 370 | $CH_3$ | $\gamma$ | CH | $H_2$ | | 2.8e+02 | 0.00 | 0.0 |

**Table D5** – *continued* (part 6)

| | Reactants | | Products | | | $\alpha$ | $\beta$ | $\gamma$ |
|---|---|---|---|---|---|---|---|---|
| 371 | $CH_3$ | $\gamma$ | $CH_3^+$ | $e^-$ | | 3.8e+02 | 0.00 | 0.0 |
| 372 | $CH_4$ | $\gamma$ | $CH_3$ | H | | 1.5e+03 | 0.00 | 0.0 |
| 373 | $CH_4$ | $\gamma$ | $CH_4^+$ | $e^-$ | | 2.2e+01 | 0.00 | 0.0 |
| 374 | $CH^+$ | $\gamma$ | C | $H^+$ | | 2.2e+02 | 0.00 | 0.0 |
| 375 | $CH_2^+$ | $\gamma$ | $CH^+$ | H | | 8.9e+01 | 0.00 | 0.0 |
| 376 | $CH_4^+$ | $\gamma$ | $CH_3^+$ | H | | 2.7e+01 | 0.00 | 0.0 |
| 377 | $C_2$ | $\gamma$ | C | C | | 1.8e+02 | 0.00 | 0.0 |
| 378 | $C_2$ | $\gamma$ | $C_2^+$ | $e^-$ | | 2.5e+02 | 0.00 | 0.0 |
| 379 | $C_2H$ | $\gamma$ | $C_2$ | H | | 1.1e+03 | 0.00 | 0.0 |
| 380 | $C_2H_2$ | $\gamma$ | $C_2H$ | H | | 3.5e+03 | 0.00 | 0.0 |
| 381 | $C_2H_2$ | $\gamma$ | $C_2H_2^+$ | $e^-$ | | 3.8e+02 | 0.00 | 0.0 |
| 382 | $C_3$ | $\gamma$ | $C_2$ | C | | 6.9e+03 | 0.00 | 0.0 |
| 383 | $C_3H$ | $\gamma$ | $C_3$ | H | | 3.0e+03 | 0.00 | 0.0 |
| 384 | $C_3H_2$ | $\gamma$ | $C_3H$ | H | | 3.4e+03 | 0.00 | 0.0 |
| 385 | $C^+$ | Gr | $Gr^+$ | C | | 4.6e-07 | 0.50 | 0.0 |
| 386 | $C^+$ | $Gr^-$ | Gr | C | | 4.6e-07 | 0.50 | 0.0 |
| 387 | C | $H_2$ | CH | H | | 1.2e-09 | 0.50 | 14100.0 |
| 388 | CH | $H_2$ | $CH_2$ | H | | 2.4e-10 | 0.00 | 1760.0 |
| 389 | $CH_2$ | $H_2$ | $CH_3$ | H | | 5.2e-11 | 0.17 | 6400.0 |
| 390 | $CH_3$ | $H_2$ | $CH_4$ | H | | 3.0e-10 | 0.00 | 5460.0 |
| 391 | $C_2$ | $H_2$ | $C_2H$ | H | | 1.6e-10 | 0.00 | 1419.0 |
| 392 | $C_2H$ | $H_2$ | $C_2H_2$ | H | | 1.1e-11 | 0.00 | 950.0 |
| 393 | $CH_3$ | H | $CH_2$ | $H_2$ | | 5.2e-11 | 0.17 | 5600.0 |
| 394 | $CH_4$ | H | $CH_3$ | $H_2$ | | 3.0e-10 | 0.00 | 6560.0 |
| 395 | $C^+$ | $H_2$ | $CH^+$ | H | | 1.5e-10 | 0.00 | 4640.0 |
| 396 | $CH_2^+$ | H | $CH^+$ | $H_2$ | | 1.2e-09 | 0.00 | 2700.0 |
| 397 | $CH_3^+$ | H | $CH_2^+$ | $H_2$ | | 7.0e-10 | 0.00 | 10560.0 |
| 398 | $CH_3^+$ | $H_2$ | $CH_4^+$ | H | | 2.0e-10 | 0.00 | 32500.0 |
| 399 | $CH_5^+$ | H | $CH_4^+$ | $H_2$ | | 4.0e-11 | 0.00 | 2200.0 |
| 400 | $C_2^+$ | $H_2$ | $H^+$ | $C_2H$ | | 1.5e-09 | 0.00 | 1260.0 |
| 401 | $C_2H_2^+$ | $H_2$ | $C_2H_3^+$ | H | | 5.0e-10 | 0.00 | 800.0 |
| 402 | $C_3H^+$ | $H_2$ | $C_3H_2^+$ | H | | 1.0e-09 | 0.00 | 500.0 |
| 403 | $C_3H_2^+$ | $H_2$ | $C_3H_3^+$ | H | | 1.0e-10 | 0.00 | 2000.0 |
| 404 | N | CRP | $N^+$ | $e^-$ | | 2.1e+00 | 0.00 | 0.0 |
| 405 | N | NH | $N_2$ | H | | 5.0e-11 | 0.10 | 0.0 |
| 406 | N | $NH_2$ | $N_2$ | H | H | 1.2e-10 | 0.00 | 0.0 |
| 407 | $N^+$ | $H_2$ | $NH^+$ | H | | 4.2e-10 | 0.00 | 41.9 |
| 408 | $N^+$ | HD | $ND^+$ | H | | 4.2e-10 | 0.00 | 0.0 |
| 409 | $N_2^+$ | $H_2$ | $N_2H^+$ | H | | 2.0e-09 | 0.24 | 0.0 |
| 410 | $NH^+$ | $H_2$ | $NH_2^+$ | H | | 1.3e-09 | 0.00 | 0.0 |
| 411 | $NH^+$ | $H_2$ | $H_3^+$ | N | | 2.2e-10 | 0.00 | 0.0 |
| 412 | $NH^+$ | H | $N^+$ | $H_2$ | | 6.5e-10 | 0.00 | 0.0 |
| 413 | $N_2H^+$ | $NH_3$ | $NH_4^+$ | $N_2$ | | 2.3e-09 | 0.00 | 0.0 |
| 414 | $N_2H^+$ | D | $N_2D^+$ | H | | 1.0e-09 | 0.00 | 0.0 |
| 415 | $NH_2^+$ | $H_2$ | $NH_3^+$ | H | | 2.7e-10 | 0.00 | 0.0 |
| 416 | $NH_3^+$ | $H_2$ | $NH_4^+$ | H | | 2.4e-12 | 0.00 | 0.0 |
| 417 | $ND^+$ | $H_2$ | $NHD^+$ | H | | 1.3e-09 | 0.00 | 0.0 |
| 418 | $ND^+$ | $H_2$ | $H_2D^+$ | N | | 2.2e-10 | 0.00 | 0.0 |
| 419 | $ND^+$ | HD | $NHD^+$ | D | | 6.3e-10 | 0.00 | 0.0 |
| 420 | $ND^+$ | HD | $ND_2^+$ | H | | 6.3e-10 | 0.00 | 0.0 |
| 421 | $N_2D^+$ | H | $N_2H^+$ | D | | 1.0e-09 | 0.00 | 170.0 |
| 422 | $N_2D^+$ | $NH_3$ | $NH_3D^+$ | $N_2$ | | 2.3e-09 | 0.00 | 0.0 |
| 423 | $NHD^+$ | $H_2$ | $NH_2D^+$ | H | | 2.7e-10 | 0.00 | 0.0 |
| 424 | $NHD^+$ | HD | $NH_2D^+$ | D | | 1.3e-10 | 0.00 | 0.0 |
| 425 | $NHD^+$ | HD | $NHD_2^+$ | H | | 1.3e-10 | 0.00 | 0.0 |
| 426 | $ND_2^+$ | $H_2$ | $NHD_2^+$ | H | | 2.7e-10 | 0.00 | 0.0 |
| 427 | $ND_2^+$ | HD | $NHD_2^+$ | D | | 1.3e-10 | 0.00 | 0.0 |
| 428 | $ND_2^+$ | HD | $ND_3^+$ | H | | 1.3e-10 | 0.00 | 0.0 |
| 429 | $NH_2D^+$ | $H_2$ | $NH_3D^+$ | H | | 2.4e-12 | 0.00 | 0.0 |
| 430 | $NH_2D^+$ | HD | $NH_3D^+$ | D | | 1.2e-12 | 0.00 | 0.0 |
| 431 | $NH_2D^+$ | HD | $NH_2D_2^+$ | H | | 1.2e-12 | 0.00 | 0.0 |
| 432 | $NHD_2^+$ | $H_2$ | $NH_2D_2^+$ | H | | 2.4e-12 | 0.00 | 0.0 |
| 433 | $NHD_2^+$ | HD | $NH_2D_2^+$ | D | | 1.2e-12 | 0.00 | 0.0 |
| 434 | $NHD_2^+$ | HD | $NHD_3^+$ | H | | 1.2e-12 | 0.00 | 0.0 |
| 435 | $ND_3^+$ | $H_2$ | $NHD_3^+$ | H | | 2.4e-12 | 0.00 | 0.0 |
| 436 | $ND_3^+$ | HD | $NHD_3^+$ | D | | 1.2e-12 | 0.00 | 0.0 |
| 437 | $ND_3^+$ | HD | $ND_4^+$ | H | | 1.2e-12 | 0.00 | 0.0 |
| 438 | N | $H_3^+$ | $NH_2^+$ | H | | 0.0e+00 | 0.00 | 0.0 |
| 439 | $N_2$ | $H_3^+$ | $N_2H^+$ | $H_2$ | | 1.3e-09 | 0.00 | 0.0 |
| 440 | NH | $He^+$ | $N^+$ | H | He | 1.1e-09 | 0.00 | 0.0 |
| 441 | NH | $H^+$ | $NH^+$ | H | | 2.1e-09 | 0.00 | 0.0 |
| 442 | NH | $H_3^+$ | $NH_2^+$ | $H_2$ | | 1.3e-09 | 0.00 | 0.0 |
| 443 | $NH_2$ | $He^+$ | $NH^+$ | H | He | 8.0e-10 | 0.00 | 0.0 |
| 444 | $NH_2$ | $He^+$ | $N^+$ | $H_2$ | He | 8.0e-10 | 0.00 | 0.0 |

**Table D5** – *continued* (part 7)

| | Reactants | | Products | | | $\alpha$ | $\beta$ | $\gamma$ |
|---|---|---|---|---|---|---|---|---|
| 445 | $NH_2$ | $H^+$ | H | $NH_2^+$ | | 2.9e-09 | 0.00 | 0.0 |
| 446 | $NH_2$ | $H_3^+$ | $NH_3^+$ | $H_2$ | | 1.8e-09 | 0.00 | 0.0 |
| 447 | $NH_3$ | $He^+$ | $NH_3^+$ | He | | 2.6e-10 | 0.00 | 0.0 |
| 448 | $NH_3$ | $He^+$ | $NH_2^+$ | H | He | 1.8e-09 | 0.00 | 0.0 |
| 449 | $NH_3$ | $He^+$ | $NH^+$ | $H_2$ | He | 1.8e-10 | 0.00 | 0.0 |
| 450 | $NH_3$ | $H^+$ | H | $NH_3^+$ | | 5.2e-09 | 0.00 | 0.0 |
| 451 | $NH_3$ | $H_3^+$ | $NH_4^+$ | $H_2$ | | 9.1e-09 | 0.00 | 0.0 |
| 452 | NH | $H_2D^+$ | $NH_2^+$ | HD | | 8.7e-10 | 0.00 | 0.0 |
| 453 | NH | $H_2D^+$ | $NHD^+$ | $H_2$ | | 4.3e-10 | 0.00 | 0.0 |
| 454 | NH | $HD_2^+$ | $NH_2^+$ | $D_2$ | | 4.3e-10 | 0.00 | 0.0 |
| 455 | NH | $HD_2^+$ | $NHD^+$ | HD | | 8.7e-10 | 0.00 | 0.0 |
| 456 | NH | $D_3^+$ | $NHD^+$ | $D_2$ | | 1.3e-09 | 0.00 | 0.0 |
| 457 | $NH_2$ | $H_2D^+$ | $NH_3^+$ | HD | | 1.2e-09 | 0.00 | 0.0 |
| 458 | $NH_2$ | $H_2D^+$ | $NH_2D^+$ | $H_2$ | | 6.0e-10 | 0.00 | 0.0 |
| 459 | $NH_2$ | $HD_2^+$ | $NH_3^+$ | $D_2$ | | 6.0e-10 | 0.00 | 0.0 |
| 460 | $NH_2$ | $HD_2^+$ | $NH_2D^+$ | HD | | 1.2e-09 | 0.00 | 0.0 |
| 461 | $NH_2$ | $D_3^+$ | $NH_2D^+$ | $D_2$ | | 1.8e-09 | 0.00 | 0.0 |
| 462 | $NH_3$ | $H_2D^+$ | $NH_4^+$ | HD | | 6.1e-09 | 0.00 | 0.0 |
| 463 | $NH_3$ | $H_2D^+$ | $NH_3D^+$ | $H_2$ | | 3.0e-09 | 0.00 | 0.0 |
| 464 | $NH_3$ | $HD_2^+$ | $NH_4^+$ | $D_2$ | | 3.0e-09 | 0.00 | 0.0 |
| 465 | $NH_3$ | $HD_2^+$ | $NH_3D^+$ | HD | | 6.1e-09 | 0.00 | 0.0 |
| 466 | $NH_3$ | $D_3^+$ | $NH_3D^+$ | $D_2$ | | 9.1e-09 | 0.00 | 0.0 |
| 467 | ND | $H_3^+$ | $NHD^+$ | $H_2$ | | 1.3e-09 | 0.00 | 0.0 |
| 468 | ND | $H_2D^+$ | $NHD^+$ | HD | | 8.7e-10 | 0.00 | 0.0 |
| 469 | ND | $H_2D^+$ | $ND_2^+$ | $H_2$ | | 4.3e-10 | 0.00 | 0.0 |
| 470 | ND | $HD_2^+$ | $NHD^+$ | $D_2$ | | 4.3e-10 | 0.00 | 0.0 |
| 471 | ND | $HD_2^+$ | $ND_2^+$ | HD | | 8.7e-10 | 0.00 | 0.0 |
| 472 | ND | $D_3^+$ | $ND_2^+$ | $D_2$ | | 1.3e-09 | 0.00 | 0.0 |
| 473 | NHD | $H_3^+$ | $NH_2D^+$ | $H_2$ | | 1.8e-09 | 0.00 | 0.0 |
| 474 | NHD | $H_2D^+$ | $NH_2D^+$ | HD | | 1.2e-09 | 0.00 | 0.0 |
| 475 | NHD | $H_2D^+$ | $NHD_2^+$ | $H_2$ | | 6.0e-10 | 0.00 | 0.0 |
| 476 | NHD | $HD_2^+$ | $NH_2D^+$ | $D_2$ | | 6.0e-10 | 0.00 | 0.0 |
| 477 | NHD | $HD_2^+$ | $NHD_2^+$ | HD | | 1.2e-09 | 0.00 | 0.0 |
| 478 | NHD | $D_3^+$ | $NHD_2^+$ | $D_2$ | | 1.8e-09 | 0.00 | 0.0 |
| 479 | $ND_2$ | $H_3^+$ | $NHD_2^+$ | $H_2$ | | 1.8e-09 | 0.00 | 0.0 |
| 480 | $ND_2$ | $H_2D^+$ | $NHD_2^+$ | HD | | 1.2e-09 | 0.00 | 0.0 |
| 481 | $ND_2$ | $H_2D^+$ | $ND_3^+$ | $H_2$ | | 6.0e-10 | 0.00 | 0.0 |
| 482 | $ND_2$ | $HD_2^+$ | $NHD_2^+$ | $D_2$ | | 6.0e-10 | 0.00 | 0.0 |
| 483 | $ND_2$ | $HD_2^+$ | $ND_3^+$ | HD | | 1.2e-09 | 0.00 | 0.0 |
| 484 | $ND_2$ | $D_3^+$ | $ND_3^+$ | $D_2$ | | 1.8e-09 | 0.00 | 0.0 |
| 485 | $NH_2D$ | $H^+$ | $NH_2D^+$ | H | | 5.2e-09 | 0.00 | 0.0 |
| 486 | $NH_2D$ | $H_3^+$ | $NH_3D^+$ | $H_2$ | | 9.1e-09 | 0.00 | 0.0 |
| 487 | $NH_2D$ | $H_2D^+$ | $NH_3D^+$ | HD | | 6.1e-09 | 0.00 | 0.0 |
| 488 | $NH_2D$ | $H_2D^+$ | $NH_2D_2^+$ | $H_2$ | | 3.0e-09 | 0.00 | 0.0 |
| 489 | $NH_2D$ | $HD_2^+$ | $NH_3D^+$ | $D_2$ | | 3.0e-09 | 0.00 | 0.0 |
| 490 | $NH_2D$ | $HD_2^+$ | $NH_2D_2^+$ | HD | | 6.1e-09 | 0.00 | 0.0 |
| 491 | $NH_2D$ | $D_3^+$ | $NH_2D_2^+$ | $D_2$ | | 9.1e-09 | 0.00 | 0.0 |
| 492 | $NHD_2$ | $H^+$ | $NHD_2^+$ | H | | 5.2e-09 | 0.00 | 0.0 |
| 493 | $NHD_2$ | $H_3^+$ | $NH_2D_2^+$ | $H_2$ | | 9.1e-09 | 0.00 | 0.0 |
| 494 | $NHD_2$ | $H_2D^+$ | $NH_2D_2^+$ | HD | | 6.1e-09 | 0.00 | 0.0 |
| 495 | $NHD_2$ | $H_2D^+$ | $NHD_3^+$ | $H_2$ | | 3.0e-09 | 0.00 | 0.0 |
| 496 | $NHD_2$ | $HD_2^+$ | $NH_2D_2^+$ | $D_2$ | | 3.0e-09 | 0.00 | 0.0 |
| 497 | $NHD_2$ | $HD_2^+$ | $NHD_3^+$ | HD | | 6.1e-09 | 0.00 | 0.0 |
| 498 | $NHD_2$ | $D_3^+$ | $NHD_3^+$ | $D_2$ | | 9.1e-09 | 0.00 | 0.0 |
| 499 | $ND_3$ | $H^+$ | $ND_3^+$ | H | | 5.2e-09 | 0.00 | 0.0 |
| 500 | $ND_3$ | $H_3^+$ | $NHD_3^+$ | $H_2$ | | 9.1e-09 | 0.00 | 0.0 |
| 501 | $ND_3$ | $H_2D^+$ | $NHD_3^+$ | HD | | 6.1e-09 | 0.00 | 0.0 |
| 502 | $ND_3$ | $H_2D^+$ | $ND_4^+$ | $H_2$ | | 3.0e-09 | 0.00 | 0.0 |
| 503 | $ND_3$ | $HD_2^+$ | $NHD_3^+$ | $D_2$ | | 3.0e-09 | 0.00 | 0.0 |
| 504 | $ND_3$ | $HD_2^+$ | $ND_4^+$ | HD | | 6.1e-09 | 0.00 | 0.0 |
| 505 | $ND_3$ | $D_3^+$ | $ND_4^+$ | $D_2$ | | 9.1e-09 | 0.00 | 0.0 |
| 506 | $N_2$ | $H_2D^+$ | $N_2H^+$ | HD | | 8.7e-10 | 0.00 | 0.0 |
| 507 | $N_2$ | $H_2D^+$ | $N_2D^+$ | $H_2$ | | 4.3e-10 | 0.00 | 0.0 |
| 508 | $N_2$ | $HD_2^+$ | $N_2H^+$ | $D_2$ | | 4.3e-10 | 0.00 | 0.0 |
| 509 | $N_2$ | $HD_2^+$ | $N_2D^+$ | HD | | 8.7e-10 | 0.00 | 0.0 |
| 510 | $N_2$ | $D_3^+$ | $N_2D^+$ | $D_2$ | | 1.3e-09 | 0.00 | 0.0 |
| 511 | $N^+$ | $e^-$ | N | $\gamma$ | | 3.8e-12 | -0.62 | 0.0 |
| 512 | $NH^+$ | $e^-$ | N | H | | 2.0e-07 | -0.50 | 0.0 |
| 513 | $NH_2^+$ | $e^-$ | NH | H | | 1.2e-07 | -0.50 | 0.0 |
| 514 | $NH_2^+$ | $e^-$ | N | $H_2$ | | 1.2e-08 | -0.50 | 0.0 |
| 515 | $NH_2^+$ | $e^-$ | N | H | H | 1.7e-07 | -0.50 | 0.0 |
| 516 | $NH_3^+$ | $e^-$ | $NH_2$ | H | | 1.5e-07 | -0.50 | 0.0 |
| 517 | $NH_3^+$ | $e^-$ | NH | H | H | 1.5e-07 | -0.50 | 0.0 |
| 518 | $NH_4^+$ | $e^-$ | $NH_2$ | $H_2$ | | 1.9e-08 | -0.60 | 0.0 |

**Table D5** – *continued* (part 8)

| | Reactants | | Products | | | $\alpha$ | $\beta$ | $\gamma$ |
|---|---|---|---|---|---|---|---|---|
| 519 | $NH_4^+$ | $e^-$ | $NH_2$ | H | H | 1.2e-07 | -0.60 | 0.0 |
| 520 | $NH_4^+$ | $e^-$ | $NH_3$ | H | | 8.0e-07 | -0.60 | 0.0 |
| 521 | $N_2^+$ | $e^-$ | N | N | | 3.6e-08 | -0.42 | 0.0 |
| 522 | $N_2H^+$ | $e^-$ | $N_2$ | H | | 2.8e-07 | -0.74 | 0.0 |
| 523 | $N_2H^+$ | $e^-$ | NH | N | | 2.1e-08 | -0.74 | 0.0 |
| 524 | $ND^+$ | $e^-$ | N | D | | 2.0e-07 | -0.50 | 0.0 |
| 525 | $NHD^+$ | $e^-$ | N | H | D | 1.5e-07 | -0.50 | 0.0 |
| 526 | $NHD^+$ | $e^-$ | NH | D | | 7.5e-08 | -0.50 | 0.0 |
| 527 | $NHD^+$ | $e^-$ | ND | H | | 7.5e-08 | -0.50 | 0.0 |
| 528 | $ND_2^+$ | $e^-$ | ND | D | | 1.5e-07 | -0.50 | 0.0 |
| 529 | $ND_2^+$ | $e^-$ | N | D | D | 1.5e-07 | -0.50 | 0.0 |
| 530 | $NH_2D^+$ | $e^-$ | $NH_2$ | D | | 1.0e-07 | -0.50 | 0.0 |
| 531 | $NH_2D^+$ | $e^-$ | NHD | H | | 2.0e-07 | -0.50 | 0.0 |
| 532 | $NHD_2^+$ | $e^-$ | $ND_2$ | H | | 1.0e-07 | -0.50 | 0.0 |
| 533 | $NHD_2^+$ | $e^-$ | NHD | D | | 2.0e-07 | -0.50 | 0.0 |
| 534 | $ND_3^+$ | $e^-$ | $ND_2$ | D | | 3.0e-07 | -0.50 | 0.0 |
| 535 | $NH_3D^+$ | $e^-$ | $NH_2$ | HD | | 2.5e-07 | -0.50 | 0.0 |
| 536 | $NH_3D^+$ | $e^-$ | NHD | $H_2$ | | 2.5e-07 | -0.50 | 0.0 |
| 537 | $NH_3D^+$ | $e^-$ | $NH_3$ | D | | 1.9e-07 | -0.50 | 0.0 |
| 538 | $NH_3D^+$ | $e^-$ | $NH_2D$ | H | | 5.7e-07 | -0.50 | 0.0 |
| 539 | $NH_2D_2^+$ | $e^-$ | $NH_2$ | $D_2$ | | 8.5e-08 | -0.50 | 0.0 |
| 540 | $NH_2D_2^+$ | $e^-$ | NHD | HD | | 3.4e-07 | -0.50 | 0.0 |
| 541 | $NH_2D_2^+$ | $e^-$ | $ND_2$ | $H_2$ | | 8.5e-08 | -0.50 | 0.0 |
| 542 | $NH_2D_2^+$ | $e^-$ | $NH_2D$ | D | | 3.8e-07 | -0.50 | 0.0 |
| 543 | $NH_2D_2^+$ | $e^-$ | $NHD_2$ | H | | 3.8e-07 | -0.50 | 0.0 |
| 544 | $NHD_3^+$ | $e^-$ | NHD | $D_2$ | | 2.5e-07 | -0.50 | 0.0 |
| 545 | $NHD_3^+$ | $e^-$ | $ND_2$ | HD | | 2.5e-07 | -0.50 | 0.0 |
| 546 | $NHD_3^+$ | $e^-$ | $ND_3$ | H | | 1.9e-07 | -0.50 | 0.0 |
| 547 | $NHD_3^+$ | $e^-$ | $NHD_2$ | D | | 5.7e-07 | -0.50 | 0.0 |
| 548 | $ND_4^+$ | $e^-$ | $ND_2$ | $D_2$ | | 5.1e-07 | -0.50 | 0.0 |
| 549 | $ND_4^+$ | $e^-$ | $ND_3$ | D | | 7.6e-07 | -0.50 | 0.0 |
| 550 | $N_2D^+$ | $e^-$ | $N_2$ | D | | 2.8e-07 | -0.74 | 0.0 |
| 551 | $N_2D^+$ | $e^-$ | ND | N | | 2.1e-08 | -0.74 | 0.0 |
| 552 | $NH^+$ | $\gamma$ | $N^+$ | H | | 2.2e+01 | 0.00 | 0.0 |
| 553 | $N_2$ | $\gamma$ | N | N | | 3.9e+01 | 0.00 | 0.0 |
| 554 | NH | $\gamma$ | N | H | | 3.7e+02 | 0.00 | 0.0 |
| 555 | NH | $\gamma$ | $NH^+$ | $e^-$ | | 7.1e+00 | 0.00 | 0.0 |
| 556 | $NH_2$ | $\gamma$ | NH | H | | 7.2e+02 | 0.00 | 0.0 |
| 557 | $NH_2$ | $\gamma$ | $NH_2^+$ | $e^-$ | | 1.4e+02 | 0.00 | 0.0 |
| 558 | $NH_3$ | $\gamma$ | NH | $H_2$ | | 1.1e+03 | 0.00 | 0.0 |
| 559 | $NH_3$ | $\gamma$ | $NH_3^+$ | $e^-$ | | 2.2e+02 | 0.00 | 0.0 |
| 560 | ND | $\gamma$ | N | D | | 3.7e+02 | 0.00 | 0.0 |
| 561 | ND | $\gamma$ | $ND^+$ | $e^-$ | | 7.1e+00 | 0.00 | 0.0 |
| 562 | NHD | $\gamma$ | NH | D | | 3.6e+02 | 0.00 | 0.0 |
| 563 | NHD | $\gamma$ | ND | H | | 3.6e+02 | 0.00 | 0.0 |
| 564 | NHD | $\gamma$ | $NHD^+$ | $e^-$ | | 1.4e+02 | 0.00 | 0.0 |
| 565 | $ND_2$ | $\gamma$ | ND | D | | 7.2e+02 | 0.00 | 0.0 |
| 566 | $ND_2$ | $\gamma$ | $ND_2^+$ | $e^-$ | | 1.4e+02 | 0.00 | 0.0 |
| 567 | $NH_2D$ | $\gamma$ | NH | HD | | 4.5e+02 | 0.00 | 0.0 |
| 568 | $NH_2D$ | $\gamma$ | ND | $H_2$ | | 2.2e+02 | 0.00 | 0.0 |
| 569 | $NH_2D$ | $\gamma$ | NHD | H | | 2.9e+02 | 0.00 | 0.0 |
| 570 | $NH_2D$ | $\gamma$ | $NH_2$ | D | | 1.4e+02 | 0.00 | 0.0 |
| 571 | $NH_2D$ | $\gamma$ | $NH_2D^+$ | $e^-$ | | 2.2e+02 | 0.00 | 0.0 |
| 572 | $NHD_2$ | $\gamma$ | ND | HD | | 4.5e+02 | 0.00 | 0.0 |
| 573 | $NHD_2$ | $\gamma$ | NH | $D_2$ | | 2.2e+02 | 0.00 | 0.0 |
| 574 | $NHD_2$ | $\gamma$ | NHD | D | | 2.9e+02 | 0.00 | 0.0 |
| 575 | $NHD_2$ | $\gamma$ | $ND_2$ | H | | 1.4e+02 | 0.00 | 0.0 |
| 576 | $NHD_2$ | $\gamma$ | $NHD_2^+$ | $e^-$ | | 2.2e+02 | 0.00 | 0.0 |
| 577 | $ND_3$ | $\gamma$ | ND | $D_2$ | | 6.7e+02 | 0.00 | 0.0 |
| 578 | $ND_3$ | $\gamma$ | $ND_2$ | D | | 4.3e+02 | 0.00 | 0.0 |
| 579 | $ND_3$ | $\gamma$ | $ND_3^+$ | $e^-$ | | 2.2e+02 | 0.00 | 0.0 |
| 580 | $NH_4^+$ | Gr | $Gr^+$ | $NH_3$ | H | 3.8e-07 | 0.50 | 0.0 |
| 581 | $NH_4^+$ | $Gr^-$ | Gr | $NH_3$ | H | 3.8e-07 | 0.50 | 0.0 |
| 582 | $N_2H^+$ | Gr | $Gr^+$ | $N_2$ | H | 1.0e-07 | 0.50 | 0.0 |
| 583 | $N_2H^+$ | Gr | $Gr^+$ | NH | N | 1.9e-07 | 0.50 | 0.0 |
| 584 | $N_2H^+$ | $Gr^-$ | Gr | $N_2$ | H | 1.0e-07 | 0.50 | 0.0 |
| 585 | $N_2H^+$ | $Gr^-$ | Gr | NH | N | 1.9e-07 | 0.50 | 0.0 |
| 586 | $N_2D^+$ | Gr | $Gr^+$ | $N_2$ | D | 3.0e-07 | 0.50 | 0.0 |
| 587 | $N_2D^+$ | $Gr^-$ | Gr | $N_2$ | D | 3.0e-07 | 0.50 | 0.0 |
| 588 | N | $H_2$ | NH | H | | 8.7e-10 | 0.50 | 14600.0 |
| 589 | NH | $H_2$ | $NH_2$ | H | | 5.2e-12 | 0.79 | 6700.0 |
| 590 | NH | H | N | $H_2$ | | 8.7e-10 | 0.50 | 2400.0 |
| 591 | $NH_2$ | $H_2$ | $NH_3$ | H | | 6.2e-11 | 0.50 | 6300.0 |
| 592 | $NH_2$ | H | NH | $H_2$ | | 5.2e-12 | 0.79 | 2200.0 |

**Table D5** – *continued* (part 9)

| | Reactants | | Products | | | $\alpha$ | $\beta$ | $\gamma$ |
|---|---|---|---|---|---|---|---|---|
| 593 | $NH_3$ | H | $NH_2$ | $H_2$ | | 6.2e-11 | 0.50 | 5700.0 |
| 594 | $NH_2^+$ | H | $NH^+$ | $H_2$ | | 1.3e-09 | 0.00 | 24000.0 |
| 595 | $NH_3^+$ | H | $NH_2^+$ | $H_2$ | | 2.2e-10 | 0.00 | 12800.0 |
| 596 | $NH_4^+$ | H | $NH_3^+$ | $H_2$ | | 1.0e-09 | 0.00 | 11000.0 |
| 597 | $NH_4^+$ | $N_2$ | $N_2H^+$ | $NH_3$ | | 2.3e-09 | 0.00 | 44000.0 |
| 598 | $N_2H^+$ | H | $N_2^+$ | $H_2$ | | 2.1e-09 | 0.00 | 30300.0 |
| 599 | $N_2H^+$ | $H_2$ | $H_3^+$ | $N_2$ | | 1.8e-09 | 0.00 | 8300.0 |
| 600 | S | $H^+$ | $S^+$ | H | | 1.0e-15 | 0.00 | 0.0 |
| 601 | S | $H_3^+$ | $SH^+$ | $H_2$ | | 2.6e-09 | 0.00 | 0.0 |
| 602 | SH | H | S | $H_2$ | | 2.5e-11 | 0.00 | 0.0 |
| 603 | SH | $He^+$ | $S^+$ | H | He | 1.7e-09 | 0.00 | 0.0 |
| 604 | SH | $H^+$ | $SH^+$ | H | | 1.6e-09 | 0.00 | 0.0 |
| 605 | SH | $H^+$ | $S^+$ | $H_2$ | | 1.6e-09 | 0.00 | 0.0 |
| 606 | SH | $H_3^+$ | $H_2S^+$ | $H_2$ | | 1.9e-09 | 0.00 | 0.0 |
| 607 | $H_2S$ | $He^+$ | $S^+$ | $H_2$ | He | 3.6e-09 | 0.00 | 0.0 |
| 608 | $H_2S$ | $He^+$ | $SH^+$ | H | He | 4.8e-10 | 0.00 | 0.0 |
| 609 | $H_2S$ | $He^+$ | He | $H_2S^+$ | | 3.1e-10 | 0.00 | 0.0 |
| 610 | $H_2S$ | $H^+$ | H | $H_2S^+$ | | 7.6e-09 | 0.00 | 0.0 |
| 611 | $H_2S$ | $H_3^+$ | $H_3S^+$ | $H_2$ | | 3.7e-09 | 0.00 | 0.0 |
| 612 | $S^+$ | Fe | $Fe^+$ | S | | 1.8e-10 | 0.00 | 0.0 |
| 613 | $SH^+$ | H | $S^+$ | $H_2$ | | 1.1e-10 | 0.00 | 0.0 |
| 614 | $SH^+$ | $H_2$ | $H_3S^+$ | $\gamma$ | | 1.0e-15 | 0.00 | 0.0 |
| 615 | $SH^+$ | S | $S_2^+$ | SH | | 9.7e-10 | 0.00 | 0.0 |
| 616 | $SH^+$ | $H_2S$ | $H_3S^+$ | S | | 5.0e-10 | 0.00 | 0.0 |
| 617 | $SH^+$ | Fe | $Fe^+$ | SH | | 1.6e-09 | 0.00 | 0.0 |
| 618 | $H_2S^+$ | H | $SH^+$ | $H_2$ | | 2.0e-10 | 0.00 | 0.0 |
| 619 | $H_2S^+$ | S | $H_2S$ | $S^+$ | | 1.1e-09 | 0.00 | 0.0 |
| 620 | $H_2S^+$ | SH | $H_2S$ | $SH^+$ | | 5.0e-10 | 0.00 | 0.0 |
| 621 | $H_2S^+$ | Fe | $H_2S$ | $Fe^+$ | | 1.8e-09 | 0.00 | 0.0 |
| 622 | $H_3S^+$ | H | $H_2S^+$ | $H_2$ | | 6.0e-11 | 0.00 | 0.0 |
| 623 | S | $H_2D^+$ | $SH^+$ | HD | | 1.7e-09 | 0.00 | 0.0 |
| 624 | S | $H_2D^+$ | $SD^+$ | $H_2$ | | 8.7e-10 | 0.00 | 0.0 |
| 625 | S | $HD_2^+$ | $SH^+$ | $D_2$ | | 8.7e-10 | 0.00 | 0.0 |
| 626 | S | $HD_2^+$ | $SD^+$ | HD | | 1.7e-09 | 0.00 | 0.0 |
| 627 | S | $D_3^+$ | $SD^+$ | $D_2$ | | 2.6e-09 | 0.00 | 0.0 |
| 628 | SH | $H_2D^+$ | $H_2S^+$ | HD | | 1.9e-09 | 0.00 | 0.0 |
| 629 | SH | $HD_2^+$ | $H_2S^+$ | $D_2$ | | 1.9e-09 | 0.00 | 0.0 |
| 630 | $H_2S$ | $H_2D^+$ | $H_3S^+$ | HD | | 3.7e-09 | 0.00 | 0.0 |
| 631 | $H_2S$ | $HD_2^+$ | $H_3S^+$ | $D_2$ | | 3.7e-09 | 0.00 | 0.0 |
| 632 | $S^+$ | Gr | $Gr^+$ | S | | 2.8e-07 | 0.50 | 0.0 |
| 633 | $S^+$ | $Gr^-$ | Gr | S | | 2.8e-07 | 0.50 | 0.0 |
| 634 | $H_3S^+$ | Gr | $Gr^+$ | $H_2S$ | H | 2.7e-07 | 0.50 | 0.0 |
| 635 | $H_3S^+$ | $Gr^-$ | Gr | $H_2S$ | H | 2.7e-07 | 0.50 | 0.0 |
| 636 | $S^+$ | $e^-$ | S | $\gamma$ | | 3.9e-12 | -0.63 | 0.0 |
| 637 | $SH^+$ | $e^-$ | S | H | | 2.0e-07 | -0.50 | 0.0 |
| 638 | $H_2S^+$ | $e^-$ | SH | H | | 1.5e-07 | -0.50 | 0.0 |
| 639 | $H_2S^+$ | $e^-$ | S | H | H | 1.5e-07 | -0.50 | 0.0 |
| 640 | $H_2S^+$ | $e^-$ | $H_2S$ | $\gamma$ | | 1.1e-10 | -0.70 | 0.0 |
| 641 | $H_3S^+$ | $e^-$ | $H_2S$ | H | | 3.0e-07 | -0.50 | 0.0 |
| 642 | $H_3S^+$ | $e^-$ | SH | $H_2$ | | 1.0e-07 | -0.50 | 0.0 |
| 643 | $SD^+$ | $e^-$ | S | D | | 2.0e-07 | -0.50 | 0.0 |
| 644 | S | $\gamma_2$ | $S^+$ | $e^-$ | | 8.0e+02 | 0.00 | 0.0 |
| 645 | SH | $\gamma_2$ | S | H | | 1.1e+03 | 0.00 | 0.0 |
| 646 | SH | $\gamma_2$ | $SH^+$ | $e^-$ | | 3.4e+01 | 0.00 | 0.0 |
| 647 | $H_2S$ | $\gamma_2$ | SH | H | | 3.4e+03 | 0.00 | 0.0 |
| 648 | $H_2S$ | $\gamma_2$ | $H_2S^+$ | $e^-$ | | 6.2e+02 | 0.00 | 0.0 |
| 649 | $SH^+$ | $\gamma_2$ | $S^+$ | H | | 4.6e+02 | 0.00 | 0.0 |
| 650 | S | $H_2$ | SH | H | | 1.0e-10 | 0.13 | 9620.0 |
| 651 | SH | $H_2$ | $H_2S$ | H | | 6.4e-12 | 0.09 | 8050.0 |
| 652 | $H_2S$ | H | SH | $H_2$ | | 6.6e-11 | 0.00 | 1350.0 |
| 653 | $S^+$ | $H_2$ | $SH^+$ | H | | 2.2e-10 | 0.00 | 9860.0 |
| 654 | $SH^+$ | $H_2$ | $H_2S^+$ | H | | 1.9e-10 | 0.00 | 8500.0 |
| 655 | $SH^+$ | $H_2S$ | $H_2S^+$ | SH | | 5.0e-10 | 0.00 | 1000.0 |
| 656 | $H_2S^+$ | $H_2$ | $H_3S^+$ | H | | 1.4e-11 | 0.00 | 2300.0 |
| 657 | O | CH | $HCO^+$ | $e^-$ | | 2.4e-14 | 0.50 | 0.0 |
| 658 | O | CH | CO | H | | 6.6e-11 | 0.00 | 0.0 |
| 659 | O | $CH_2$ | CO | H | H | 1.0e-10 | 0.00 | 0.0 |
| 660 | O | $CH_2$ | CO | $H_2$ | | 4.0e-11 | 0.00 | 0.0 |
| 661 | O | $CH_3$ | CO | $H_2$ | H | 1.8e-10 | 0.50 | 0.0 |
| 662 | O | $C_2$ | CO | C | | 2.0e-10 | -0.12 | 0.0 |
| 663 | O | $C_2H$ | CO | CH | | 1.0e-10 | 0.00 | 250.0 |
| 664 | O | $C_3$ | CO | $C_2$ | | 5.0e-11 | 0.50 | 0.0 |
| 665 | O | $C_3H$ | $C_2H$ | CO | | 5.0e-11 | 0.50 | 0.0 |
| 666 | O | $C_3H_2$ | $C_2H_2$ | CO | | 5.0e-11 | 0.50 | 0.0 |

**Table D5** – *continued* (part 10)

| | Reactants | Products | | | $\alpha$ | $\beta$ | $\gamma$ |
|---|---|---|---|---|---|---|---|
| 667 | O | NH | NO | H | | 6.6e-11 | 0.00 | 0.0 |
| 668 | O | $NH_2$ | NH | OH | 7.0e-12 | -0.10 | 0.0 |
| 669 | O | CN | CO | N | 5.0e-11 | 0.00 | 0.0 |
| 670 | O | HNC | CO | NH | 2.0e-10 | 0.50 | 200.0 |
| 671 | O | SH | OH | S | 1.7e-11 | 0.67 | 956.0 |
| 672 | O | SH | SO | H | 1.6e-10 | 0.00 | 0.0 |
| 673 | O | CS | CO | S | 2.6e-10 | 0.00 | 760.0 |
| 674 | O | $CH_3^+$ | $HCO^+$ | $H_2$ | 3.1e-10 | 0.00 | 0.0 |
| 675 | O | $CH_3^+$ | $H_3^+$ | CO | 1.3e-11 | 0.00 | 0.0 |
| 676 | O | $CH_5^+$ | $H_3O^+$ | $CH_2$ | 2.2e-10 | 0.00 | 0.0 |
| 677 | O | $HCO_2^+$ | $HCO^+$ | $O_2$ | 1.0e-09 | 0.00 | 0.0 |
| 678 | O | $SH^+$ | $SO^+$ | H | 2.9e-10 | 0.00 | 0.0 |
| 679 | O | $SH^+$ | $S^+$ | OH | 2.9e-10 | 0.00 | 0.0 |
| 680 | O | $H_2S^+$ | $SH^+$ | OH | 3.1e-10 | 0.00 | 0.0 |
| 681 | O | $H_2S^+$ | $SO^+$ | $H_2$ | 3.1e-10 | 0.00 | 0.0 |
| 682 | O | $HCS^+$ | $HCO^+$ | S | 1.0e-09 | 0.00 | 0.0 |
| 683 | C | OH | CO | H | 3.1e-11 | -0.36 | 0.0 |
| 684 | C | $O_2$ | CO | O | 3.3e-11 | 0.50 | 0.0 |
| 685 | C | NH | CN | H | 1.2e-10 | 0.00 | 0.0 |
| 686 | C | $NH_2$ | HCN | H | 3.0e-11 | -0.20 | -6.0 |
| 687 | C | $NH_2$ | HNC | H | 3.0e-11 | -0.20 | -6.0 |
| 688 | C | NO | CN | O | 6.0e-11 | -0.16 | 0.0 |
| 689 | C | NO | CO | N | 9.0e-11 | -0.16 | 0.0 |
| 690 | C | SH | CS | H | 2.0e-11 | 0.00 | 0.0 |
| 691 | C | SO | CO | S | 7.2e-11 | 0.00 | 0.0 |
| 692 | C | SO | CS | O | 1.7e-10 | 0.00 | 0.0 |
| 693 | C | $H_3O^+$ | $HCO^+$ | $H_2$ | 1.0e-11 | 0.00 | 0.0 |
| 694 | C | $HCO^+$ | $CH^+$ | CO | 1.1e-09 | 0.00 | 0.0 |
| 695 | C | $O_2^+$ | $CO^+$ | O | 5.2e-11 | 0.00 | 0.0 |
| 696 | C | $O_2^+$ | $C^+$ | $O_2$ | 5.2e-11 | 0.00 | 0.0 |
| 697 | C | $SH^+$ | $CS^+$ | H | 9.9e-10 | 0.00 | 0.0 |
| 698 | C | $H_2S^+$ | $HCS^+$ | H | 1.0e-09 | 0.00 | 0.0 |
| 699 | C | $H_2DO^+$ | $DCO^+$ | $H_2$ | 1.0e-11 | 0.00 | 0.0 |
| 700 | C | $HD_2O^+$ | $DCO^+$ | HD | 1.0e-11 | 0.00 | 0.0 |
| 701 | C | $D_3O^+$ | $DCO^+$ | $D_2$ | 1.0e-11 | 0.00 | 0.0 |
| 702 | N | CH | CN | H | 1.4e-10 | 0.41 | 0.0 |
| 703 | N | CN | $N_2$ | C | 8.8e-11 | 0.42 | 0.0 |
| 704 | N | NO | $N_2$ | O | 7.3e-11 | 0.44 | 12.7 |
| 705 | N | $CH_2$ | HCN | H | 5.0e-11 | 0.17 | 0.0 |
| 706 | N | $CH_2$ | HNC | H | 3.0e-11 | 0.17 | 0.0 |
| 707 | N | $CH_3$ | HCN | $H_2$ | 1.3e-11 | 0.50 | 0.0 |
| 708 | N | OH | NO | H | 5.0e-11 | 0.00 | 6.0 |
| 709 | N | $O_2^+$ | $NO^+$ | O | 7.8e-11 | 0.00 | 0.0 |
| 710 | N | $CH_2^+$ | $HCN^+$ | H | 9.4e-10 | 0.00 | 0.0 |
| 711 | N | $C_2H^+$ | $C_2N^+$ | H | 8.3e-10 | 0.00 | 0.0 |
| 712 | N | $CH_3^+$ | $HCN^+$ | $H_2$ | 6.7e-11 | 0.00 | 0.0 |
| 713 | N | $CH_3^+$ | $HCNH^+$ | H | 6.7e-11 | 0.00 | 0.0 |
| 714 | N | $C_2H_2^+$ | $CH^+$ | HCN | 2.5e-11 | 0.00 | 0.0 |
| 715 | N | SO | NO | S | 1.7e-11 | 0.50 | 750.0 |
| 716 | S | CH | CS | H | 1.1e-12 | 0.00 | 0.0 |
| 717 | S | OH | SO | H | 1.0e-10 | 0.00 | 100.0 |
| 718 | S | $O_2$ | SO | O | 5.2e-12 | 0.00 | 265.0 |
| 719 | S | $CH^+$ | $S^+$ | CH | 4.7e-10 | 0.00 | 0.0 |
| 720 | S | $CH^+$ | $SH^+$ | C | 4.7e-10 | 0.00 | 0.0 |
| 721 | S | $CH^+$ | $CS^+$ | H | 4.7e-10 | 0.00 | 0.0 |
| 722 | S | $CH_3^+$ | $HCS^+$ | $H_2$ | 1.4e-09 | 0.00 | 0.0 |
| 723 | S | $CH_5^+$ | $SH^+$ | $CH_4$ | 1.3e-09 | 0.00 | 0.0 |
| 724 | S | $HCO^+$ | $SH^+$ | CO | 3.3e-10 | 0.00 | 0.0 |
| 725 | S | $O_2^+$ | $SO^+$ | O | 5.4e-10 | 0.00 | 0.0 |
| 726 | S | $O_2^+$ | $S^+$ | $O_2$ | 5.4e-10 | 0.00 | 0.0 |
| 727 | S | $HNO^+$ | $SH^+$ | NO | 1.1e-09 | 0.00 | 0.0 |
| 728 | S | $N_2H^+$ | $SH^+$ | $N_2$ | 1.1e-09 | 0.00 | 0.0 |
| 729 | CO | $He^+$ | $C^+$ | O | He | 1.5e-09 | 0.00 | 0.0 |
| 730 | SO | $He^+$ | $O^+$ | S | He | 8.3e-10 | 0.00 | 0.0 |
| 731 | SO | $He^+$ | $S^+$ | O | He | 8.3e-10 | 0.00 | 0.0 |
| 732 | NO | $He^+$ | $N^+$ | O | He | 1.4e-09 | 0.00 | 0.0 |
| 733 | NO | $He^+$ | $O^+$ | N | He | 2.2e-10 | 0.00 | 0.0 |
| 734 | CN | $He^+$ | $C^+$ | N | He | 8.8e-10 | 0.00 | 0.0 |
| 735 | CN | $He^+$ | $N^+$ | C | He | 8.8e-10 | 0.00 | 0.0 |
| 736 | CS | $He^+$ | $C^+$ | S | He | 1.3e-09 | 0.00 | 0.0 |
| 737 | CS | $He^+$ | $S^+$ | C | He | 1.3e-09 | 0.00 | 0.0 |
| 738 | $N_2$ | $He^+$ | $N^+$ | N | He | 7.9e-10 | 0.00 | 0.0 |
| 739 | $N_2$ | $He^+$ | $N_2^+$ | He | | 4.1e-10 | 0.00 | 0.0 |
| 740 | HCN | $He^+$ | $CN^+$ | H | He | 1.5e-09 | 0.00 | 0.0 |

**Table D5** – *continued* (part 11)

| | Reactants | | Products | | | $\alpha$ | $\beta$ | $\gamma$ |
|---|---|---|---|---|---|---|---|---|
| 741 | HCN | He$^+$ | CH$^+$ | N | He | 6.2e-10 | 0.00 | 0.0 |
| 742 | HCN | He$^+$ | C$^+$ | NH | He | 7.8e-10 | 0.00 | 0.0 |
| 743 | HCN | He$^+$ | N$^+$ | CH | He | 2.5e-10 | 0.00 | 0.0 |
| 744 | HNC | He$^+$ | CN$^+$ | H | He | 1.6e-09 | 0.00 | 0.0 |
| 745 | HNC | He$^+$ | C$^+$ | NH | He | 1.6e-09 | 0.00 | 0.0 |
| 746 | CO$_2$ | He$^+$ | CO$^+$ | O | He | 7.7e-10 | 0.00 | 0.0 |
| 747 | CO$_2$ | He$^+$ | O$^+$ | CO | He | 1.8e-10 | 0.00 | 0.0 |
| 748 | CO$_2$ | He$^+$ | C$^+$ | O$_2$ | He | 4.0e-11 | 0.00 | 0.0 |
| 749 | SO$_2$ | He$^+$ | S$^+$ | O$_2$ | He | 8.6e-10 | 0.00 | 0.0 |
| 750 | SO$_2$ | He$^+$ | SO$^+$ | O | He | 3.4e-09 | 0.00 | 0.0 |
| 751 | OCS | He$^+$ | CS$^+$ | O | He | 7.6e-10 | 0.00 | 0.0 |
| 752 | OCS | He$^+$ | S$^+$ | CO | He | 7.6e-10 | 0.00 | 0.0 |
| 753 | OCS | He$^+$ | CO$^+$ | S | He | 7.6e-10 | 0.00 | 0.0 |
| 754 | OCS | He$^+$ | O$^+$ | CS | He | 7.6e-11 | 0.00 | 0.0 |
| 755 | CH | SH$^+$ | CH$_2^+$ | S | | 5.8e-10 | 0.00 | 0.0 |
| 756 | H$_2$O | CH$_5^+$ | H$_3$O$^+$ | CH$_4$ | | 3.7e-09 | 0.00 | 0.0 |
| 757 | H$_2$O | C$_2$H$_2^+$ | H$_3$O$^+$ | C$_2$H | | 2.2e-10 | 0.00 | 0.0 |
| 758 | H$_2$O | C$_2$H$_3^+$ | H$_3$O$^+$ | C$_2$H$_2$ | | 1.1e-09 | 0.00 | 0.0 |
| 759 | H$_2$O | C$_3$H$^+$ | HCO$^+$ | C$_2$H$_2$ | | 2.5e-10 | 0.00 | 0.0 |
| 760 | H$_2$O | C$_3$H$^+$ | C$_2$H$_3^+$ | CO | | 2.0e-10 | 0.00 | 0.0 |
| 761 | H$_2$O | NH$_3^+$ | NH$_4^+$ | OH | | 2.5e-10 | 0.00 | 0.0 |
| 762 | H$_2$O | N$_2$H$^+$ | H$_3$O$^+$ | N$_2$ | | 2.6e-09 | 0.00 | 0.0 |
| 763 | H$_2$O | HNO$^+$ | H$_3$O$^+$ | NO | | 2.3e-09 | 0.00 | 0.0 |
| 764 | H$_2$O | SH$^+$ | H$_3$O$^+$ | S | | 6.3e-10 | 0.00 | 0.0 |
| 765 | H$_2$O | H$_2$S$^+$ | H$_3$O$^+$ | SH | | 8.1e-10 | 0.00 | 0.0 |
| 766 | NH$_3$ | SH$^+$ | SH | NH$_3^+$ | | 5.3e-10 | 0.00 | 0.0 |
| 767 | NH$_3$ | SH$^+$ | NH$_4^+$ | S | | 9.8e-10 | 0.00 | 0.0 |
| 768 | NH$_3$ | H$_2$S$^+$ | NH$_4^+$ | SH | | 1.4e-09 | 0.00 | 0.0 |
| 769 | NH$_3$ | H$_2$S$^+$ | H$_2$S | NH$_3^+$ | | 3.4e-10 | 0.00 | 0.0 |
| 770 | NH$_3$ | H$_3$S$^+$ | NH$_4^+$ | H$_2$S | | 1.9e-09 | 0.00 | 0.0 |
| 771 | NH$_3$ | SO$^+$ | NH$_3^+$ | SO | | 1.3e-09 | 0.00 | 0.0 |
| 772 | NH$_3$ | O$_2^+$ | NH$_3^+$ | O$_2$ | | 2.0e-09 | 0.00 | 0.0 |
| 773 | CO | OH | CO$_2$ | H | | 4.4e-13 | -1.15 | 390.0 |
| 774 | CO | H$_2^+$ | HCO$^+$ | H | | 2.2e-09 | 0.00 | 0.0 |
| 775 | CO | H$_2^+$ | CO$^+$ | H$_2$ | | 6.4e-10 | 0.00 | 0.0 |
| 776 | CO | H$_3^+$ | HCO$^+$ | H$_2$ | | 1.7e-09 | 0.00 | 0.0 |
| 777 | CO | CH$_5^+$ | HCO$^+$ | CH$_4$ | | 9.9e-10 | 0.00 | 0.0 |
| 778 | CO | H$_2$D$^+$ | HCO$^+$ | HD | | 1.1e-09 | 0.00 | 0.0 |
| 779 | CO | H$_2$D$^+$ | DCO$^+$ | H$_2$ | | 5.7e-10 | 0.00 | 0.0 |
| 780 | CO | HD$_2^+$ | HCO$^+$ | D$_2$ | | 5.7e-10 | 0.00 | 0.0 |
| 781 | CO | HD$_2^+$ | DCO$^+$ | HD | | 1.1e-09 | 0.00 | 0.0 |
| 782 | CO | D$_3^+$ | DCO$^+$ | D$_2$ | | 1.7e-09 | 0.00 | 0.0 |
| 783 | SO | OH | SO$_2$ | H | | 2.0e-10 | -0.17 | 0.0 |
| 784 | SO | H$^+$ | SO$^+$ | H | | 3.2e-09 | 0.00 | 0.0 |
| 785 | SO | H$_3^+$ | HSO$^+$ | H$_2$ | | 1.9e-09 | 0.00 | 0.0 |
| 786 | SO | CH$^+$ | OH$^+$ | CS | | 1.0e-09 | 0.00 | 0.0 |
| 787 | SO | CH$^+$ | SH$^+$ | CO | | 1.0e-09 | 0.00 | 0.0 |
| 788 | SO | CH$_3^+$ | HOCS$^+$ | H$_2$ | | 9.5e-10 | 0.00 | 0.0 |
| 789 | SO | HCO$^+$ | HSO$^+$ | CO | | 7.5e-10 | 0.00 | 0.0 |
| 790 | NO | CH | HCN | O | | 1.2e-11 | -0.13 | 0.0 |
| 791 | NO | H$^+$ | NO$^+$ | H | | 1.9e-09 | 0.00 | 0.0 |
| 792 | NO | H$_3^+$ | HNO$^+$ | H$_2$ | | 1.1e-09 | 0.00 | 0.0 |
| 793 | NO | HCO$_2^+$ | HNO$^+$ | CO$_2$ | | 1.0e-10 | 0.00 | 0.0 |
| 794 | NO | O$_2^+$ | NO$^+$ | O$_2$ | | 4.4e-10 | 0.00 | 0.0 |
| 795 | NO | SH$^+$ | NO$^+$ | SH | | 3.3e-10 | 0.00 | 0.0 |
| 796 | NO | H$_2$S$^+$ | NO$^+$ | H$_2$S | | 3.7e-10 | 0.00 | 0.0 |
| 797 | NO | H$_2$D$^+$ | HNO$^+$ | HD | | 7.3e-10 | 0.00 | 0.0 |
| 798 | NO | HD$_2^+$ | HNO$^+$ | D$_2$ | | 3.7e-10 | 0.00 | 0.0 |
| 799 | CN | NH$_3$ | NH$_2$ | HCN | | 2.8e-11 | -0.85 | 0.0 |
| 800 | CN | H$_3^+$ | HCN$^+$ | H$_2$ | | 1.0e-09 | 0.00 | 0.0 |
| 801 | CN | H$_3^+$ | HCNH$^+$ | H | | 1.0e-09 | 0.00 | 0.0 |
| 802 | CN | H$_3$O$^+$ | HCNH$^+$ | OH | | 4.5e-09 | 0.00 | 0.0 |
| 803 | CN | H$_2$D$^+$ | HCNH$^+$ | D | | 1.0e-09 | 0.00 | 0.0 |
| 804 | CN | H$_2$D$^+$ | HCN$^+$ | HD | | 1.0e-09 | 0.00 | 0.0 |
| 805 | CS | OH | OCS | H | | 1.7e-10 | 0.00 | 0.0 |
| 806 | CS | OH | CO | SH | | 3.0e-11 | 0.00 | 0.0 |
| 807 | CS | H$^+$ | CS$^+$ | H | | 4.9e-09 | 0.00 | 0.0 |
| 808 | CS | H$_3^+$ | HCS$^+$ | H$_2$ | | 2.9e-09 | 0.00 | 0.0 |
| 809 | CS | H$_2$D$^+$ | HCS$^+$ | HD | | 2.9e-09 | 0.00 | 0.0 |
| 810 | CS | HD$_2^+$ | HCS$^+$ | D$_2$ | | 2.9e-09 | 0.00 | 0.0 |
| 811 | SO | H$_2$D$^+$ | HSO$^+$ | HD | | 1.9e-09 | 0.00 | 0.0 |
| 812 | SO | HD$_2^+$ | HSO$^+$ | D$_2$ | | 1.9e-09 | 0.00 | 0.0 |
| 813 | HCN | H$^+$ | HCN$^+$ | H | | 1.1e-08 | 0.00 | 0.0 |
| 814 | HCN | H$_3^+$ | HCNH$^+$ | H$_2$ | | 9.5e-09 | 0.00 | 0.0 |

**Table D5** – *continued* (part 12)

| | Reactants | | Products | | $\alpha$ | $\beta$ | $\gamma$ |
|---|---|---|---|---|---|---|---|
| 815 | HCN | $H_3O^+$ | $HCNH^+$ | $H_2O$ | 4.5e-09 | 0.00 | 0.0 |
| 816 | HCN | $H_3S^+$ | $HCNH^+$ | $H_2S$ | 1.9e-09 | 0.00 | 0.0 |
| 817 | HCN | $HCO^+$ | $HCNH^+$ | CO | 3.7e-09 | 0.00 | 0.0 |
| 818 | HCN | $H_2D^+$ | $HCNH^+$ | HD | 9.5e-09 | 0.00 | 0.0 |
| 819 | HNC | H | HCN | H | 1.0e-15 | 0.00 | 0.0 |
| 820 | HNC | $H^+$ | $H^+$ | HCN | 1.0e-09 | 0.00 | 0.0 |
| 821 | HNC | $H_3^+$ | $HCNH^+$ | $H_2$ | 9.5e-09 | 0.00 | 0.0 |
| 822 | HNC | $H_2D^+$ | $HCNH^+$ | HD | 9.5e-09 | 0.00 | 0.0 |
| 823 | HNC | $H_3O^+$ | $HCNH^+$ | $H_2O$ | 4.5e-09 | 0.00 | 0.0 |
| 824 | HNC | $HCO^+$ | $HCNH^+$ | CO | 3.7e-09 | 0.00 | 0.0 |
| 825 | HNC | $CH_5^+$ | $C_2H_3^+$ | $NH_3$ | 1.0e-09 | 0.00 | 0.0 |
| 826 | $CO_2$ | $H^+$ | $HCO^+$ | O | 4.2e-09 | 0.00 | 0.0 |
| 827 | $CO_2$ | $H_3^+$ | $HCO_2^+$ | $H_2$ | 2.0e-09 | 0.00 | 0.0 |
| 828 | $CO_2$ | $H_2D^+$ | $HCO_2^+$ | HD | 1.3e-09 | 0.00 | 0.0 |
| 829 | $CO_2$ | $H_2D^+$ | $DCO_2^+$ | $H_2$ | 6.7e-10 | 0.00 | 0.0 |
| 830 | $CO_2$ | $HD_2^+$ | $HCO_2^+$ | $D_2$ | 6.7e-10 | 0.00 | 0.0 |
| 831 | $CO_2$ | $HD_2^+$ | $DCO_2^+$ | HD | 1.3e-09 | 0.00 | 0.0 |
| 832 | $CO_2$ | $D_3^+$ | $DCO_2^+$ | $D_2$ | 2.0e-09 | 0.00 | 0.0 |
| 833 | $CO_2$ | $N_2D^+$ | $DCO_2^+$ | $N_2$ | 1.4e-09 | 0.00 | 0.0 |
| 834 | $SO_2$ | $H_3^+$ | $HSO_2^+$ | $H_2$ | 1.3e-09 | 0.00 | 0.0 |
| 835 | $SO_2$ | $H_2D^+$ | $HSO_2^+$ | HD | 1.3e-09 | 0.00 | 0.0 |
| 836 | $SO_2$ | $HD_2^+$ | $HSO_2^+$ | $D_2$ | 1.3e-09 | 0.00 | 0.0 |
| 837 | OCS | $H^+$ | $SH^+$ | CO | 5.9e-09 | 0.00 | 0.0 |
| 838 | OCS | $H_3^+$ | $HOCS^+$ | $H_2$ | 1.9e-09 | 0.00 | 0.0 |
| 839 | OCS | $HCO^+$ | $HOCS^+$ | CO | 1.1e-09 | 0.00 | 0.0 |
| 840 | $H_2S$ | $H_3O^+$ | $H_3S^+$ | $H_2O$ | 1.9e-09 | 0.00 | 0.0 |
| 841 | SH | $HCO^+$ | $H_2S^+$ | CO | 8.2e-10 | 0.00 | 0.0 |
| 842 | CS | $HCO^+$ | $HCS^+$ | CO | 1.2e-09 | 0.00 | 0.0 |
| 843 | $H_2S$ | $HCO^+$ | $H_3S^+$ | CO | 1.6e-09 | 0.00 | 0.0 |
| 844 | $H_2S$ | $O_2^+$ | $H_2S^+$ | $O_2$ | 1.4e-09 | 0.00 | 0.0 |
| 845 | $H_2S$ | $NH_3^+$ | $NH_4^+$ | SH | 6.0e-10 | 0.00 | 0.0 |
| 846 | OCS | $H_2D^+$ | $HOCS^+$ | HD | 1.9e-09 | 0.00 | 0.0 |
| 847 | OCS | $HD_2^+$ | $HOCS^+$ | $D_2$ | 1.9e-09 | 0.00 | 0.0 |
| 848 | S | $N_2D^+$ | $SD^+$ | $N_2$ | 1.1e-09 | 0.00 | 0.0 |
| 849 | $H_2O$ | $N_2D^+$ | $H_2DO^+$ | $N_2$ | 2.6e-09 | 0.00 | 0.0 |
| 850 | CO | $N_2D^+$ | $DCO^+$ | $N_2$ | 8.8e-10 | 0.00 | 0.0 |
| 851 | $C^+$ | S | $S^+$ | C | 5.5e-12 | 0.86 | 681.0 |
| 852 | $C^+$ | OH | $CO^+$ | H | 8.0e-10 | 0.00 | 0.0 |
| 853 | $C^+$ | OH | $H^+$ | CO | 8.0e-10 | 0.00 | 0.0 |
| 854 | $C^+$ | $H_2O$ | $HCO^+$ | H | 2.4e-09 | -0.63 | 0.0 |
| 855 | $C^+$ | $O_2$ | $O^+$ | CO | 5.1e-10 | 0.00 | 0.0 |
| 856 | $C^+$ | $O_2$ | $CO^+$ | O | 3.1e-10 | 0.00 | 0.0 |
| 857 | $C^+$ | $CO_2$ | $CO^+$ | CO | 1.1e-09 | 0.00 | 0.0 |
| 858 | $C^+$ | NH | $CN^+$ | H | 7.8e-10 | 0.00 | 0.0 |
| 859 | $C^+$ | $NH_2$ | $HCN^+$ | H | 1.1e-09 | 0.00 | 0.0 |
| 860 | $C^+$ | $NH_3$ | $NH_3^+$ | C | 5.3e-10 | 0.00 | 0.0 |
| 861 | $C^+$ | $NH_3$ | $H_2NC^+$ | H | 7.8e-10 | 0.00 | 0.0 |
| 862 | $C^+$ | $NH_3$ | $HCNH^+$ | H | 7.8e-10 | 0.00 | 0.0 |
| 863 | $C^+$ | $NH_3$ | $HCN^+$ | $H_2$ | 2.1e-10 | 0.00 | 0.0 |
| 864 | $C^+$ | HCN | $C_2N^+$ | H | 3.4e-09 | 0.00 | 0.0 |
| 865 | $C^+$ | HNC | $C_2N^+$ | H | 3.4e-09 | 0.00 | 0.0 |
| 866 | $C^+$ | NO | $NO^+$ | C | 3.4e-09 | 0.00 | 0.0 |
| 867 | $C^+$ | NO | $N^+$ | CO | 9.0e-10 | 0.00 | 0.0 |
| 868 | $C^+$ | SH | $CS^+$ | H | 1.1e-09 | 0.00 | 0.0 |
| 869 | $C^+$ | $H_2S$ | $HCS^+$ | H | 1.3e-09 | 0.00 | 0.0 |
| 870 | $C^+$ | $H_2S$ | $H_2S^+$ | C | 4.2e-10 | 0.00 | 0.0 |
| 871 | $C^+$ | SO | $S^+$ | CO | 2.6e-10 | 0.00 | 0.0 |
| 872 | $C^+$ | SO | $CS^+$ | O | 2.6e-10 | 0.00 | 0.0 |
| 873 | $C^+$ | SO | $SO^+$ | C | 2.6e-10 | 0.00 | 0.0 |
| 874 | $C^+$ | SO | $CO^+$ | S | 2.6e-10 | 0.00 | 0.0 |
| 875 | $C^+$ | $SO_2$ | $SO^+$ | CO | 2.3e-09 | 0.00 | 0.0 |
| 876 | $C^+$ | CS | $CS^+$ | C | 1.6e-09 | 0.00 | 700.0 |
| 877 | $C^+$ | OCS | $CS^+$ | CO | 1.6e-09 | 0.00 | 0.0 |
| 878 | $C^+$ | OD | $CO^+$ | D | 8.0e-10 | 0.00 | 0.0 |
| 879 | $C^+$ | OD | $D^+$ | CO | 8.0e-10 | 0.00 | 0.0 |
| 880 | $C^+$ | HDO | $DCO^+$ | H | 1.2e-09 | -0.63 | 0.0 |
| 881 | $C^+$ | HDO | $HCO^+$ | D | 1.2e-09 | -0.63 | 0.0 |
| 882 | $C^+$ | $D_2O$ | $DCO^+$ | D | 2.4e-09 | -0.63 | 0.0 |
| 883 | $N^+$ | $O_2$ | $O_2^+$ | N | 2.8e-10 | 0.00 | 0.0 |
| 884 | $N^+$ | $O_2$ | $NO^+$ | O | 2.4e-10 | 0.00 | 0.0 |
| 885 | $N^+$ | $O_2$ | $O^+$ | NO | 3.3e-11 | 0.00 | 0.0 |
| 886 | $N^+$ | CO | $CO^+$ | N | 8.3e-10 | 0.00 | 0.0 |
| 887 | $N^+$ | CO | $NO^+$ | C | 1.5e-10 | 0.00 | 0.0 |
| 888 | $N^+$ | NO | $NO^+$ | N | 4.5e-10 | 0.00 | 0.0 |

**Table D5** – *continued* (part 13)

| | Reactants | | Products | | | $\alpha$ | $\beta$ | $\gamma$ |
|---|---|---|---|---|---|---|---|---|
| 889 | N$^+$ | NO | N$_2^+$ | O | | 7.9e-11 | 0.00 | 0.0 |
| 890 | S$^+$ | CH | CS$^+$ | H | | 6.2e-10 | 0.00 | 0.0 |
| 891 | S$^+$ | CH$_2$ | HCS$^+$ | H | | 1.0e-11 | 0.00 | 0.0 |
| 892 | S$^+$ | OH | SO$^+$ | H | | 6.1e-10 | 0.00 | 0.0 |
| 893 | S$^+$ | NO | NO$^+$ | S | | 3.2e-10 | 0.00 | 0.0 |
| 894 | S$^+$ | NH$_3$ | NH$_3^+$ | S | | 1.6e-09 | 0.00 | 0.0 |
| 895 | S$^+$ | O$_2$ | SO$^+$ | O | | 2.3e-11 | 0.00 | 0.0 |
| 896 | S$^+$ | NH$_2$D | NH$_2$D$^+$ | S | | 1.6e-09 | 0.00 | 0.0 |
| 897 | S$^+$ | NHD$_2$ | NHD$_2^+$ | S | | 1.6e-09 | 0.00 | 0.0 |
| 898 | S$^+$ | ND$_3$ | ND$_3^+$ | S | | 1.6e-09 | 0.00 | 0.0 |
| 899 | CO$^+$ | H$_2$ | HCO$^+$ | H | | 1.3e-09 | 0.00 | 0.0 |
| 900 | CO$^+$ | H | H$^+$ | CO | | 7.5e-10 | 0.00 | 0.0 |
| 901 | HCO$^+$ | CH | CH$_2^+$ | CO | | 6.3e-10 | 0.00 | 0.0 |
| 902 | HCO$^+$ | CH$_2$ | CH$_3^+$ | CO | | 8.6e-10 | 0.00 | 0.0 |
| 903 | HCO$^+$ | H$_2$O | H$_3$O$^+$ | CO | | 2.5e-09 | 0.00 | 0.0 |
| 904 | HCO$^+$ | OH | HCO$_2^+$ | H | | 1.0e-09 | 0.00 | 0.0 |
| 905 | HCO$^+$ | C$_2$H | C$_2$H$_2^+$ | CO | | 7.8e-10 | 0.00 | 0.0 |
| 906 | HCO$^+$ | C$_2$H$_2$ | C$_2$H$_3^+$ | CO | | 1.4e-09 | 0.00 | 0.0 |
| 907 | HCO$^+$ | C$_3$H | C$_3$H$_2^+$ | CO | | 1.4e-09 | 0.00 | 0.0 |
| 908 | HCO$^+$ | C$_3$H$_2$ | C$_3$H$_3^+$ | CO | | 1.4e-09 | 0.00 | 0.0 |
| 909 | HCO$^+$ | NH | NH$_2^+$ | CO | | 6.4e-10 | 0.00 | 0.0 |
| 910 | HCO$^+$ | NH$_2$ | NH$_3^+$ | CO | | 8.9e-10 | 0.00 | 0.0 |
| 911 | HCO$^+$ | NH$_3$ | NH$_4^+$ | CO | | 1.9e-09 | 0.00 | 0.0 |
| 912 | HCO$^+$ | Fe | Fe$^+$ | CO | H | 1.9e-09 | 0.00 | 0.0 |
| 913 | HCO$_2^+$ | CO | HCO$^+$ | CO$_2$ | | 1.0e-09 | 0.00 | 0.0 |
| 914 | HCO$_2^+$ | CH$_4$ | CH$_5^+$ | CO$_2$ | | 7.8e-10 | 0.00 | 0.0 |
| 915 | H$_3$O$^+$ | CH | CH$_2^+$ | H$_2$O | | 6.8e-10 | 0.00 | 0.0 |
| 916 | H$_3$O$^+$ | CH$_2$ | CH$_3^+$ | H$_2$O | | 9.4e-10 | 0.00 | 0.0 |
| 917 | H$_3$O$^+$ | C$_3$H | C$_3$H$_2^+$ | H$_2$O | | 2.0e-09 | 0.00 | 0.0 |
| 918 | H$_3$O$^+$ | C$_3$H$_2$ | C$_3$H$_3^+$ | H$_2$O | | 3.0e-09 | 0.00 | 0.0 |
| 919 | H$_3$O$^+$ | NH$_3$ | NH$_4^+$ | H$_2$O | | 2.2e-09 | 0.00 | 0.0 |
| 920 | CN$^+$ | H$_2$ | HCN$^+$ | H | | 1.0e-09 | 0.00 | 0.0 |
| 921 | HCN$^+$ | H$_2$ | HCNH$^+$ | H | | 9.8e-10 | 0.00 | 0.0 |
| 922 | HCNH$^+$ | CH | CH$_2^+$ | HCN | | 3.1e-10 | 0.00 | 0.0 |
| 923 | HCNH$^+$ | CH | CH$_2^+$ | HNC | | 3.1e-10 | 0.00 | 0.0 |
| 924 | HCNH$^+$ | CH$_2$ | CH$_3^+$ | HCN | | 4.3e-10 | 0.00 | 0.0 |
| 925 | HCNH$^+$ | CH$_2$ | CH$_3^+$ | HNC | | 4.3e-10 | 0.00 | 0.0 |
| 926 | HCNH$^+$ | NH$_2$ | NH$_3^+$ | HCN | | 4.5e-10 | 0.00 | 0.0 |
| 927 | HCNH$^+$ | NH$_2$ | NH$_3^+$ | HNC | | 4.5e-10 | 0.00 | 0.0 |
| 928 | HCNH$^+$ | NH$_3$ | NH$_4^+$ | HCN | | 1.1e-09 | 0.00 | 0.0 |
| 929 | HCNH$^+$ | NH$_3$ | NH$_4^+$ | HNC | | 1.1e-09 | 0.00 | 0.0 |
| 930 | HCNH$^+$ | H$_2$S | H$_3$S$^+$ | HCN | | 1.7e-10 | 0.00 | 0.0 |
| 931 | HCNH$^+$ | H$_2$S | H$_3$S$^+$ | HNC | | 1.7e-10 | 0.00 | 0.0 |
| 932 | N$_2$H$^+$ | CO | HCO$^+$ | N$_2$ | | 8.8e-10 | 0.00 | 0.0 |
| 933 | N$_2$H$^+$ | CO$_2$ | HCO$_2^+$ | N$_2$ | | 1.4e-09 | 0.00 | 0.0 |
| 934 | N$_2$H$^+$ | NO | HNO$^+$ | N$_2$ | | 3.4e-10 | 0.00 | 0.0 |
| 935 | C$_2$N$^+$ | NH$_3$ | N$_2$H$^+$ | C$_2$H$_2$ | | 1.9e-09 | 0.00 | 0.0 |
| 936 | NO$^+$ | Fe | Fe$^+$ | NO | | 1.0e-09 | 0.00 | 0.0 |
| 937 | HNO$^+$ | C | CH$^+$ | NO | | 1.0e-09 | 0.00 | 0.0 |
| 938 | HNO$^+$ | CO | HCO$^+$ | NO | | 1.0e-10 | 0.00 | 0.0 |
| 939 | HNO$^+$ | CO$_2$ | HCO$_2^+$ | NO | | 1.0e-10 | 0.00 | 0.0 |
| 940 | HNO$^+$ | OH | H$_2$O$^+$ | NO | | 6.2e-10 | 0.00 | 0.0 |
| 941 | SO$^+$ | Fe | Fe$^+$ | SO | | 1.6e-09 | 0.00 | 0.0 |
| 942 | CS$^+$ | H$_2$ | HCS$^+$ | H | | 4.8e-10 | 0.00 | 0.0 |
| 943 | HCO$^+$ | Gr | Gr$^+$ | CO | H | 3.0e-07 | 0.50 | 0.0 |
| 944 | HCS$^+$ | Gr | Gr$^+$ | CS | H | 2.4e-07 | 0.50 | 0.0 |
| 945 | HCO$^+$ | Gr$^-$ | Gr | CO | H | 3.0e-07 | 0.50 | 0.0 |
| 946 | HCS$^+$ | Gr$^-$ | Gr | CS | H | 2.4e-07 | 0.50 | 0.0 |
| 947 | CO$^+$ | e$^-$ | C | O | | 1.0e-07 | -0.46 | 0.0 |
| 948 | HCO$^+$ | e$^-$ | CO | H | | 2.4e-07 | -0.69 | 0.0 |
| 949 | HCO$_2^+$ | e$^-$ | CO$_2$ | H | | 2.2e-07 | -0.50 | 0.0 |
| 950 | HCO$_2^+$ | e$^-$ | CO | OH | | 1.2e-07 | -0.50 | 0.0 |
| 951 | CN$^+$ | e$^-$ | C | N | | 1.8e-07 | -0.50 | 0.0 |
| 952 | C$_2$N$^+$ | e$^-$ | C$_2$ | N | | 1.0e-07 | -0.50 | 0.0 |
| 953 | C$_2$N$^+$ | e$^-$ | CN | C | | 2.0e-07 | -0.50 | 0.0 |
| 954 | HCN$^+$ | e$^-$ | CN | H | | 1.5e-07 | -0.50 | 0.0 |
| 955 | HCN$^+$ | e$^-$ | CH | N | | 1.5e-07 | -0.50 | 0.0 |
| 956 | HCNH$^+$ | e$^-$ | HCN | H | | 9.6e-08 | -0.65 | 0.0 |
| 957 | HCNH$^+$ | e$^-$ | HNC | H | | 9.6e-08 | -0.65 | 0.0 |
| 958 | HCNH$^+$ | e$^-$ | CN | H | H | 9.1e-08 | -0.65 | 0.0 |
| 959 | H$_2$NC$^+$ | e$^-$ | HNC | H | | 1.8e-07 | -0.50 | 0.0 |
| 960 | H$_2$NC$^+$ | e$^-$ | CN | H | H | 1.8e-08 | -0.50 | 0.0 |
| 961 | NO$^+$ | e$^-$ | N | O | | 4.3e-07 | -0.37 | 0.0 |
| 962 | HNO$^+$ | e$^-$ | NO | H | | 3.0e-07 | -0.50 | 0.0 |

**Table D5** – *continued* (part 14)

| | Reactants | | Products | | | $\alpha$ | $\beta$ | $\gamma$ |
|---|---|---|---|---|---|---|---|---|
| 963 | $CS^+$ | $e^-$ | C | S | | 2.0e-07 | -0.50 | 0.0 |
| 964 | $HCS^+$ | $e^-$ | CS | H | | 7.0e-07 | -0.50 | 0.0 |
| 965 | $SO^+$ | $e^-$ | S | O | | 2.0e-07 | -0.50 | 0.0 |
| 966 | $HSO^+$ | $e^-$ | SO | H | | 2.0e-07 | -0.50 | 0.0 |
| 967 | $HSO_2^+$ | $e^-$ | SO | H | O | 1.0e-07 | -0.50 | 0.0 |
| 968 | $HSO_2^+$ | $e^-$ | SO | OH | | 1.0e-07 | -0.50 | 0.0 |
| 969 | $HOCS^+$ | $e^-$ | OH | CS | | 2.0e-07 | -0.50 | 0.0 |
| 970 | $HOCS^+$ | $e^-$ | OCS | H | | 2.0e-07 | -0.50 | 0.0 |
| 971 | $Fe^+$ | $e^-$ | Fe | $\gamma$ | | 3.7e-12 | -0.65 | 0.0 |
| 972 | $DCO^+$ | $e^-$ | CO | D | | 2.4e-07 | -0.69 | 0.0 |
| 973 | $DCO_2^+$ | $e^-$ | $CO_2$ | D | | 2.2e-07 | -0.50 | 0.0 |
| 974 | $DCO_2^+$ | $e^-$ | CO | OD | | 1.2e-07 | -0.50 | 0.0 |
| 975 | $CO_2$ | $\eta$ | CO | O | | 6.0e+02 | 0.00 | 0.0 |
| 976 | CO | $\eta$ | C | O | | 4.6e+01 | 0.00 | 0.0 |
| 977 | CO | $\eta$ | $CO^+$ | $e^-$ | | 1.4e+01 | 0.00 | 0.0 |
| 978 | $H_2CO$ | $\eta$ | CO | $H_2$ | | 1.3e+03 | 0.00 | 0.0 |
| 979 | $CH_3OH$ | $\eta$ | $CH_3$ | OH | | 1.6e+03 | 0.00 | 0.0 |
| 980 | $HCO^+$ | $\eta$ | $CO^+$ | H | | 3.3e+00 | 0.00 | 0.0 |
| 981 | $CO^+$ | $\eta$ | $C^+$ | O | | 7.7e+01 | 0.00 | 0.0 |
| 982 | CN | $\eta$ | C | N | | 4.5e+02 | 0.00 | 0.0 |
| 983 | CN | $\eta$ | $CN^+$ | $e^-$ | | 8.3e+00 | 0.00 | 0.0 |
| 984 | HCN | $\eta$ | CN | H | | 2.0e+03 | 0.00 | 0.0 |
| 985 | HCN | $\eta$ | $HCN^+$ | $e^-$ | | 1.4e+00 | 0.00 | 0.0 |
| 986 | HNC | $\eta$ | CN | H | | 2.0e+03 | 0.00 | 0.0 |
| 987 | NO | $\eta$ | N | O | | 3.0e+02 | 0.00 | 0.0 |
| 988 | NO | $\eta$ | $NO^+$ | $e^-$ | | 2.4e+02 | 0.00 | 0.0 |
| 989 | SO | $\eta$ | S | O | | 5.5e+03 | 0.00 | 0.0 |
| 990 | SO | $\eta$ | $SO^+$ | $e^-$ | | 4.5e+02 | 0.00 | 0.0 |
| 991 | CS | $\eta$ | S | C | | 1.9e+03 | 0.00 | 0.0 |
| 992 | CS | $\eta$ | $CS^+$ | $e^-$ | | 2.0e+01 | 0.00 | 0.0 |
| 993 | OCS | $\eta$ | CO | S | | 5.2e+03 | 0.00 | 0.0 |
| 994 | $SO_2$ | $\eta$ | SO | O | | 2.7e+03 | 0.00 | 0.0 |
| 995 | O | $NH_3$ | $NH_2$ | OH | | 2.5e-12 | 0.00 | 3020.0 |
| 996 | O | NO | $O_2$ | N | | 7.5e-13 | 1.00 | 16000.0 |
| 997 | O | $N_2H^+$ | $OH^+$ | $N_2$ | | 1.4e-10 | 0.00 | 1400.0 |
| 998 | O | $H_2S$ | SH | OH | | 1.4e-11 | 0.00 | 1920.0 |
| 999 | O | SO | S | $O_2$ | | 6.6e-13 | 0.00 | 2760.0 |
| 1000 | O | $SO_2$ | SO | $O_2$ | | 9.3e-11 | -0.46 | 9140.0 |
| 1001 | O | OCS | SO | CO | | 2.6e-11 | 0.00 | 2250.0 |
| 1002 | C | SH | CH | S | | 1.2e-11 | 0.58 | 5880.0 |
| 1003 | N | $O_2$ | NO | O | | 3.3e-12 | 1.00 | 3150.0 |
| 1004 | N | $C_2H_2^+$ | $CH^+$ | HNC | | 2.5e-11 | 0.00 | 2600.0 |
| 1005 | S | CH | SH | C | | 1.7e-11 | 0.50 | 4000.0 |
| 1006 | S | $H_3O^+$ | $SH^+$ | $H_2O$ | | 3.2e-10 | 0.00 | 4930.0 |
| 1007 | OH | $NH_3$ | $NH_2$ | $H_2O$ | | 3.5e-12 | 0.00 | 925.0 |
| 1008 | OH | $NH_4^+$ | $NH_3^+$ | $H_2O$ | | 2.5e-10 | 0.00 | 3400.0 |
| 1009 | OH | $H_2S$ | SH | $H_2O$ | | 6.1e-12 | 0.00 | 80.0 |
| 1010 | OH | $S^+$ | $SH^+$ | O | | 2.9e-10 | 0.00 | 8820.0 |
| 1011 | OH | $SH^+$ | $H_2S^+$ | O | | 3.1e-10 | 0.00 | 7500.0 |
| 1012 | OH | $SH^+$ | $H_2O^+$ | S | | 4.3e-10 | 0.00 | 9200.0 |
| 1013 | CO | H | OH | C | | 1.1e-10 | 0.50 | 77700.0 |
| 1014 | CO | $NH_2^+$ | $HCO^+$ | NH | | 6.4e-10 | 0.00 | 6100.0 |
| 1015 | CO | $N^+$ | $C^+$ | NO | | 9.0e-10 | 0.00 | 15400.0 |
| 1016 | CO | SH | OCS | H | | 5.9e-14 | 1.12 | 8330.0 |
| 1017 | CN | $H_2$ | HCN | H | | 5.7e-13 | 2.45 | 1130.0 |
| 1018 | CN | $H^+$ | $CN^+$ | H | | 2.1e-09 | 0.00 | 6150.0 |
| 1019 | HCN | $CH_5^+$ | $C_2H_3^+$ | $NH_3$ | | 1.0e-09 | 0.00 | 5120.0 |
| 1020 | HCN | $H^+$ | $H^+$ | HNC | | 1.0e-09 | 0.00 | 7850.0 |
| 1021 | HNC | $O_2$ | $CO_2$ | NH | | 2.0e-11 | 0.50 | 2000.0 |
| 1022 | SO | H | OH | S | | 5.9e-10 | -0.31 | 11100.0 |
| 1023 | SO | $O_2$ | $SO_2$ | O | | 1.4e-12 | 0.00 | 2820.0 |
| 1024 | $SO_2$ | H | SO | OH | | 9.3e-09 | -0.74 | 14700.0 |
| 1025 | OCS | H | SH | CO | | 1.7e-11 | 0.00 | 2000.0 |
| 1026 | $HCO^+$ | H | $CO^+$ | $H_2$ | | 1.3e-09 | 0.00 | 24500.0 |
| 1027 | $HCO^+$ | $CH_3$ | $CH_4^+$ | CO | | 1.4e-09 | 0.00 | 9060.0 |
| 1028 | $HCO^+$ | $CH_4$ | $CH_5^+$ | CO | | 9.9e-10 | 0.00 | 4920.0 |
| 1029 | $HCO^+$ | $O_2$ | $HCO_2^+$ | O | | 1.0e-09 | 0.00 | 1450.0 |
| 1030 | $HCO^+$ | $CO_2$ | $HCO_2^+$ | CO | | 1.0e-09 | 0.00 | 5000.0 |
| 1031 | $HCO^+$ | $N_2$ | $N_2H^+$ | CO | | 8.8e-10 | 0.00 | 11200.0 |
| 1032 | $HCO_2^+$ | H | $HCO^+$ | OH | | 1.0e-09 | 0.00 | 7500.0 |
| 1033 | $HCO_2^+$ | $N_2$ | $N_2H^+$ | $CO_2$ | | 1.4e-09 | 0.00 | 6400.0 |
| 1034 | $H_3O^+$ | $C_2H$ | $C_2H_2^+$ | $H_2O$ | | 2.2e-09 | 0.00 | 4100.0 |
| 1035 | $H_3O^+$ | $C_2H_2$ | $C_2H_3^+$ | $H_2O$ | | 1.0e-09 | 0.00 | 7330.0 |
| 1036 | $HCN^+$ | H | $CN^+$ | $H_2$ | | 1.0e-09 | 0.00 | 15800.0 |

**Table D5** – *continued* (part 15)

| | Reactants | | Products | | | $\alpha$ | $\beta$ | $\gamma$ |
|---|---|---|---|---|---|---|---|---|
| 1037 | HCNH$^+$ | H | HCN$^+$ | H$_2$ | | 9.8e-10 | 0.00 | 34400.0 |
| 1038 | HCNH$^+$ | H$_2$O | H$_3$O$^+$ | HCN | | 4.5e-09 | 0.00 | 2460.0 |
| 1039 | HCNH$^+$ | H$_2$O | H$_3$O$^+$ | HNC | | 4.5e-09 | 0.00 | 10300.0 |
| 1040 | SO$^+$ | H | S$^+$ | OH | | 6.1e-10 | 0.00 | 11380.0 |
| 1041 | H$_3^+$ | HD | H$_2$D$^+$ | H$_2$ | | 1.7e-09 | 0.00 | 0.0 |
| 1042 | H$_3^+$ | D$_2$ | HD$_2^+$ | H$_2$ | | 8.0e-10 | 0.00 | 0.0 |
| 1043 | H$_3^+$ | D$_2$ | H$_2$D$^+$ | HD | | 8.0e-10 | 0.00 | 0.0 |
| 1044 | H$_2$D$^+$ | H$_2$ | H$_3^+$ | HD | | 9.4e-11 | -0.79 | 154.6 |
| 1045 | H$_2$D$^+$ | HD | HD$_2^+$ | H$_2$ | | 1.3e-09 | 0.00 | 0.0 |
| 1046 | H$_2$D$^+$ | HD | H$_3^+$ | D$_2$ | | 1.1e-11 | -0.49 | 106.9 |
| 1047 | H$_2$D$^+$ | D$_2$ | D$_3^+$ | H$_2$ | | 3.0e-10 | -0.00 | 0.0 |
| 1048 | H$_2$D$^+$ | D$_2$ | HD$_2^+$ | HD | | 1.0e-09 | 0.00 | -0.0 |
| 1049 | HD$_2^+$ | H$_2$ | H$_2$D$^+$ | HD | | 3.3e-10 | -0.55 | 144.3 |
| 1050 | HD$_2^+$ | H$_2$ | H$_3^+$ | D$_2$ | | 3.5e-11 | -0.82 | 249.9 |
| 1051 | HD$_2^+$ | HD | H$_2$D$^+$ | D$_2$ | | 6.8e-11 | -0.27 | 113.0 |
| 1052 | HD$_2^+$ | HD | D$_3^+$ | H$_2$ | | 6.3e-10 | 0.00 | 0.0 |
| 1053 | HD$_2^+$ | D$_2$ | D$_3^+$ | HD | | 1.0e-09 | 0.00 | 0.0 |
| 1054 | D$_3^+$ | H$_2$ | H$_2$D$^+$ | D$_2$ | | 1.5e-10 | -0.85 | 315.3 |
| 1055 | D$_3^+$ | H$_2$ | HD$_2^+$ | HD | | 9.2e-10 | -0.59 | 197.8 |
| 1056 | D$_3^+$ | HD | HD$_2^+$ | D$_2$ | | 4.2e-10 | -0.85 | 259.0 |
| 1057 | H$_3^+$ | e$^-$ | H$_2$ | H | | 2.3e-08 | -0.52 | 0.0 |
| 1058 | H$_3^+$ | e$^-$ | H | H | H | 4.4e-08 | -0.52 | 0.0 |
| 1059 | H$_2$D$^+$ | e$^-$ | H | H | D | 5.6e-07 | 0.44 | -2.8 |
| 1060 | H$_2$D$^+$ | e$^-$ | HD | H | | 1.3e-07 | 0.44 | -2.8 |
| 1061 | H$_2$D$^+$ | e$^-$ | H$_2$ | D | | 5.2e-09 | 0.44 | -2.8 |
| 1062 | HD$_2^+$ | e$^-$ | D | D | H | 9.4e-08 | 0.66 | -12.5 |
| 1063 | HD$_2^+$ | e$^-$ | HD | D | | 1.2e-08 | 0.66 | -12.5 |
| 1064 | HD$_2^+$ | e$^-$ | D$_2$ | H | | 1.1e-08 | 0.66 | -12.5 |
| 1065 | D$_2^+$ | e$^-$ | D | D | | 2.3e-08 | -0.69 | 0.0 |
| 1066 | D$_3^+$ | e$^-$ | D | D | D | 2.2e-08 | -0.73 | 0.0 |
| 1067 | D$_3^+$ | e$^-$ | D$_2$ | D | | 1.3e-08 | -0.73 | 0.0 |
| 1068 | Gr | $\gamma_2$ | Gr$^+$ | e$^-$ | | 6.3e+07 | 0.00 | 0.0 |
| 1069 | Gr$^-$ | $\gamma_2$ | Gr | e$^-$ | | 4.2e+08 | 0.00 | 0.0 |
| 1070 | Gr | e$^-$ | Gr$^-$ | $\gamma$ | | 6.9e-05 | 0.50 | 0.0 |
| 1071 | Gr$^+$ | e$^-$ | Gr | $\gamma$ | | 6.9e-05 | 0.50 | 0.0 |
| 1072 | Fe | $\gamma_2$ | Fe$^+$ | e$^-$ | | 4.8e+02 | 0.00 | 0.0 |

**Table D6.** The separated version of the `UGAN` chemical network (total: 3266 reactions split into 43 Tables).

| # | Reactants | | Products | | | $\alpha$ | $\beta$ | $\gamma$ |
|---|---|---|---|---|---|---|---|---|
| 1 | H | H | $oH_2$ | | | 7.5e-01 | | |
| 2 | H | H | $pH_2$ | | | 2.5e-01 | | |
| 3 | H | D | HD | | | 1.0e+00 | | |
| 4 | D | D | $oD_2$ | | | 6.6e-01 | | |
| 5 | D | D | $pD_2$ | | | 3.3e-01 | | |
| 6 | H | CRP | $H^+$ | $e^-$ | | 4.6e-01 | 0.00 | 0.0 |
| 7 | D | CRP | $D^+$ | $e^-$ | | 4.6e-01 | 0.00 | 0.0 |
| 8 | He | CRP | $He^+$ | $e^-$ | | 5.0e-01 | 0.00 | 0.0 |
| 9 | $oH_2$ | CRP | H | H | | 1.0e-01 | 0.00 | 0.0 |
| 10 | $pH_2$ | CRP | H | H | | 1.0e-01 | 0.00 | 0.0 |
| 11 | $oH_2$ | CRP | $H^+$ | H | $e^-$ | 2.3e-02 | 0.00 | 0.0 |
| 12 | $pH_2$ | CRP | $H^+$ | H | $e^-$ | 2.3e-02 | 0.00 | 0.0 |
| 13 | $oH_2$ | CRP | $oH_2^+$ | $e^-$ | | 9.8e-01 | 0.00 | 0.0 |
| 14 | $pH_2$ | CRP | $pH_2^+$ | $e^-$ | | 9.8e-01 | 0.00 | 0.0 |
| 15 | HD | CRP | $H^+$ | D | $e^-$ | 2.0e-02 | 0.00 | 0.0 |
| 16 | HD | CRP | $D^+$ | H | $e^-$ | 2.0e-02 | 0.00 | 0.0 |
| 17 | HD | CRP | H | D | | 1.5e+00 | 0.00 | 0.0 |
| 18 | HD | CRP | $HD^+$ | $e^-$ | | 9.6e-01 | 0.00 | 0.0 |
| 19 | $oD_2$ | CRP | $D^+$ | D | $e^-$ | 4.0e-02 | 0.00 | 0.0 |
| 20 | $pD_2$ | CRP | $D^+$ | D | $e^-$ | 4.0e-02 | 0.00 | 0.0 |
| 21 | $oD_2$ | CRP | D | D | | 1.5e+00 | 0.00 | 0.0 |
| 22 | $pD_2$ | CRP | D | D | | 1.5e+00 | 0.00 | 0.0 |
| 23 | $oD_2$ | CRP | $oD_2^+$ | $e^-$ | | 9.6e-01 | 0.00 | 0.0 |
| 24 | $pD_2$ | CRP | $pD_2^+$ | $e^-$ | | 9.6e-01 | 0.00 | 0.0 |
| 25 | H | $oH_2^+$ | $oH_2$ | $H^+$ | | 6.4e-10 | 0.00 | 0.0 |
| 26 | H | $pH_2^+$ | $pH_2$ | $H^+$ | | 6.4e-10 | 0.00 | 0.0 |
| 27 | H | $HD^+$ | $H^+$ | HD | | 6.4e-10 | 0.00 | 0.0 |
| 28 | H | $HD^+$ | $oH_2D^+$ | $\gamma$ | | 9.0e-18 | 1.80 | -20.0 |
| 29 | H | $HD^+$ | $pH_2D^+$ | $\gamma$ | | 3.0e-18 | 1.80 | -20.0 |
| 30 | H | $D^+$ | $H^+$ | D | | 1.0e-09 | 0.00 | 0.0 |
| 31 | H | $D^+$ | $HD^+$ | $\gamma$ | | 3.9e-19 | 1.80 | -20.0 |
| 32 | H | $oD_2^+$ | $HD^+$ | D | | 1.0e-09 | 0.00 | 430.0 |
| 33 | H | $pD_2^+$ | $HD^+$ | D | | 1.0e-09 | 0.00 | 430.0 |
| 34 | H | $oD_2^+$ | $oD_2$ | $H^+$ | | 6.4e-10 | 0.00 | 0.0 |
| 35 | H | $pD_2^+$ | $pD_2$ | $H^+$ | | 6.4e-10 | 0.00 | 0.0 |
| 36 | $oH_2$ | $He^+$ | $H^+$ | H | He | 3.3e-15 | 0.00 | 0.0 |
| 37 | $pH_2$ | $He^+$ | $H^+$ | H | He | 3.3e-15 | 0.00 | 0.0 |
| 38 | $oH_2$ | $He^+$ | $oH_2^+$ | He | | 9.6e-15 | 0.00 | 0.0 |
| 39 | $pH_2$ | $He^+$ | $pH_2^+$ | He | | 9.6e-15 | 0.00 | 0.0 |
| 40 | $oH_2$ | $H^+$ | $oH_2^+$ | H | | 6.4e-10 | 0.00 | 21300.0 |
| 41 | $pH_2$ | $H^+$ | $pH_2^+$ | H | | 6.4e-10 | 0.00 | 21300.0 |
| 42 | $oH_2$ | $oH_2^+$ | $oH_3^+$ | H | | 1.4e-09 | 0.00 | 0.0 |
| 43 | $oH_2$ | $oH_2^+$ | $pH_3^+$ | H | | 7.0e-10 | 0.00 | 0.0 |
| 44 | $oH_2$ | $pH_2^+$ | $oH_3^+$ | H | | 7.0e-10 | 0.00 | 0.0 |
| 45 | $oH_2$ | $pH_2^+$ | $pH_3^+$ | H | | 1.4e-09 | 0.00 | 0.0 |
| 46 | $pH_2$ | $oH_2^+$ | $oH_3^+$ | H | | 7.0e-10 | 0.00 | 0.0 |
| 47 | $pH_2$ | $oH_2^+$ | $pH_3^+$ | H | | 1.4e-09 | 0.00 | 0.0 |
| 48 | $pH_2$ | $pH_2^+$ | $pH_3^+$ | H | | 2.1e-09 | 0.00 | 0.0 |
| 49 | $oH_2$ | $HD^+$ | $oH_2D^+$ | H | | 8.7e-10 | 0.00 | 0.0 |
| 50 | $oH_2$ | $HD^+$ | $pH_2D^+$ | H | | 1.7e-10 | 0.00 | 0.0 |
| 51 | $pH_2$ | $HD^+$ | $oH_2D^+$ | H | | 5.3e-10 | 0.00 | 0.0 |
| 52 | $pH_2$ | $HD^+$ | $pH_2D^+$ | H | | 5.3e-10 | 0.00 | 0.0 |
| 53 | $oH_2$ | $HD^+$ | $oH_3^+$ | D | | 7.0e-10 | 0.00 | 0.0 |
| 54 | $oH_2$ | $HD^+$ | $pH_3^+$ | D | | 3.5e-10 | 0.00 | 0.0 |
| 55 | $pH_2$ | $HD^+$ | $pH_3^+$ | D | | 1.1e-09 | 0.00 | 0.0 |
| 56 | $oH_2$ | $D^+$ | $H^+$ | HD | | 2.1e-09 | 0.00 | 0.0 |
| 57 | $pH_2$ | $D^+$ | $H^+$ | HD | | 2.1e-09 | 0.00 | 0.0 |
| 58 | $oH_2$ | $oD_2^+$ | $oH_2D^+$ | D | | 1.1e-09 | 0.00 | 0.0 |
| 59 | $oH_2$ | $pD_2^+$ | $oH_2D^+$ | D | | 1.1e-09 | 0.00 | 0.0 |
| 60 | $pH_2$ | $oD_2^+$ | $pH_2D^+$ | D | | 1.1e-09 | 0.00 | 0.0 |
| 61 | $pH_2$ | $pD_2^+$ | $pH_2D^+$ | D | | 1.1e-09 | 0.00 | 0.0 |
| 62 | $oH_2$ | $oD_2^+$ | $oHD_2^+$ | H | | 1.1e-09 | 0.00 | 0.0 |
| 63 | $oH_2$ | $pD_2^+$ | $pHD_2^+$ | H | | 1.1e-09 | 0.00 | 0.0 |
| 64 | $pH_2$ | $oD_2^+$ | $oHD_2^+$ | H | | 1.1e-09 | 0.00 | 0.0 |
| 65 | $pH_2$ | $pD_2^+$ | $pHD_2^+$ | H | | 1.1e-09 | 0.00 | 0.0 |
| 66 | HD | $He^+$ | $H^+$ | D | He | 5.5e-14 | -0.24 | 0.0 |
| 67 | HD | $He^+$ | $D^+$ | H | He | 5.5e-14 | -0.24 | 0.0 |
| 68 | HD | $oH_2^+$ | $oH_2D^+$ | H | | 8.7e-10 | 0.00 | 0.0 |
| 69 | HD | $oH_2^+$ | $pH_2D^+$ | H | | 1.7e-10 | 0.00 | 0.0 |
| 70 | HD | $pH_2^+$ | $oH_2D^+$ | H | | 5.3e-10 | 0.00 | 0.0 |
| 71 | HD | $pH_2^+$ | $pH_2D^+$ | H | | 5.3e-10 | 0.00 | 0.0 |
| 72 | HD | $oH_2^+$ | $oH_3^+$ | D | | 7.0e-10 | 0.00 | 0.0 |
| 73 | HD | $oH_2^+$ | $pH_3^+$ | D | | 3.5e-10 | 0.00 | 0.0 |
| 74 | HD | $pH_2^+$ | $pH_3^+$ | D | | 1.1e-09 | 0.00 | 0.0 |

**Table D6** – *continued* (part 2)

| # | Reactants | | Products | | | $\alpha$ | $\beta$ | $\gamma$ |
|---|---|---|---|---|---|---|---|---|
| 75 | HD | HD$^+$ | oH$_2$D$^+$ | D | | 7.9e-10 | 0.00 | 0.0 |
| 76 | HD | HD$^+$ | pH$_2$D$^+$ | D | | 2.6e-10 | 0.00 | 0.0 |
| 77 | HD | HD$^+$ | oHD$_2^+$ | H | | 7.0e-10 | 0.00 | 0.0 |
| 78 | HD | HD$^+$ | pHD$_2^+$ | H | | 3.5e-10 | 0.00 | 0.0 |
| 79 | HD | D$^+$ | oD$_2$ | H$^+$ | | 6.7e-10 | 0.00 | 0.0 |
| 80 | HD | D$^+$ | pD$_2$ | H$^+$ | | 3.3e-10 | 0.00 | 0.0 |
| 81 | HD | oD$_2^+$ | oHD$_2^+$ | D | | 8.2e-10 | 0.00 | 0.0 |
| 82 | HD | oD$_2^+$ | pHD$_2^+$ | D | | 2.3e-10 | 0.00 | 0.0 |
| 83 | HD | pD$_2^+$ | oHD$_2^+$ | D | | 4.7e-10 | 0.00 | 0.0 |
| 84 | HD | pD$_2^+$ | pHD$_2^+$ | D | | 5.8e-10 | 0.00 | 0.0 |
| 85 | HD | oD$_2^+$ | mD$_3^+$ | H | | 5.8e-10 | 0.00 | 0.0 |
| 86 | HD | oD$_2^+$ | oD$_3^+$ | H | | 4.7e-10 | 0.00 | 0.0 |
| 87 | HD | pD$_2^+$ | pD$_3^+$ | H | | 1.2e-10 | 0.00 | 0.0 |
| 88 | HD | pD$_2^+$ | oD$_3^+$ | H | | 9.3e-10 | 0.00 | 0.0 |
| 89 | D | H$^+$ | D$^+$ | H | | 1.0e-09 | 0.00 | 41.0 |
| 90 | D | H$^+$ | HD$^+$ | $\gamma$ | | 3.9e-19 | 1.80 | -20.0 |
| 91 | D | oH$_2^+$ | oH$_2$ | D$^+$ | | 6.4e-10 | 0.00 | 0.0 |
| 92 | D | pH$_2^+$ | pH$_2$ | D$^+$ | | 6.4e-10 | 0.00 | 0.0 |
| 93 | D | oH$_2^+$ | oH$_2$D$^+$ | $\gamma$ | | 7.0e-18 | 1.80 | -20.0 |
| 94 | D | pH$_2^+$ | pH$_2$D$^+$ | $\gamma$ | | 7.0e-18 | 1.80 | -20.0 |
| 95 | D | oH$_2^+$ | HD$^+$ | H | | 1.0e-09 | 0.00 | 0.0 |
| 96 | D | pH$_2^+$ | HD$^+$ | H | | 1.0e-09 | 0.00 | 0.0 |
| 97 | D | oH$_3^+$ | oH$_2$D$^+$ | H | | 1.0e-09 | 0.00 | 0.0 |
| 98 | D | pH$_3^+$ | oH$_2$D$^+$ | H | | 5.0e-10 | 0.00 | 0.0 |
| 99 | D | pH$_3^+$ | pH$_2$D$^+$ | H | | 5.0e-10 | 0.00 | 0.0 |
| 100 | D | HD$^+$ | oD$_2^+$ | H | | 6.7e-10 | 0.00 | 0.0 |
| 101 | D | HD$^+$ | pD$_2^+$ | H | | 3.3e-10 | 0.00 | 0.0 |
| 102 | D | HD$^+$ | D$^+$ | HD | | 6.4e-10 | 0.00 | 0.0 |
| 103 | D | oH$_2$D$^+$ | oHD$_2^+$ | H | | 6.7e-10 | 0.00 | 0.0 |
| 104 | D | oH$_2$D$^+$ | pHD$_2^+$ | H | | 3.3e-10 | 0.00 | 0.0 |
| 105 | D | pH$_2$D$^+$ | oHD$_2^+$ | H | | 6.7e-10 | 0.00 | 0.0 |
| 106 | D | pH$_2$D$^+$ | pHD$_2^+$ | H | | 3.3e-10 | 0.00 | 0.0 |
| 107 | D | oHD$_2^+$ | mD$_3^+$ | H | | 5.6e-10 | 0.00 | 0.0 |
| 108 | D | oHD$_2^+$ | oD$_3^+$ | H | | 4.4e-10 | 0.00 | 0.0 |
| 109 | D | pHD$_2^+$ | pD$_3^+$ | H | | 1.1e-10 | 0.00 | 0.0 |
| 110 | D | pHD$_2^+$ | oD$_3^+$ | H | | 8.9e-10 | 0.00 | 0.0 |
| 111 | D | oD$_2^+$ | oD$_2$ | D$^+$ | | 6.4e-10 | 0.00 | 0.0 |
| 112 | D | pD$_2^+$ | pD$_2$ | D$^+$ | | 6.4e-10 | 0.00 | 0.0 |
| 113 | oD$_2$ | He$^+$ | D$^+$ | D | He | 1.1e-13 | -0.24 | 0.0 |
| 114 | pD$_2$ | He$^+$ | D$^+$ | D | He | 1.1e-13 | -0.24 | 0.0 |
| 115 | oD$_2$ | He$^+$ | oD$_2^+$ | He | | 2.5e-14 | 0.00 | 0.0 |
| 116 | pD$_2$ | He$^+$ | pD$_2^+$ | He | | 2.5e-14 | 0.00 | 0.0 |
| 117 | oD$_2$ | H$^+$ | D$^+$ | HD | | 2.1e-09 | 0.00 | 405.0 |
| 118 | pD$_2$ | H$^+$ | D$^+$ | HD | | 2.1e-09 | 0.00 | 405.0 |
| 119 | oD$_2$ | oH$_2^+$ | oH$_2$D$^+$ | D | | 1.1e-09 | 0.00 | 0.0 |
| 120 | pD$_2$ | oH$_2^+$ | oH$_2$D$^+$ | D | | 1.1e-09 | 0.00 | 0.0 |
| 121 | oD$_2$ | pH$_2^+$ | pH$_2$D$^+$ | D | | 1.1e-09 | 0.00 | 0.0 |
| 122 | pD$_2$ | pH$_2^+$ | pH$_2$D$^+$ | D | | 1.1e-09 | 0.00 | 0.0 |
| 123 | oD$_2$ | oH$_2^+$ | oHD$_2^+$ | H | | 1.1e-09 | 0.00 | 0.0 |
| 124 | pD$_2$ | oH$_2^+$ | pHD$_2^+$ | H | | 1.1e-09 | 0.00 | 0.0 |
| 125 | oD$_2$ | pH$_2^+$ | oHD$_2^+$ | H | | 1.1e-09 | 0.00 | 0.0 |
| 126 | pD$_2$ | pH$_2^+$ | pHD$_2^+$ | H | | 1.1e-09 | 0.00 | 0.0 |
| 127 | oD$_2$ | HD$^+$ | oHD$_2^+$ | D | | 8.2e-10 | 0.00 | 0.0 |
| 128 | oD$_2$ | HD$^+$ | pHD$_2^+$ | D | | 2.3e-10 | 0.00 | 0.0 |
| 129 | pD$_2$ | HD$^+$ | oHD$_2^+$ | D | | 4.7e-10 | 0.00 | 0.0 |
| 130 | pD$_2$ | HD$^+$ | pHD$_2^+$ | D | | 5.8e-10 | 0.00 | 0.0 |
| 131 | oD$_2$ | HD$^+$ | mD$_3^+$ | H | | 5.8e-10 | 0.00 | 0.0 |
| 132 | oD$_2$ | HD$^+$ | oD$_3^+$ | H | | 4.7e-10 | 0.00 | 0.0 |
| 133 | pD$_2$ | HD$^+$ | pD$_3^+$ | H | | 1.2e-10 | 0.00 | 0.0 |
| 134 | pD$_2$ | HD$^+$ | oD$_3^+$ | H | | 9.3e-10 | 0.00 | 0.0 |
| 135 | oD$_2$ | oD$_2^+$ | mD$_3^+$ | D | | 1.2e-09 | 0.00 | 0.0 |
| 136 | oD$_2$ | oD$_2^+$ | oD$_3^+$ | D | | 9.3e-10 | 0.00 | 0.0 |
| 137 | oD$_2$ | pD$_2^+$ | mD$_3^+$ | D | | 5.8e-10 | 0.00 | 0.0 |
| 138 | oD$_2$ | pD$_2^+$ | pD$_3^+$ | D | | 1.2e-10 | 0.00 | 0.0 |
| 139 | oD$_2$ | pD$_2^+$ | oD$_3^+$ | D | | 1.4e-09 | 0.00 | 0.0 |
| 140 | pD$_2$ | oD$_2^+$ | mD$_3^+$ | D | | 5.8e-10 | 0.00 | 0.0 |
| 141 | pD$_2$ | oD$_2^+$ | pD$_3^+$ | D | | 1.2e-10 | 0.00 | 0.0 |
| 142 | pD$_2$ | oD$_2^+$ | oD$_3^+$ | D | | 1.4e-09 | 0.00 | 0.0 |
| 143 | pD$_2$ | pD$_2^+$ | pD$_3^+$ | D | | 2.3e-10 | 0.00 | 0.0 |
| 144 | pD$_2$ | pD$_2^+$ | oD$_3^+$ | D | | 1.9e-09 | 0.00 | 0.0 |
| 145 | Fe | H$^+$ | Fe$^+$ | H | | 7.4e-09 | 0.00 | 0.0 |
| 146 | Fe | oH$_3^+$ | oH$_2$ | H | Fe$^+$ | 4.9e-09 | 0.00 | 0.0 |
| 147 | Fe | pH$_3^+$ | oH$_2$ | H | Fe$^+$ | 2.5e-09 | 0.00 | 0.0 |
| 148 | Fe | pH$_3^+$ | pH$_2$ | H | Fe$^+$ | 2.5e-09 | 0.00 | 0.0 |

**Table D6** – *continued* (part 3)

| # | Reactants | | Products | | | | | $\alpha$ | $\beta$ | $\gamma$ |
|---|---|---|---|---|---|---|---|---|---|---|
| 149 | Fe | $oH_2D^+$ | $oH_2$ | $Fe^+$ | D | | | 4.9e-09 | 0.00 | 0.0 |
| 150 | Fe | $pH_2D^+$ | $pH_2$ | $Fe^+$ | D | | | 4.9e-09 | 0.00 | 0.0 |
| 151 | Fe | $oHD_2^+$ | $oD_2$ | $Fe^+$ | H | | | 1.6e-09 | 0.00 | 0.0 |
| 152 | Fe | $pHD_2^+$ | $pD_2$ | $Fe^+$ | H | | | 1.6e-09 | 0.00 | 0.0 |
| 153 | Fe | $oHD_2^+$ | HD | D | $Fe^+$ | | | 3.3e-09 | 0.00 | 0.0 |
| 154 | Fe | $pHD_2^+$ | HD | D | $Fe^+$ | | | 3.3e-09 | 0.00 | 0.0 |
| 155 | Fe | $mD_3^+$ | $oD_2$ | D | $Fe^+$ | | | 4.9e-09 | 0.00 | 0.0 |
| 156 | Fe | $pD_3^+$ | $pD_2$ | D | $Fe^+$ | | | 4.9e-09 | 0.00 | 0.0 |
| 157 | Fe | $oD_3^+$ | $oD_2$ | D | $Fe^+$ | | | 2.5e-09 | 0.00 | 0.0 |
| 158 | Fe | $oD_3^+$ | $pD_2$ | D | $Fe^+$ | | | 2.5e-09 | 0.00 | 0.0 |
| 159 | $H^+$ | $e^-$ | H | $\gamma$ | | | | 3.6e-12 | -0.75 | 0.0 |
| 160 | $oH_2^+$ | $e^-$ | H | H | | | | 2.5e-07 | -0.50 | 0.0 |
| 161 | $pH_2^+$ | $e^-$ | H | H | | | | 2.5e-07 | -0.50 | 0.0 |
| 162 | $oH_2^+$ | $e^-$ | $oH_2$ | $\gamma$ | | | | 2.2e-07 | -0.40 | 0.0 |
| 163 | $pH_2^+$ | $e^-$ | $pH_2$ | $\gamma$ | | | | 2.2e-07 | -0.40 | 0.0 |
| 164 | $HD^+$ | $e^-$ | H | D | | | | 9.0e-09 | -0.50 | 0.0 |
| 165 | $D^+$ | $e^-$ | D | $\gamma$ | | | | 3.6e-12 | -0.75 | 0.0 |
| 166 | $He^+$ | $e^-$ | He | $\gamma$ | | | | 4.5e-12 | -0.67 | 0.0 |
| 167 | $oH_2^+$ | $\eta$ | $H^+$ | H | | | | 6.1e+02 | 0.00 | 0.0 |
| 168 | $pH_2^+$ | $\eta$ | $H^+$ | H | | | | 6.1e+02 | 0.00 | 0.0 |
| 169 | $H^+$ | $Gr^-$ | Gr | H | | | | 1.6e-06 | 0.50 | 0.0 |
| 170 | $oH_3^+$ | $Gr^-$ | $oH_2$ | H | Gr | | | 4.6e-07 | 0.50 | 0.0 |
| 171 | $pH_3^+$ | $Gr^-$ | $oH_2$ | H | Gr | | | 2.3e-07 | 0.50 | 0.0 |
| 172 | $pH_3^+$ | $Gr^-$ | $pH_2$ | H | Gr | | | 2.3e-07 | 0.50 | 0.0 |
| 173 | $oH_3^+$ | $Gr^-$ | H | H | H | Gr | | 4.6e-07 | 0.50 | 0.0 |
| 174 | $pH_3^+$ | $Gr^-$ | H | H | H | Gr | | 4.6e-07 | 0.50 | 0.0 |
| 175 | $He^+$ | $Gr^-$ | Gr | He | | | | 8.0e-07 | 0.50 | 0.0 |
| 176 | $H^+$ | Gr | $Gr^+$ | H | | | | 1.6e-06 | 0.50 | 0.0 |
| 177 | $oH_3^+$ | Gr | $oH_2$ | H | $Gr^+$ | | | 4.6e-07 | 0.50 | 0.0 |
| 178 | $pH_3^+$ | Gr | $oH_2$ | H | $Gr^+$ | | | 2.3e-07 | 0.50 | 0.0 |
| 179 | $pH_3^+$ | Gr | $pH_2$ | H | $Gr^+$ | | | 2.3e-07 | 0.50 | 0.0 |
| 180 | $oH_3^+$ | Gr | H | H | H | $Gr^+$ | | 4.6e-07 | 0.50 | 0.0 |
| 181 | $pH_3^+$ | Gr | H | H | H | $Gr^+$ | | 4.6e-07 | 0.50 | 0.0 |
| 182 | $He^+$ | Gr | $Gr^+$ | He | | | | 8.0e-07 | 0.50 | 0.0 |
| 183 | $D^+$ | $Gr^-$ | Gr | D | | | | 1.1e-06 | 0.50 | 0.0 |
| 184 | $oH_2D^+$ | $Gr^-$ | $oH_2$ | Gr | D | | | 1.3e-07 | 0.50 | 0.0 |
| 185 | $pH_2D^+$ | $Gr^-$ | $pH_2$ | Gr | D | | | 1.3e-07 | 0.50 | 0.0 |
| 186 | $oH_2D^+$ | $Gr^-$ | HD | H | Gr | | | 2.7e-07 | 0.50 | 0.0 |
| 187 | $pH_2D^+$ | $Gr^-$ | HD | H | Gr | | | 2.7e-07 | 0.50 | 0.0 |
| 188 | $oH_2D^+$ | $Gr^-$ | H | H | Gr | D | | 4.0e-07 | 0.50 | 0.0 |
| 189 | $pH_2D^+$ | $Gr^-$ | H | H | Gr | D | | 4.0e-07 | 0.50 | 0.0 |
| 190 | $D^+$ | Gr | $Gr^+$ | D | | | | 1.1e-06 | 0.50 | 0.0 |
| 191 | $oH_2D^+$ | Gr | $oH_2$ | $Gr^+$ | D | | | 1.3e-07 | 0.50 | 0.0 |
| 192 | $pH_2D^+$ | Gr | $pH_2$ | $Gr^+$ | D | | | 1.3e-07 | 0.50 | 0.0 |
| 193 | $oH_2D^+$ | Gr | HD | H | $Gr^+$ | | | 2.7e-07 | 0.50 | 0.0 |
| 194 | $pH_2D^+$ | Gr | HD | H | $Gr^+$ | | | 2.7e-07 | 0.50 | 0.0 |
| 195 | $oH_2D^+$ | Gr | H | H | $Gr^+$ | D | | 4.0e-07 | 0.50 | 0.0 |
| 196 | $pH_2D^+$ | Gr | H | H | $Gr^+$ | D | | 4.0e-07 | 0.50 | 0.0 |
| 197 | $oHD_2^+$ | $Gr^-$ | HD | D | Gr | | | 2.4e-07 | 0.50 | 0.0 |
| 198 | $pHD_2^+$ | $Gr^-$ | HD | D | Gr | | | 2.4e-07 | 0.50 | 0.0 |
| 199 | $oHD_2^+$ | $Gr^-$ | $oD_2$ | Gr | H | | | 1.2e-07 | 0.50 | 0.0 |
| 200 | $pHD_2^+$ | $Gr^-$ | $pD_2$ | Gr | H | | | 1.2e-07 | 0.50 | 0.0 |
| 201 | $oHD_2^+$ | $Gr^-$ | D | D | Gr | H | | 3.6e-07 | 0.50 | 0.0 |
| 202 | $pHD_2^+$ | $Gr^-$ | D | D | Gr | H | | 3.6e-07 | 0.50 | 0.0 |
| 203 | $oHD_2^+$ | Gr | HD | D | $Gr^+$ | | | 2.4e-07 | 0.50 | 0.0 |
| 204 | $pHD_2^+$ | Gr | HD | D | $Gr^+$ | | | 2.4e-07 | 0.50 | 0.0 |
| 205 | $oHD_2^+$ | Gr | $oD_2$ | $Gr^+$ | H | | | 1.2e-07 | 0.50 | 0.0 |
| 206 | $pHD_2^+$ | Gr | $pD_2$ | $Gr^+$ | H | | | 1.2e-07 | 0.50 | 0.0 |
| 207 | $oHD_2^+$ | Gr | D | D | $Gr^+$ | H | | 3.6e-07 | 0.50 | 0.0 |
| 208 | $pHD_2^+$ | Gr | D | D | $Gr^+$ | H | | 3.6e-07 | 0.50 | 0.0 |
| 209 | $mD_3^+$ | $Gr^-$ | $oD_2$ | D | Gr | | | 3.3e-07 | 0.50 | 0.0 |
| 210 | $pD_3^+$ | $Gr^-$ | $pD_2$ | D | Gr | | | 3.3e-07 | 0.50 | 0.0 |
| 211 | $oD_3^+$ | $Gr^-$ | $oD_2$ | D | Gr | | | 1.6e-07 | 0.50 | 0.0 |
| 212 | $oD_3^+$ | $Gr^-$ | $pD_2$ | D | Gr | | | 1.6e-07 | 0.50 | 0.0 |
| 213 | $mD_3^+$ | $Gr^-$ | D | D | D | Gr | | 3.3e-07 | 0.50 | 0.0 |
| 214 | $pD_3^+$ | $Gr^-$ | D | D | D | Gr | | 3.3e-07 | 0.50 | 0.0 |
| 215 | $oD_3^+$ | $Gr^-$ | D | D | D | Gr | | 3.3e-07 | 0.50 | 0.0 |
| 216 | $mD_3^+$ | Gr | $oD_2$ | D | $Gr^+$ | | | 3.3e-07 | 0.50 | 0.0 |
| 217 | $pD_3^+$ | Gr | $pD_2$ | D | $Gr^+$ | | | 3.3e-07 | 0.50 | 0.0 |
| 218 | $oD_3^+$ | Gr | $oD_2$ | D | $Gr^+$ | | | 1.6e-07 | 0.50 | 0.0 |
| 219 | $oD_3^+$ | Gr | $pD_2$ | D | $Gr^+$ | | | 1.6e-07 | 0.50 | 0.0 |
| 220 | $mD_3^+$ | Gr | D | D | D | $Gr^+$ | | 3.3e-07 | 0.50 | 0.0 |
| 221 | $pD_3^+$ | Gr | D | D | D | $Gr^+$ | | 3.3e-07 | 0.50 | 0.0 |
| 222 | $oD_3^+$ | Gr | D | D | D | $Gr^+$ | | 3.3e-07 | 0.50 | 0.0 |

**Table D6** – *continued* (part 4)

| # | Reactants | | Products | | | $\alpha$ | $\beta$ | $\gamma$ |
|---|---|---|---|---|---|---|---|---|
| 223 | O | CRP | O$^+$ | e$^-$ | | 2.8e+00 | 0.00 | 0.0 |
| 224 | O | OH | O$_2$ | H | | 4.0e-11 | 0.00 | 0.0 |
| 225 | O$^+$ | oH$_2$ | OH$^+$ | H | | 1.2e-09 | 0.00 | 0.0 |
| 226 | O$^+$ | pH$_2$ | OH$^+$ | H | | 1.2e-09 | 0.00 | 0.0 |
| 227 | O$^+$ | H | H$^+$ | O | | 6.0e-10 | 0.00 | 0.0 |
| 228 | O | H$^+$ | O$^+$ | H | | 6.0e-10 | 0.00 | 227.0 |
| 229 | O | oH$_2^+$ | OH$^+$ | H | | 1.5e-09 | 0.00 | 0.0 |
| 230 | O | pH$_2^+$ | OH$^+$ | H | | 1.5e-09 | 0.00 | 0.0 |
| 231 | O | oH$_3^+$ | oH$_2$ | OH$^+$ | | 8.0e-10 | -0.16 | 1.4 |
| 232 | O | pH$_3^+$ | oH$_2$ | OH$^+$ | | 4.0e-10 | -0.16 | 1.4 |
| 233 | O | pH$_3^+$ | pH$_2$ | OH$^+$ | | 4.0e-10 | -0.16 | 1.4 |
| 234 | O | oH$_3^+$ | oH$_2$O$^+$ | H | | 3.4e-10 | -0.16 | 1.4 |
| 235 | O | pH$_3^+$ | oH$_2$O$^+$ | H | | 1.7e-10 | -0.16 | 1.4 |
| 236 | O | pH$_3^+$ | pH$_2$O$^+$ | H | | 1.7e-10 | -0.16 | 1.4 |
| 237 | O | oH$_2$D$^+$ | OH$^+$ | HD | | 5.3e-10 | 0.00 | 0.0 |
| 238 | O | pH$_2$D$^+$ | OH$^+$ | HD | | 5.3e-10 | 0.00 | 0.0 |
| 239 | O | oH$_2$D$^+$ | oH$_2$ | OD$^+$ | | 2.7e-10 | 0.00 | 0.0 |
| 240 | O | pH$_2$D$^+$ | pH$_2$ | OD$^+$ | | 2.7e-10 | 0.00 | 0.0 |
| 241 | O | oHD$_2^+$ | oD$_2$ | OH$^+$ | | 2.7e-10 | 0.00 | 0.0 |
| 242 | O | pHD$_2^+$ | pD$_2$ | OH$^+$ | | 2.7e-10 | 0.00 | 0.0 |
| 243 | O | oHD$_2^+$ | OD$^+$ | HD | | 5.3e-10 | 0.00 | 0.0 |
| 244 | O | pHD$_2^+$ | OD$^+$ | HD | | 5.3e-10 | 0.00 | 0.0 |
| 245 | O | mD$_3^+$ | oD$_2$ | OD$^+$ | | 8.0e-10 | 0.00 | 0.0 |
| 246 | O | pD$_3^+$ | pD$_2$ | OD$^+$ | | 8.0e-10 | 0.00 | 0.0 |
| 247 | O | oD$_3^+$ | oD$_2$ | OD$^+$ | | 4.0e-10 | 0.00 | 0.0 |
| 248 | O | oD$_3^+$ | pD$_2$ | OD$^+$ | | 4.0e-10 | 0.00 | 0.0 |
| 249 | O$_2$ | He$^+$ | O$^+$ | O | He | 1.0e-09 | 0.00 | 0.0 |
| 250 | O$_2$ | H$^+$ | O$_2^+$ | H | | 1.2e-09 | 0.00 | 0.0 |
| 251 | OH | He$^+$ | OH$^+$ | He | | 5.5e-10 | 0.00 | 0.0 |
| 252 | OH | He$^+$ | O$^+$ | H | He | 5.5e-10 | 0.00 | 0.0 |
| 253 | OH | H$^+$ | OH$^+$ | H | | 2.1e-09 | 0.00 | 0.0 |
| 254 | OH | oH$_2^+$ | oH$_2$ | OH$^+$ | | 7.6e-10 | 0.00 | 0.0 |
| 255 | OH | pH$_2^+$ | pH$_2$ | OH$^+$ | | 7.6e-10 | 0.00 | 0.0 |
| 256 | OH | oH$_3^+$ | oH$_2$O$^+$ | oH$_2$ | | 9.8e-10 | 0.00 | 0.0 |
| 257 | OH | oH$_3^+$ | oH$_2$O$^+$ | pH$_2$ | | 1.6e-10 | 0.00 | 0.0 |
| 258 | OH | oH$_3^+$ | pH$_2$O$^+$ | oH$_2$ | | 1.6e-10 | 0.00 | 0.0 |
| 259 | OH | pH$_3^+$ | oH$_2$O$^+$ | oH$_2$ | | 4.9e-10 | 0.00 | 0.0 |
| 260 | OH | pH$_3^+$ | oH$_2$O$^+$ | pH$_2$ | | 3.3e-10 | 0.00 | 0.0 |
| 261 | OH | pH$_3^+$ | pH$_2$O$^+$ | oH$_2$ | | 3.3e-10 | 0.00 | 0.0 |
| 262 | OH | pH$_3^+$ | pH$_2$O$^+$ | pH$_2$ | | 1.6e-10 | 0.00 | 0.0 |
| 263 | OH | oH$_2$D$^+$ | oH$_2$O$^+$ | HD | | 7.2e-10 | 0.00 | 0.0 |
| 264 | OH | oH$_2$D$^+$ | pH$_2$O$^+$ | HD | | 1.4e-10 | 0.00 | 0.0 |
| 265 | OH | pH$_2$D$^+$ | oH$_2$O$^+$ | HD | | 4.3e-10 | 0.00 | 0.0 |
| 266 | OH | pH$_2$D$^+$ | pH$_2$O$^+$ | HD | | 4.3e-10 | 0.00 | 0.0 |
| 267 | OH | oH$_2$D$^+$ | oH$_2$ | HDO$^+$ | | 3.6e-10 | 0.00 | 0.0 |
| 268 | OH | oH$_2$D$^+$ | pH$_2$ | HDO$^+$ | | 7.2e-11 | 0.00 | 0.0 |
| 269 | OH | pH$_2$D$^+$ | oH$_2$ | HDO$^+$ | | 2.2e-10 | 0.00 | 0.0 |
| 270 | OH | pH$_2$D$^+$ | pH$_2$ | HDO$^+$ | | 2.2e-10 | 0.00 | 0.0 |
| 271 | OH | oHD$_2^+$ | oD$_2$ | oH$_2$O$^+$ | | 3.2e-10 | 0.00 | 0.0 |
| 272 | OH | pHD$_2^+$ | pD$_2$ | oH$_2$O$^+$ | | 3.2e-10 | 0.00 | 0.0 |
| 273 | OH | oHD$_2^+$ | oD$_2$ | pH$_2$O$^+$ | | 1.1e-10 | 0.00 | 0.0 |
| 274 | OH | pHD$_2^+$ | pD$_2$ | pH$_2$O$^+$ | | 1.1e-10 | 0.00 | 0.0 |
| 275 | OH | oHD$_2^+$ | HDO$^+$ | HD | | 8.7e-10 | 0.00 | 0.0 |
| 276 | OH | pHD$_2^+$ | HDO$^+$ | HD | | 8.7e-10 | 0.00 | 0.0 |
| 277 | OH | mD$_3^+$ | oD$_2$ | HDO$^+$ | | 1.3e-09 | 0.00 | 0.0 |
| 278 | OH | pD$_3^+$ | pD$_2$ | HDO$^+$ | | 1.3e-09 | 0.00 | 0.0 |
| 279 | OH | oD$_3^+$ | oD$_2$ | HDO$^+$ | | 6.5e-10 | 0.00 | 0.0 |
| 280 | OH | oD$_3^+$ | pD$_2$ | HDO$^+$ | | 6.5e-10 | 0.00 | 0.0 |
| 281 | OD | He$^+$ | OD$^+$ | He | | 5.5e-10 | 0.00 | 0.0 |
| 282 | OD | He$^+$ | O$^+$ | D | He | 5.5e-10 | 0.00 | 0.0 |
| 283 | OD | H$^+$ | OD$^+$ | H | | 2.1e-09 | 0.00 | 0.0 |
| 284 | OD | oH$_3^+$ | oH$_2$ | HDO$^+$ | | 1.3e-09 | 0.00 | 0.0 |
| 285 | OD | pH$_3^+$ | oH$_2$ | HDO$^+$ | | 6.5e-10 | 0.00 | 0.0 |
| 286 | OD | pH$_3^+$ | pH$_2$ | HDO$^+$ | | 6.5e-10 | 0.00 | 0.0 |
| 287 | OD | oH$_2$D$^+$ | HDO$^+$ | HD | | 8.7e-10 | 0.00 | 0.0 |
| 288 | OD | pH$_2$D$^+$ | HDO$^+$ | HD | | 8.7e-10 | 0.00 | 0.0 |
| 289 | OD | oH$_2$D$^+$ | oD$_2$O$^+$ | oH$_2$ | | 2.9e-10 | 0.00 | 0.0 |
| 290 | OD | oH$_2$D$^+$ | pD$_2$O$^+$ | oH$_2$ | | 1.4e-10 | 0.00 | 0.0 |
| 291 | OD | pH$_2$D$^+$ | oD$_2$O$^+$ | pH$_2$ | | 2.9e-10 | 0.00 | 0.0 |
| 292 | OD | pH$_2$D$^+$ | pD$_2$O$^+$ | pH$_2$ | | 1.4e-10 | 0.00 | 0.0 |
| 293 | OD | oHD$_2^+$ | oD$_2$ | HDO$^+$ | | 3.4e-10 | 0.00 | 0.0 |
| 294 | OD | oHD$_2^+$ | pD$_2$ | HDO$^+$ | | 9.6e-11 | 0.00 | 0.0 |
| 295 | OD | pHD$_2^+$ | oD$_2$ | HDO$^+$ | | 1.9e-10 | 0.00 | 0.0 |
| 296 | OD | pHD$_2^+$ | pD$_2$ | HDO$^+$ | | 2.4e-10 | 0.00 | 0.0 |

**Table D6** – *continued* (part 5)

| # | Reactants | | Products | | | $\alpha$ | $\beta$ | $\gamma$ |
|---|---|---|---|---|---|---|---|---|
| 297 | OD | $oHD_2^+$ | $oD_2O^+$ | HD | | 6.7e-10 | 0.00 | 0.0 |
| 298 | OD | $oHD_2^+$ | $pD_2O^+$ | HD | | 1.9e-10 | 0.00 | 0.0 |
| 299 | OD | $pHD_2^+$ | $oD_2O^+$ | HD | | 3.9e-10 | 0.00 | 0.0 |
| 300 | OD | $pHD_2^+$ | $pD_2O^+$ | HD | | 4.8e-10 | 0.00 | 0.0 |
| 301 | OD | $mD_3^+$ | $oD_2O^+$ | $oD_2$ | | 8.7e-11 | 0.00 | 0.0 |
| 302 | OD | $mD_3^+$ | $oD_2O^+$ | $pD_2$ | | 2.2e-11 | 0.00 | 0.0 |
| 303 | OD | $mD_3^+$ | $pD_2O^+$ | $oD_2$ | | 2.2e-11 | 0.00 | 0.0 |
| 304 | OD | $pD_3^+$ | $oD_2O^+$ | $pD_2$ | | 4.3e-11 | 0.00 | 0.0 |
| 305 | OD | $pD_3^+$ | $pD_2O^+$ | $oD_2$ | | 4.3e-11 | 0.00 | 0.0 |
| 306 | OD | $pD_3^+$ | $pD_2O^+$ | $pD_2$ | | 4.3e-11 | 0.00 | 0.0 |
| 307 | OD | $oD_3^+$ | $oD_2O^+$ | $oD_2$ | | 4.3e-11 | 0.00 | 0.0 |
| 308 | OD | $oD_3^+$ | $oD_2O^+$ | $pD_2$ | | 3.2e-11 | 0.00 | 0.0 |
| 309 | OD | $oD_3^+$ | $pD_2O^+$ | $oD_2$ | | 3.2e-11 | 0.00 | 0.0 |
| 310 | OD | $oD_3^+$ | $pD_2O^+$ | $pD_2$ | | 2.2e-11 | 0.00 | 0.0 |
| 311 | $oH_2O$ | $He^+$ | $OH^+$ | H | He | 2.3e-10 | -0.94 | 0.0 |
| 312 | $pH_2O$ | $He^+$ | $OH^+$ | H | He | 2.3e-10 | -0.94 | 0.0 |
| 313 | $oH_2O$ | $He^+$ | $oH_2O^+$ | He | | 4.9e-11 | -0.94 | 0.0 |
| 314 | $pH_2O$ | $He^+$ | $pH_2O^+$ | He | | 4.9e-11 | -0.94 | 0.0 |
| 315 | $oH_2O$ | $He^+$ | $H^+$ | OH | He | 1.6e-10 | -0.94 | 0.0 |
| 316 | $pH_2O$ | $He^+$ | $H^+$ | OH | He | 1.6e-10 | -0.94 | 0.0 |
| 317 | $oH_2O$ | $H^+$ | $oH_2O^+$ | H | | 8.2e-09 | 0.00 | 0.0 |
| 318 | $pH_2O$ | $H^+$ | $pH_2O^+$ | H | | 8.2e-09 | 0.00 | 0.0 |
| 319 | $oH_2O$ | $oH_2^+$ | $oH_2$ | $oH_2O^+$ | | 3.9e-09 | 0.00 | 0.0 |
| 320 | $oH_2O$ | $pH_2^+$ | $oH_2$ | $pH_2O^+$ | | 3.9e-09 | 0.00 | 0.0 |
| 321 | $pH_2O$ | $oH_2^+$ | $pH_2$ | $oH_2O^+$ | | 3.9e-09 | 0.00 | 0.0 |
| 322 | $pH_2O$ | $pH_2^+$ | $pH_2$ | $pH_2O^+$ | | 3.9e-09 | 0.00 | 0.0 |
| 323 | $oH_2O$ | $oH_2^+$ | $oH_3O^+$ | H | | 2.3e-09 | 0.00 | 0.0 |
| 324 | $oH_2O$ | $oH_2^+$ | $pH_3O^+$ | H | | 1.1e-09 | 0.00 | 0.0 |
| 325 | $oH_2O$ | $pH_2^+$ | $oH_3O^+$ | H | | 1.1e-09 | 0.00 | 0.0 |
| 326 | $oH_2O$ | $pH_2^+$ | $pH_3O^+$ | H | | 2.3e-09 | 0.00 | 0.0 |
| 327 | $pH_2O$ | $oH_2^+$ | $pH_3O^+$ | H | | 1.1e-09 | 0.00 | 0.0 |
| 328 | $pH_2O$ | $oH_2^+$ | $pH_3O^+$ | H | | 2.3e-09 | 0.00 | 0.0 |
| 329 | $pH_2O$ | $pH_2^+$ | $pH_3O^+$ | H | | 3.4e-09 | 0.00 | 0.0 |
| 330 | $oH_2O$ | $oH_3^+$ | $oH_3O^+$ | $oH_2$ | | 2.7e-09 | 0.00 | 0.0 |
| 331 | $oH_2O$ | $oH_3^+$ | $oH_3O^+$ | $pH_2$ | | 3.6e-10 | 0.00 | 0.0 |
| 332 | $oH_2O$ | $oH_3^+$ | $pH_3O^+$ | $oH_2$ | | 1.0e-09 | 0.00 | 0.0 |
| 333 | $oH_2O$ | $oH_3^+$ | $pH_3O^+$ | $pH_2$ | | 2.9e-10 | 0.00 | 0.0 |
| 334 | $pH_2O$ | $oH_3^+$ | $oH_3O^+$ | $oH_2$ | | 1.1e-09 | 0.00 | 0.0 |
| 335 | $pH_2O$ | $oH_3^+$ | $oH_3O^+$ | $pH_2$ | | 1.1e-09 | 0.00 | 0.0 |
| 336 | $oH_2O$ | $pH_3^+$ | $oH_3O^+$ | $oH_2$ | | 2.1e-09 | 0.00 | 0.0 |
| 337 | $oH_2O$ | $pH_3^+$ | $oH_3O^+$ | $pH_2$ | | 1.0e-09 | 0.00 | 0.0 |
| 338 | $oH_2O$ | $pH_3^+$ | $pH_3O^+$ | $pH_2$ | | 7.2e-10 | 0.00 | 0.0 |
| 339 | $oH_2O$ | $pH_3^+$ | $pH_3O^+$ | $oH_2$ | | 2.0e-09 | 0.00 | 0.0 |
| 340 | $oH_2O$ | $pH_3^+$ | $pH_3O^+$ | $pH_2$ | | 5.7e-10 | 0.00 | 0.0 |
| 341 | $pH_2O$ | $pH_3^+$ | $oH_3O^+$ | $oH_2$ | | 8.6e-10 | 0.00 | 0.0 |
| 342 | $pH_2O$ | $pH_3^+$ | $pH_3O^+$ | $oH_2$ | | 1.7e-09 | 0.00 | 0.0 |
| 343 | $pH_2O$ | $pH_3^+$ | $pH_3O^+$ | $pH_2$ | | 1.7e-09 | 0.00 | 0.0 |
| 344 | $oH_2O$ | $oH_2D^+$ | $oH_3O^+$ | HD | | 1.9e-09 | 0.00 | 0.0 |
| 345 | $oH_2O$ | $oH_2D^+$ | $pH_3O^+$ | HD | | 9.6e-10 | 0.00 | 0.0 |
| 346 | $oH_2O$ | $pH_2D^+$ | $oH_3O^+$ | HD | | 9.6e-10 | 0.00 | 0.0 |
| 347 | $oH_2O$ | $pH_2D^+$ | $pH_3O^+$ | HD | | 1.9e-09 | 0.00 | 0.0 |
| 348 | $pH_2O$ | $oH_2D^+$ | $oH_3O^+$ | HD | | 9.6e-10 | 0.00 | 0.0 |
| 349 | $pH_2O$ | $oH_2D^+$ | $pH_3O^+$ | HD | | 1.9e-09 | 0.00 | 0.0 |
| 350 | $pH_2O$ | $pH_2D^+$ | $pH_3O^+$ | HD | | 2.9e-09 | 0.00 | 0.0 |
| 351 | $oH_2O$ | $oHD_2^+$ | $oH_2DO^+$ | HD | | 2.4e-09 | 0.00 | 0.0 |
| 352 | $oH_2O$ | $pHD_2^+$ | $oH_2DO^+$ | HD | | 2.4e-09 | 0.00 | 0.0 |
| 353 | $oH_2O$ | $oHD_2^+$ | $pH_2DO^+$ | HD | | 4.8e-10 | 0.00 | 0.0 |
| 354 | $oH_2O$ | $pHD_2^+$ | $pH_2DO^+$ | HD | | 4.8e-10 | 0.00 | 0.0 |
| 355 | $pH_2O$ | $oHD_2^+$ | $oH_2DO^+$ | HD | | 1.4e-09 | 0.00 | 0.0 |
| 356 | $pH_2O$ | $pHD_2^+$ | $oH_2DO^+$ | HD | | 1.4e-09 | 0.00 | 0.0 |
| 357 | $pH_2O$ | $oHD_2^+$ | $pH_2DO^+$ | HD | | 1.4e-09 | 0.00 | 0.0 |
| 358 | $pH_2O$ | $pHD_2^+$ | $pH_2DO^+$ | HD | | 1.4e-09 | 0.00 | 0.0 |
| 359 | $oH_2O$ | $oHD_2^+$ | $oD_2$ | $oH_3O^+$ | | 9.5e-10 | 0.00 | 0.0 |
| 360 | $oH_2O$ | $pHD_2^+$ | $pD_2$ | $oH_3O^+$ | | 9.5e-10 | 0.00 | 0.0 |
| 361 | $oH_2O$ | $oHD_2^+$ | $oD_2$ | $pH_3O^+$ | | 4.8e-10 | 0.00 | 0.0 |
| 362 | $oH_2O$ | $pHD_2^+$ | $pD_2$ | $pH_3O^+$ | | 4.8e-10 | 0.00 | 0.0 |
| 363 | $pH_2O$ | $oHD_2^+$ | $oD_2$ | $pH_3O^+$ | | 1.4e-09 | 0.00 | 0.0 |
| 364 | $pH_2O$ | $pHD_2^+$ | $pD_2$ | $pH_3O^+$ | | 1.4e-09 | 0.00 | 0.0 |
| 365 | $oH_2O$ | $mD_3^+$ | $oD_2$ | $oH_2DO^+$ | | 4.3e-09 | 0.00 | 0.0 |
| 366 | $oH_2O$ | $pD_3^+$ | $pD_2$ | $oH_2DO^+$ | | 4.3e-09 | 0.00 | 0.0 |
| 367 | $oH_2O$ | $oD_3^+$ | $oD_2$ | $oH_2DO^+$ | | 2.1e-09 | 0.00 | 0.0 |
| 368 | $oH_2O$ | $oD_3^+$ | $pD_2$ | $oH_2DO^+$ | | 2.1e-09 | 0.00 | 0.0 |
| 369 | $pH_2O$ | $mD_3^+$ | $oD_2$ | $pH_2DO^+$ | | 4.3e-09 | 0.00 | 0.0 |
| 370 | $pH_2O$ | $pD_3^+$ | $pD_2$ | $pH_2DO^+$ | | 4.3e-09 | 0.00 | 0.0 |

**Table D6** – *continued* (part 6)

| # | Reactants | | Products | | | $\alpha$ | $\beta$ | $\gamma$ |
|---|---|---|---|---|---|---|---|---|
| 371 | $pH_2O$ | $oD_3^+$ | $oD_2$ | $pH_2DO^+$ | | 2.1e-09 | 0.00 | 0.0 |
| 372 | $pH_2O$ | $oD_3^+$ | $pD_2$ | $pH_2DO^+$ | | 2.1e-09 | 0.00 | 0.0 |
| 373 | $oH_2O$ | $oH_2D^+$ | $oH_2DO^+$ | $oH_2$ | | 1.0e-09 | 0.00 | 0.0 |
| 374 | $oH_2O$ | $oH_2D^+$ | $oH_2DO^+$ | $pH_2$ | | 1.6e-10 | 0.00 | 0.0 |
| 375 | $oH_2O$ | $oH_2D^+$ | $pH_2DO^+$ | $oH_2$ | | 1.6e-10 | 0.00 | 0.0 |
| 376 | $oH_2O$ | $oH_2D^+$ | $pH_2DO^+$ | $pH_2$ | | 7.9e-11 | 0.00 | 0.0 |
| 377 | $oH_2O$ | $pH_2D^+$ | $oH_2DO^+$ | $oH_2$ | | 4.8e-10 | 0.00 | 0.0 |
| 378 | $oH_2O$ | $pH_2D^+$ | $oH_2DO^+$ | $pH_2$ | | 4.8e-10 | 0.00 | 0.0 |
| 379 | $oH_2O$ | $pH_2D^+$ | $pH_2DO^+$ | $oH_2$ | | 4.8e-10 | 0.00 | 0.0 |
| 380 | $pH_2O$ | $oH_2D^+$ | $oH_2DO^+$ | $oH_2$ | | 4.8e-10 | 0.00 | 0.0 |
| 381 | $pH_2O$ | $oH_2D^+$ | $oH_2DO^+$ | $pH_2$ | | 4.8e-10 | 0.00 | 0.0 |
| 382 | $pH_2O$ | $oH_2D^+$ | $pH_2DO^+$ | $oH_2$ | | 4.8e-10 | 0.00 | 0.0 |
| 383 | $pH_2O$ | $pH_2D^+$ | $oH_2DO^+$ | $oH_2$ | | 7.2e-10 | 0.00 | 0.0 |
| 384 | $pH_2O$ | $pH_2D^+$ | $pH_2DO^+$ | $pH_2$ | | 7.2e-10 | 0.00 | 0.0 |
| 385 | HDO | $He^+$ | $OD^+$ | H | He | 2.3e-10 | -0.94 | 0.0 |
| 386 | HDO | $He^+$ | $HDO^+$ | He | | 4.9e-11 | -0.94 | 0.0 |
| 387 | HDO | $He^+$ | $H^+$ | OD | He | 1.6e-10 | -0.94 | 0.0 |
| 388 | HDO | $H^+$ | $HDO^+$ | H | | 8.2e-09 | 0.00 | 0.0 |
| 389 | HDO | $oH_3^+$ | $oH_2DO^+$ | $oH_2$ | | 3.2e-09 | 0.00 | 0.0 |
| 390 | HDO | $oH_3^+$ | $oH_2DO^+$ | $pH_2$ | | 5.4e-10 | 0.00 | 0.0 |
| 391 | HDO | $oH_3^+$ | $pH_2DO^+$ | $oH_2$ | | 5.4e-10 | 0.00 | 0.0 |
| 392 | HDO | $pH_3^+$ | $oH_2DO^+$ | $oH_2$ | | 1.6e-09 | 0.00 | 0.0 |
| 393 | HDO | $pH_3^+$ | $oH_2DO^+$ | $pH_2$ | | 1.1e-09 | 0.00 | 0.0 |
| 394 | HDO | $pH_3^+$ | $pH_2DO^+$ | $oH_2$ | | 1.1e-09 | 0.00 | 0.0 |
| 395 | HDO | $pH_3^+$ | $pH_2DO^+$ | $pH_2$ | | 5.4e-10 | 0.00 | 0.0 |
| 396 | HDO | $oH_2D^+$ | $oH_2DO^+$ | HD | | 2.4e-09 | 0.00 | 0.0 |
| 397 | HDO | $oH_2D^+$ | $pH_2DO^+$ | HD | | 4.8e-10 | 0.00 | 0.0 |
| 398 | HDO | $pH_2D^+$ | $oH_2DO^+$ | HD | | 1.4e-09 | 0.00 | 0.0 |
| 399 | HDO | $pH_2D^+$ | $pH_2DO^+$ | HD | | 1.4e-09 | 0.00 | 0.0 |
| 400 | HDO | $oH_2D^+$ | $oHD_2O^+$ | $oH_2$ | | 7.9e-10 | 0.00 | 0.0 |
| 401 | HDO | $oH_2D^+$ | $pHD_2O^+$ | $oH_2$ | | 4.0e-10 | 0.00 | 0.0 |
| 402 | HDO | $oH_2D^+$ | $oHD_2O^+$ | $pH_2$ | | 1.6e-10 | 0.00 | 0.0 |
| 403 | HDO | $oH_2D^+$ | $pHD_2O^+$ | $pH_2$ | | 7.9e-11 | 0.00 | 0.0 |
| 404 | HDO | $pH_2D^+$ | $oHD_2O^+$ | $oH_2$ | | 4.8e-10 | 0.00 | 0.0 |
| 405 | HDO | $pH_2D^+$ | $pHD_2O^+$ | $oH_2$ | | 2.4e-10 | 0.00 | 0.0 |
| 406 | HDO | $pH_2D^+$ | $oHD_2O^+$ | $pH_2$ | | 4.8e-10 | 0.00 | 0.0 |
| 407 | HDO | $pH_2D^+$ | $pHD_2O^+$ | $pH_2$ | | 2.4e-10 | 0.00 | 0.0 |
| 408 | HDO | $oHD_2^+$ | $oHD_2O^+$ | HD | | 2.2e-09 | 0.00 | 0.0 |
| 409 | HDO | $oHD_2^+$ | $pHD_2O^+$ | HD | | 6.4e-10 | 0.00 | 0.0 |
| 410 | HDO | $pHD_2^+$ | $oHD_2O^+$ | HD | | 1.3e-09 | 0.00 | 0.0 |
| 411 | HDO | $pHD_2^+$ | $pHD_2O^+$ | HD | | 1.6e-09 | 0.00 | 0.0 |
| 412 | HDO | $oHD_2^+$ | $oD_2$ | $oH_2DO^+$ | | 8.3e-10 | 0.00 | 0.0 |
| 413 | HDO | $oHD_2^+$ | $pD_2$ | $oH_2DO^+$ | | 2.4e-10 | 0.00 | 0.0 |
| 414 | HDO | $pHD_2^+$ | $oD_2$ | $oH_2DO^+$ | | 4.8e-10 | 0.00 | 0.0 |
| 415 | HDO | $pHD_2^+$ | $pD_2$ | $oH_2DO^+$ | | 6.0e-10 | 0.00 | 0.0 |
| 416 | HDO | $oHD_2^+$ | $oD_2$ | $pH_2DO^+$ | | 2.8e-10 | 0.00 | 0.0 |
| 417 | HDO | $oHD_2^+$ | $pD_2$ | $pH_2DO^+$ | | 7.9e-11 | 0.00 | 0.0 |
| 418 | HDO | $pHD_2^+$ | $oD_2$ | $pH_2DO^+$ | | 1.6e-10 | 0.00 | 0.0 |
| 419 | HDO | $pHD_2^+$ | $pD_2$ | $pH_2DO^+$ | | 2.0e-10 | 0.00 | 0.0 |
| 420 | HDO | $mD_3^+$ | $oHD_2O^+$ | $oD_2$ | | 2.9e-09 | 0.00 | 0.0 |
| 421 | HDO | $mD_3^+$ | $oH_2DO^+$ | $pD_2$ | | 7.2e-10 | 0.00 | 0.0 |
| 422 | HDO | $mD_3^+$ | $pH_2DO^+$ | $oD_2$ | | 7.2e-10 | 0.00 | 0.0 |
| 423 | HDO | $pD_3^+$ | $oHD_2O^+$ | $pD_2$ | | 1.4e-09 | 0.00 | 0.0 |
| 424 | HDO | $pD_3^+$ | $pHD_2O^+$ | $oD_2$ | | 1.4e-09 | 0.00 | 0.0 |
| 425 | HDO | $pD_3^+$ | $pHD_2O^+$ | $pD_2$ | | 1.4e-09 | 0.00 | 0.0 |
| 426 | HDO | $oD_3^+$ | $oHD_2O^+$ | $oD_2$ | | 1.4e-09 | 0.00 | 0.0 |
| 427 | HDO | $oD_3^+$ | $oHD_2O^+$ | $pD_2$ | | 1.1e-09 | 0.00 | 0.0 |
| 428 | HDO | $oD_3^+$ | $pHD_2O^+$ | $oD_2$ | | 1.1e-09 | 0.00 | 0.0 |
| 429 | HDO | $oD_3^+$ | $pHD_2O^+$ | $pD_2$ | | 7.2e-10 | 0.00 | 0.0 |
| 430 | $oD_2O$ | $He^+$ | $OD^+$ | D | He | 2.3e-10 | -0.94 | 0.0 |
| 431 | $pD_2O$ | $He^+$ | $OD^+$ | D | He | 2.3e-10 | -0.94 | 0.0 |
| 432 | $oD_2O$ | $He^+$ | $oD_2O^+$ | He | | 4.9e-11 | -0.94 | 0.0 |
| 433 | $pD_2O$ | $He^+$ | $pD_2O^+$ | He | | 4.9e-11 | -0.94 | 0.0 |
| 434 | $oD_2O$ | $He^+$ | $D^+$ | OD | He | 1.6e-10 | -0.94 | 0.0 |
| 435 | $pD_2O$ | $He^+$ | $D^+$ | OD | He | 1.6e-10 | -0.94 | 0.0 |
| 436 | $oD_2O$ | $H^+$ | $oD_2O^+$ | H | | 8.2e-09 | 0.00 | 0.0 |
| 437 | $pD_2O$ | $H^+$ | $pD_2O^+$ | H | | 8.2e-09 | 0.00 | 0.0 |
| 438 | $oD_2O$ | $oH_3^+$ | $oHD_2O^+$ | $oH_2$ | | 4.3e-09 | 0.00 | 0.0 |
| 439 | $pD_2O$ | $oH_3^+$ | $pHD_2O^+$ | $oH_2$ | | 4.3e-09 | 0.00 | 0.0 |
| 440 | $oD_2O$ | $pH_3^+$ | $oHD_2O^+$ | $oH_2$ | | 2.1e-09 | 0.00 | 0.0 |
| 441 | $pD_2O$ | $pH_3^+$ | $pHD_2O^+$ | $oH_2$ | | 2.1e-09 | 0.00 | 0.0 |
| 442 | $oD_2O$ | $pH_3^+$ | $oHD_2O^+$ | $pH_2$ | | 2.1e-09 | 0.00 | 0.0 |
| 443 | $pD_2O$ | $pH_3^+$ | $pHD_2O^+$ | $pH_2$ | | 2.1e-09 | 0.00 | 0.0 |
| 444 | $oD_2O$ | $oH_2D^+$ | $oHD_2O^+$ | HD | | 2.2e-09 | 0.00 | 0.0 |

**Table D6** – *continued* (part 7)

| # | Reactants | | Products | | $\alpha$ | $\beta$ | $\gamma$ |
|---|---|---|---|---|---|---|---|
| 445 | $oD_2O$ | $oH_2D^+$ | $pHD_2O^+$ | HD | 6.4e-10 | 0.00 | 0.0 |
| 446 | $pD_2O$ | $oH_2D^+$ | $oHD_2O^+$ | HD | 1.3e-09 | 0.00 | 0.0 |
| 447 | $pD_2O$ | $oH_2D^+$ | $pHD_2O^+$ | HD | 1.6e-09 | 0.00 | 0.0 |
| 448 | $oD_2O$ | $pH_2D^+$ | $oHD_2O^+$ | HD | 2.2e-09 | 0.00 | 0.0 |
| 449 | $oD_2O$ | $pH_2D^+$ | $pHD_2O^+$ | HD | 6.4e-10 | 0.00 | 0.0 |
| 450 | $pD_2O$ | $pH_2D^+$ | $oHD_2O^+$ | HD | 1.3e-09 | 0.00 | 0.0 |
| 451 | $pD_2O$ | $pH_2D^+$ | $pHD_2O^+$ | HD | 1.6e-09 | 0.00 | 0.0 |
| 452 | $oD_2O$ | $oH_2D^+$ | $mD_3O^+$ | $oH_2$ | 7.9e-10 | 0.00 | 0.0 |
| 453 | $oD_2O$ | $oH_2D^+$ | $oD_3O^+$ | $oH_2$ | 6.4e-10 | 0.00 | 0.0 |
| 454 | $pD_2O$ | $oH_2D^+$ | $pD_3O^+$ | $oH_2$ | 1.6e-09 | 0.00 | 0.0 |
| 455 | $pD_2O$ | $oH_2D^+$ | $oD_3O^+$ | $oH_2$ | 1.3e-09 | 0.00 | 0.0 |
| 456 | $oD_2O$ | $pH_2D^+$ | $mD_3O^+$ | $pH_2$ | 7.9e-10 | 0.00 | 0.0 |
| 457 | $oD_2O$ | $pH_2D^+$ | $oD_3O^+$ | $pH_2$ | 6.4e-10 | 0.00 | 0.0 |
| 458 | $pD_2O$ | $pH_2D^+$ | $pD_3O^+$ | $pH_2$ | 1.6e-09 | 0.00 | 0.0 |
| 459 | $pD_2O$ | $pH_2D^+$ | $oD_3O^+$ | $pH_2$ | 1.3e-09 | 0.00 | 0.0 |
| 460 | $oD_2O$ | $mD_3^+$ | $mD_3O^+$ | $oD_2$ | 2.1e-09 | 0.00 | 0.0 |
| 461 | $oD_2O$ | $mD_3^+$ | $mD_3O^+$ | $pD_2$ | 4.3e-10 | 0.00 | 0.0 |
| 462 | $oD_2O$ | $mD_3^+$ | $oD_3O^+$ | $oD_2$ | 1.3e-09 | 0.00 | 0.0 |
| 463 | $oD_2O$ | $mD_3^+$ | $oD_3O^+$ | $pD_2$ | 4.3e-10 | 0.00 | 0.0 |
| 464 | $pD_2O$ | $mD_3^+$ | $mD_3O^+$ | $oD_2$ | 8.6e-10 | 0.00 | 0.0 |
| 465 | $pD_2O$ | $mD_3^+$ | $mD_3O^+$ | $pD_2$ | 1.0e-09 | 0.00 | 0.0 |
| 466 | $pD_2O$ | $mD_3^+$ | $pD_3O^+$ | $oD_2$ | 1.4e-10 | 0.00 | 0.0 |
| 467 | $pD_2O$ | $mD_3^+$ | $oD_3O^+$ | $oD_2$ | 2.0e-09 | 0.00 | 0.0 |
| 468 | $pD_2O$ | $mD_3^+$ | $oD_3O^+$ | $pD_2$ | 2.9e-10 | 0.00 | 0.0 |
| 469 | $oD_2O$ | $pD_3^+$ | $mD_3O^+$ | $pD_2$ | 7.2e-10 | 0.00 | 0.0 |
| 470 | $oD_2O$ | $pD_3^+$ | $pD_3O^+$ | $oD_2$ | 7.2e-10 | 0.00 | 0.0 |
| 471 | $oD_2O$ | $pD_3^+$ | $oD_3O^+$ | $oD_2$ | 1.4e-09 | 0.00 | 0.0 |
| 472 | $oD_2O$ | $pD_3^+$ | $oD_3O^+$ | $pD_2$ | 1.4e-09 | 0.00 | 0.0 |
| 473 | $pD_2O$ | $pD_3^+$ | $pD_3O^+$ | $pD_2$ | 8.6e-10 | 0.00 | 0.0 |
| 474 | $pD_2O$ | $pD_3^+$ | $oD_3O^+$ | $oD_2$ | 1.7e-09 | 0.00 | 0.0 |
| 475 | $pD_2O$ | $pD_3^+$ | $oD_3O^+$ | $pD_2$ | 1.7e-09 | 0.00 | 0.0 |
| 476 | $oD_2O$ | $oD_3^+$ | $mD_3O^+$ | $oD_2$ | 8.1e-10 | 0.00 | 0.0 |
| 477 | $oD_2O$ | $oD_3^+$ | $mD_3O^+$ | $pD_2$ | 6.3e-10 | 0.00 | 0.0 |
| 478 | $oD_2O$ | $oD_3^+$ | $pD_3O^+$ | $oD_2$ | 9.0e-11 | 0.00 | 0.0 |
| 479 | $oD_2O$ | $oD_3^+$ | $pD_3O^+$ | $pD_2$ | 5.4e-11 | 0.00 | 0.0 |
| 480 | $oD_2O$ | $oD_3^+$ | $oD_3O^+$ | $oD_2$ | 1.9e-09 | 0.00 | 0.0 |
| 481 | $oD_2O$ | $oD_3^+$ | $oD_3O^+$ | $pD_2$ | 8.2e-10 | 0.00 | 0.0 |
| 482 | $pD_2O$ | $oD_3^+$ | $mD_3O^+$ | $oD_2$ | 5.4e-10 | 0.00 | 0.0 |
| 483 | $pD_2O$ | $oD_3^+$ | $mD_3O^+$ | $pD_2$ | 1.8e-10 | 0.00 | 0.0 |
| 484 | $pD_2O$ | $oD_3^+$ | $pD_3O^+$ | $oD_2$ | 1.8e-10 | 0.00 | 0.0 |
| 485 | $pD_2O$ | $oD_3^+$ | $pD_3O^+$ | $pD_2$ | 1.1e-10 | 0.00 | 0.0 |
| 486 | $pD_2O$ | $oD_3^+$ | $oD_3O^+$ | $oD_2$ | 1.6e-09 | 0.00 | 0.0 |
| 487 | $pD_2O$ | $oD_3^+$ | $oD_3O^+$ | $pD_2$ | 1.6e-09 | 0.00 | 0.0 |
| 488 | $oD_2O$ | $oHD_2^+$ | $mD_3O^+$ | HD | 1.6e-09 | 0.00 | 0.0 |
| 489 | $oD_2O$ | $oHD_2^+$ | $oD_3O^+$ | HD | 1.3e-09 | 0.00 | 0.0 |
| 490 | $oD_2O$ | $pHD_2^+$ | $mD_3O^+$ | HD | 8.0e-10 | 0.00 | 0.0 |
| 491 | $oD_2O$ | $pHD_2^+$ | $pD_3O^+$ | HD | 1.6e-10 | 0.00 | 0.0 |
| 492 | $oD_2O$ | $pHD_2^+$ | $oD_3O^+$ | HD | 1.9e-09 | 0.00 | 0.0 |
| 493 | $pD_2O$ | $oHD_2^+$ | $mD_3O^+$ | HD | 8.0e-10 | 0.00 | 0.0 |
| 494 | $pD_2O$ | $oHD_2^+$ | $pD_3O^+$ | HD | 1.6e-10 | 0.00 | 0.0 |
| 495 | $pD_2O$ | $oHD_2^+$ | $oD_3O^+$ | HD | 1.9e-09 | 0.00 | 0.0 |
| 496 | $pD_2O$ | $pHD_2^+$ | $pD_3O^+$ | HD | 3.2e-10 | 0.00 | 0.0 |
| 497 | $pD_2O$ | $pHD_2^+$ | $oD_3O^+$ | HD | 2.6e-09 | 0.00 | 0.0 |
| 498 | $oD_2O$ | $oHD_2^+$ | $oHD_2O^+$ | $oD_2$ | 9.1e-10 | 0.00 | 0.0 |
| 499 | $oD_2O$ | $oHD_2^+$ | $oHD_2O^+$ | $pD_2$ | 2.0e-10 | 0.00 | 0.0 |
| 500 | $oD_2O$ | $oHD_2^+$ | $pHD_2O^+$ | $oD_2$ | 2.0e-10 | 0.00 | 0.0 |
| 501 | $oD_2O$ | $oHD_2^+$ | $pHD_2O^+$ | $pD_2$ | 1.2e-10 | 0.00 | 0.0 |
| 502 | $oD_2O$ | $pHD_2^+$ | $oHD_2O^+$ | $oD_2$ | 4.0e-10 | 0.00 | 0.0 |
| 503 | $oD_2O$ | $pHD_2^+$ | $oHD_2O^+$ | $pD_2$ | 4.8e-10 | 0.00 | 0.0 |
| 504 | $oD_2O$ | $pHD_2^+$ | $pHD_2O^+$ | $oD_2$ | 4.8e-10 | 0.00 | 0.0 |
| 505 | $oD_2O$ | $pHD_2^+$ | $pHD_2O^+$ | $pD_2$ | 7.9e-11 | 0.00 | 0.0 |
| 506 | $pD_2O$ | $oHD_2^+$ | $oHD_2O^+$ | $oD_2$ | 4.0e-10 | 0.00 | 0.0 |
| 507 | $pD_2O$ | $oHD_2^+$ | $oHD_2O^+$ | $pD_2$ | 4.8e-10 | 0.00 | 0.0 |
| 508 | $pD_2O$ | $oHD_2^+$ | $pHD_2O^+$ | $oD_2$ | 4.8e-10 | 0.00 | 0.0 |
| 509 | $pD_2O$ | $oHD_2^+$ | $pHD_2O^+$ | $pD_2$ | 7.9e-11 | 0.00 | 0.0 |
| 510 | $pD_2O$ | $oHD_2^+$ | $oHD_2O^+$ | $oD_2$ | 4.8e-10 | 0.00 | 0.0 |
| 511 | $pD_2O$ | $oHD_2^+$ | $oHD_2O^+$ | $pD_2$ | 1.6e-10 | 0.00 | 0.0 |
| 512 | $pD_2O$ | $pHD_2^+$ | $pHD_2O^+$ | $oD_2$ | 1.6e-10 | 0.00 | 0.0 |
| 513 | $pD_2O$ | $pHD_2^+$ | $pHD_2O^+$ | $pD_2$ | 6.4e-10 | 0.00 | 0.0 |
| 514 | $OH^+$ | $oH_2$ | $oH_2O^+$ | H | 8.4e-10 | 0.00 | 0.0 |
| 515 | $OH^+$ | $oH_2$ | $pH_2O^+$ | H | 1.7e-10 | 0.00 | 0.0 |
| 516 | $OH^+$ | $pH_2$ | $oH_2O^+$ | H | 5.1e-10 | 0.00 | 0.0 |
| 517 | $OH^+$ | $pH_2$ | $pH_2O^+$ | H | 5.1e-10 | 0.00 | 0.0 |
| 518 | $OH^+$ | HD | $HDO^+$ | H | 5.1e-10 | 0.00 | 0.0 |

*Chemical models of collapsing prestellar sources* 41**Table D6** – *continued* (part 8)

| # | Reactants | | Products | | | $\alpha$ | $\beta$ | $\gamma$ |
|---|---|---|---|---|---|---|---|---|
| 519 | OH$^+$ | HD | oH$_2$O$^+$ | D | | 3.8e-10 | 0.00 | 0.0 |
| 520 | OH$^+$ | HD | pH$_2$O$^+$ | D | | 1.3e-10 | 0.00 | 0.0 |
| 521 | OD$^+$ | oH$_2$ | HDO$^+$ | H | | 1.0e-09 | 0.00 | 0.0 |
| 522 | OD$^+$ | pH$_2$ | HDO$^+$ | H | | 1.0e-09 | 0.00 | 0.0 |
| 523 | OD$^+$ | HD | oD$_2$O$^+$ | H | | 3.4e-10 | 0.00 | 0.0 |
| 524 | OD$^+$ | HD | pD$_2$O$^+$ | H | | 1.7e-10 | 0.00 | 0.0 |
| 525 | OD$^+$ | HD | HDO$^+$ | D | | 5.1e-10 | 0.00 | 0.0 |
| 526 | OD$^+$ | oD$_2$ | oD$_2$O$^+$ | D | | 8.6e-10 | 0.00 | 0.0 |
| 527 | OD$^+$ | oD$_2$ | pD$_2$O$^+$ | D | | 2.4e-10 | 0.00 | 0.0 |
| 528 | OD$^+$ | pD$_2$ | oD$_2$O$^+$ | D | | 4.9e-10 | 0.00 | 0.0 |
| 529 | OD$^+$ | pD$_2$ | pD$_2$O$^+$ | D | | 6.1e-10 | 0.00 | 0.0 |
| 530 | oH$_2$O$^+$ | oH$_2$ | oH$_3$O$^+$ | H | | 5.5e-10 | 0.00 | 0.0 |
| 531 | oH$_2$O$^+$ | oH$_2$ | pH$_3$O$^+$ | H | | 2.8e-10 | 0.00 | 0.0 |
| 532 | oH$_2$O$^+$ | pH$_2$ | oH$_3$O$^+$ | H | | 2.8e-10 | 0.00 | 0.0 |
| 533 | oH$_2$O$^+$ | pH$_2$ | pH$_3$O$^+$ | H | | 5.5e-10 | 0.00 | 0.0 |
| 534 | pH$_2$O$^+$ | oH$_2$ | oH$_3$O$^+$ | H | | 2.8e-10 | 0.00 | 0.0 |
| 535 | pH$_2$O$^+$ | oH$_2$ | pH$_3$O$^+$ | H | | 5.5e-10 | 0.00 | 0.0 |
| 536 | pH$_2$O$^+$ | pH$_2$ | pH$_3$O$^+$ | H | | 8.3e-10 | 0.00 | 0.0 |
| 537 | oH$_2$O$^+$ | HD | oH$_2$DO$^+$ | H | | 3.5e-10 | 0.00 | 0.0 |
| 538 | oH$_2$O$^+$ | HD | pH$_2$DO$^+$ | H | | 6.9e-11 | 0.00 | 0.0 |
| 539 | pH$_2$O$^+$ | HD | oH$_2$DO$^+$ | H | | 2.1e-10 | 0.00 | 0.0 |
| 540 | pH$_2$O$^+$ | HD | pH$_2$DO$^+$ | H | | 2.1e-10 | 0.00 | 0.0 |
| 541 | oH$_2$O$^+$ | HD | oH$_3$O$^+$ | D | | 2.8e-10 | 0.00 | 0.0 |
| 542 | oH$_2$O$^+$ | HD | pH$_3$O$^+$ | D | | 1.4e-10 | 0.00 | 0.0 |
| 543 | pH$_2$O$^+$ | HD | pH$_3$O$^+$ | D | | 4.2e-10 | 0.00 | 0.0 |
| 544 | HDO$^+$ | oH$_2$ | oH$_2$DO$^+$ | H | | 6.9e-10 | 0.00 | 0.0 |
| 545 | HDO$^+$ | oH$_2$ | pH$_2$DO$^+$ | H | | 1.4e-10 | 0.00 | 0.0 |
| 546 | HDO$^+$ | pH$_2$ | oH$_2$DO$^+$ | H | | 4.2e-10 | 0.00 | 0.0 |
| 547 | HDO$^+$ | pH$_2$ | pH$_2$DO$^+$ | H | | 4.2e-10 | 0.00 | 0.0 |
| 548 | HDO$^+$ | HD | oHD$_2$O$^+$ | H | | 2.8e-10 | 0.00 | 0.0 |
| 549 | HDO$^+$ | HD | pHD$_2$O$^+$ | H | | 1.4e-10 | 0.00 | 0.0 |
| 550 | HDO$^+$ | HD | oH$_2$DO$^+$ | D | | 3.1e-10 | 0.00 | 0.0 |
| 551 | HDO$^+$ | HD | pH$_2$DO$^+$ | D | | 1.0e-10 | 0.00 | 0.0 |
| 552 | HDO$^+$ | oD$_2$ | oHD$_2$O$^+$ | D | | 6.5e-10 | 0.00 | 0.0 |
| 553 | HDO$^+$ | oD$_2$ | pHD$_2$O$^+$ | D | | 1.8e-10 | 0.00 | 0.0 |
| 554 | HDO$^+$ | pD$_2$ | oHD$_2$O$^+$ | D | | 3.7e-10 | 0.00 | 0.0 |
| 555 | HDO$^+$ | pD$_2$ | pHD$_2$O$^+$ | D | | 4.6e-10 | 0.00 | 0.0 |
| 556 | oD$_2$O$^+$ | oD$_2$ | mD$_3$O$^+$ | D | | 4.6e-10 | 0.00 | 0.0 |
| 557 | oD$_2$O$^+$ | oD$_2$ | oD$_3$O$^+$ | D | | 3.7e-10 | 0.00 | 0.0 |
| 558 | oD$_2$O$^+$ | pD$_2$ | mD$_3$O$^+$ | D | | 2.3e-10 | 0.00 | 0.0 |
| 559 | oD$_2$O$^+$ | pD$_2$ | pD$_3$O$^+$ | D | | 4.6e-11 | 0.00 | 0.0 |
| 560 | oD$_2$O$^+$ | pD$_2$ | oD$_3$O$^+$ | D | | 5.5e-10 | 0.00 | 0.0 |
| 561 | pD$_2$O$^+$ | oD$_2$ | mD$_3$O$^+$ | D | | 2.3e-10 | 0.00 | 0.0 |
| 562 | pD$_2$O$^+$ | oD$_2$ | pD$_3$O$^+$ | D | | 4.6e-11 | 0.00 | 0.0 |
| 563 | pD$_2$O$^+$ | oD$_2$ | oD$_3$O$^+$ | D | | 5.5e-10 | 0.00 | 0.0 |
| 564 | pD$_2$O$^+$ | pD$_2$ | pD$_3$O$^+$ | D | | 9.2e-11 | 0.00 | 0.0 |
| 565 | pD$_2$O$^+$ | pD$_2$ | oD$_3$O$^+$ | D | | 7.4e-10 | 0.00 | 0.0 |
| 566 | oD$_2$O$^+$ | oH$_2$ | oHD$_2$O$^+$ | H | | 8.3e-10 | 0.00 | 0.0 |
| 567 | pD$_2$O$^+$ | oH$_2$ | pHD$_2$O$^+$ | H | | 8.3e-10 | 0.00 | 0.0 |
| 568 | oD$_2$O$^+$ | pH$_2$ | oHD$_2$O$^+$ | H | | 8.3e-10 | 0.00 | 0.0 |
| 569 | pD$_2$O$^+$ | pH$_2$ | pHD$_2$O$^+$ | H | | 8.3e-10 | 0.00 | 0.0 |
| 570 | oD$_2$O$^+$ | HD | oHD$_2$O$^+$ | D | | 3.2e-10 | 0.00 | 0.0 |
| 571 | oD$_2$O$^+$ | HD | pHD$_2$O$^+$ | D | | 9.2e-11 | 0.00 | 0.0 |
| 572 | pD$_2$O$^+$ | HD | oHD$_2$O$^+$ | D | | 1.8e-10 | 0.00 | 0.0 |
| 573 | pD$_2$O$^+$ | HD | pHD$_2$O$^+$ | D | | 2.3e-10 | 0.00 | 0.0 |
| 574 | oD$_2$O$^+$ | HD | mD$_3$O$^+$ | H | | 2.3e-10 | 0.00 | 0.0 |
| 575 | oD$_2$O$^+$ | HD | oD$_3$O$^+$ | H | | 1.8e-10 | 0.00 | 0.0 |
| 576 | pD$_2$O$^+$ | HD | pD$_3$O$^+$ | H | | 4.6e-11 | 0.00 | 0.0 |
| 577 | pD$_2$O$^+$ | HD | oD$_3$O$^+$ | H | | 3.7e-10 | 0.00 | 0.0 |
| 578 | oH$_3$O$^+$ | Fe | oH$_2$O | H | Fe$^+$ | 3.1e-09 | 0.00 | 0.0 |
| 579 | pH$_3$O$^+$ | Fe | oH$_2$O | H | Fe$^+$ | 1.6e-09 | 0.00 | 0.0 |
| 580 | pH$_3$O$^+$ | Fe | pH$_2$O | H | Fe$^+$ | 1.6e-09 | 0.00 | 0.0 |
| 581 | O$_2^+$ | Fe | Fe$^+$ | O$_2$ | | 1.1e-09 | 0.00 | 0.0 |
| 582 | O$^+$ | e$^-$ | O | $\gamma$ | | 3.4e-12 | -0.64 | 0.0 |
| 583 | O$_2^+$ | e$^-$ | O | O | | 2.0e-07 | -0.70 | 0.0 |
| 584 | OH$^+$ | e$^-$ | O | H | | 3.7e-08 | -0.50 | 0.0 |
| 585 | oH$_2$O$^+$ | e$^-$ | OH | H | | 7.8e-08 | -0.50 | 0.0 |
| 586 | pH$_2$O$^+$ | e$^-$ | OH | H | | 7.8e-08 | -0.50 | 0.0 |
| 587 | oH$_2$O$^+$ | e$^-$ | oH$_2$ | O | | 3.4e-08 | -0.50 | 0.0 |
| 588 | pH$_2$O$^+$ | e$^-$ | pH$_2$ | O | | 3.4e-08 | -0.50 | 0.0 |
| 589 | oH$_2$O$^+$ | e$^-$ | H | H | O | 1.5e-07 | -0.50 | 0.0 |
| 590 | pH$_2$O$^+$ | e$^-$ | H | H | O | 1.5e-07 | -0.50 | 0.0 |
| 591 | oH$_3$O$^+$ | e$^-$ | OH | H | H | 2.6e-07 | -0.50 | 0.0 |
| 592 | pH$_3$O$^+$ | e$^-$ | OH | H | H | 2.6e-07 | -0.50 | 0.0 |

MNRAS **000**, 1–**??** (2017)

**Table D6** – *continued* (part 9)

| # | Reactants | | Products | | | $\alpha$ | $\beta$ | $\gamma$ |
|---|---|---|---|---|---|---|---|---|
| 593 | $oH_3O^+$ | $e^-$ | $oH_2O$ | H | | 1.1e-07 | -0.50 | 0.0 |
| 594 | $pH_3O^+$ | $e^-$ | $oH_2O$ | H | | 5.4e-08 | -0.50 | 0.0 |
| 595 | $pH_3O^+$ | $e^-$ | $pH_2O$ | H | | 5.4e-08 | -0.50 | 0.0 |
| 596 | $oH_3O^+$ | $e^-$ | $oH_2$ | OH | | 6.0e-08 | -0.50 | 0.0 |
| 597 | $pH_3O^+$ | $e^-$ | $oH_2$ | OH | | 3.0e-08 | -0.50 | 0.0 |
| 598 | $pH_3O^+$ | $e^-$ | $pH_2$ | OH | | 3.0e-08 | -0.50 | 0.0 |
| 599 | $oH_3O^+$ | $e^-$ | $oH_2$ | H | O | 5.6e-09 | -0.50 | 0.0 |
| 600 | $pH_3O^+$ | $e^-$ | $oH_2$ | H | O | 2.8e-09 | -0.50 | 0.0 |
| 601 | $pH_3O^+$ | $e^-$ | $pH_2$ | H | O | 2.8e-09 | -0.50 | 0.0 |
| 602 | $OD^+$ | $e^-$ | O | D | | 3.7e-08 | -0.50 | 0.0 |
| 603 | $HDO^+$ | $e^-$ | OD | H | | 3.1e-08 | -0.50 | 0.0 |
| 604 | $HDO^+$ | $e^-$ | OH | D | | 1.5e-08 | -0.50 | 0.0 |
| 605 | $HDO^+$ | $e^-$ | O | HD | | 1.5e-08 | -0.50 | 0.0 |
| 606 | $HDO^+$ | $e^-$ | O | H | D | 8.9e-08 | -0.50 | 0.0 |
| 607 | $oD_2O^+$ | $e^-$ | OD | D | | 7.8e-08 | -0.50 | 0.0 |
| 608 | $pD_2O^+$ | $e^-$ | OD | D | | 7.8e-08 | -0.50 | 0.0 |
| 609 | $oD_2O^+$ | $e^-$ | $oD_2$ | O | | 3.4e-08 | -0.50 | 0.0 |
| 610 | $pD_2O^+$ | $e^-$ | $pD_2$ | O | | 3.4e-08 | -0.50 | 0.0 |
| 611 | $oD_2O^+$ | $e^-$ | D | D | O | 1.5e-07 | -0.50 | 0.0 |
| 612 | $pD_2O^+$ | $e^-$ | D | D | O | 1.5e-07 | -0.50 | 0.0 |
| 613 | $oH_2DO^+$ | $e^-$ | H | H | OD | 8.6e-08 | -0.50 | 0.0 |
| 614 | $pH_2DO^+$ | $e^-$ | H | H | OD | 8.6e-08 | -0.50 | 0.0 |
| 615 | $oH_2DO^+$ | $e^-$ | OH | H | D | 1.7e-07 | -0.50 | 0.0 |
| 616 | $pH_2DO^+$ | $e^-$ | OH | H | D | 1.7e-07 | -0.50 | 0.0 |
| 617 | $oH_2DO^+$ | $e^-$ | $oH_2O$ | D | | 3.6e-08 | -0.50 | 0.0 |
| 618 | $pH_2DO^+$ | $e^-$ | $pH_2O$ | D | | 3.6e-08 | -0.50 | 0.0 |
| 619 | $oH_2DO^+$ | $e^-$ | HDO | H | | 7.2e-08 | -0.50 | 0.0 |
| 620 | $pH_2DO^+$ | $e^-$ | HDO | H | | 7.2e-08 | -0.50 | 0.0 |
| 621 | $oH_2DO^+$ | $e^-$ | $oH_2$ | OD | | 2.0e-08 | -0.50 | 0.0 |
| 622 | $pH_2DO^+$ | $e^-$ | $pH_2$ | OD | | 2.0e-08 | -0.50 | 0.0 |
| 623 | $oH_2DO^+$ | $e^-$ | OH | HD | | 4.0e-08 | -0.50 | 0.0 |
| 624 | $pH_2DO^+$ | $e^-$ | OH | HD | | 4.0e-08 | -0.50 | 0.0 |
| 625 | $oH_2DO^+$ | $e^-$ | HD | H | O | 3.7e-09 | -0.50 | 0.0 |
| 626 | $pH_2DO^+$ | $e^-$ | HD | H | O | 3.7e-09 | -0.50 | 0.0 |
| 627 | $oH_2DO^+$ | $e^-$ | $oH_2$ | D | O | 1.9e-09 | -0.50 | 0.0 |
| 628 | $pH_2DO^+$ | $e^-$ | $pH_2$ | D | O | 1.9e-09 | -0.50 | 0.0 |
| 629 | $oHD_2O^+$ | $e^-$ | D | D | OH | 8.6e-08 | -0.50 | 0.0 |
| 630 | $pHD_2O^+$ | $e^-$ | D | D | OH | 8.6e-08 | -0.50 | 0.0 |
| 631 | $oHD_2O^+$ | $e^-$ | OD | D | H | 1.7e-07 | -0.50 | 0.0 |
| 632 | $pHD_2O^+$ | $e^-$ | OD | D | H | 1.7e-07 | -0.50 | 0.0 |
| 633 | $oHD_2O^+$ | $e^-$ | $oD_2O$ | H | | 3.6e-08 | -0.50 | 0.0 |
| 634 | $pHD_2O^+$ | $e^-$ | $pD_2O$ | H | | 3.6e-08 | -0.50 | 0.0 |
| 635 | $oHD_2O^+$ | $e^-$ | HDO | D | | 7.2e-08 | -0.50 | 0.0 |
| 636 | $pHD_2O^+$ | $e^-$ | HDO | D | | 7.2e-08 | -0.50 | 0.0 |
| 637 | $oHD_2O^+$ | $e^-$ | $oD_2$ | OH | | 2.0e-08 | -0.50 | 0.0 |
| 638 | $pHD_2O^+$ | $e^-$ | $pD_2$ | OH | | 2.0e-08 | -0.50 | 0.0 |
| 639 | $oHD_2O^+$ | $e^-$ | OD | HD | | 4.0e-08 | -0.50 | 0.0 |
| 640 | $pHD_2O^+$ | $e^-$ | OD | HD | | 4.0e-08 | -0.50 | 0.0 |
| 641 | $oHD_2O^+$ | $e^-$ | HD | D | O | 3.7e-09 | -0.50 | 0.0 |
| 642 | $pHD_2O^+$ | $e^-$ | HD | D | O | 3.7e-09 | -0.50 | 0.0 |
| 643 | $oHD_2O^+$ | $e^-$ | $oD_2$ | H | O | 1.9e-09 | -0.50 | 0.0 |
| 644 | $pHD_2O^+$ | $e^-$ | $pD_2$ | H | O | 1.9e-09 | -0.50 | 0.0 |
| 645 | $mD_3O^+$ | $e^-$ | OD | D | D | 2.6e-07 | -0.50 | 0.0 |
| 646 | $pD_3O^+$ | $e^-$ | OD | D | D | 2.6e-07 | -0.50 | 0.0 |
| 647 | $oD_3O^+$ | $e^-$ | OD | D | D | 2.6e-07 | -0.50 | 0.0 |
| 648 | $mD_3O^+$ | $e^-$ | $oD_2O$ | D | | 1.1e-07 | -0.50 | 0.0 |
| 649 | $pD_3O^+$ | $e^-$ | $pD_2O$ | D | | 1.1e-07 | -0.50 | 0.0 |
| 650 | $oD_3O^+$ | $e^-$ | $oD_2O$ | D | | 5.4e-08 | -0.50 | 0.0 |
| 651 | $oD_3O^+$ | $e^-$ | $pD_2O$ | D | | 5.4e-08 | -0.50 | 0.0 |
| 652 | $mD_3O^+$ | $e^-$ | $oD_2$ | OD | | 6.0e-08 | -0.50 | 0.0 |
| 653 | $pD_3O^+$ | $e^-$ | $pD_2$ | OD | | 6.0e-08 | -0.50 | 0.0 |
| 654 | $oD_3O^+$ | $e^-$ | $oD_2$ | OD | | 3.0e-08 | -0.50 | 0.0 |
| 655 | $oD_3O^+$ | $e^-$ | $pD_2$ | OD | | 3.0e-08 | -0.50 | 0.0 |
| 656 | $mD_3O^+$ | $e^-$ | $oD_2$ | D | O | 5.6e-09 | -0.50 | 0.0 |
| 657 | $pD_3O^+$ | $e^-$ | $pD_2$ | D | O | 5.6e-09 | -0.50 | 0.0 |
| 658 | $oD_3O^+$ | $e^-$ | $oD_2$ | D | O | 2.8e-09 | -0.50 | 0.0 |
| 659 | $oD_3O^+$ | $e^-$ | $pD_2$ | D | O | 2.8e-09 | -0.50 | 0.0 |
| 660 | OH | $\gamma$ | O | H | | 4.7e+02 | 0.00 | 0.0 |
| 661 | $OH^+$ | $\gamma$ | $O^+$ | H | | 8.6e+00 | 0.00 | 0.0 |
| 662 | $oH_2O$ | $\gamma$ | OH | H | | 1.0e+03 | 0.00 | 0.0 |
| 663 | $pH_2O$ | $\gamma$ | OH | H | | 1.0e+03 | 0.00 | 0.0 |
| 664 | $oH_2O$ | $\gamma$ | $oH_2O^+$ | $e^-$ | | 2.3e+01 | 0.00 | 0.0 |
| 665 | $pH_2O$ | $\gamma$ | $pH_2O^+$ | $e^-$ | | 2.3e+01 | 0.00 | 0.0 |
| 666 | $O_2$ | $\gamma$ | O | O | | 7.8e+02 | 0.00 | 0.0 |

**Table D6** – *continued* (part 10)

| # | Reactants | | Products | | | α | β | γ |
|---|---|---|---|---|---|---|---|---|
| 667 | $O_2$ | $\gamma_2$ | $O_2^+$ | $e^-$ | | 2.8e+01 | 0.00 | 0.0 |
| 668 | $O_2^+$ | $\gamma_2$ | $O^+$ | $O$ | | 7.0e+01 | 0.00 | 0.0 |
| 669 | $oH_3O^+$ | Gr | $oH_2O$ | H | $Gr^+$ | 3.7e-07 | 0.50 | 0.0 |
| 670 | $pH_3O^+$ | Gr | $oH_2O$ | H | $Gr^+$ | 1.8e-07 | 0.50 | 0.0 |
| 671 | $pH_3O^+$ | Gr | $pH_2O$ | H | $Gr^+$ | 1.8e-07 | 0.50 | 0.0 |
| 672 | $oH_3O^+$ | $Gr^-$ | $oH_2O$ | H | Gr | 3.7e-07 | 0.50 | 0.0 |
| 673 | $pH_3O^+$ | $Gr^-$ | $oH_2O$ | H | Gr | 1.8e-07 | 0.50 | 0.0 |
| 674 | $pH_3O^+$ | $Gr^-$ | $pH_2O$ | H | Gr | 1.8e-07 | 0.50 | 0.0 |
| 675 | $C^+$ | H | $CH^+$ | $\gamma$ | | 7.0e-17 | 0.00 | 0.0 |
| 676 | $C^+$ | $oH_2$ | $oCH_2^+$ | $\gamma$ | | 2.0e-16 | -1.30 | 23.0 |
| 677 | $C^+$ | $pH_2$ | $pCH_2^+$ | $\gamma$ | | 2.0e-16 | -1.30 | 23.0 |
| 678 | $C^+$ | CH | $C_2^+$ | H | | 3.8e-10 | 0.00 | 0.0 |
| 679 | $C^+$ | CH | $CH^+$ | C | | 3.8e-10 | 0.00 | 0.0 |
| 680 | $C^+$ | $oCH_2$ | $oCH_2^+$ | C | | 5.2e-10 | 0.00 | 0.0 |
| 681 | $C^+$ | $pCH_2$ | $pCH_2^+$ | C | | 5.2e-10 | 0.00 | 0.0 |
| 682 | $C^+$ | $oCH_2$ | $C_2H^+$ | H | | 5.2e-10 | 0.00 | 0.0 |
| 683 | $C^+$ | $pCH_2$ | $C_2H^+$ | H | | 5.2e-10 | 0.00 | 0.0 |
| 684 | $C^+$ | $mCH_4$ | $C_2H_3^+$ | H | | 9.8e-10 | 0.00 | 0.0 |
| 685 | $C^+$ | $pCH_4$ | $C_2H_3^+$ | H | | 9.8e-10 | 0.00 | 0.0 |
| 686 | $C^+$ | $oCH_4$ | $C_2H_3^+$ | H | | 9.8e-10 | 0.00 | 0.0 |
| 687 | $C^+$ | $C_2H$ | $C_3^+$ | H | | 1.0e-09 | 0.00 | 0.0 |
| 688 | $C^+$ | $C_2H_2$ | $C_3H^+$ | H | | 2.2e-09 | 0.00 | 0.0 |
| 689 | $C^+$ | Fe | $Fe^+$ | C | | 2.6e-09 | 0.00 | 0.0 |
| 690 | $CH^+$ | H | $oH_2$ | $C^+$ | | 1.1e-10 | 0.00 | 0.0 |
| 691 | $CH^+$ | H | $pH_2$ | $C^+$ | | 3.7e-11 | 0.00 | 0.0 |
| 692 | $CH^+$ | $oH_2$ | $oCH_2^+$ | H | | 1.0e-09 | 0.00 | 0.0 |
| 693 | $CH^+$ | $oH_2$ | $pCH_2^+$ | H | | 2.0e-10 | 0.00 | 0.0 |
| 694 | $CH^+$ | $pH_2$ | $oCH_2^+$ | H | | 6.0e-10 | 0.00 | 0.0 |
| 695 | $CH^+$ | $pH_2$ | $pCH_2^+$ | H | | 6.0e-10 | 0.00 | 0.0 |
| 696 | $oCH_2^+$ | $oH_2$ | $oCH_3^+$ | H | | 4.7e-10 | 0.00 | 0.0 |
| 697 | $oCH_2^+$ | $oH_2$ | $pCH_3^+$ | H | | 2.3e-10 | 0.00 | 0.0 |
| 698 | $oCH_2^+$ | $pH_2$ | $oCH_3^+$ | H | | 2.3e-10 | 0.00 | 0.0 |
| 699 | $oCH_2^+$ | $pH_2$ | $pCH_3^+$ | H | | 4.7e-10 | 0.00 | 0.0 |
| 700 | $pCH_2^+$ | $oH_2$ | $oCH_3^+$ | H | | 2.3e-10 | 0.00 | 0.0 |
| 701 | $pCH_2^+$ | $oH_2$ | $pCH_3^+$ | H | | 4.7e-10 | 0.00 | 0.0 |
| 702 | $pCH_2^+$ | $pH_2$ | $pCH_3^+$ | H | | 7.0e-10 | 0.00 | 0.0 |
| 703 | $oCH_3^+$ | $oH_2$ | $pCH_5^+$ | $\gamma$ | | 1.9e-16 | -2.30 | 21.5 |
| 704 | $oCH_3^+$ | $oH_2$ | $oCH_5^+$ | $\gamma$ | | 1.3e-16 | -2.30 | 21.5 |
| 705 | $oCH_3^+$ | $oH_2$ | $mCH_5^+$ | $\gamma$ | | 6.3e-17 | -2.30 | 21.5 |
| 706 | $oCH_3^+$ | $pH_2$ | $oCH_5^+$ | $\gamma$ | | 3.8e-16 | -2.30 | 21.5 |
| 707 | $pCH_3^+$ | $oH_2$ | $oCH_5^+$ | $\gamma$ | | 2.5e-16 | -2.30 | 21.5 |
| 708 | $pCH_3^+$ | $oH_2$ | $mCH_5^+$ | $\gamma$ | | 1.3e-16 | -2.30 | 21.5 |
| 709 | $pCH_3^+$ | $pH_2$ | $mCH_5^+$ | $\gamma$ | | 3.8e-16 | -2.30 | 21.5 |
| 710 | $mCH_4^+$ | H | $oCH_3^+$ | $oH_2$ | | 1.4e-10 | 0.00 | 0.0 |
| 711 | $mCH_4^+$ | H | $oCH_3^+$ | $pH_2$ | | 2.0e-11 | 0.00 | 0.0 |
| 712 | $mCH_4^+$ | H | $pCH_3^+$ | $oH_2$ | | 4.0e-11 | 0.00 | 0.0 |
| 713 | $pCH_4^+$ | H | $oCH_3^+$ | $oH_2$ | | 4.0e-11 | 0.00 | 0.0 |
| 714 | $pCH_4^+$ | H | $pCH_3^+$ | $oH_2$ | | 8.0e-11 | 0.00 | 0.0 |
| 715 | $pCH_4^+$ | H | $pCH_3^+$ | $pH_2$ | | 8.0e-11 | 0.00 | 0.0 |
| 716 | $oCH_4^+$ | H | $oCH_3^+$ | $oH_2$ | | 4.7e-11 | 0.00 | 0.0 |
| 717 | $oCH_4^+$ | H | $oCH_3^+$ | $pH_2$ | | 3.3e-11 | 0.00 | 0.0 |
| 718 | $oCH_4^+$ | H | $pCH_3^+$ | $oH_2$ | | 9.3e-11 | 0.00 | 0.0 |
| 719 | $oCH_4^+$ | H | $pCH_3^+$ | $pH_2$ | | 2.7e-11 | 0.00 | 0.0 |
| 720 | $mCH_4^+$ | $oH_2$ | $pCH_5^+$ | H | | 2.1e-11 | 0.00 | 0.0 |
| 721 | $mCH_4^+$ | $oH_2$ | $oCH_5^+$ | H | | 1.4e-11 | 0.00 | 0.0 |
| 722 | $mCH_4^+$ | $oH_2$ | $mCH_5^+$ | H | | 4.4e-12 | 0.00 | 0.0 |
| 723 | $mCH_4^+$ | $pH_2$ | $pCH_5^+$ | H | | 8.0e-12 | 0.00 | 0.0 |
| 724 | $mCH_4^+$ | $pH_2$ | $oCH_5^+$ | H | | 3.2e-11 | 0.00 | 0.0 |
| 725 | $pCH_4^+$ | $oH_2$ | $oCH_5^+$ | H | | 1.8e-11 | 0.00 | 0.0 |
| 726 | $pCH_4^+$ | $oH_2$ | $mCH_5^+$ | H | | 2.2e-11 | 0.00 | 0.0 |
| 727 | $pCH_4^+$ | $pH_2$ | $mCH_5^+$ | H | | 4.0e-11 | 0.00 | 0.0 |
| 728 | $oCH_4^+$ | $oH_2$ | $pCH_5^+$ | H | | 4.4e-12 | 0.00 | 0.0 |
| 729 | $oCH_4^+$ | $oH_2$ | $oCH_5^+$ | H | | 2.4e-11 | 0.00 | 0.0 |
| 730 | $oCH_4^+$ | $oH_2$ | $mCH_5^+$ | H | | 1.2e-11 | 0.00 | 0.0 |
| 731 | $oCH_4^+$ | $pH_2$ | $oCH_5^+$ | H | | 1.8e-11 | 0.00 | 0.0 |
| 732 | $oCH_4^+$ | $pH_2$ | $mCH_5^+$ | H | | 2.2e-11 | 0.00 | 0.0 |
| 733 | $C_2^+$ | $oH_2$ | $C_2H^+$ | H | | 1.4e-09 | 0.00 | 0.0 |
| 734 | $C_2^+$ | $pH_2$ | $C_2H^+$ | H | | 1.4e-09 | 0.00 | 0.0 |
| 735 | $C_2H^+$ | $oH_2$ | $C_2H_2^+$ | H | | 1.7e-09 | 0.00 | 0.0 |
| 736 | $C_2H^+$ | $pH_2$ | $C_2H_2^+$ | H | | 1.7e-09 | 0.00 | 0.0 |
| 737 | $C_3^+$ | $oH_2$ | $C_3H^+$ | H | | 3.0e-10 | 0.00 | 0.0 |
| 738 | $C_3^+$ | $pH_2$ | $C_3H^+$ | H | | 3.0e-10 | 0.00 | 0.0 |
| 739 | $C_3H^+$ | $oH_2$ | $C_3H_3^+$ | $\gamma$ | | 3.0e-13 | -1.00 | 0.0 |
| 740 | $C_3H^+$ | $pH_2$ | $C_3H_3^+$ | $\gamma$ | | 3.0e-13 | -1.00 | 0.0 |

**Table D6** – *continued* (part 11)

| # | Reactants | | Products | | | $\alpha$ | $\beta$ | $\gamma$ |
|---|---|---|---|---|---|---|---|---|
| 741 | C | CRP | C$^+$ | e$^-$ | | 1.8e+00 | 0.00 | 0.0 |
| 742 | C | H | CH | $\gamma$ | | 1.0e-17 | 0.00 | 0.0 |
| 743 | C | oH$_2^+$ | CH$^+$ | H | | 2.4e-09 | 0.00 | 0.0 |
| 744 | C | pH$_2^+$ | CH$^+$ | H | | 2.4e-09 | 0.00 | 0.0 |
| 745 | C | oH$_3^+$ | oH$_2$ | CH$^+$ | | 2.0e-09 | 0.00 | 0.0 |
| 746 | C | pH$_3^+$ | oH$_2$ | CH$^+$ | | 1.0e-09 | 0.00 | 0.0 |
| 747 | C | pH$_3^+$ | pH$_2$ | CH$^+$ | | 1.0e-09 | 0.00 | 0.0 |
| 748 | CH | H | oH$_2$ | C | | 9.3e-11 | 0.26 | 0.0 |
| 749 | CH | H | pH$_2$ | C | | 3.1e-11 | 0.26 | 0.0 |
| 750 | CH | He$^+$ | C$^+$ | H | He | 1.1e-09 | 0.00 | 0.0 |
| 751 | CH | H$^+$ | H | CH$^+$ | | 1.9e-09 | 0.00 | 0.0 |
| 752 | CH | oH$_2^+$ | oH$_2$ | CH$^+$ | | 7.1e-10 | 0.00 | 0.0 |
| 753 | CH | pH$_2^+$ | pH$_2$ | CH$^+$ | | 7.1e-10 | 0.00 | 0.0 |
| 754 | CH | oH$_2^+$ | oCH$_2^+$ | H | | 5.9e-10 | 0.00 | 0.0 |
| 755 | CH | oH$_2^+$ | pCH$_2^+$ | H | | 1.2e-10 | 0.00 | 0.0 |
| 756 | CH | pH$_2^+$ | oCH$_2^+$ | H | | 3.6e-10 | 0.00 | 0.0 |
| 757 | CH | pH$_2^+$ | pCH$_2^+$ | H | | 3.6e-10 | 0.00 | 0.0 |
| 758 | CH | oH$_3^+$ | oCH$_2^+$ | oH$_2$ | | 9.0e-10 | 0.00 | 0.0 |
| 759 | CH | oH$_3^+$ | oCH$_2^+$ | pH$_2$ | | 1.5e-10 | 0.00 | 0.0 |
| 760 | CH | oH$_3^+$ | pCH$_2^+$ | oH$_2$ | | 1.5e-10 | 0.00 | 0.0 |
| 761 | CH | pH$_3^+$ | oCH$_2^+$ | oH$_2$ | | 4.5e-10 | 0.00 | 0.0 |
| 762 | CH | pH$_3^+$ | oCH$_2^+$ | pH$_2$ | | 3.0e-10 | 0.00 | 0.0 |
| 763 | CH | pH$_3^+$ | pCH$_2^+$ | oH$_2$ | | 3.0e-10 | 0.00 | 0.0 |
| 764 | CH | pH$_3^+$ | pCH$_2^+$ | pH$_2$ | | 1.5e-10 | 0.00 | 0.0 |
| 765 | CH | oH$_2$D$^+$ | oCH$_2^+$ | HD | | 1.0e-09 | 0.00 | 0.0 |
| 766 | CH | oH$_2$D$^+$ | pCH$_2^+$ | HD | | 2.0e-10 | 0.00 | 0.0 |
| 767 | CH | pH$_2$D$^+$ | oCH$_2^+$ | HD | | 6.0e-10 | 0.00 | 0.0 |
| 768 | CH | pH$_2$D$^+$ | pCH$_2^+$ | HD | | 6.0e-10 | 0.00 | 0.0 |
| 769 | CH | oHD$_2^+$ | oD$_2$ | oCH$_2^+$ | | 9.0e-10 | 0.00 | 0.0 |
| 770 | CH | pHD$_2^+$ | pD$_2$ | oCH$_2^+$ | | 9.0e-10 | 0.00 | 0.0 |
| 771 | CH | oHD$_2^+$ | oD$_2$ | pCH$_2^+$ | | 3.0e-10 | 0.00 | 0.0 |
| 772 | CH | pHD$_2^+$ | pD$_2$ | pCH$_2^+$ | | 3.0e-10 | 0.00 | 0.0 |
| 773 | oCH$_2$ | H | oH$_2$ | CH | | 1.8e-10 | 0.00 | 0.0 |
| 774 | oCH$_2$ | H | pH$_2$ | CH | | 3.7e-11 | 0.00 | 0.0 |
| 775 | pCH$_2$ | H | oH$_2$ | CH | | 1.1e-10 | 0.00 | 0.0 |
| 776 | pCH$_2$ | H | pH$_2$ | CH | | 1.1e-10 | 0.00 | 0.0 |
| 777 | oCH$_2$ | He$^+$ | oH$_2$ | C$^+$ | He | 7.5e-10 | 0.00 | 0.0 |
| 778 | pCH$_2$ | He$^+$ | pH$_2$ | C$^+$ | He | 7.5e-10 | 0.00 | 0.0 |
| 779 | oCH$_2$ | He$^+$ | CH$^+$ | H | He | 7.5e-10 | 0.00 | 0.0 |
| 780 | pCH$_2$ | He$^+$ | CH$^+$ | H | He | 7.5e-10 | 0.00 | 0.0 |
| 781 | oCH$_2$ | H$^+$ | oH$_2$ | CH$^+$ | | 1.2e-09 | 0.00 | 0.0 |
| 782 | oCH$_2$ | H$^+$ | pH$_2$ | CH$^+$ | | 2.3e-10 | 0.00 | 0.0 |
| 783 | pCH$_2$ | H$^+$ | oH$_2$ | CH$^+$ | | 7.0e-10 | 0.00 | 0.0 |
| 784 | pCH$_2$ | H$^+$ | pH$_2$ | CH$^+$ | | 7.0e-10 | 0.00 | 0.0 |
| 785 | oCH$_2$ | H$^+$ | oCH$_2^+$ | H | | 1.4e-09 | 0.00 | 0.0 |
| 786 | pCH$_2$ | H$^+$ | pCH$_2^+$ | H | | 1.4e-09 | 0.00 | 0.0 |
| 787 | oCH$_2$ | oH$_2^+$ | oCH$_3^+$ | H | | 6.7e-10 | 0.00 | 0.0 |
| 788 | oCH$_2$ | oH$_2^+$ | pCH$_3^+$ | H | | 3.3e-10 | 0.00 | 0.0 |
| 789 | oCH$_2$ | pH$_2^+$ | oCH$_3^+$ | H | | 3.3e-10 | 0.00 | 0.0 |
| 790 | oCH$_2$ | pH$_2^+$ | pCH$_3^+$ | H | | 6.7e-10 | 0.00 | 0.0 |
| 791 | pCH$_2$ | oH$_2^+$ | oCH$_3^+$ | H | | 3.3e-10 | 0.00 | 0.0 |
| 792 | pCH$_2$ | oH$_2^+$ | pCH$_3^+$ | H | | 6.7e-10 | 0.00 | 0.0 |
| 793 | pCH$_2$ | pH$_2^+$ | pCH$_3^+$ | H | | 1.0e-09 | 0.00 | 0.0 |
| 794 | oCH$_2$ | oH$_2^+$ | oH$_2$ | oCH$_2^+$ | | 1.0e-09 | 0.00 | 0.0 |
| 795 | oCH$_2$ | pH$_2^+$ | oH$_2$ | pCH$_2^+$ | | 1.0e-09 | 0.00 | 0.0 |
| 796 | pCH$_2$ | oH$_2^+$ | pH$_2$ | oCH$_2^+$ | | 1.0e-09 | 0.00 | 0.0 |
| 797 | pCH$_2$ | pH$_2^+$ | pH$_2$ | pCH$_2^+$ | | 1.0e-09 | 0.00 | 0.0 |
| 798 | oCH$_2$ | oH$_3^+$ | oCH$_3^+$ | oH$_2$ | | 1.0e-09 | 0.00 | 0.0 |
| 799 | oCH$_2$ | oH$_3^+$ | oCH$_3^+$ | pH$_2$ | | 1.4e-10 | 0.00 | 0.0 |
| 800 | oCH$_2$ | oH$_3^+$ | pCH$_3^+$ | oH$_2$ | | 4.0e-10 | 0.00 | 0.0 |
| 801 | oCH$_2$ | oH$_3^+$ | pCH$_3^+$ | pH$_2$ | | 1.1e-10 | 0.00 | 0.0 |
| 802 | pCH$_2$ | oH$_3^+$ | oCH$_3^+$ | oH$_2$ | | 4.2e-10 | 0.00 | 0.0 |
| 803 | pCH$_2$ | oH$_3^+$ | oCH$_3^+$ | pH$_2$ | | 4.2e-10 | 0.00 | 0.0 |
| 804 | pCH$_2$ | oH$_3^+$ | pCH$_3^+$ | oH$_2$ | | 8.5e-10 | 0.00 | 0.0 |
| 805 | oCH$_2$ | pH$_3^+$ | oCH$_3^+$ | oH$_2$ | | 4.0e-10 | 0.00 | 0.0 |
| 806 | oCH$_2$ | pH$_3^+$ | oCH$_3^+$ | pH$_2$ | | 2.8e-10 | 0.00 | 0.0 |
| 807 | oCH$_2$ | pH$_3^+$ | pCH$_3^+$ | oH$_2$ | | 7.9e-10 | 0.00 | 0.0 |
| 808 | oCH$_2$ | pH$_3^+$ | pCH$_3^+$ | pH$_2$ | | 2.3e-10 | 0.00 | 0.0 |
| 809 | pCH$_2$ | pH$_3^+$ | oCH$_3^+$ | oH$_2$ | | 3.4e-10 | 0.00 | 0.0 |
| 810 | pCH$_2$ | pH$_3^+$ | pCH$_3^+$ | oH$_2$ | | 6.8e-10 | 0.00 | 0.0 |
| 811 | pCH$_2$ | pH$_3^+$ | pCH$_3^+$ | pH$_2$ | | 6.8e-10 | 0.00 | 0.0 |
| 812 | oCH$_2$ | oH$_2$D$^+$ | oCH$_3^+$ | HD | | 1.1e-09 | 0.00 | 0.0 |
| 813 | oCH$_2$ | oH$_2$D$^+$ | pCH$_3^+$ | HD | | 5.7e-10 | 0.00 | 0.0 |
| 814 | oCH$_2$ | pH$_2$D$^+$ | oCH$_3^+$ | HD | | 5.7e-10 | 0.00 | 0.0 |

**Table D6** – *continued* (part 12)

| # | Reactants | | Products | | | $\alpha$ | $\beta$ | $\gamma$ |
|---|---|---|---|---|---|---|---|---|
| 815 | $oCH_2$ | $pH_2D^+$ | $pCH_3^+$ | HD | | 1.1e-09 | 0.00 | 0.0 |
| 816 | $pCH_2$ | $oH_2D^+$ | $oCH_3^+$ | HD | | 5.7e-10 | 0.00 | 0.0 |
| 817 | $pCH_2$ | $oH_2D^+$ | $pCH_3^+$ | HD | | 1.1e-09 | 0.00 | 0.0 |
| 818 | $pCH_2$ | $pH_2D^+$ | $pCH_3^+$ | HD | | 1.7e-09 | 0.00 | 0.0 |
| 819 | $oCH_2$ | $oHD_2^+$ | $oD_2$ | $oCH_3^+$ | | 1.1e-09 | 0.00 | 0.0 |
| 820 | $oCH_2$ | $pHD_2^+$ | $pD_2$ | $oCH_3^+$ | | 1.1e-09 | 0.00 | 0.0 |
| 821 | $oCH_2$ | $oHD_2^+$ | $oD_2$ | $pCH_3^+$ | | 5.7e-10 | 0.00 | 0.0 |
| 822 | $oCH_2$ | $pHD_2^+$ | $pD_2$ | $pCH_3^+$ | | 5.7e-10 | 0.00 | 0.0 |
| 823 | $pCH_2$ | $oHD_2^+$ | $oD_2$ | $pCH_3^+$ | | 1.7e-09 | 0.00 | 0.0 |
| 824 | $pCH_2$ | $pHD_2^+$ | $pD_2$ | $pCH_3^+$ | | 1.7e-09 | 0.00 | 0.0 |
| 825 | $oCH_3$ | $He^+$ | $oH_2$ | $CH^+$ | He | 9.0e-10 | 0.00 | 0.0 |
| 826 | $pCH_3$ | $He^+$ | $oH_2$ | $CH^+$ | He | 4.5e-10 | 0.00 | 0.0 |
| 827 | $pCH_3$ | $He^+$ | $pH_2$ | $CH^+$ | He | 4.5e-10 | 0.00 | 0.0 |
| 828 | $oCH_3$ | $He^+$ | $oCH_2^+$ | H | He | 9.0e-10 | 0.00 | 0.0 |
| 829 | $pCH_3$ | $He^+$ | $oCH_2^+$ | H | He | 4.5e-10 | 0.00 | 0.0 |
| 830 | $pCH_3$ | $He^+$ | $pCH_2^+$ | H | He | 4.5e-10 | 0.00 | 0.0 |
| 831 | $oCH_3$ | $H^+$ | $oCH_3^+$ | H | | 3.4e-09 | 0.00 | 0.0 |
| 832 | $pCH_3$ | $H^+$ | $pCH_3^+$ | H | | 3.4e-09 | 0.00 | 0.0 |
| 833 | $oCH_3$ | $oH_3^+$ | $mCH_4^+$ | $oH_2$ | | 1.1e-09 | 0.00 | 0.0 |
| 834 | $oCH_3$ | $oH_3^+$ | $mCH_4^+$ | $pH_2$ | | 1.3e-10 | 0.00 | 0.0 |
| 835 | $oCH_3$ | $oH_3^+$ | $pCH_4^+$ | $oH_2$ | | 8.7e-11 | 0.00 | 0.0 |
| 836 | $oCH_3$ | $oH_3^+$ | $pCH_4^+$ | $pH_2$ | | 5.2e-11 | 0.00 | 0.0 |
| 837 | $oCH_3$ | $oH_3^+$ | $oCH_4^+$ | $oH_2$ | | 6.0e-10 | 0.00 | 0.0 |
| 838 | $oCH_3$ | $oH_3^+$ | $oCH_4^+$ | $pH_2$ | | 1.3e-10 | 0.00 | 0.0 |
| 839 | $oCH_3$ | $pH_3^+$ | $mCH_4^+$ | $oH_2$ | | 3.5e-10 | 0.00 | 0.0 |
| 840 | $oCH_3$ | $pH_3^+$ | $mCH_4^+$ | $pH_2$ | | 2.6e-10 | 0.00 | 0.0 |
| 841 | $oCH_3$ | $pH_3^+$ | $pCH_4^+$ | $oH_2$ | | 1.7e-10 | 0.00 | 0.0 |
| 842 | $oCH_3$ | $pH_3^+$ | $oCH_4^+$ | $oH_2$ | | 1.1e-09 | 0.00 | 0.0 |
| 843 | $oCH_3$ | $pH_3^+$ | $oCH_4^+$ | $pH_2$ | | 2.6e-10 | 0.00 | 0.0 |
| 844 | $pCH_3$ | $oH_3^+$ | $mCH_4^+$ | $oH_2$ | | 3.5e-10 | 0.00 | 0.0 |
| 845 | $pCH_3$ | $oH_3^+$ | $mCH_4^+$ | $pH_2$ | | 2.6e-10 | 0.00 | 0.0 |
| 846 | $pCH_3$ | $oH_3^+$ | $pCH_4^+$ | $oH_2$ | | 1.7e-10 | 0.00 | 0.0 |
| 847 | $pCH_3$ | $oH_3^+$ | $oCH_4^+$ | $oH_2$ | | 1.1e-09 | 0.00 | 0.0 |
| 848 | $pCH_3$ | $oH_3^+$ | $oCH_4^+$ | $pH_2$ | | 2.6e-10 | 0.00 | 0.0 |
| 849 | $pCH_3$ | $pH_3^+$ | $mCH_4^+$ | $oH_2$ | | 1.7e-10 | 0.00 | 0.0 |
| 850 | $pCH_3$ | $pH_3^+$ | $pCH_4^+$ | $oH_2$ | | 3.5e-10 | 0.00 | 0.0 |
| 851 | $pCH_3$ | $pH_3^+$ | $pCH_4^+$ | $pH_2$ | | 2.1e-10 | 0.00 | 0.0 |
| 852 | $pCH_3$ | $pH_3^+$ | $oCH_4^+$ | $oH_2$ | | 8.4e-10 | 0.00 | 0.0 |
| 853 | $pCH_3$ | $pH_3^+$ | $oCH_4^+$ | $pH_2$ | | 5.3e-10 | 0.00 | 0.0 |
| 854 | $oCH_3$ | $oH_2D^+$ | $mCH_4^+$ | HD | | 1.2e-09 | 0.00 | 0.0 |
| 855 | $oCH_3$ | $oH_2D^+$ | $pCH_4^+$ | HD | | 1.4e-10 | 0.00 | 0.0 |
| 856 | $oCH_3$ | $oH_2D^+$ | $oCH_4^+$ | HD | | 7.4e-10 | 0.00 | 0.0 |
| 857 | $oCH_3$ | $pH_2D^+$ | $mCH_4^+$ | HD | | 5.3e-10 | 0.00 | 0.0 |
| 858 | $oCH_3$ | $pH_2D^+$ | $oCH_4^+$ | HD | | 1.6e-09 | 0.00 | 0.0 |
| 859 | $pCH_3$ | $oH_2D^+$ | $mCH_4^+$ | HD | | 3.5e-10 | 0.00 | 0.0 |
| 860 | $pCH_3$ | $oH_2D^+$ | $pCH_4^+$ | HD | | 2.8e-10 | 0.00 | 0.0 |
| 861 | $pCH_3$ | $oH_2D^+$ | $oCH_4^+$ | HD | | 1.5e-09 | 0.00 | 0.0 |
| 862 | $pCH_3$ | $pH_2D^+$ | $pCH_4^+$ | HD | | 8.4e-10 | 0.00 | 0.0 |
| 863 | $pCH_3$ | $pH_2D^+$ | $oCH_4^+$ | HD | | 1.3e-09 | 0.00 | 0.0 |
| 864 | $oCH_3$ | $oHD_2^+$ | $oD_2$ | $mCH_4^+$ | | 1.3e-09 | 0.00 | 0.0 |
| 865 | $oCH_3$ | $pHD_2^+$ | $pD_2$ | $mCH_4^+$ | | 1.3e-09 | 0.00 | 0.0 |
| 866 | $oCH_3$ | $oHD_2^+$ | $oD_2$ | $oCH_4^+$ | | 7.9e-10 | 0.00 | 0.0 |
| 867 | $oCH_3$ | $pHD_2^+$ | $pD_2$ | $oCH_4^+$ | | 7.9e-10 | 0.00 | 0.0 |
| 868 | $pCH_3$ | $oHD_2^+$ | $oD_2$ | $pCH_4^+$ | | 5.3e-10 | 0.00 | 0.0 |
| 869 | $pCH_3$ | $pHD_2^+$ | $pD_2$ | $pCH_4^+$ | | 5.3e-10 | 0.00 | 0.0 |
| 870 | $pCH_3$ | $oHD_2^+$ | $oD_2$ | $oCH_4^+$ | | 1.6e-09 | 0.00 | 0.0 |
| 871 | $pCH_3$ | $pHD_2^+$ | $pD_2$ | $oCH_4^+$ | | 1.6e-09 | 0.00 | 0.0 |
| 872 | $mCH_4$ | $He^+$ | $oCH_3$ | $H^+$ | He | 4.0e-10 | 0.00 | 0.0 |
| 873 | $pCH_4$ | $He^+$ | $pCH_3$ | $H^+$ | He | 4.0e-10 | 0.00 | 0.0 |
| 874 | $oCH_4$ | $He^+$ | $oCH_3$ | $H^+$ | He | 1.3e-10 | 0.00 | 0.0 |
| 875 | $oCH_4$ | $He^+$ | $pCH_3$ | $H^+$ | He | 2.7e-10 | 0.00 | 0.0 |
| 876 | $mCH_4$ | $He^+$ | $oCH_2^+$ | $oH_2$ | He | 8.5e-10 | 0.00 | 0.0 |
| 877 | $pCH_4$ | $He^+$ | $oCH_2^+$ | $oH_2$ | He | 4.2e-10 | 0.00 | 0.0 |
| 878 | $pCH_4$ | $He^+$ | $pCH_2^+$ | $pH_2$ | He | 4.2e-10 | 0.00 | 0.0 |
| 879 | $oCH_4$ | $He^+$ | $oCH_2^+$ | $oH_2$ | He | 2.8e-10 | 0.00 | 0.0 |
| 880 | $oCH_4$ | $He^+$ | $oCH_2^+$ | $pH_2$ | He | 2.8e-10 | 0.00 | 0.0 |
| 881 | $oCH_4$ | $He^+$ | $pCH_2^+$ | $oH_2$ | He | 2.8e-10 | 0.00 | 0.0 |
| 882 | $mCH_4$ | $He^+$ | $oCH_3^+$ | H | He | 8.0e-11 | 0.00 | 0.0 |
| 883 | $pCH_4$ | $He^+$ | $pCH_3^+$ | H | He | 8.0e-11 | 0.00 | 0.0 |
| 884 | $oCH_4$ | $He^+$ | $oCH_3^+$ | H | He | 2.7e-11 | 0.00 | 0.0 |
| 885 | $oCH_4$ | $He^+$ | $pCH_3^+$ | H | He | 5.3e-11 | 0.00 | 0.0 |
| 886 | $mCH_4$ | $He^+$ | $mCH_4^+$ | He | | 1.6e-11 | 0.00 | 0.0 |
| 887 | $pCH_4$ | $He^+$ | $pCH_4^+$ | He | | 1.6e-11 | 0.00 | 0.0 |
| 888 | $oCH_4$ | $He^+$ | $oCH_4^+$ | He | | 1.6e-11 | 0.00 | 0.0 |

**Table D6** – *continued* (part 13)

| # | Reactants | | Products | | | $\alpha$ | $\beta$ | $\gamma$ |
|---|---|---|---|---|---|---|---|---|
| 889 | $mCH_4$ | $H^+$ | $oCH_3^+$ | $oH_2$ | | 1.6e-09 | 0.00 | 0.0 |
| 890 | $mCH_4$ | $H^+$ | $oCH_3^+$ | $pH_2$ | | 2.3e-10 | 0.00 | 0.0 |
| 891 | $mCH_4$ | $H^+$ | $pCH_3^+$ | $oH_2$ | | 4.6e-10 | 0.00 | 0.0 |
| 892 | $pCH_4$ | $H^+$ | $oCH_3^+$ | $oH_2$ | | 4.6e-10 | 0.00 | 0.0 |
| 893 | $pCH_4$ | $H^+$ | $pCH_3^+$ | $oH_2$ | | 9.1e-10 | 0.00 | 0.0 |
| 894 | $pCH_4$ | $H^+$ | $pCH_3^+$ | $pH_2$ | | 9.1e-10 | 0.00 | 0.0 |
| 895 | $oCH_4$ | $H^+$ | $oCH_3^+$ | $oH_2$ | | 5.3e-10 | 0.00 | 0.0 |
| 896 | $oCH_4$ | $H^+$ | $oCH_3^+$ | $pH_2$ | | 3.8e-10 | 0.00 | 0.0 |
| 897 | $oCH_4$ | $H^+$ | $pCH_3^+$ | $oH_2$ | | 1.1e-09 | 0.00 | 0.0 |
| 898 | $oCH_4$ | $H^+$ | $pCH_3^+$ | $pH_2$ | | 3.0e-10 | 0.00 | 0.0 |
| 899 | $mCH_4$ | $H^+$ | $mCH_4^+$ | $H$ | | 1.5e-09 | 0.00 | 0.0 |
| 900 | $pCH_4$ | $H^+$ | $pCH_4^+$ | $H$ | | 1.5e-09 | 0.00 | 0.0 |
| 901 | $oCH_4$ | $H^+$ | $oCH_4^+$ | $H$ | | 1.5e-09 | 0.00 | 0.0 |
| 902 | $mCH_4$ | $oH_3^+$ | $pCH_5^+$ | $oH_2$ | | 8.8e-10 | 0.00 | 0.0 |
| 903 | $mCH_4$ | $oH_3^+$ | $pCH_5^+$ | $pH_2$ | | 9.5e-11 | 0.00 | 0.0 |
| 904 | $mCH_4$ | $oH_3^+$ | $oCH_5^+$ | $oH_2$ | | 5.4e-10 | 0.00 | 0.0 |
| 905 | $mCH_4$ | $oH_3^+$ | $oCH_5^+$ | $pH_2$ | | 1.1e-10 | 0.00 | 0.0 |
| 906 | $mCH_4$ | $oH_3^+$ | $mCH_5^+$ | $oH_2$ | | 2.0e-10 | 0.00 | 0.0 |
| 907 | $mCH_4$ | $oH_3^+$ | $mCH_5^+$ | $pH_2$ | | 6.8e-11 | 0.00 | 0.0 |
| 908 | $pCH_4$ | $oH_3^+$ | $pCH_5^+$ | $oH_2$ | | 1.4e-10 | 0.00 | 0.0 |
| 909 | $pCH_4$ | $oH_3^+$ | $oCH_5^+$ | $oH_2$ | | 5.4e-10 | 0.00 | 0.0 |
| 910 | $pCH_4$ | $oH_3^+$ | $oCH_5^+$ | $pH_2$ | | 5.4e-10 | 0.00 | 0.0 |
| 911 | $pCH_4$ | $oH_3^+$ | $mCH_5^+$ | $oH_2$ | | 6.8e-10 | 0.00 | 0.0 |
| 912 | $oCH_4$ | $oH_3^+$ | $pCH_5^+$ | $oH_2$ | | 2.0e-10 | 0.00 | 0.0 |
| 913 | $oCH_4$ | $oH_3^+$ | $pCH_5^+$ | $pH_2$ | | 1.6e-10 | 0.00 | 0.0 |
| 914 | $oCH_4$ | $oH_3^+$ | $oCH_5^+$ | $oH_2$ | | 9.0e-10 | 0.00 | 0.0 |
| 915 | $oCH_4$ | $oH_3^+$ | $oCH_5^+$ | $pH_2$ | | 1.8e-10 | 0.00 | 0.0 |
| 916 | $oCH_4$ | $oH_3^+$ | $mCH_5^+$ | $oH_2$ | | 3.4e-10 | 0.00 | 0.0 |
| 917 | $oCH_4$ | $oH_3^+$ | $mCH_5^+$ | $pH_2$ | | 1.1e-10 | 0.00 | 0.0 |
| 918 | $mCH_4$ | $pH_3^+$ | $pCH_5^+$ | $oH_2$ | | 2.4e-10 | 0.00 | 0.0 |
| 919 | $mCH_4$ | $pH_3^+$ | $pCH_5^+$ | $pH_2$ | | 1.9e-10 | 0.00 | 0.0 |
| 920 | $mCH_4$ | $pH_3^+$ | $oCH_5^+$ | $oH_2$ | | 9.8e-10 | 0.00 | 0.0 |
| 921 | $mCH_4$ | $pH_3^+$ | $oCH_5^+$ | $pH_2$ | | 2.2e-10 | 0.00 | 0.0 |
| 922 | $mCH_4$ | $pH_3^+$ | $mCH_5^+$ | $oH_2$ | | 2.7e-10 | 0.00 | 0.0 |
| 923 | $pCH_4$ | $pH_3^+$ | $oCH_5^+$ | $oH_2$ | | 5.4e-10 | 0.00 | 0.0 |
| 924 | $pCH_4$ | $pH_3^+$ | $mCH_5^+$ | $oH_2$ | | 6.8e-10 | 0.00 | 0.0 |
| 925 | $pCH_4$ | $pH_3^+$ | $mCH_5^+$ | $pH_2$ | | 6.8e-10 | 0.00 | 0.0 |
| 926 | $oCH_4$ | $pH_3^+$ | $pCH_5^+$ | $oH_2$ | | 6.2e-11 | 0.00 | 0.0 |
| 927 | $oCH_4$ | $pH_3^+$ | $oCH_5^+$ | $oH_2$ | | 5.4e-10 | 0.00 | 0.0 |
| 928 | $oCH_4$ | $pH_3^+$ | $oCH_5^+$ | $pH_2$ | | 2.5e-10 | 0.00 | 0.0 |
| 929 | $oCH_4$ | $pH_3^+$ | $mCH_5^+$ | $oH_2$ | | 6.8e-10 | 0.00 | 0.0 |
| 930 | $oCH_4$ | $pH_3^+$ | $mCH_5^+$ | $pH_2$ | | 3.7e-10 | 0.00 | 0.0 |
| 931 | $mCH_4$ | $oH_2D^+$ | $pCH_5^+$ | $HD$ | | 1.0e-09 | 0.00 | 0.0 |
| 932 | $mCH_4$ | $oH_2D^+$ | $oCH_5^+$ | $HD$ | | 6.8e-10 | 0.00 | 0.0 |
| 933 | $mCH_4$ | $oH_2D^+$ | $mCH_5^+$ | $HD$ | | 2.1e-10 | 0.00 | 0.0 |
| 934 | $mCH_4$ | $pH_2D^+$ | $pCH_5^+$ | $HD$ | | 3.8e-10 | 0.00 | 0.0 |
| 935 | $mCH_4$ | $pH_2D^+$ | $oCH_5^+$ | $HD$ | | 1.5e-09 | 0.00 | 0.0 |
| 936 | $pCH_4$ | $oH_2D^+$ | $oCH_5^+$ | $HD$ | | 8.4e-10 | 0.00 | 0.0 |
| 937 | $pCH_4$ | $oH_2D^+$ | $mCH_5^+$ | $HD$ | | 1.1e-09 | 0.00 | 0.0 |
| 938 | $pCH_4$ | $pH_2D^+$ | $mCH_5^+$ | $HD$ | | 1.9e-09 | 0.00 | 0.0 |
| 939 | $oCH_4$ | $oH_2D^+$ | $pCH_5^+$ | $HD$ | | 2.1e-10 | 0.00 | 0.0 |
| 940 | $oCH_4$ | $oH_2D^+$ | $oCH_5^+$ | $HD$ | | 1.1e-09 | 0.00 | 0.0 |
| 941 | $oCH_4$ | $oH_2D^+$ | $mCH_5^+$ | $HD$ | | 5.6e-10 | 0.00 | 0.0 |
| 942 | $oCH_4$ | $pH_2D^+$ | $oCH_5^+$ | $HD$ | | 8.4e-10 | 0.00 | 0.0 |
| 943 | $oCH_4$ | $pH_2D^+$ | $mCH_5^+$ | $HD$ | | 1.1e-09 | 0.00 | 0.0 |
| 944 | $mCH_4$ | $oHD_2^+$ | $oD_2$ | $pCH_5^+$ | | 1.1e-09 | 0.00 | 0.0 |
| 945 | $mCH_4$ | $pHD_2^+$ | $pD_2$ | $pCH_5^+$ | | 1.1e-09 | 0.00 | 0.0 |
| 946 | $mCH_4$ | $oHD_2^+$ | $oD_2$ | $oCH_5^+$ | | 7.6e-10 | 0.00 | 0.0 |
| 947 | $mCH_4$ | $pHD_2^+$ | $pD_2$ | $oCH_5^+$ | | 7.6e-10 | 0.00 | 0.0 |
| 948 | $pCH_4$ | $oHD_2^+$ | $oD_2$ | $mCH_5^+$ | | 1.9e-09 | 0.00 | 0.0 |
| 949 | $pCH_4$ | $pHD_2^+$ | $pD_2$ | $mCH_5^+$ | | 1.9e-09 | 0.00 | 0.0 |
| 950 | $oCH_4$ | $oHD_2^+$ | $oD_2$ | $oCH_5^+$ | | 1.3e-09 | 0.00 | 0.0 |
| 951 | $oCH_4$ | $pHD_2^+$ | $pD_2$ | $oCH_5^+$ | | 1.3e-09 | 0.00 | 0.0 |
| 952 | $oCH_4$ | $oHD_2^+$ | $oD_2$ | $mCH_5^+$ | | 6.3e-10 | 0.00 | 0.0 |
| 953 | $oCH_4$ | $pHD_2^+$ | $pD_2$ | $mCH_5^+$ | | 6.3e-10 | 0.00 | 0.0 |
| 954 | $C$ | $oH_2D^+$ | $CH^+$ | $HD$ | | 2.0e-09 | 0.00 | 0.0 |
| 955 | $C$ | $pH_2D^+$ | $CH^+$ | $HD$ | | 2.0e-09 | 0.00 | 0.0 |
| 956 | $C$ | $oHD_2^+$ | $oD_2$ | $CH^+$ | | 2.0e-09 | 0.00 | 0.0 |
| 957 | $C$ | $pHD_2^+$ | $pD_2$ | $CH^+$ | | 2.0e-09 | 0.00 | 0.0 |
| 958 | $C_2$ | $H^+$ | $C_2^+$ | $H$ | | 3.1e-09 | 0.00 | 0.0 |
| 959 | $C_2H$ | $He^+$ | $C^+$ | $CH$ | $He$ | 5.1e-10 | 0.00 | 0.0 |
| 960 | $C_2H$ | $He^+$ | $CH^+$ | $C$ | $He$ | 5.1e-10 | 0.00 | 0.0 |
| 961 | $C_2H$ | $He^+$ | $C_2^+$ | $H$ | $He$ | 5.1e-10 | 0.00 | 0.0 |
| 962 | $C_2H$ | $H^+$ | $oH_2$ | $C_2^+$ | | 1.1e-09 | 0.00 | 0.0 |

**Table D6** – *continued* (part 14)

| # | Reactants | | Products | | | $\alpha$ | $\beta$ | $\gamma$ |
|---|---|---|---|---|---|---|---|---|
| 963 | $C_2H$ | $H^+$ | $pH_2$ | $C_2^+$ | | 3.7e-10 | 0.00 | 0.0 |
| 964 | $C_2H$ | $H^+$ | $C_2H^+$ | H | | 1.5e-09 | 0.00 | 0.0 |
| 965 | $C_2H$ | $oH_2D^+$ | $C_2H_2^+$ | HD | | 1.7e-09 | 0.00 | 0.0 |
| 966 | $C_2H$ | $pH_2D^+$ | $C_2H_2^+$ | HD | | 1.7e-09 | 0.00 | 0.0 |
| 967 | $C_2H$ | $oHD_2^+$ | $oD_2$ | $C_2H_2^+$ | | 1.7e-09 | 0.00 | 0.0 |
| 968 | $C_2H$ | $pHD_2^+$ | $pD_2$ | $C_2H_2^+$ | | 1.7e-09 | 0.00 | 0.0 |
| 969 | $C_3H$ | $He^+$ | $C_3^+$ | H | He | 2.0e-09 | 0.00 | 0.0 |
| 970 | $C_3H$ | $H^+$ | $oH_2$ | $C_3^+$ | | 1.5e-09 | 0.00 | 0.0 |
| 971 | $C_3H$ | $H^+$ | $pH_2$ | $C_3^+$ | | 5.0e-10 | 0.00 | 0.0 |
| 972 | $C_3H$ | $H^+$ | $C_3H^+$ | H | | 2.0e-09 | 0.00 | 0.0 |
| 973 | $C_3H$ | $oH_2D^+$ | $C_3H_2^+$ | HD | | 2.0e-09 | 0.00 | 0.0 |
| 974 | $C_3H$ | $pH_2D^+$ | $C_3H_2^+$ | HD | | 2.0e-09 | 0.00 | 0.0 |
| 975 | $C_3H$ | $oHD_2^+$ | $oD_2$ | $C_3H_2^+$ | | 2.0e-09 | 0.00 | 0.0 |
| 976 | $C_3H$ | $pHD_2^+$ | $pD_2$ | $C_3H_2^+$ | | 2.0e-09 | 0.00 | 0.0 |
| 977 | $C_2H_2$ | $He^+$ | $CH^+$ | CH | He | 7.7e-10 | 0.00 | 0.0 |
| 978 | $C_2H_2$ | $He^+$ | $C_2H^+$ | H | He | 8.7e-10 | 0.00 | 0.0 |
| 979 | $C_2H_2$ | $He^+$ | $C_2H_2^+$ | He | | 2.4e-10 | 0.00 | 0.0 |
| 980 | $C_2H_2$ | $H^+$ | $C_2H_2^+$ | H | | 2.0e-09 | 0.00 | 0.0 |
| 981 | $C_3H_2$ | $H^+$ | $C_3H_2^+$ | H | | 2.0e-09 | 0.00 | 0.0 |
| 982 | $C_3H_2$ | $He^+$ | $C_3H^+$ | H | He | 1.0e-09 | 0.00 | 0.0 |
| 983 | $C^+$ | $e^-$ | C | $\gamma$ | | 4.4e-12 | -0.61 | 0.0 |
| 984 | $CH^+$ | $e^-$ | C | H | | 1.5e-07 | -0.42 | 0.0 |
| 985 | $oCH_2^+$ | $e^-$ | $oH_2$ | C | | 1.2e-07 | -0.50 | 0.0 |
| 986 | $pCH_2^+$ | $e^-$ | $pH_2$ | C | | 1.2e-07 | -0.50 | 0.0 |
| 987 | $oCH_2^+$ | $e^-$ | CH | H | | 1.2e-07 | -0.50 | 0.0 |
| 988 | $pCH_2^+$ | $e^-$ | CH | H | | 1.2e-07 | -0.50 | 0.0 |
| 989 | $oCH_3^+$ | $e^-$ | $oH_2$ | H | C | 3.0e-07 | -0.30 | 0.0 |
| 990 | $pCH_3^+$ | $e^-$ | $oH_2$ | H | C | 1.5e-07 | -0.30 | 0.0 |
| 991 | $pCH_3^+$ | $e^-$ | $pH_2$ | H | C | 1.5e-07 | -0.30 | 0.0 |
| 992 | $oCH_3^+$ | $e^-$ | CH | H | H | 1.6e-07 | -0.30 | 0.0 |
| 993 | $pCH_3^+$ | $e^-$ | CH | H | H | 1.6e-07 | -0.30 | 0.0 |
| 994 | $oCH_3^+$ | $e^-$ | $oH_2$ | CH | | 1.4e-07 | -0.30 | 0.0 |
| 995 | $pCH_3^+$ | $e^-$ | $oH_2$ | CH | | 7.0e-08 | -0.30 | 0.0 |
| 996 | $pCH_3^+$ | $e^-$ | $pH_2$ | CH | | 7.0e-08 | -0.30 | 0.0 |
| 997 | $oCH_3^+$ | $e^-$ | $oCH_2$ | H | | 4.0e-07 | -0.30 | 0.0 |
| 998 | $pCH_3^+$ | $e^-$ | $oCH_2$ | H | | 2.0e-07 | -0.30 | 0.0 |
| 999 | $pCH_3^+$ | $e^-$ | $pCH_2$ | H | | 2.0e-07 | -0.30 | 0.0 |
| 1000 | $mCH_4^+$ | $e^-$ | $oCH_3$ | H | | 3.0e-07 | -0.50 | 0.0 |
| 1001 | $pCH_4^+$ | $e^-$ | $pCH_3$ | H | | 3.0e-07 | -0.50 | 0.0 |
| 1002 | $oCH_4^+$ | $e^-$ | $oCH_3$ | H | | 1.0e-07 | -0.50 | 0.0 |
| 1003 | $oCH_4^+$ | $e^-$ | $pCH_3$ | H | | 2.0e-07 | -0.50 | 0.0 |
| 1004 | $pCH_5^+$ | $e^-$ | $oCH_3$ | $oH_2$ | | 8.7e-08 | -0.30 | 0.0 |
| 1005 | $oCH_5^+$ | $e^-$ | $oCH_3$ | $oH_2$ | | 2.2e-08 | -0.30 | 0.0 |
| 1006 | $oCH_5^+$ | $e^-$ | $oCH_3$ | $pH_2$ | | 2.2e-08 | -0.30 | 0.0 |
| 1007 | $oCH_5^+$ | $e^-$ | $pCH_3$ | $oH_2$ | | 4.4e-08 | -0.30 | 0.0 |
| 1008 | $mCH_5^+$ | $e^-$ | $oCH_3$ | $oH_2$ | | 1.8e-08 | -0.30 | 0.0 |
| 1009 | $mCH_5^+$ | $e^-$ | $pCH_3$ | $oH_2$ | | 3.5e-08 | -0.30 | 0.0 |
| 1010 | $mCH_5^+$ | $e^-$ | $pCH_3$ | $pH_2$ | | 3.5e-08 | -0.30 | 0.0 |
| 1011 | $pCH_5^+$ | $e^-$ | $mCH_4$ | H | | 8.7e-08 | -0.30 | 0.0 |
| 1012 | $oCH_5^+$ | $e^-$ | $mCH_4$ | H | | 2.2e-08 | -0.30 | 0.0 |
| 1013 | $oCH_5^+$ | $e^-$ | $oCH_4$ | H | | 6.6e-08 | -0.30 | 0.0 |
| 1014 | $mCH_5^+$ | $e^-$ | $pCH_4$ | H | | 3.5e-08 | -0.30 | 0.0 |
| 1015 | $mCH_5^+$ | $e^-$ | $oCH_4$ | H | | 5.3e-08 | -0.30 | 0.0 |
| 1016 | $C_2^+$ | $e^-$ | C | C | | 3.0e-07 | -0.50 | 0.0 |
| 1017 | $C_2H^+$ | $e^-$ | $C_2$ | H | | 1.4e-07 | -0.50 | 0.0 |
| 1018 | $C_2H^+$ | $e^-$ | CH | C | | 1.4e-07 | -0.50 | 0.0 |
| 1019 | $C_2H_2^+$ | $e^-$ | $C_2H$ | H | | 1.5e-07 | -0.50 | 0.0 |
| 1020 | $C_2H_2^+$ | $e^-$ | CH | CH | | 1.5e-07 | -0.50 | 0.0 |
| 1021 | $C_2H_3^+$ | $e^-$ | $C_2H_2$ | H | | 3.0e-08 | -0.50 | 0.0 |
| 1022 | $C_3^+$ | $e^-$ | $C_2$ | C | | 3.0e-07 | -0.50 | 0.0 |
| 1023 | $C_3H^+$ | $e^-$ | $C_2$ | CH | | 1.5e-07 | -0.50 | 0.0 |
| 1024 | $C_3H^+$ | $e^-$ | $C_2H$ | C | | 1.5e-07 | -0.50 | 0.0 |
| 1025 | $C_3H_2^+$ | $e^-$ | $C_3H$ | H | | 1.5e-07 | -0.50 | 0.0 |
| 1026 | $C_3H_2^+$ | $e^-$ | $C_2H$ | CH | | 1.5e-07 | -0.50 | 0.0 |
| 1027 | $C_3H_3^+$ | $e^-$ | $C_3H_2$ | H | | 1.5e-07 | -0.50 | 0.0 |
| 1028 | $C_3H_3^+$ | $e^-$ | $C_2H_2$ | CH | | 1.5e-07 | -0.50 | 0.0 |
| 1029 | C | $\gamma$ | $C^+$ | $e^-$ | | 2.6e+02 | 0.00 | 0.0 |
| 1030 | CH | $\gamma$ | C | H | | 1.1e+03 | 0.00 | 0.0 |
| 1031 | CH | $\gamma$ | $CH^+$ | $e^-$ | | 5.8e+02 | 0.00 | 0.0 |
| 1032 | $oCH_2$ | $\gamma$ | $oH_2$ | C | | 2.9e+02 | 0.00 | 0.0 |
| 1033 | $pCH_2$ | $\gamma$ | $pH_2$ | C | | 2.9e+02 | 0.00 | 0.0 |
| 1034 | $oCH_3$ | $\gamma$ | $oH_2$ | CH | | 2.8e+02 | 0.00 | 0.0 |
| 1035 | $pCH_3$ | $\gamma$ | $oH_2$ | CH | | 1.4e+02 | 0.00 | 0.0 |
| 1036 | $pCH_3$ | $\gamma$ | $pH_2$ | CH | | 1.4e+02 | 0.00 | 0.0 |

**Table D6** – *continued* (part 15)

| # | Reactants | | Products | | | $\alpha$ | $\beta$ | $\gamma$ |
|---|---|---|---|---|---|---|---|---|
| 1037 | $oCH_3$ | $\gamma_2$ | $oCH_3^+$ | $e^-$ | | 3.8e+02 | 0.00 | 0.0 |
| 1038 | $pCH_3$ | $\gamma_2$ | $pCH_3^+$ | $e^-$ | | 3.8e+02 | 0.00 | 0.0 |
| 1039 | $mCH_4$ | $\gamma_2$ | $oCH_3$ | H | | 1.5e+03 | 0.00 | 0.0 |
| 1040 | $pCH_4$ | $\gamma_2$ | $pCH_3$ | H | | 1.5e+03 | 0.00 | 0.0 |
| 1041 | $oCH_4$ | $\gamma_2$ | $oCH_3$ | H | | 5.0e+02 | 0.00 | 0.0 |
| 1042 | $oCH_4$ | $\gamma_2$ | $pCH_3$ | H | | 1.0e+03 | 0.00 | 0.0 |
| 1043 | $mCH_4$ | $\gamma_2$ | $mCH_4^+$ | $e^-$ | | 2.2e+01 | 0.00 | 0.0 |
| 1044 | $pCH_4$ | $\gamma_2$ | $pCH_4^+$ | $e^-$ | | 2.2e+01 | 0.00 | 0.0 |
| 1045 | $oCH_4$ | $\gamma_2$ | $oCH_4^+$ | $e^-$ | | 2.2e+01 | 0.00 | 0.0 |
| 1046 | $CH^+$ | $\gamma_2$ | C | $H^+$ | | 2.2e+02 | 0.00 | 0.0 |
| 1047 | $oCH_2^+$ | $\gamma_2$ | $CH^+$ | H | | 8.9e+01 | 0.00 | 0.0 |
| 1048 | $pCH_2^+$ | $\gamma_2$ | $CH^+$ | H | | 8.9e+01 | 0.00 | 0.0 |
| 1049 | $mCH_4^+$ | $\gamma_2$ | $oCH_3^+$ | H | | 2.7e+01 | 0.00 | 0.0 |
| 1050 | $pCH_4^+$ | $\gamma_2$ | $pCH_3^+$ | H | | 2.7e+01 | 0.00 | 0.0 |
| 1051 | $oCH_4^+$ | $\gamma_2$ | $oCH_3^+$ | H | | 9.0e+00 | 0.00 | 0.0 |
| 1052 | $oCH_4^+$ | $\gamma_2$ | $pCH_3^+$ | H | | 1.8e+01 | 0.00 | 0.0 |
| 1053 | $C_2$ | $\gamma_2$ | C | C | | 1.8e+02 | 0.00 | 0.0 |
| 1054 | $C_2$ | $\gamma_2$ | $C_2^+$ | $e^-$ | | 2.5e+02 | 0.00 | 0.0 |
| 1055 | $C_2H$ | $\gamma_2$ | $C_2$ | H | | 1.1e+03 | 0.00 | 0.0 |
| 1056 | $C_2H_2$ | $\gamma_2$ | $C_2H$ | H | | 3.5e+03 | 0.00 | 0.0 |
| 1057 | $C_2H_2$ | $\gamma_2$ | $C_2H_2^+$ | $e^-$ | | 3.8e+02 | 0.00 | 0.0 |
| 1058 | $C_3$ | $\gamma_2$ | $C_2$ | C | | 6.9e+03 | 0.00 | 0.0 |
| 1059 | $C_3H$ | $\gamma_2$ | $C_3$ | H | | 3.0e+03 | 0.00 | 0.0 |
| 1060 | $C_3H_2$ | $\gamma_2$ | $C_3H$ | H | | 3.4e+03 | 0.00 | 0.0 |
| 1061 | $C^+$ | Gr | $Gr^+$ | C | | 4.6e-07 | 0.50 | 0.0 |
| 1062 | $C^+$ | $Gr^-$ | Gr | C | | 4.6e-07 | 0.50 | 0.0 |
| 1063 | $C^+$ | $mCH_4$ | $C_2H_2^+$ | $oH_2$ | | 3.3e-10 | 0.00 | 0.0 |
| 1064 | $C^+$ | $pCH_4$ | $C_2H_2^+$ | $oH_2$ | | 1.6e-10 | 0.00 | 0.0 |
| 1065 | $C^+$ | $pCH_4$ | $C_2H_2^+$ | $pH_2$ | | 1.6e-10 | 0.00 | 0.0 |
| 1066 | $C^+$ | $oCH_4$ | $C_2H_2^+$ | $oH_2$ | | 2.2e-10 | 0.00 | 0.0 |
| 1067 | $C^+$ | $oCH_4$ | $C_2H_2^+$ | $pH_2$ | | 1.1e-10 | 0.00 | 0.0 |
| 1068 | $C_2H$ | $oH_3^+$ | $C_2H_2^+$ | $oH_2$ | | 1.5e-09 | 0.00 | 0.0 |
| 1069 | $C_2H$ | $oH_3^+$ | $C_2H_2^+$ | $pH_2$ | | 2.1e-10 | 0.00 | 0.0 |
| 1070 | $C_2H$ | $pH_3^+$ | $C_2H_2^+$ | $oH_2$ | | 1.1e-09 | 0.00 | 0.0 |
| 1071 | $C_2H$ | $pH_3^+$ | $C_2H_2^+$ | $pH_2$ | | 6.4e-10 | 0.00 | 0.0 |
| 1072 | $C_3H$ | $oH_3^+$ | $C_3H_2^+$ | $oH_2$ | | 1.7e-09 | 0.00 | 0.0 |
| 1073 | $C_3H$ | $oH_3^+$ | $C_3H_2^+$ | $pH_2$ | | 2.5e-10 | 0.00 | 0.0 |
| 1074 | $C_3H$ | $pH_3^+$ | $C_3H_2^+$ | $oH_2$ | | 1.3e-09 | 0.00 | 0.0 |
| 1075 | $C_3H$ | $pH_3^+$ | $C_3H_2^+$ | $pH_2$ | | 7.5e-10 | 0.00 | 0.0 |
| 1076 | $C_3H^+$ | $oH_2$ | $C_3H_2^+$ | H | | 1.0e-09 | 0.00 | 500.0 |
| 1077 | $C_3H^+$ | $pH_2$ | $C_3H_2^+$ | H | | 1.0e-09 | 0.00 | 500.0 |
| 1078 | N | CRP | $N^+$ | $e^-$ | | 2.1e+00 | 0.00 | 0.0 |
| 1079 | N | NH | $N_2$ | H | | 5.0e-11 | 0.10 | 0.0 |
| 1080 | N | $oNH_2$ | H | H | $N_2$ | 1.2e-10 | 0.00 | 0.0 |
| 1081 | N | $pNH_2$ | H | H | $N_2$ | 1.2e-10 | 0.00 | 0.0 |
| 1082 | $N^+$ | HD | $ND^+$ | H | | 4.2e-10 | 0.00 | 0.0 |
| 1083 | $N_2^+$ | $oH_2$ | $N_2H^+$ | H | | 2.0e-09 | 0.24 | 0.0 |
| 1084 | $N_2^+$ | $pH_2$ | $N_2H^+$ | H | | 2.0e-09 | 0.24 | 0.0 |
| 1085 | $NH^+$ | $oH_2$ | $oNH_2^+$ | H | | 1.1e-09 | 0.00 | 0.0 |
| 1086 | $NH^+$ | $oH_2$ | $pNH_2^+$ | H | | 2.1e-10 | 0.00 | 0.0 |
| 1087 | $NH^+$ | $pH_2$ | $oNH_2^+$ | H | | 6.4e-10 | 0.00 | 0.0 |
| 1088 | $NH^+$ | $pH_2$ | $pNH_2^+$ | H | | 6.4e-10 | 0.00 | 0.0 |
| 1089 | $NH^+$ | $oH_2$ | $oH_3^+$ | N | | 1.5e-10 | 0.00 | 0.0 |
| 1090 | $NH^+$ | $oH_2$ | $pH_3^+$ | N | | 7.5e-11 | 0.00 | 0.0 |
| 1091 | $NH^+$ | $pH_2$ | $pH_3^+$ | N | | 2.2e-10 | 0.00 | 0.0 |
| 1092 | $NH^+$ | H | $oH_2$ | $N^+$ | | 4.9e-10 | 0.00 | 0.0 |
| 1093 | $NH^+$ | H | $pH_2$ | $N^+$ | | 1.6e-10 | 0.00 | 0.0 |
| 1094 | $N_2H^+$ | $oNH_3$ | $mNH_4^+$ | $N_2$ | | 1.4e-09 | 0.00 | 0.0 |
| 1095 | $N_2H^+$ | $oNH_3$ | $oNH_4^+$ | $N_2$ | | 8.6e-10 | 0.00 | 0.0 |
| 1096 | $N_2H^+$ | $pNH_3$ | $pNH_4^+$ | $N_2$ | | 5.7e-10 | 0.00 | 0.0 |
| 1097 | $N_2H^+$ | $pNH_3$ | $oNH_4^+$ | $N_2$ | | 1.7e-09 | 0.00 | 0.0 |
| 1098 | $N_2H^+$ | D | $N_2D^+$ | H | | 1.0e-09 | 0.00 | 0.0 |
| 1099 | $oNH_2^+$ | $oH_2$ | $oNH_3^+$ | H | | 1.8e-10 | 0.00 | 0.0 |
| 1100 | $oNH_2^+$ | $oH_2$ | $pNH_3^+$ | H | | 9.0e-11 | 0.00 | 0.0 |
| 1101 | $oNH_2^+$ | $pH_2$ | $oNH_3^+$ | H | | 9.0e-11 | 0.00 | 0.0 |
| 1102 | $oNH_2^+$ | $pH_2$ | $pNH_3^+$ | H | | 1.8e-10 | 0.00 | 0.0 |
| 1103 | $pNH_2^+$ | $oH_2$ | $oNH_3^+$ | H | | 9.0e-11 | 0.00 | 0.0 |
| 1104 | $pNH_2^+$ | $oH_2$ | $pNH_3^+$ | H | | 1.8e-10 | 0.00 | 0.0 |
| 1105 | $pNH_2^+$ | $pH_2$ | $pNH_3^+$ | H | | 2.7e-10 | 0.00 | 0.0 |
| 1106 | $oNH_3^+$ | $oH_2$ | $mNH_4^+$ | H | | 1.4e-12 | 0.00 | 0.0 |
| 1107 | $oNH_3^+$ | $oH_2$ | $pNH_4^+$ | H | | 1.6e-13 | 0.00 | 0.0 |
| 1108 | $oNH_3^+$ | $oH_2$ | $oNH_4^+$ | H | | 8.4e-13 | 0.00 | 0.0 |
| 1109 | $oNH_3^+$ | $pH_2$ | $mNH_4^+$ | H | | 6.0e-13 | 0.00 | 0.0 |
| 1110 | $oNH_3^+$ | $pH_2$ | $oNH_4^+$ | H | | 1.8e-12 | 0.00 | 0.0 |

**Table D6** – *continued* (part 16)

| # | Reactants | | Products | | $\alpha$ | $\beta$ | $\gamma$ |
|---|---|---|---|---|---|---|---|
| 1111 | $pNH_3^+$ | $oH_2$ | $mNH_4^+$ | H | 4.0e-13 | 0.00 | 0.0 |
| 1112 | $pNH_3^+$ | $oH_2$ | $pNH_4^+$ | H | 3.2e-13 | 0.00 | 0.0 |
| 1113 | $pNH_3^+$ | $oH_2$ | $oNH_4^+$ | H | 1.7e-12 | 0.00 | 0.0 |
| 1114 | $pNH_3^+$ | $pH_2$ | $pNH_4^+$ | H | 9.6e-13 | 0.00 | 0.0 |
| 1115 | $pNH_3^+$ | $pH_2$ | $oNH_4^+$ | H | 1.4e-12 | 0.00 | 0.0 |
| 1116 | $ND^+$ | $oH_2$ | $NHD^+$ | H | 1.3e-09 | 0.00 | 0.0 |
| 1117 | $ND^+$ | $pH_2$ | $NHD^+$ | H | 1.3e-09 | 0.00 | 0.0 |
| 1118 | $ND^+$ | $oH_2$ | $oH_2D^+$ | N | 2.2e-10 | 0.00 | 0.0 |
| 1119 | $ND^+$ | $pH_2$ | $pH_2D^+$ | N | 2.2e-10 | 0.00 | 0.0 |
| 1120 | $ND^+$ | HD | $NHD^+$ | D | 6.3e-10 | 0.00 | 0.0 |
| 1121 | $ND^+$ | HD | $oND_2^+$ | H | 4.2e-10 | 0.00 | 0.0 |
| 1122 | $ND^+$ | HD | $pND_2^+$ | H | 2.1e-10 | 0.00 | 0.0 |
| 1123 | $N_2D^+$ | H | $N_2H^+$ | D | 1.0e-09 | 0.00 | 170.0 |
| 1124 | $N_2D^+$ | $oNH_3$ | $oNH_3D^+$ | $N_2$ | 2.3e-09 | 0.00 | 0.0 |
| 1125 | $N_2D^+$ | $pNH_3$ | $pNH_3D^+$ | $N_2$ | 2.3e-09 | 0.00 | 0.0 |
| 1126 | $NHD^+$ | $oH_2$ | $oNH_2D^+$ | H | 2.2e-10 | 0.00 | 0.0 |
| 1127 | $NHD^+$ | $oH_2$ | $pNH_2D^+$ | H | 4.5e-11 | 0.00 | 0.0 |
| 1128 | $NHD^+$ | $pH_2$ | $oNH_2D^+$ | H | 1.3e-10 | 0.00 | 0.0 |
| 1129 | $NHD^+$ | $pH_2$ | $pNH_2D^+$ | H | 1.3e-10 | 0.00 | 0.0 |
| 1130 | $NHD^+$ | HD | $oNH_2D^+$ | D | 1.0e-10 | 0.00 | 0.0 |
| 1131 | $NHD^+$ | HD | $pNH_2D^+$ | D | 3.4e-11 | 0.00 | 0.0 |
| 1132 | $NHD^+$ | HD | $oNHD_2^+$ | H | 9.0e-11 | 0.00 | 0.0 |
| 1133 | $NHD^+$ | HD | $pNHD_2^+$ | H | 4.5e-11 | 0.00 | 0.0 |
| 1134 | $oND_2^+$ | $oH_2$ | $oNHD_2^+$ | H | 2.7e-10 | 0.00 | 0.0 |
| 1135 | $pND_2^+$ | $oH_2$ | $pNHD_2^+$ | H | 2.7e-10 | 0.00 | 0.0 |
| 1136 | $oND_2^+$ | $pH_2$ | $oNHD_2^+$ | H | 2.7e-10 | 0.00 | 0.0 |
| 1137 | $pND_2^+$ | $pH_2$ | $pNHD_2^+$ | H | 2.7e-10 | 0.00 | 0.0 |
| 1138 | $oND_2^+$ | HD | $oNHD_2^+$ | D | 1.0e-10 | 0.00 | 0.0 |
| 1139 | $oND_2^+$ | HD | $pNHD_2^+$ | D | 3.0e-11 | 0.00 | 0.0 |
| 1140 | $pND_2^+$ | HD | $oNHD_2^+$ | D | 6.0e-11 | 0.00 | 0.0 |
| 1141 | $pND_2^+$ | HD | $pNHD_2^+$ | D | 7.5e-11 | 0.00 | 0.0 |
| 1142 | $oND_2^+$ | HD | $mND_3^+$ | H | 7.5e-11 | 0.00 | 0.0 |
| 1143 | $oND_2^+$ | HD | $oND_3^+$ | H | 6.0e-11 | 0.00 | 0.0 |
| 1144 | $pND_2^+$ | HD | $pND_3^+$ | H | 1.5e-11 | 0.00 | 0.0 |
| 1145 | $pND_2^+$ | HD | $oND_3^+$ | H | 1.2e-10 | 0.00 | 0.0 |
| 1146 | $oNH_2D^+$ | $oH_2$ | $oNH_3D^+$ | H | 1.6e-12 | 0.00 | 0.0 |
| 1147 | $oNH_2D^+$ | $oH_2$ | $pNH_3D^+$ | H | 8.0e-13 | 0.00 | 0.0 |
| 1148 | $oNH_2D^+$ | $pH_2$ | $oNH_3D^+$ | H | 8.0e-13 | 0.00 | 0.0 |
| 1149 | $oNH_2D^+$ | $pH_2$ | $pNH_3D^+$ | H | 1.6e-12 | 0.00 | 0.0 |
| 1150 | $pNH_2D^+$ | $oH_2$ | $oNH_3D^+$ | H | 8.0e-13 | 0.00 | 0.0 |
| 1151 | $pNH_2D^+$ | $oH_2$ | $pNH_3D^+$ | H | 1.6e-12 | 0.00 | 0.0 |
| 1152 | $pNH_2D^+$ | $pH_2$ | $pNH_3D^+$ | H | 2.4e-12 | 0.00 | 0.0 |
| 1153 | $oNH_2D^+$ | HD | $oNH_3D^+$ | D | 8.0e-13 | 0.00 | 0.0 |
| 1154 | $oNH_2D^+$ | HD | $pNH_3D^+$ | D | 4.0e-13 | 0.00 | 0.0 |
| 1155 | $pNH_2D^+$ | HD | $pNH_3D^+$ | D | 1.2e-12 | 0.00 | 0.0 |
| 1156 | $oNH_2D^+$ | HD | $ooNH_2D_2^+$ | H | 6.7e-13 | 0.00 | 0.0 |
| 1157 | $oNH_2D^+$ | HD | $poNH_2D_2^+$ | H | 3.3e-13 | 0.00 | 0.0 |
| 1158 | $oNH_2D^+$ | HD | $opNH_2D_2^+$ | H | 1.3e-13 | 0.00 | 0.0 |
| 1159 | $oNH_2D^+$ | HD | $ppNH_2D_2^+$ | H | 6.7e-14 | 0.00 | 0.0 |
| 1160 | $pNH_2D^+$ | HD | $ooNH_2D_2^+$ | H | 4.0e-13 | 0.00 | 0.0 |
| 1161 | $pNH_2D^+$ | HD | $poNH_2D_2^+$ | H | 2.0e-13 | 0.00 | 0.0 |
| 1162 | $pNH_2D^+$ | HD | $opNH_2D_2^+$ | H | 4.0e-13 | 0.00 | 0.0 |
| 1163 | $pNH_2D^+$ | HD | $ppNH_2D_2^+$ | H | 2.0e-13 | 0.00 | 0.0 |
| 1164 | $oNHD_2^+$ | $oH_2$ | $ooNH_2D_2^+$ | H | 2.0e-12 | 0.00 | 0.0 |
| 1165 | $pNHD_2^+$ | $oH_2$ | $poNH_2D_2^+$ | H | 2.0e-12 | 0.00 | 0.0 |
| 1166 | $oNHD_2^+$ | $oH_2$ | $opNH_2D_2^+$ | H | 4.0e-13 | 0.00 | 0.0 |
| 1167 | $pNHD_2^+$ | $oH_2$ | $ppNH_2D_2^+$ | H | 4.0e-13 | 0.00 | 0.0 |
| 1168 | $oNHD_2^+$ | $pH_2$ | $ooNH_2D_2^+$ | H | 1.2e-12 | 0.00 | 0.0 |
| 1169 | $pNHD_2^+$ | $pH_2$ | $poNH_2D_2^+$ | H | 1.2e-12 | 0.00 | 0.0 |
| 1170 | $oNHD_2^+$ | $pH_2$ | $opNH_2D_2^+$ | H | 1.2e-12 | 0.00 | 0.0 |
| 1171 | $pNHD_2^+$ | $pH_2$ | $ppNH_2D_2^+$ | H | 1.2e-12 | 0.00 | 0.0 |
| 1172 | $oNHD_2^+$ | HD | $ooNH_2D_2^+$ | D | 7.0e-13 | 0.00 | 0.0 |
| 1173 | $oNHD_2^+$ | HD | $poNH_2D_2^+$ | D | 2.0e-13 | 0.00 | 0.0 |
| 1174 | $pNHD_2^+$ | HD | $ooNH_2D_2^+$ | D | 4.0e-13 | 0.00 | 0.0 |
| 1175 | $pNHD_2^+$ | HD | $poNH_2D_2^+$ | D | 5.0e-13 | 0.00 | 0.0 |
| 1176 | $oNHD_2^+$ | HD | $opNH_2D_2^+$ | D | 2.3e-13 | 0.00 | 0.0 |
| 1177 | $oNHD_2^+$ | HD | $ppNH_2D_2^+$ | D | 6.7e-14 | 0.00 | 0.0 |
| 1178 | $pNHD_2^+$ | HD | $opNH_2D_2^+$ | D | 1.3e-13 | 0.00 | 0.0 |
| 1179 | $pNHD_2^+$ | HD | $ppNH_2D_2^+$ | D | 1.7e-13 | 0.00 | 0.0 |
| 1180 | $oNHD_2^+$ | HD | $mNHD_3^+$ | H | 6.7e-13 | 0.00 | 0.0 |
| 1181 | $oNHD_2^+$ | HD | $oNHD_3^+$ | H | 5.3e-13 | 0.00 | 0.0 |
| 1182 | $pNHD_2^+$ | HD | $pNHD_3^+$ | H | 1.3e-13 | 0.00 | 0.0 |
| 1183 | $pNHD_2^+$ | HD | $oNHD_3^+$ | H | 1.1e-12 | 0.00 | 0.0 |
| 1184 | $mND_3^+$ | $oH_2$ | $mNHD_3^+$ | H | 2.4e-12 | 0.00 | 0.0 |

**Table D6** – *continued* (part 17)

| # | Reactants | | Products | | | $\alpha$ | $\beta$ | $\gamma$ |
|---|---|---|---|---|---|---|---|---|
| 1185 | $pND_3^+$ | $oH_2$ | $pNHD_3^+$ | H | | 2.4e-12 | 0.00 | 0.0 |
| 1186 | $oND_3^+$ | $oH_2$ | $oNHD_3^+$ | H | | 2.4e-12 | 0.00 | 0.0 |
| 1187 | $mND_3^+$ | $pH_2$ | $mNHD_3^+$ | H | | 2.4e-12 | 0.00 | 0.0 |
| 1188 | $pND_3^+$ | $pH_2$ | $pNHD_3^+$ | H | | 2.4e-12 | 0.00 | 0.0 |
| 1189 | $oND_3^+$ | $pH_2$ | $oNHD_3^+$ | H | | 2.4e-12 | 0.00 | 0.0 |
| 1190 | $mND_3^+$ | HD | $mNHD_3^+$ | D | | 8.0e-13 | 0.00 | 0.0 |
| 1191 | $mND_3^+$ | HD | $oNHD_3^+$ | D | | 4.0e-13 | 0.00 | 0.0 |
| 1192 | $pND_3^+$ | HD | $pNHD_3^+$ | D | | 4.0e-13 | 0.00 | 0.0 |
| 1193 | $pND_3^+$ | HD | $oNHD_3^+$ | D | | 8.0e-13 | 0.00 | 0.0 |
| 1194 | $oND_3^+$ | HD | $mNHD_3^+$ | D | | 2.5e-13 | 0.00 | 0.0 |
| 1195 | $oND_3^+$ | HD | $pNHD_3^+$ | D | | 5.0e-14 | 0.00 | 0.0 |
| 1196 | $oND_3^+$ | HD | $oNHD_3^+$ | D | | 9.0e-13 | 0.00 | 0.0 |
| 1197 | $mND_3^+$ | HD | $lND_4^+$ | H | | 6.0e-13 | 0.00 | 0.0 |
| 1198 | $mND_3^+$ | HD | $oND_4^+$ | H | | 6.0e-13 | 0.00 | 0.0 |
| 1199 | $pND_3^+$ | HD | $pND_4^+$ | H | | 1.2e-12 | 0.00 | 0.0 |
| 1200 | $oND_3^+$ | HD | $mND_4^+$ | H | | 3.0e-13 | 0.00 | 0.0 |
| 1201 | $oND_3^+$ | HD | $oND_4^+$ | H | | 7.5e-13 | 0.00 | 0.0 |
| 1202 | $oND_3^+$ | HD | $pND_4^+$ | H | | 1.5e-13 | 0.00 | 0.0 |
| 1203 | N | $oH_3^+$ | $oNH_2^+$ | H | | 0.0e+00 | 0.00 | 0.0 |
| 1204 | N | $pH_3^+$ | $oNH_2^+$ | H | | 0.0e+00 | 0.00 | 0.0 |
| 1205 | N | $pH_3^+$ | $pNH_2^+$ | H | | 0.0e+00 | 0.00 | 0.0 |
| 1206 | $N_2$ | $oH_3^+$ | $oH_2$ | $N_2H^+$ | | 1.3e-09 | 0.00 | 0.0 |
| 1207 | $N_2$ | $pH_3^+$ | $oH_2$ | $N_2H^+$ | | 6.5e-10 | 0.00 | 0.0 |
| 1208 | $N_2$ | $pH_3^+$ | $pH_2$ | $N_2H^+$ | | 6.5e-10 | 0.00 | 0.0 |
| 1209 | NH | $He^+$ | $N^+$ | H | He | 1.1e-09 | 0.00 | 0.0 |
| 1210 | NH | $H^+$ | $NH^+$ | H | | 2.1e-09 | 0.00 | 0.0 |
| 1211 | NH | $oH_3^+$ | $oNH_2^+$ | $oH_2$ | | 9.8e-10 | 0.00 | 0.0 |
| 1212 | NH | $oH_3^+$ | $oNH_2^+$ | $pH_2$ | | 1.6e-10 | 0.00 | 0.0 |
| 1213 | NH | $oH_3^+$ | $pNH_2^+$ | $oH_2$ | | 1.6e-10 | 0.00 | 0.0 |
| 1214 | NH | $pH_3^+$ | $oNH_2^+$ | $oH_2$ | | 4.9e-10 | 0.00 | 0.0 |
| 1215 | NH | $pH_3^+$ | $oNH_2^+$ | $pH_2$ | | 3.3e-10 | 0.00 | 0.0 |
| 1216 | NH | $pH_3^+$ | $pNH_2^+$ | $oH_2$ | | 3.3e-10 | 0.00 | 0.0 |
| 1217 | NH | $pH_3^+$ | $pNH_2^+$ | $pH_2$ | | 1.6e-10 | 0.00 | 0.0 |
| 1218 | $oNH_2$ | $He^+$ | $NH^+$ | H | He | 8.0e-10 | 0.00 | 0.0 |
| 1219 | $pNH_2$ | $He^+$ | $NH^+$ | H | He | 8.0e-10 | 0.00 | 0.0 |
| 1220 | $oNH_2$ | $He^+$ | $oH_2$ | $N^+$ | He | 8.0e-10 | 0.00 | 0.0 |
| 1221 | $pNH_2$ | $He^+$ | $pH_2$ | $N^+$ | He | 8.0e-10 | 0.00 | 0.0 |
| 1222 | $oNH_2$ | $H^+$ | $oNH_2^+$ | H | | 2.9e-09 | 0.00 | 0.0 |
| 1223 | $pNH_2$ | $H^+$ | $pNH_2^+$ | H | | 2.9e-09 | 0.00 | 0.0 |
| 1224 | $oNH_2$ | $oH_3^+$ | $oNH_3^+$ | $oH_2$ | | 1.1e-09 | 0.00 | 0.0 |
| 1225 | $oNH_2$ | $oH_3^+$ | $oNH_3^+$ | $pH_2$ | | 1.5e-10 | 0.00 | 0.0 |
| 1226 | $oNH_2$ | $oH_3^+$ | $pNH_3^+$ | $oH_2$ | | 4.2e-10 | 0.00 | 0.0 |
| 1227 | $oNH_2$ | $oH_3^+$ | $pNH_3^+$ | $pH_2$ | | 1.2e-10 | 0.00 | 0.0 |
| 1228 | $pNH_2$ | $oH_3^+$ | $oNH_3^+$ | $oH_2$ | | 4.5e-10 | 0.00 | 0.0 |
| 1229 | $pNH_2$ | $oH_3^+$ | $oNH_3^+$ | $pH_2$ | | 4.5e-10 | 0.00 | 0.0 |
| 1230 | $pNH_2$ | $oH_3^+$ | $pNH_3^+$ | $oH_2$ | | 9.0e-10 | 0.00 | 0.0 |
| 1231 | $oNH_2$ | $pH_3^+$ | $oNH_3^+$ | $oH_2$ | | 4.2e-10 | 0.00 | 0.0 |
| 1232 | $oNH_2$ | $pH_3^+$ | $oNH_3^+$ | $pH_2$ | | 3.0e-10 | 0.00 | 0.0 |
| 1233 | $oNH_2$ | $pH_3^+$ | $pNH_3^+$ | $oH_2$ | | 8.4e-10 | 0.00 | 0.0 |
| 1234 | $oNH_2$ | $pH_3^+$ | $pNH_3^+$ | $pH_2$ | | 2.4e-10 | 0.00 | 0.0 |
| 1235 | $pNH_2$ | $pH_3^+$ | $oNH_3^+$ | $oH_2$ | | 3.6e-10 | 0.00 | 0.0 |
| 1236 | $pNH_2$ | $pH_3^+$ | $pNH_3^+$ | $oH_2$ | | 7.2e-10 | 0.00 | 0.0 |
| 1237 | $pNH_2$ | $pH_3^+$ | $pNH_3^+$ | $pH_2$ | | 7.2e-10 | 0.00 | 0.0 |
| 1238 | $oNH_3$ | $He^+$ | $oNH_3^+$ | He | | 2.6e-10 | 0.00 | 0.0 |
| 1239 | $pNH_3$ | $He^+$ | $pNH_3^+$ | He | | 2.6e-10 | 0.00 | 0.0 |
| 1240 | $oNH_3$ | $He^+$ | $oNH_2^+$ | H | He | 1.8e-09 | 0.00 | 0.0 |
| 1241 | $pNH_3$ | $He^+$ | $oNH_2^+$ | H | He | 8.8e-10 | 0.00 | 0.0 |
| 1242 | $pNH_3$ | $He^+$ | $pNH_2^+$ | H | He | 8.8e-10 | 0.00 | 0.0 |
| 1243 | $oNH_3$ | $He^+$ | $oH_2$ | $NH^+$ | He | 1.8e-10 | 0.00 | 0.0 |
| 1244 | $pNH_3$ | $He^+$ | $oH_2$ | $NH^+$ | He | 8.8e-11 | 0.00 | 0.0 |
| 1245 | $pNH_3$ | $He^+$ | $pH_2$ | $NH^+$ | He | 8.8e-11 | 0.00 | 0.0 |
| 1246 | $oNH_3$ | $H^+$ | $oNH_3^+$ | H | | 5.2e-09 | 0.00 | 0.0 |
| 1247 | $pNH_3$ | $H^+$ | $pNH_3^+$ | H | | 5.2e-09 | 0.00 | 0.0 |
| 1248 | $oNH_3$ | $oH_3^+$ | $mNH_4^+$ | $oH_2$ | | 4.7e-09 | 0.00 | 0.0 |
| 1249 | $oNH_3$ | $oH_3^+$ | $mNH_4^+$ | $pH_2$ | | 5.7e-10 | 0.00 | 0.0 |
| 1250 | $oNH_3$ | $oH_3^+$ | $pNH_4^+$ | $oH_2$ | | 3.8e-10 | 0.00 | 0.0 |
| 1251 | $oNH_3$ | $oH_3^+$ | $pNH_4^+$ | $pH_2$ | | 2.3e-10 | 0.00 | 0.0 |
| 1252 | $oNH_3$ | $oH_3^+$ | $oNH_4^+$ | $oH_2$ | | 2.6e-09 | 0.00 | 0.0 |
| 1253 | $oNH_3$ | $oH_3^+$ | $oNH_4^+$ | $pH_2$ | | 5.7e-10 | 0.00 | 0.0 |
| 1254 | $oNH_3$ | $pH_3^+$ | $mNH_4^+$ | $oH_2$ | | 1.5e-09 | 0.00 | 0.0 |
| 1255 | $oNH_3$ | $pH_3^+$ | $mNH_4^+$ | $pH_2$ | | 1.1e-09 | 0.00 | 0.0 |
| 1256 | $oNH_3$ | $pH_3^+$ | $pNH_4^+$ | $oH_2$ | | 7.6e-10 | 0.00 | 0.0 |
| 1257 | $oNH_3$ | $pH_3^+$ | $oNH_4^+$ | $oH_2$ | | 4.6e-09 | 0.00 | 0.0 |
| 1258 | $oNH_3$ | $pH_3^+$ | $oNH_4^+$ | $pH_2$ | | 1.1e-09 | 0.00 | 0.0 |

**Table D6** – *continued* (part 18)

| # | Reactants | | Products | | $\alpha$ | $\beta$ | $\gamma$ |
|---|---|---|---|---|---|---|---|
| 1259 | $pNH_3$ | $oH_3^+$ | $mNH_4^+$ | $oH_2$ | 1.5e-09 | 0.00 | 0.0 |
| 1260 | $pNH_3$ | $oH_3^+$ | $mNH_4^+$ | $pH_2$ | 1.1e-09 | 0.00 | 0.0 |
| 1261 | $pNH_3$ | $oH_3^+$ | $pNH_4^+$ | $oH_2$ | 7.6e-10 | 0.00 | 0.0 |
| 1262 | $pNH_3$ | $oH_3^+$ | $oNH_4^+$ | $oH_2$ | 4.6e-09 | 0.00 | 0.0 |
| 1263 | $pNH_3$ | $oH_3^+$ | $oNH_4^+$ | $pH_2$ | 1.1e-09 | 0.00 | 0.0 |
| 1264 | $pNH_3$ | $pH_3^+$ | $mNH_4^+$ | $oH_2$ | 7.6e-10 | 0.00 | 0.0 |
| 1265 | $pNH_3$ | $pH_3^+$ | $pNH_4^+$ | $oH_2$ | 1.5e-09 | 0.00 | 0.0 |
| 1266 | $pNH_3$ | $pH_3^+$ | $pNH_4^+$ | $pH_2$ | 9.1e-10 | 0.00 | 0.0 |
| 1267 | $pNH_3$ | $pH_3^+$ | $oNH_4^+$ | $oH_2$ | 3.6e-09 | 0.00 | 0.0 |
| 1268 | $pNH_3$ | $pH_3^+$ | $oNH_4^+$ | $pH_2$ | 2.3e-09 | 0.00 | 0.0 |
| 1269 | NH | $oH_2D^+$ | $oNH_2^+$ | HD | 7.2e-10 | 0.00 | 0.0 |
| 1270 | NH | $oH_2D^+$ | $pNH_2^+$ | HD | 1.4e-10 | 0.00 | 0.0 |
| 1271 | NH | $pH_2D^+$ | $oNH_2^+$ | HD | 4.3e-10 | 0.00 | 0.0 |
| 1272 | NH | $pH_2D^+$ | $pNH_2^+$ | HD | 4.3e-10 | 0.00 | 0.0 |
| 1273 | NH | $oH_2D^+$ | $oH_2$ | $NHD^+$ | 3.6e-10 | 0.00 | 0.0 |
| 1274 | NH | $oH_2D^+$ | $pH_2$ | $NHD^+$ | 7.2e-11 | 0.00 | 0.0 |
| 1275 | NH | $pH_2D^+$ | $oH_2$ | $NHD^+$ | 2.2e-10 | 0.00 | 0.0 |
| 1276 | NH | $pH_2D^+$ | $pH_2$ | $NHD^+$ | 2.2e-10 | 0.00 | 0.0 |
| 1277 | NH | $oHD_2^+$ | $oD_2$ | $oNH_2^+$ | 3.2e-10 | 0.00 | 0.0 |
| 1278 | NH | $pHD_2^+$ | $pD_2$ | $oNH_2^+$ | 3.2e-10 | 0.00 | 0.0 |
| 1279 | NH | $oHD_2^+$ | $oD_2$ | $pNH_2^+$ | 1.1e-10 | 0.00 | 0.0 |
| 1280 | NH | $pHD_2^+$ | $pD_2$ | $pNH_2^+$ | 1.1e-10 | 0.00 | 0.0 |
| 1281 | NH | $oHD_2^+$ | $NHD^+$ | HD | 8.7e-10 | 0.00 | 0.0 |
| 1282 | NH | $pHD_2^+$ | $NHD^+$ | HD | 8.7e-10 | 0.00 | 0.0 |
| 1283 | NH | $mD_3^+$ | $oD_2$ | $NHD^+$ | 1.3e-09 | 0.00 | 0.0 |
| 1284 | NH | $pD_3^+$ | $pD_2$ | $NHD^+$ | 1.3e-09 | 0.00 | 0.0 |
| 1285 | NH | $oD_3^+$ | $oD_2$ | $NHD^+$ | 6.5e-10 | 0.00 | 0.0 |
| 1286 | NH | $oD_3^+$ | $pD_2$ | $NHD^+$ | 6.5e-10 | 0.00 | 0.0 |
| 1287 | $oNH_2$ | $oH_2D^+$ | $oNH_3^+$ | HD | 8.0e-10 | 0.00 | 0.0 |
| 1288 | $oNH_2$ | $oH_2D^+$ | $pNH_3^+$ | HD | 4.0e-10 | 0.00 | 0.0 |
| 1289 | $oNH_2$ | $pH_2D^+$ | $oNH_3^+$ | HD | 4.0e-10 | 0.00 | 0.0 |
| 1290 | $oNH_2$ | $pH_2D^+$ | $pNH_3^+$ | HD | 8.0e-10 | 0.00 | 0.0 |
| 1291 | $pNH_2$ | $oH_2D^+$ | $oNH_3^+$ | HD | 4.0e-10 | 0.00 | 0.0 |
| 1292 | $pNH_2$ | $oH_2D^+$ | $pNH_3^+$ | HD | 8.0e-10 | 0.00 | 0.0 |
| 1293 | $pNH_2$ | $pH_2D^+$ | $pNH_3^+$ | HD | 1.2e-09 | 0.00 | 0.0 |
| 1294 | $oNH_2$ | $oH_2D^+$ | $oNH_2D^+$ | $oH_2$ | 4.3e-10 | 0.00 | 0.0 |
| 1295 | $oNH_2$ | $oH_2D^+$ | $oNH_2D^+$ | $pH_2$ | 6.7e-11 | 0.00 | 0.0 |
| 1296 | $oNH_2$ | $oH_2D^+$ | $pNH_2D^+$ | $oH_2$ | 6.7e-11 | 0.00 | 0.0 |
| 1297 | $oNH_2$ | $oH_2D^+$ | $pNH_2D^+$ | $pH_2$ | 3.3e-11 | 0.00 | 0.0 |
| 1298 | $oNH_2$ | $pH_2D^+$ | $oNH_2D^+$ | $oH_2$ | 2.0e-10 | 0.00 | 0.0 |
| 1299 | $oNH_2$ | $pH_2D^+$ | $oNH_2D^+$ | $pH_2$ | 2.0e-10 | 0.00 | 0.0 |
| 1300 | $oNH_2$ | $pH_2D^+$ | $pNH_2D^+$ | $oH_2$ | 2.0e-10 | 0.00 | 0.0 |
| 1301 | $pNH_2$ | $oH_2D^+$ | $oNH_2D^+$ | $oH_2$ | 2.0e-10 | 0.00 | 0.0 |
| 1302 | $pNH_2$ | $oH_2D^+$ | $oNH_2D^+$ | $pH_2$ | 2.0e-10 | 0.00 | 0.0 |
| 1303 | $pNH_2$ | $oH_2D^+$ | $pNH_2D^+$ | $oH_2$ | 2.0e-10 | 0.00 | 0.0 |
| 1304 | $pNH_2$ | $pH_2D^+$ | $oNH_2D^+$ | $oH_2$ | 3.0e-10 | 0.00 | 0.0 |
| 1305 | $pNH_2$ | $pH_2D^+$ | $pNH_2D^+$ | $pH_2$ | 3.0e-10 | 0.00 | 0.0 |
| 1306 | $oNH_2$ | $oHD_2^+$ | $oD_2$ | $oNH_3^+$ | 4.0e-10 | 0.00 | 0.0 |
| 1307 | $oNH_2$ | $pHD_2^+$ | $pD_2$ | $oNH_3^+$ | 4.0e-10 | 0.00 | 0.0 |
| 1308 | $oNH_2$ | $oHD_2^+$ | $oD_2$ | $pNH_3^+$ | 2.0e-10 | 0.00 | 0.0 |
| 1309 | $oNH_2$ | $pHD_2^+$ | $pD_2$ | $pNH_3^+$ | 2.0e-10 | 0.00 | 0.0 |
| 1310 | $pNH_2$ | $oHD_2^+$ | $oD_2$ | $pNH_3^+$ | 6.0e-10 | 0.00 | 0.0 |
| 1311 | $pNH_2$ | $pHD_2^+$ | $pD_2$ | $pNH_3^+$ | 6.0e-10 | 0.00 | 0.0 |
| 1312 | $oNH_2$ | $oHD_2^+$ | $oNH_2D^+$ | HD | 1.0e-09 | 0.00 | 0.0 |
| 1313 | $oNH_2$ | $pHD_2^+$ | $oNH_2D^+$ | HD | 1.0e-09 | 0.00 | 0.0 |
| 1314 | $oNH_2$ | $oHD_2^+$ | $pNH_2D^+$ | HD | 2.0e-10 | 0.00 | 0.0 |
| 1315 | $oNH_2$ | $pHD_2^+$ | $pNH_2D^+$ | HD | 2.0e-10 | 0.00 | 0.0 |
| 1316 | $pNH_2$ | $oHD_2^+$ | $oNH_2D^+$ | HD | 6.0e-10 | 0.00 | 0.0 |
| 1317 | $pNH_2$ | $pHD_2^+$ | $oNH_2D^+$ | HD | 6.0e-10 | 0.00 | 0.0 |
| 1318 | $pNH_2$ | $oHD_2^+$ | $pNH_2D^+$ | HD | 6.0e-10 | 0.00 | 0.0 |
| 1319 | $pNH_2$ | $pHD_2^+$ | $pNH_2D^+$ | HD | 6.0e-10 | 0.00 | 0.0 |
| 1320 | $oNH_2$ | $mD_3^+$ | $oD_2$ | $oNH_2D^+$ | 1.8e-09 | 0.00 | 0.0 |
| 1321 | $oNH_2$ | $pD_3^+$ | $pD_2$ | $oNH_2D^+$ | 1.8e-09 | 0.00 | 0.0 |
| 1322 | $oNH_2$ | $oD_3^+$ | $oD_2$ | $oNH_2D^+$ | 9.0e-10 | 0.00 | 0.0 |
| 1323 | $oNH_2$ | $oD_3^+$ | $pD_2$ | $oNH_2D^+$ | 9.0e-10 | 0.00 | 0.0 |
| 1324 | $pNH_2$ | $mD_3^+$ | $oD_2$ | $pNH_2D^+$ | 1.8e-09 | 0.00 | 0.0 |
| 1325 | $pNH_2$ | $pD_3^+$ | $pD_2$ | $pNH_2D^+$ | 1.8e-09 | 0.00 | 0.0 |
| 1326 | $pNH_2$ | $oD_3^+$ | $oD_2$ | $pNH_2D^+$ | 9.0e-10 | 0.00 | 0.0 |
| 1327 | $pNH_2$ | $oD_3^+$ | $pD_2$ | $pNH_2D^+$ | 9.0e-10 | 0.00 | 0.0 |
| 1328 | $oNH_3$ | $oH_2D^+$ | $mNH_4^+$ | HD | 3.5e-09 | 0.00 | 0.0 |
| 1329 | $oNH_3$ | $oH_2D^+$ | $pNH_4^+$ | HD | 4.0e-10 | 0.00 | 0.0 |
| 1330 | $oNH_3$ | $oH_2D^+$ | $oNH_4^+$ | HD | 2.1e-09 | 0.00 | 0.0 |
| 1331 | $oNH_3$ | $pH_2D^+$ | $mNH_4^+$ | HD | 1.5e-09 | 0.00 | 0.0 |
| 1332 | $oNH_3$ | $pH_2D^+$ | $oNH_4^+$ | HD | 4.5e-09 | 0.00 | 0.0 |

**Table D6** – *continued* (part 19)

| # | Reactants | | Products | | $\alpha$ | $\beta$ | $\gamma$ |
|---|---|---|---|---|---|---|---|
| 1333 | $pNH_3$ | $oH_2D^+$ | $mNH_4^+$ | HD | 1.0e-09 | 0.00 | 0.0 |
| 1334 | $pNH_3$ | $oH_2D^+$ | $pNH_4^+$ | HD | 8.1e-10 | 0.00 | 0.0 |
| 1335 | $pNH_3$ | $oH_2D^+$ | $oNH_4^+$ | HD | 4.2e-09 | 0.00 | 0.0 |
| 1336 | $pNH_3$ | $pH_2D^+$ | $pNH_4^+$ | HD | 2.4e-09 | 0.00 | 0.0 |
| 1337 | $pNH_3$ | $pH_2D^+$ | $oNH_4^+$ | HD | 3.6e-09 | 0.00 | 0.0 |
| 1338 | $oNH_3$ | $oH_2D^+$ | $oNH_3D^+$ | $oH_2$ | 1.9e-09 | 0.00 | 0.0 |
| 1339 | $oNH_3$ | $oH_2D^+$ | $oNH_3D^+$ | $pH_2$ | 2.5e-10 | 0.00 | 0.0 |
| 1340 | $oNH_3$ | $oH_2D^+$ | $pNH_3D^+$ | $oH_2$ | 7.1e-10 | 0.00 | 0.0 |
| 1341 | $oNH_3$ | $oH_2D^+$ | $pNH_3D^+$ | $pH_2$ | 2.0e-10 | 0.00 | 0.0 |
| 1342 | $oNH_3$ | $pH_2D^+$ | $oNH_3D^+$ | $oH_2$ | 7.6e-10 | 0.00 | 0.0 |
| 1343 | $oNH_3$ | $pH_2D^+$ | $oNH_3D^+$ | $pH_2$ | 7.6e-10 | 0.00 | 0.0 |
| 1344 | $oNH_3$ | $pH_2D^+$ | $pNH_3D^+$ | $oH_2$ | 1.5e-09 | 0.00 | 0.0 |
| 1345 | $pNH_3$ | $oH_2D^+$ | $oNH_3D^+$ | $oH_2$ | 7.1e-10 | 0.00 | 0.0 |
| 1346 | $pNH_3$ | $oH_2D^+$ | $oNH_3D^+$ | $pH_2$ | 5.1e-10 | 0.00 | 0.0 |
| 1347 | $pNH_3$ | $oH_2D^+$ | $pNH_3D^+$ | $oH_2$ | 1.4e-09 | 0.00 | 0.0 |
| 1348 | $pNH_3$ | $oH_2D^+$ | $pNH_3D^+$ | $pH_2$ | 4.0e-10 | 0.00 | 0.0 |
| 1349 | $pNH_3$ | $pH_2D^+$ | $oNH_3D^+$ | $oH_2$ | 6.1e-10 | 0.00 | 0.0 |
| 1350 | $pNH_3$ | $pH_2D^+$ | $pNH_3D^+$ | $oH_2$ | 1.2e-09 | 0.00 | 0.0 |
| 1351 | $pNH_3$ | $pH_2D^+$ | $pNH_3D^+$ | $pH_2$ | 1.2e-09 | 0.00 | 0.0 |
| 1352 | $oNH_3$ | $oHD_2^+$ | $oD_2$ | $mNH_4^+$ | 1.9e-09 | 0.00 | 0.0 |
| 1353 | $oNH_3$ | $pHD_2^+$ | $pD_2$ | $mNH_4^+$ | 1.9e-09 | 0.00 | 0.0 |
| 1354 | $oNH_3$ | $oHD_2^+$ | $oD_2$ | $oNH_4^+$ | 1.1e-09 | 0.00 | 0.0 |
| 1355 | $oNH_3$ | $pHD_2^+$ | $pD_2$ | $oNH_4^+$ | 1.1e-09 | 0.00 | 0.0 |
| 1356 | $pNH_3$ | $oHD_2^+$ | $oD_2$ | $pNH_4^+$ | 7.6e-10 | 0.00 | 0.0 |
| 1357 | $pNH_3$ | $pHD_2^+$ | $pD_2$ | $pNH_4^+$ | 7.6e-10 | 0.00 | 0.0 |
| 1358 | $pNH_3$ | $oHD_2^+$ | $oD_2$ | $oNH_4^+$ | 2.3e-09 | 0.00 | 0.0 |
| 1359 | $pNH_3$ | $pHD_2^+$ | $pD_2$ | $oNH_4^+$ | 2.3e-09 | 0.00 | 0.0 |
| 1360 | $oNH_3$ | $oHD_2^+$ | $oNH_3D^+$ | HD | 4.5e-09 | 0.00 | 0.0 |
| 1361 | $oNH_3$ | $pHD_2^+$ | $oNH_3D^+$ | HD | 4.5e-09 | 0.00 | 0.0 |
| 1362 | $oNH_3$ | $oHD_2^+$ | $pNH_3D^+$ | HD | 1.5e-09 | 0.00 | 0.0 |
| 1363 | $oNH_3$ | $pHD_2^+$ | $pNH_3D^+$ | HD | 1.5e-09 | 0.00 | 0.0 |
| 1364 | $pNH_3$ | $oHD_2^+$ | $oNH_3D^+$ | HD | 1.5e-09 | 0.00 | 0.0 |
| 1365 | $pNH_3$ | $pHD_2^+$ | $oNH_3D^+$ | HD | 1.5e-09 | 0.00 | 0.0 |
| 1366 | $pNH_3$ | $oHD_2^+$ | $pNH_3D^+$ | HD | 4.5e-09 | 0.00 | 0.0 |
| 1367 | $pNH_3$ | $pHD_2^+$ | $pNH_3D^+$ | HD | 4.5e-09 | 0.00 | 0.0 |
| 1368 | $oNH_3$ | $mD_3^+$ | $oD_2$ | $oNH_3D^+$ | 9.1e-09 | 0.00 | 0.0 |
| 1369 | $oNH_3$ | $pD_3^+$ | $pD_2$ | $oNH_3D^+$ | 9.1e-09 | 0.00 | 0.0 |
| 1370 | $oNH_3$ | $oD_3^+$ | $oD_2$ | $oNH_3D^+$ | 4.6e-09 | 0.00 | 0.0 |
| 1371 | $oNH_3$ | $oD_3^+$ | $pD_2$ | $oNH_3D^+$ | 4.6e-09 | 0.00 | 0.0 |
| 1372 | $pNH_3$ | $mD_3^+$ | $oD_2$ | $pNH_3D^+$ | 9.1e-09 | 0.00 | 0.0 |
| 1373 | $pNH_3$ | $pD_3^+$ | $pD_2$ | $pNH_3D^+$ | 9.1e-09 | 0.00 | 0.0 |
| 1374 | $pNH_3$ | $oD_3^+$ | $oD_2$ | $pNH_3D^+$ | 4.6e-09 | 0.00 | 0.0 |
| 1375 | $pNH_3$ | $oD_3^+$ | $pD_2$ | $pNH_3D^+$ | 4.6e-09 | 0.00 | 0.0 |
| 1376 | ND | $oH_3^+$ | $oH_2$ | $NHD^+$ | 1.3e-09 | 0.00 | 0.0 |
| 1377 | ND | $pH_3^+$ | $oH_2$ | $NHD^+$ | 6.5e-10 | 0.00 | 0.0 |
| 1378 | ND | $pH_3^+$ | $pH_2$ | $NHD^+$ | 6.5e-10 | 0.00 | 0.0 |
| 1379 | ND | $oH_2D^+$ | $NHD^+$ | HD | 8.7e-10 | 0.00 | 0.0 |
| 1380 | ND | $pH_2D^+$ | $NHD^+$ | HD | 8.7e-10 | 0.00 | 0.0 |
| 1381 | ND | $oH_2D^+$ | $oND_2^+$ | $oH_2$ | 2.9e-10 | 0.00 | 0.0 |
| 1382 | ND | $oH_2D^+$ | $pND_2^+$ | $oH_2$ | 1.4e-10 | 0.00 | 0.0 |
| 1383 | ND | $pH_2D^+$ | $oND_2^+$ | $pH_2$ | 2.9e-10 | 0.00 | 0.0 |
| 1384 | ND | $pH_2D^+$ | $pND_2^+$ | $pH_2$ | 1.4e-10 | 0.00 | 0.0 |
| 1385 | ND | $oHD_2^+$ | $oD_2$ | $NHD^+$ | 3.4e-10 | 0.00 | 0.0 |
| 1386 | ND | $oHD_2^+$ | $pD_2$ | $NHD^+$ | 9.6e-11 | 0.00 | 0.0 |
| 1387 | ND | $pHD_2^+$ | $oD_2$ | $NHD^+$ | 1.9e-10 | 0.00 | 0.0 |
| 1388 | ND | $pHD_2^+$ | $pD_2$ | $NHD^+$ | 2.4e-10 | 0.00 | 0.0 |
| 1389 | ND | $oHD_2^+$ | $oND_2^+$ | HD | 6.7e-10 | 0.00 | 0.0 |
| 1390 | ND | $oHD_2^+$ | $pND_2^+$ | HD | 1.9e-10 | 0.00 | 0.0 |
| 1391 | ND | $pHD_2^+$ | $oND_2^+$ | HD | 3.9e-10 | 0.00 | 0.0 |
| 1392 | ND | $pHD_2^+$ | $pND_2^+$ | HD | 4.8e-10 | 0.00 | 0.0 |
| 1393 | ND | $mD_3^+$ | $oND_2^+$ | $oD_2$ | 8.7e-10 | 0.00 | 0.0 |
| 1394 | ND | $mD_3^+$ | $oND_2^+$ | $pD_2$ | 2.2e-10 | 0.00 | 0.0 |
| 1395 | ND | $mD_3^+$ | $pND_2^+$ | $oD_2$ | 2.2e-10 | 0.00 | 0.0 |
| 1396 | ND | $pD_3^+$ | $oND_2^+$ | $pD_2$ | 4.3e-10 | 0.00 | 0.0 |
| 1397 | ND | $pD_3^+$ | $pND_2^+$ | $oD_2$ | 4.3e-10 | 0.00 | 0.0 |
| 1398 | ND | $pD_3^+$ | $pND_2^+$ | $pD_2$ | 4.3e-10 | 0.00 | 0.0 |
| 1399 | ND | $oD_3^+$ | $oND_2^+$ | $oD_2$ | 4.3e-10 | 0.00 | 0.0 |
| 1400 | ND | $oD_3^+$ | $oND_2^+$ | $pD_2$ | 3.3e-10 | 0.00 | 0.0 |
| 1401 | ND | $oD_3^+$ | $pND_2^+$ | $oD_2$ | 3.3e-10 | 0.00 | 0.0 |
| 1402 | ND | $oD_3^+$ | $pND_2^+$ | $pD_2$ | 2.2e-10 | 0.00 | 0.0 |
| 1403 | NHD | $oH_3^+$ | $oNH_2D^+$ | $oH_2$ | 1.4e-09 | 0.00 | 0.0 |
| 1404 | NHD | $oH_3^+$ | $oNH_2D^+$ | $pH_2$ | 2.2e-10 | 0.00 | 0.0 |
| 1405 | NHD | $oH_3^+$ | $pNH_2D^+$ | $oH_2$ | 2.2e-10 | 0.00 | 0.0 |
| 1406 | NHD | $pH_3^+$ | $oNH_2D^+$ | $oH_2$ | 6.8e-10 | 0.00 | 0.0 |

**Table D6** – *continued* (part 20)

| # | Reactants | | Products | | α | β | γ |
|---|---|---|---|---|---|---|---|
| 1407 | NHD | $pH_3^+$ | $oNH_2D^+$ | $pH_2$ | 4.5e-10 | 0.00 | 0.0 |
| 1408 | NHD | $pH_3^+$ | $pNH_2D^+$ | $oH_2$ | 4.5e-10 | 0.00 | 0.0 |
| 1409 | NHD | $pH_3^+$ | $pNH_2D^+$ | $pH_2$ | 2.2e-10 | 0.00 | 0.0 |
| 1410 | NHD | $oH_2D^+$ | $oNH_2D^+$ | HD | 1.0e-09 | 0.00 | 0.0 |
| 1411 | NHD | $oH_2D^+$ | $pNH_2D^+$ | HD | 2.0e-10 | 0.00 | 0.0 |
| 1412 | NHD | $pH_2D^+$ | $oNH_2D^+$ | HD | 6.0e-10 | 0.00 | 0.0 |
| 1413 | NHD | $pH_2D^+$ | $pNH_2D^+$ | HD | 6.0e-10 | 0.00 | 0.0 |
| 1414 | NHD | $oH_2D^+$ | $oNHD_2^+$ | $oH_2$ | 3.3e-10 | 0.00 | 0.0 |
| 1415 | NHD | $oH_2D^+$ | $pNHD_2^+$ | $oH_2$ | 1.7e-10 | 0.00 | 0.0 |
| 1416 | NHD | $oH_2D^+$ | $oNHD_2^+$ | $pH_2$ | 6.7e-11 | 0.00 | 0.0 |
| 1417 | NHD | $oH_2D^+$ | $pNHD_2^+$ | $pH_2$ | 3.3e-11 | 0.00 | 0.0 |
| 1418 | NHD | $pH_2D^+$ | $oNHD_2^+$ | $oH_2$ | 2.0e-10 | 0.00 | 0.0 |
| 1419 | NHD | $pH_2D^+$ | $pNHD_2^+$ | $oH_2$ | 1.0e-10 | 0.00 | 0.0 |
| 1420 | NHD | $pH_2D^+$ | $oNHD_2^+$ | $pH_2$ | 2.0e-10 | 0.00 | 0.0 |
| 1421 | NHD | $pH_2D^+$ | $pNHD_2^+$ | $pH_2$ | 1.0e-10 | 0.00 | 0.0 |
| 1422 | NHD | $oHD_2^+$ | $oD_2$ | $oNH_2D^+$ | 3.5e-10 | 0.00 | 0.0 |
| 1423 | NHD | $oHD_2^+$ | $pD_2$ | $oNH_2D^+$ | 1.0e-10 | 0.00 | 0.0 |
| 1424 | NHD | $pHD_2^+$ | $oD_2$ | $oNH_2D^+$ | 2.0e-10 | 0.00 | 0.0 |
| 1425 | NHD | $pHD_2^+$ | $pD_2$ | $oNH_2D^+$ | 2.5e-10 | 0.00 | 0.0 |
| 1426 | NHD | $oHD_2^+$ | $oD_2$ | $pNH_2D^+$ | 1.2e-10 | 0.00 | 0.0 |
| 1427 | NHD | $oHD_2^+$ | $pD_2$ | $pNH_2D^+$ | 3.3e-11 | 0.00 | 0.0 |
| 1428 | NHD | $pHD_2^+$ | $oD_2$ | $pNH_2D^+$ | 6.7e-11 | 0.00 | 0.0 |
| 1429 | NHD | $pHD_2^+$ | $pD_2$ | $pNH_2D^+$ | 8.3e-11 | 0.00 | 0.0 |
| 1430 | NHD | $oHD_2^+$ | $oNHD_2^+$ | HD | 9.3e-10 | 0.00 | 0.0 |
| 1431 | NHD | $oHD_2^+$ | $pNHD_2^+$ | HD | 2.7e-10 | 0.00 | 0.0 |
| 1432 | NHD | $pHD_2^+$ | $oNHD_2^+$ | HD | 5.3e-10 | 0.00 | 0.0 |
| 1433 | NHD | $pHD_2^+$ | $pNHD_2^+$ | HD | 6.7e-10 | 0.00 | 0.0 |
| 1434 | NHD | $mD_3^+$ | $oNHD_2^+$ | $oD_2$ | 1.2e-09 | 0.00 | 0.0 |
| 1435 | NHD | $mD_3^+$ | $oNHD_2^+$ | $pD_2$ | 3.0e-10 | 0.00 | 0.0 |
| 1436 | NHD | $mD_3^+$ | $pNHD_2^+$ | $oD_2$ | 3.0e-10 | 0.00 | 0.0 |
| 1437 | NHD | $pD_3^+$ | $oNHD_2^+$ | $pD_2$ | 6.0e-10 | 0.00 | 0.0 |
| 1438 | NHD | $pD_3^+$ | $pNHD_2^+$ | $oD_2$ | 6.0e-10 | 0.00 | 0.0 |
| 1439 | NHD | $pD_3^+$ | $pNHD_2^+$ | $pD_2$ | 6.0e-10 | 0.00 | 0.0 |
| 1440 | NHD | $oD_3^+$ | $oNHD_2^+$ | $oD_2$ | 6.0e-10 | 0.00 | 0.0 |
| 1441 | NHD | $oD_3^+$ | $oNHD_2^+$ | $pD_2$ | 4.5e-10 | 0.00 | 0.0 |
| 1442 | NHD | $oD_3^+$ | $pNHD_2^+$ | $oD_2$ | 4.5e-10 | 0.00 | 0.0 |
| 1443 | NHD | $oD_3^+$ | $pNHD_2^+$ | $pD_2$ | 3.0e-10 | 0.00 | 0.0 |
| 1444 | $oND_2$ | $oH_3^+$ | $oNHD_2^+$ | $oH_2$ | 1.8e-09 | 0.00 | 0.0 |
| 1445 | $pND_2$ | $oH_3^+$ | $pNHD_2^+$ | $oH_2$ | 1.8e-09 | 0.00 | 0.0 |
| 1446 | $oND_2$ | $pH_3^+$ | $oNHD_2^+$ | $oH_2$ | 9.0e-10 | 0.00 | 0.0 |
| 1447 | $pND_2$ | $pH_3^+$ | $pNHD_2^+$ | $oH_2$ | 9.0e-10 | 0.00 | 0.0 |
| 1448 | $oND_2$ | $pH_3^+$ | $oNHD_2^+$ | $pH_2$ | 9.0e-10 | 0.00 | 0.0 |
| 1449 | $pND_2$ | $pH_3^+$ | $pNHD_2^+$ | $pH_2$ | 9.0e-10 | 0.00 | 0.0 |
| 1450 | $oND_2$ | $oH_2D^+$ | $oNHD_2^+$ | HD | 9.3e-10 | 0.00 | 0.0 |
| 1451 | $oND_2$ | $oH_2D^+$ | $pNHD_2^+$ | HD | 2.7e-10 | 0.00 | 0.0 |
| 1452 | $pND_2$ | $oH_2D^+$ | $oNHD_2^+$ | HD | 5.3e-10 | 0.00 | 0.0 |
| 1453 | $pND_2$ | $oH_2D^+$ | $pNHD_2^+$ | HD | 6.7e-10 | 0.00 | 0.0 |
| 1454 | $oND_2$ | $pH_2D^+$ | $oNHD_2^+$ | HD | 9.3e-10 | 0.00 | 0.0 |
| 1455 | $oND_2$ | $pH_2D^+$ | $pNHD_2^+$ | HD | 2.7e-10 | 0.00 | 0.0 |
| 1456 | $pND_2$ | $pH_2D^+$ | $oNHD_2^+$ | HD | 5.3e-10 | 0.00 | 0.0 |
| 1457 | $pND_2$ | $pH_2D^+$ | $pNHD_2^+$ | HD | 6.7e-10 | 0.00 | 0.0 |
| 1458 | $oND_2$ | $oH_2D^+$ | $mND_3^+$ | $oH_2$ | 3.3e-10 | 0.00 | 0.0 |
| 1459 | $oND_2$ | $oH_2D^+$ | $oND_3^+$ | $oH_2$ | 2.7e-10 | 0.00 | 0.0 |
| 1460 | $pND_2$ | $oH_2D^+$ | $pND_3^+$ | $oH_2$ | 6.7e-11 | 0.00 | 0.0 |
| 1461 | $pND_2$ | $oH_2D^+$ | $oND_3^+$ | $oH_2$ | 5.3e-10 | 0.00 | 0.0 |
| 1462 | $oND_2$ | $pH_2D^+$ | $mND_3^+$ | $pH_2$ | 3.3e-10 | 0.00 | 0.0 |
| 1463 | $oND_2$ | $pH_2D^+$ | $oND_3^+$ | $pH_2$ | 2.7e-10 | 0.00 | 0.0 |
| 1464 | $pND_2$ | $pH_2D^+$ | $pND_3^+$ | $pH_2$ | 6.7e-11 | 0.00 | 0.0 |
| 1465 | $pND_2$ | $pH_2D^+$ | $oND_3^+$ | $pH_2$ | 5.3e-10 | 0.00 | 0.0 |
| 1466 | $oND_2$ | $oHD_2^+$ | $oNHD_2^+$ | $oD_2$ | 3.8e-10 | 0.00 | 0.0 |
| 1467 | $oND_2$ | $oHD_2^+$ | $oNHD_2^+$ | $pD_2$ | 8.3e-11 | 0.00 | 0.0 |
| 1468 | $oND_2$ | $oHD_2^+$ | $pNHD_2^+$ | $oD_2$ | 8.3e-11 | 0.00 | 0.0 |
| 1469 | $oND_2$ | $oHD_2^+$ | $pNHD_2^+$ | $pD_2$ | 5.0e-11 | 0.00 | 0.0 |
| 1470 | $oND_2$ | $pHD_2^+$ | $oNHD_2^+$ | $oD_2$ | 1.7e-10 | 0.00 | 0.0 |
| 1471 | $oND_2$ | $pHD_2^+$ | $oNHD_2^+$ | $pD_2$ | 2.0e-10 | 0.00 | 0.0 |
| 1472 | $oND_2$ | $pHD_2^+$ | $pNHD_2^+$ | $oD_2$ | 2.0e-10 | 0.00 | 0.0 |
| 1473 | $oND_2$ | $pHD_2^+$ | $pNHD_2^+$ | $pD_2$ | 3.3e-11 | 0.00 | 0.0 |
| 1474 | $pND_2$ | $oHD_2^+$ | $oNHD_2^+$ | $oD_2$ | 1.7e-10 | 0.00 | 0.0 |
| 1475 | $pND_2$ | $oHD_2^+$ | $oNHD_2^+$ | $pD_2$ | 2.0e-10 | 0.00 | 0.0 |
| 1476 | $pND_2$ | $oHD_2^+$ | $pNHD_2^+$ | $oD_2$ | 2.0e-10 | 0.00 | 0.0 |
| 1477 | $pND_2$ | $oHD_2^+$ | $pNHD_2^+$ | $pD_2$ | 3.3e-11 | 0.00 | 0.0 |
| 1478 | $pND_2$ | $pHD_2^+$ | $oNHD_2^+$ | $oD_2$ | 2.0e-10 | 0.00 | 0.0 |
| 1479 | $pND_2$ | $pHD_2^+$ | $oNHD_2^+$ | $pD_2$ | 6.7e-11 | 0.00 | 0.0 |
| 1480 | $pND_2$ | $pHD_2^+$ | $pNHD_2^+$ | $oD_2$ | 6.7e-11 | 0.00 | 0.0 |

**Table D6** – *continued* (part 21)

| # | Reactants | | Products | | $\alpha$ | $\beta$ | $\gamma$ |
|---|---|---|---|---|---|---|---|
| 1481 | $pND_2$ | $pHD_2^+$ | $pNHD_2^+$ | $pD_2$ | 2.7e-10 | 0.00 | 0.0 |
| 1482 | $oND_2$ | $oHD_2^+$ | $mND_3^+$ | HD | 6.7e-10 | 0.00 | 0.0 |
| 1483 | $oND_2$ | $oHD_2^+$ | $oND_3^+$ | HD | 5.3e-10 | 0.00 | 0.0 |
| 1484 | $oND_2$ | $pHD_2^+$ | $mND_3^+$ | HD | 3.3e-10 | 0.00 | 0.0 |
| 1485 | $oND_2$ | $pHD_2^+$ | $pND_3^+$ | HD | 6.7e-11 | 0.00 | 0.0 |
| 1486 | $oND_2$ | $pHD_2^+$ | $oND_3^+$ | HD | 8.0e-10 | 0.00 | 0.0 |
| 1487 | $pND_2$ | $oHD_2^+$ | $mND_3^+$ | HD | 3.3e-10 | 0.00 | 0.0 |
| 1488 | $pND_2$ | $oHD_2^+$ | $pND_3^+$ | HD | 6.7e-11 | 0.00 | 0.0 |
| 1489 | $pND_2$ | $oHD_2^+$ | $oND_3^+$ | HD | 8.0e-10 | 0.00 | 0.0 |
| 1490 | $pND_2$ | $pHD_2^+$ | $pND_3^+$ | HD | 1.3e-10 | 0.00 | 0.0 |
| 1491 | $pND_2$ | $pHD_2^+$ | $oND_3^+$ | HD | 1.1e-09 | 0.00 | 0.0 |
| 1492 | $oND_2$ | $mD_3^+$ | $mND_3^+$ | $oD_2$ | 9.0e-10 | 0.00 | 0.0 |
| 1493 | $oND_2$ | $mD_3^+$ | $mND_3^+$ | $pD_2$ | 1.8e-10 | 0.00 | 0.0 |
| 1494 | $oND_2$ | $mD_3^+$ | $oND_3^+$ | $oD_2$ | 5.4e-10 | 0.00 | 0.0 |
| 1495 | $oND_2$ | $mD_3^+$ | $oND_3^+$ | $pD_2$ | 1.8e-10 | 0.00 | 0.0 |
| 1496 | $pND_2$ | $mD_3^+$ | $mND_3^+$ | $oD_2$ | 3.6e-10 | 0.00 | 0.0 |
| 1497 | $pND_2$ | $mD_3^+$ | $mND_3^+$ | $pD_2$ | 4.2e-10 | 0.00 | 0.0 |
| 1498 | $pND_2$ | $mD_3^+$ | $pND_3^+$ | $oD_2$ | 6.0e-11 | 0.00 | 0.0 |
| 1499 | $pND_2$ | $mD_3^+$ | $oND_3^+$ | $oD_2$ | 8.4e-10 | 0.00 | 0.0 |
| 1500 | $pND_2$ | $mD_3^+$ | $oND_3^+$ | $pD_2$ | 1.2e-10 | 0.00 | 0.0 |
| 1501 | $oND_2$ | $pD_3^+$ | $mND_3^+$ | $pD_2$ | 3.0e-10 | 0.00 | 0.0 |
| 1502 | $oND_2$ | $pD_3^+$ | $pND_3^+$ | $oD_2$ | 3.0e-10 | 0.00 | 0.0 |
| 1503 | $oND_2$ | $pD_3^+$ | $oND_3^+$ | $oD_2$ | 6.0e-10 | 0.00 | 0.0 |
| 1504 | $oND_2$ | $pD_3^+$ | $oND_3^+$ | $pD_2$ | 6.0e-10 | 0.00 | 0.0 |
| 1505 | $pND_2$ | $pD_3^+$ | $pND_3^+$ | $pD_2$ | 3.6e-10 | 0.00 | 0.0 |
| 1506 | $pND_2$ | $pD_3^+$ | $oND_3^+$ | $oD_2$ | 7.2e-10 | 0.00 | 0.0 |
| 1507 | $pND_2$ | $pD_3^+$ | $oND_3^+$ | $pD_2$ | 7.2e-10 | 0.00 | 0.0 |
| 1508 | $oND_2$ | $oD_3^+$ | $mND_3^+$ | $oD_2$ | 3.4e-10 | 0.00 | 0.0 |
| 1509 | $oND_2$ | $oD_3^+$ | $mND_3^+$ | $pD_2$ | 2.6e-10 | 0.00 | 0.0 |
| 1510 | $oND_2$ | $oD_3^+$ | $pND_3^+$ | $oD_2$ | 3.7e-11 | 0.00 | 0.0 |
| 1511 | $oND_2$ | $oD_3^+$ | $pND_3^+$ | $pD_2$ | 2.2e-11 | 0.00 | 0.0 |
| 1512 | $oND_2$ | $oD_3^+$ | $oND_3^+$ | $oD_2$ | 8.0e-10 | 0.00 | 0.0 |
| 1513 | $oND_2$ | $oD_3^+$ | $oND_3^+$ | $pD_2$ | 3.4e-10 | 0.00 | 0.0 |
| 1514 | $pND_2$ | $oD_3^+$ | $mND_3^+$ | $oD_2$ | 2.2e-10 | 0.00 | 0.0 |
| 1515 | $pND_2$ | $oD_3^+$ | $mND_3^+$ | $pD_2$ | 7.5e-11 | 0.00 | 0.0 |
| 1516 | $pND_2$ | $oD_3^+$ | $pND_3^+$ | $oD_2$ | 7.5e-11 | 0.00 | 0.0 |
| 1517 | $pND_2$ | $oD_3^+$ | $pND_3^+$ | $pD_2$ | 4.5e-11 | 0.00 | 0.0 |
| 1518 | $pND_2$ | $oD_3^+$ | $oND_3^+$ | $oD_2$ | 6.9e-10 | 0.00 | 0.0 |
| 1519 | $pND_2$ | $oD_3^+$ | $oND_3^+$ | $pD_2$ | 6.9e-10 | 0.00 | 0.0 |
| 1520 | $oNH_2D$ | $H^+$ | $oNH_2D^+$ | H | 5.2e-09 | 0.00 | 0.0 |
| 1521 | $pNH_2D$ | $H^+$ | $pNH_2D^+$ | H | 5.2e-09 | 0.00 | 0.0 |
| 1522 | $oNH_2D$ | $oH_3^+$ | $oNH_3D^+$ | $oH_2$ | 5.6e-09 | 0.00 | 0.0 |
| 1523 | $oNH_2D$ | $oH_3^+$ | $oNH_3D^+$ | $pH_2$ | 7.6e-10 | 0.00 | 0.0 |
| 1524 | $oNH_2D$ | $oH_3^+$ | $pNH_3D^+$ | $oH_2$ | 2.1e-09 | 0.00 | 0.0 |
| 1525 | $oNH_2D$ | $oH_3^+$ | $pNH_3D^+$ | $pH_2$ | 6.1e-10 | 0.00 | 0.0 |
| 1526 | $pNH_2D$ | $oH_3^+$ | $oNH_3D^+$ | $oH_2$ | 2.3e-09 | 0.00 | 0.0 |
| 1527 | $pNH_2D$ | $oH_3^+$ | $oNH_3D^+$ | $pH_2$ | 2.3e-09 | 0.00 | 0.0 |
| 1528 | $pNH_2D$ | $oH_3^+$ | $pNH_3D^+$ | $oH_2$ | 4.6e-09 | 0.00 | 0.0 |
| 1529 | $oNH_2D$ | $pH_3^+$ | $oNH_3D^+$ | $oH_2$ | 2.1e-09 | 0.00 | 0.0 |
| 1530 | $oNH_2D$ | $pH_3^+$ | $oNH_3D^+$ | $pH_2$ | 1.5e-09 | 0.00 | 0.0 |
| 1531 | $oNH_2D$ | $pH_3^+$ | $pNH_3D^+$ | $oH_2$ | 4.2e-09 | 0.00 | 0.0 |
| 1532 | $oNH_2D$ | $pH_3^+$ | $pNH_3D^+$ | $pH_2$ | 1.2e-09 | 0.00 | 0.0 |
| 1533 | $pNH_2D$ | $pH_3^+$ | $oNH_3D^+$ | $oH_2$ | 1.8e-09 | 0.00 | 0.0 |
| 1534 | $pNH_2D$ | $pH_3^+$ | $pNH_3D^+$ | $oH_2$ | 3.6e-09 | 0.00 | 0.0 |
| 1535 | $pNH_2D$ | $pH_3^+$ | $pNH_3D^+$ | $pH_2$ | 3.6e-09 | 0.00 | 0.0 |
| 1536 | $oNH_2D$ | $oH_2D^+$ | $oNH_3D^+$ | HD | 4.0e-09 | 0.00 | 0.0 |
| 1537 | $oNH_2D$ | $oH_2D^+$ | $pNH_3D^+$ | HD | 2.0e-09 | 0.00 | 0.0 |
| 1538 | $oNH_2D$ | $pH_2D^+$ | $oNH_3D^+$ | HD | 2.0e-09 | 0.00 | 0.0 |
| 1539 | $oNH_2D$ | $pH_2D^+$ | $pNH_3D^+$ | HD | 4.0e-09 | 0.00 | 0.0 |
| 1540 | $pNH_2D$ | $oH_2D^+$ | $oNH_3D^+$ | HD | 2.0e-09 | 0.00 | 0.0 |
| 1541 | $pNH_2D$ | $oH_2D^+$ | $pNH_3D^+$ | HD | 4.0e-09 | 0.00 | 0.0 |
| 1542 | $pNH_2D$ | $pH_2D^+$ | $pNH_3D^+$ | HD | 6.1e-09 | 0.00 | 0.0 |
| 1543 | $oNH_2D$ | $oH_2D^+$ | $ooNH_2D_2^+$ | $oH_2$ | 1.5e-09 | 0.00 | 0.0 |
| 1544 | $oNH_2D$ | $oH_2D^+$ | $poNH_2D_2^+$ | $oH_2$ | 7.3e-10 | 0.00 | 0.0 |
| 1545 | $oNH_2D$ | $oH_2D^+$ | $ooNH_2D_2^+$ | $pH_2$ | 2.2e-10 | 0.00 | 0.0 |
| 1546 | $oNH_2D$ | $oH_2D^+$ | $poNH_2D_2^+$ | $pH_2$ | 1.1e-10 | 0.00 | 0.0 |
| 1547 | $oNH_2D$ | $oH_2D^+$ | $opNH_2D_2^+$ | $oH_2$ | 2.2e-10 | 0.00 | 0.0 |
| 1548 | $oNH_2D$ | $oH_2D^+$ | $ppNH_2D_2^+$ | $oH_2$ | 1.1e-10 | 0.00 | 0.0 |
| 1549 | $oNH_2D$ | $oH_2D^+$ | $opNH_2D_2^+$ | $pH_2$ | 1.1e-10 | 0.00 | 0.0 |
| 1550 | $oNH_2D$ | $oH_2D^+$ | $ppNH_2D_2^+$ | $pH_2$ | 5.6e-11 | 0.00 | 0.0 |
| 1551 | $oNH_2D$ | $pH_2D^+$ | $ooNH_2D_2^+$ | $oH_2$ | 6.7e-10 | 0.00 | 0.0 |
| 1552 | $oNH_2D$ | $pH_2D^+$ | $poNH_2D_2^+$ | $oH_2$ | 3.4e-10 | 0.00 | 0.0 |
| 1553 | $oNH_2D$ | $pH_2D^+$ | $ooNH_2D_2^+$ | $pH_2$ | 6.7e-10 | 0.00 | 0.0 |
| 1554 | $oNH_2D$ | $pH_2D^+$ | $poNH_2D_2^+$ | $pH_2$ | 3.4e-10 | 0.00 | 0.0 |

**Table D6** – *continued* (part 22)

| # | Reactants | | Products | | $\alpha$ | $\beta$ | $\gamma$ |
|---|---|---|---|---|---|---|---|
| 1555 | $oNH_2D$ | $pH_2D^+$ | $opNH_2D_2^+$ | $oH_2$ | 6.7e-10 | 0.00 | 0.0 |
| 1556 | $oNH_2D$ | $pH_2D^+$ | $ppNH_2D_2^+$ | $oH_2$ | 3.4e-10 | 0.00 | 0.0 |
| 1557 | $pNH_2D$ | $oH_2D^+$ | $ooNH_2D_2^+$ | $oH_2$ | 6.7e-10 | 0.00 | 0.0 |
| 1558 | $pNH_2D$ | $oH_2D^+$ | $poNH_2D_2^+$ | $oH_2$ | 3.4e-10 | 0.00 | 0.0 |
| 1559 | $pNH_2D$ | $oH_2D^+$ | $ooNH_2D_2^+$ | $pH_2$ | 6.7e-10 | 0.00 | 0.0 |
| 1560 | $pNH_2D$ | $oH_2D^+$ | $poNH_2D_2^+$ | $pH_2$ | 3.4e-10 | 0.00 | 0.0 |
| 1561 | $pNH_2D$ | $oH_2D^+$ | $opNH_2D_2^+$ | $oH_2$ | 6.7e-10 | 0.00 | 0.0 |
| 1562 | $pNH_2D$ | $oH_2D^+$ | $ppNH_2D_2^+$ | $oH_2$ | 3.4e-10 | 0.00 | 0.0 |
| 1563 | $pNH_2D$ | $pH_2D^+$ | $ooNH_2D_2^+$ | $oH_2$ | 1.0e-09 | 0.00 | 0.0 |
| 1564 | $pNH_2D$ | $pH_2D^+$ | $poNH_2D_2^+$ | $oH_2$ | 5.1e-10 | 0.00 | 0.0 |
| 1565 | $pNH_2D$ | $pH_2D^+$ | $opNH_2D_2^+$ | $pH_2$ | 1.0e-09 | 0.00 | 0.0 |
| 1566 | $pNH_2D$ | $pH_2D^+$ | $ppNH_2D_2^+$ | $pH_2$ | 5.1e-10 | 0.00 | 0.0 |
| 1567 | $oNH_2D$ | $oHD_2^+$ | $oD_2$ | $oNH_3D^+$ | 1.6e-09 | 0.00 | 0.0 |
| 1568 | $oNH_2D$ | $oHD_2^+$ | $pD_2$ | $oNH_3D^+$ | 4.5e-10 | 0.00 | 0.0 |
| 1569 | $oNH_2D$ | $pHD_2^+$ | $oD_2$ | $oNH_3D^+$ | 9.0e-10 | 0.00 | 0.0 |
| 1570 | $oNH_2D$ | $pHD_2^+$ | $pD_2$ | $oNH_3D^+$ | 1.1e-09 | 0.00 | 0.0 |
| 1571 | $oNH_2D$ | $oHD_2^+$ | $oD_2$ | $pNH_3D^+$ | 7.9e-10 | 0.00 | 0.0 |
| 1572 | $oNH_2D$ | $oHD_2^+$ | $pD_2$ | $pNH_3D^+$ | 2.2e-10 | 0.00 | 0.0 |
| 1573 | $oNH_2D$ | $pHD_2^+$ | $oD_2$ | $pNH_3D^+$ | 4.5e-10 | 0.00 | 0.0 |
| 1574 | $oNH_2D$ | $pHD_2^+$ | $pD_2$ | $pNH_3D^+$ | 5.6e-10 | 0.00 | 0.0 |
| 1575 | $pNH_2D$ | $oHD_2^+$ | $oD_2$ | $pNH_3D^+$ | 2.4e-09 | 0.00 | 0.0 |
| 1576 | $pNH_2D$ | $oHD_2^+$ | $pD_2$ | $pNH_3D^+$ | 6.7e-10 | 0.00 | 0.0 |
| 1577 | $pNH_2D$ | $pHD_2^+$ | $oD_2$ | $pNH_3D^+$ | 1.3e-09 | 0.00 | 0.0 |
| 1578 | $pNH_2D$ | $pHD_2^+$ | $pD_2$ | $pNH_3D^+$ | 1.7e-09 | 0.00 | 0.0 |
| 1579 | $oNH_2D$ | $oHD_2^+$ | $ooNH_2D_2^+$ | $HD$ | 3.9e-09 | 0.00 | 0.0 |
| 1580 | $oNH_2D$ | $oHD_2^+$ | $poNH_2D_2^+$ | $HD$ | 1.1e-09 | 0.00 | 0.0 |
| 1581 | $oNH_2D$ | $pHD_2^+$ | $ooNH_2D_2^+$ | $HD$ | 2.2e-09 | 0.00 | 0.0 |
| 1582 | $oNH_2D$ | $pHD_2^+$ | $poNH_2D_2^+$ | $HD$ | 2.8e-09 | 0.00 | 0.0 |
| 1583 | $oNH_2D$ | $oHD_2^+$ | $opNH_2D_2^+$ | $HD$ | 7.9e-10 | 0.00 | 0.0 |
| 1584 | $oNH_2D$ | $oHD_2^+$ | $ppNH_2D_2^+$ | $HD$ | 2.2e-10 | 0.00 | 0.0 |
| 1585 | $oNH_2D$ | $pHD_2^+$ | $opNH_2D_2^+$ | $HD$ | 4.5e-10 | 0.00 | 0.0 |
| 1586 | $oNH_2D$ | $pHD_2^+$ | $ppNH_2D_2^+$ | $HD$ | 5.6e-10 | 0.00 | 0.0 |
| 1587 | $pNH_2D$ | $oHD_2^+$ | $ooNH_2D_2^+$ | $HD$ | 2.4e-09 | 0.00 | 0.0 |
| 1588 | $pNH_2D$ | $oHD_2^+$ | $poNH_2D_2^+$ | $HD$ | 6.7e-10 | 0.00 | 0.0 |
| 1589 | $pNH_2D$ | $pHD_2^+$ | $ooNH_2D_2^+$ | $HD$ | 1.3e-09 | 0.00 | 0.0 |
| 1590 | $pNH_2D$ | $pHD_2^+$ | $poNH_2D_2^+$ | $HD$ | 1.7e-09 | 0.00 | 0.0 |
| 1591 | $pNH_2D$ | $oHD_2^+$ | $opNH_2D_2^+$ | $HD$ | 2.4e-09 | 0.00 | 0.0 |
| 1592 | $pNH_2D$ | $oHD_2^+$ | $ppNH_2D_2^+$ | $HD$ | 6.7e-10 | 0.00 | 0.0 |
| 1593 | $pNH_2D$ | $pHD_2^+$ | $opNH_2D_2^+$ | $HD$ | 1.3e-09 | 0.00 | 0.0 |
| 1594 | $pNH_2D$ | $pHD_2^+$ | $ppNH_2D_2^+$ | $HD$ | 1.7e-09 | 0.00 | 0.0 |
| 1595 | $oNH_2D$ | $mD_3^+$ | $ooNH_2D_2^+$ | $oD_2$ | 6.1e-09 | 0.00 | 0.0 |
| 1596 | $oNH_2D$ | $mD_3^+$ | $ooNH_2D_2^+$ | $pD_2$ | 1.5e-09 | 0.00 | 0.0 |
| 1597 | $oNH_2D$ | $mD_3^+$ | $poNH_2D_2^+$ | $oD_2$ | 1.5e-09 | 0.00 | 0.0 |
| 1598 | $oNH_2D$ | $pD_3^+$ | $ooNH_2D_2^+$ | $pD_2$ | 3.0e-09 | 0.00 | 0.0 |
| 1599 | $oNH_2D$ | $pD_3^+$ | $poNH_2D_2^+$ | $oD_2$ | 3.0e-09 | 0.00 | 0.0 |
| 1600 | $oNH_2D$ | $pD_3^+$ | $poNH_2D_2^+$ | $pD_2$ | 3.0e-09 | 0.00 | 0.0 |
| 1601 | $oNH_2D$ | $oD_3^+$ | $ooNH_2D_2^+$ | $oD_2$ | 3.0e-09 | 0.00 | 0.0 |
| 1602 | $oNH_2D$ | $oD_3^+$ | $ooNH_2D_2^+$ | $pD_2$ | 2.3e-09 | 0.00 | 0.0 |
| 1603 | $oNH_2D$ | $oD_3^+$ | $poNH_2D_2^+$ | $oD_2$ | 2.3e-09 | 0.00 | 0.0 |
| 1604 | $oNH_2D$ | $oD_3^+$ | $poNH_2D_2^+$ | $pD_2$ | 1.5e-09 | 0.00 | 0.0 |
| 1605 | $pNH_2D$ | $mD_3^+$ | $opNH_2D_2^+$ | $oD_2$ | 6.1e-09 | 0.00 | 0.0 |
| 1606 | $pNH_2D$ | $mD_3^+$ | $opNH_2D_2^+$ | $pD_2$ | 1.5e-09 | 0.00 | 0.0 |
| 1607 | $pNH_2D$ | $mD_3^+$ | $ppNH_2D_2^+$ | $oD_2$ | 1.5e-09 | 0.00 | 0.0 |
| 1608 | $pNH_2D$ | $pD_3^+$ | $opNH_2D_2^+$ | $pD_2$ | 3.0e-09 | 0.00 | 0.0 |
| 1609 | $pNH_2D$ | $pD_3^+$ | $ppNH_2D_2^+$ | $oD_2$ | 3.0e-09 | 0.00 | 0.0 |
| 1610 | $pNH_2D$ | $pD_3^+$ | $ppNH_2D_2^+$ | $pD_2$ | 3.0e-09 | 0.00 | 0.0 |
| 1611 | $pNH_2D$ | $oD_3^+$ | $opNH_2D_2^+$ | $oD_2$ | 3.0e-09 | 0.00 | 0.0 |
| 1612 | $pNH_2D$ | $oD_3^+$ | $opNH_2D_2^+$ | $pD_2$ | 2.3e-09 | 0.00 | 0.0 |
| 1613 | $pNH_2D$ | $oD_3^+$ | $ppNH_2D_2^+$ | $oD_2$ | 2.3e-09 | 0.00 | 0.0 |
| 1614 | $pNH_2D$ | $oD_3^+$ | $ppNH_2D_2^+$ | $pD_2$ | 1.5e-09 | 0.00 | 0.0 |
| 1615 | $oNHD_2$ | $H^+$ | $oNHD_2^+$ | $H$ | 5.2e-09 | 0.00 | 0.0 |
| 1616 | $pNHD_2$ | $H^+$ | $pNHD_2^+$ | $H$ | 5.2e-09 | 0.00 | 0.0 |
| 1617 | $oNHD_2$ | $oH_3^+$ | $ooNH_2D_2^+$ | $oH_2$ | 6.8e-09 | 0.00 | 0.0 |
| 1618 | $pNHD_2$ | $oH_3^+$ | $poNH_2D_2^+$ | $oH_2$ | 6.8e-09 | 0.00 | 0.0 |
| 1619 | $oNHD_2$ | $oH_3^+$ | $ooNH_2D_2^+$ | $pH_2$ | 1.1e-09 | 0.00 | 0.0 |
| 1620 | $pNHD_2$ | $oH_3^+$ | $poNH_2D_2^+$ | $pH_2$ | 1.1e-09 | 0.00 | 0.0 |
| 1621 | $oNHD_2$ | $oH_3^+$ | $opNH_2D_2^+$ | $oH_2$ | 1.1e-09 | 0.00 | 0.0 |
| 1622 | $pNHD_2$ | $oH_3^+$ | $ppNH_2D_2^+$ | $oH_2$ | 1.1e-09 | 0.00 | 0.0 |
| 1623 | $oNHD_2$ | $pH_3^+$ | $ooNH_2D_2^+$ | $oH_2$ | 3.4e-09 | 0.00 | 0.0 |
| 1624 | $pNHD_2$ | $pH_3^+$ | $poNH_2D_2^+$ | $oH_2$ | 3.4e-09 | 0.00 | 0.0 |
| 1625 | $oNHD_2$ | $pH_3^+$ | $ooNH_2D_2^+$ | $pH_2$ | 2.3e-09 | 0.00 | 0.0 |
| 1626 | $pNHD_2$ | $pH_3^+$ | $poNH_2D_2^+$ | $pH_2$ | 2.3e-09 | 0.00 | 0.0 |
| 1627 | $oNHD_2$ | $pH_3^+$ | $opNH_2D_2^+$ | $oH_2$ | 2.3e-09 | 0.00 | 0.0 |
| 1628 | $pNHD_2$ | $pH_3^+$ | $ppNH_2D_2^+$ | $oH_2$ | 2.3e-09 | 0.00 | 0.0 |

**Table D6** – *continued* (part 23)

| # | Reactants | | Products | | $\alpha$ | $\beta$ | $\gamma$ |
|---|---|---|---|---|---|---|---|
| 1629 | $oNHD_2$ | $pH_3^+$ | $opNH_2D_2^+$ | $pH_2$ | 1.1e-09 | 0.00 | 0.0 |
| 1630 | $pNHD_2$ | $pH_3^+$ | $ppNH_2D_2^+$ | $pH_2$ | 1.1e-09 | 0.00 | 0.0 |
| 1631 | $oNHD_2$ | $oH_2D^+$ | $ooNH_2D_2^+$ | HD | 3.9e-09 | 0.00 | 0.0 |
| 1632 | $oNHD_2$ | $oH_2D^+$ | $poNH_2D_2^+$ | HD | 1.1e-09 | 0.00 | 0.0 |
| 1633 | $pNHD_2$ | $oH_2D^+$ | $ooNH_2D_2^+$ | HD | 2.2e-09 | 0.00 | 0.0 |
| 1634 | $pNHD_2$ | $oH_2D^+$ | $poNH_2D_2^+$ | HD | 2.8e-09 | 0.00 | 0.0 |
| 1635 | $oNHD_2$ | $oH_2D^+$ | $opNH_2D_2^+$ | HD | 7.9e-10 | 0.00 | 0.0 |
| 1636 | $oNHD_2$ | $oH_2D^+$ | $ppNH_2D_2^+$ | HD | 2.2e-10 | 0.00 | 0.0 |
| 1637 | $pNHD_2$ | $oH_2D^+$ | $opNH_2D_2^+$ | HD | 4.5e-10 | 0.00 | 0.0 |
| 1638 | $pNHD_2$ | $oH_2D^+$ | $ppNH_2D_2^+$ | HD | 5.6e-10 | 0.00 | 0.0 |
| 1639 | $oNHD_2$ | $pH_2D^+$ | $ooNH_2D_2^+$ | HD | 2.4e-09 | 0.00 | 0.0 |
| 1640 | $oNHD_2$ | $pH_2D^+$ | $poNH_2D_2^+$ | HD | 6.7e-10 | 0.00 | 0.0 |
| 1641 | $pNHD_2$ | $pH_2D^+$ | $ooNH_2D_2^+$ | HD | 1.3e-09 | 0.00 | 0.0 |
| 1642 | $pNHD_2$ | $pH_2D^+$ | $poNH_2D_2^+$ | HD | 1.7e-09 | 0.00 | 0.0 |
| 1643 | $oNHD_2$ | $pH_2D^+$ | $opNH_2D_2^+$ | HD | 2.4e-09 | 0.00 | 0.0 |
| 1644 | $oNHD_2$ | $pH_2D^+$ | $ppNH_2D_2^+$ | HD | 6.7e-10 | 0.00 | 0.0 |
| 1645 | $pNHD_2$ | $pH_2D^+$ | $opNH_2D_2^+$ | HD | 1.3e-09 | 0.00 | 0.0 |
| 1646 | $pNHD_2$ | $pH_2D^+$ | $ppNH_2D_2^+$ | HD | 1.7e-09 | 0.00 | 0.0 |
| 1647 | $oNHD_2$ | $oH_2D^+$ | $mNHD_3^+$ | $oH_2$ | 1.4e-09 | 0.00 | 0.0 |
| 1648 | $oNHD_2$ | $oH_2D^+$ | $oNHD_3^+$ | $oH_2$ | 1.1e-09 | 0.00 | 0.0 |
| 1649 | $pNHD_2$ | $oH_2D^+$ | $pNHD_3^+$ | $oH_2$ | 2.8e-10 | 0.00 | 0.0 |
| 1650 | $pNHD_2$ | $oH_2D^+$ | $oNHD_3^+$ | $oH_2$ | 2.2e-09 | 0.00 | 0.0 |
| 1651 | $oNHD_2$ | $oH_2D^+$ | $mNHD_3^+$ | $pH_2$ | 2.8e-10 | 0.00 | 0.0 |
| 1652 | $oNHD_2$ | $oH_2D^+$ | $oNHD_3^+$ | $pH_2$ | 2.2e-10 | 0.00 | 0.0 |
| 1653 | $pNHD_2$ | $oH_2D^+$ | $pNHD_3^+$ | $pH_2$ | 5.6e-11 | 0.00 | 0.0 |
| 1654 | $pNHD_2$ | $oH_2D^+$ | $oNHD_3^+$ | $pH_2$ | 4.5e-10 | 0.00 | 0.0 |
| 1655 | $oNHD_2$ | $pH_2D^+$ | $mNHD_3^+$ | $oH_2$ | 8.4e-10 | 0.00 | 0.0 |
| 1656 | $oNHD_2$ | $pH_2D^+$ | $oNHD_3^+$ | $oH_2$ | 6.7e-10 | 0.00 | 0.0 |
| 1657 | $pNHD_2$ | $pH_2D^+$ | $pNHD_3^+$ | $oH_2$ | 1.7e-10 | 0.00 | 0.0 |
| 1658 | $pNHD_2$ | $pH_2D^+$ | $oNHD_3^+$ | $oH_2$ | 1.3e-09 | 0.00 | 0.0 |
| 1659 | $oNHD_2$ | $pH_2D^+$ | $mNHD_3^+$ | $pH_2$ | 8.4e-10 | 0.00 | 0.0 |
| 1660 | $oNHD_2$ | $pH_2D^+$ | $oNHD_3^+$ | $pH_2$ | 6.7e-10 | 0.00 | 0.0 |
| 1661 | $pNHD_2$ | $pH_2D^+$ | $pNHD_3^+$ | $pH_2$ | 1.7e-10 | 0.00 | 0.0 |
| 1662 | $pNHD_2$ | $pH_2D^+$ | $oNHD_3^+$ | $pH_2$ | 1.3e-09 | 0.00 | 0.0 |
| 1663 | $oNHD_2$ | $oHD_2^+$ | $ooNH_2D_2^+$ | $oD_2$ | 1.5e-09 | 0.00 | 0.0 |
| 1664 | $oNHD_2$ | $oHD_2^+$ | $ooNH_2D_2^+$ | $pD_2$ | 3.2e-10 | 0.00 | 0.0 |
| 1665 | $oNHD_2$ | $oHD_2^+$ | $poNH_2D_2^+$ | $oD_2$ | 3.2e-10 | 0.00 | 0.0 |
| 1666 | $oNHD_2$ | $oHD_2^+$ | $poNH_2D_2^+$ | $pD_2$ | 1.9e-10 | 0.00 | 0.0 |
| 1667 | $oNHD_2$ | $pHD_2^+$ | $ooNH_2D_2^+$ | $oD_2$ | 6.3e-10 | 0.00 | 0.0 |
| 1668 | $oNHD_2$ | $pHD_2^+$ | $ooNH_2D_2^+$ | $pD_2$ | 7.6e-10 | 0.00 | 0.0 |
| 1669 | $oNHD_2$ | $pHD_2^+$ | $poNH_2D_2^+$ | $oD_2$ | 7.6e-10 | 0.00 | 0.0 |
| 1670 | $oNHD_2$ | $pHD_2^+$ | $poNH_2D_2^+$ | $pD_2$ | 1.3e-10 | 0.00 | 0.0 |
| 1671 | $pNHD_2$ | $oHD_2^+$ | $ooNH_2D_2^+$ | $oD_2$ | 6.3e-10 | 0.00 | 0.0 |
| 1672 | $pNHD_2$ | $oHD_2^+$ | $ooNH_2D_2^+$ | $pD_2$ | 7.6e-10 | 0.00 | 0.0 |
| 1673 | $pNHD_2$ | $oHD_2^+$ | $poNH_2D_2^+$ | $oD_2$ | 7.6e-10 | 0.00 | 0.0 |
| 1674 | $pNHD_2$ | $oHD_2^+$ | $poNH_2D_2^+$ | $pD_2$ | 1.3e-10 | 0.00 | 0.0 |
| 1675 | $pNHD_2$ | $pHD_2^+$ | $ooNH_2D_2^+$ | $oD_2$ | 7.6e-10 | 0.00 | 0.0 |
| 1676 | $pNHD_2$ | $pHD_2^+$ | $ooNH_2D_2^+$ | $pD_2$ | 2.5e-10 | 0.00 | 0.0 |
| 1677 | $pNHD_2$ | $pHD_2^+$ | $poNH_2D_2^+$ | $oD_2$ | 2.5e-10 | 0.00 | 0.0 |
| 1678 | $pNHD_2$ | $pHD_2^+$ | $poNH_2D_2^+$ | $pD_2$ | 1.0e-09 | 0.00 | 0.0 |
| 1679 | $oNHD_2$ | $oHD_2^+$ | $opNH_2D_2^+$ | $oD_2$ | 4.8e-10 | 0.00 | 0.0 |
| 1680 | $oNHD_2$ | $oHD_2^+$ | $opNH_2D_2^+$ | $pD_2$ | 1.1e-10 | 0.00 | 0.0 |
| 1681 | $oNHD_2$ | $oHD_2^+$ | $ppNH_2D_2^+$ | $oD_2$ | 1.1e-10 | 0.00 | 0.0 |
| 1682 | $oNHD_2$ | $oHD_2^+$ | $ppNH_2D_2^+$ | $pD_2$ | 6.3e-11 | 0.00 | 0.0 |
| 1683 | $oNHD_2$ | $pHD_2^+$ | $opNH_2D_2^+$ | $oD_2$ | 2.1e-10 | 0.00 | 0.0 |
| 1684 | $oNHD_2$ | $pHD_2^+$ | $opNH_2D_2^+$ | $pD_2$ | 2.5e-10 | 0.00 | 0.0 |
| 1685 | $oNHD_2$ | $pHD_2^+$ | $ppNH_2D_2^+$ | $oD_2$ | 2.5e-10 | 0.00 | 0.0 |
| 1686 | $oNHD_2$ | $pHD_2^+$ | $ppNH_2D_2^+$ | $pD_2$ | 4.2e-11 | 0.00 | 0.0 |
| 1687 | $pNHD_2$ | $oHD_2^+$ | $opNH_2D_2^+$ | $oD_2$ | 2.1e-10 | 0.00 | 0.0 |
| 1688 | $pNHD_2$ | $oHD_2^+$ | $opNH_2D_2^+$ | $pD_2$ | 2.5e-10 | 0.00 | 0.0 |
| 1689 | $pNHD_2$ | $oHD_2^+$ | $ppNH_2D_2^+$ | $oD_2$ | 2.5e-10 | 0.00 | 0.0 |
| 1690 | $pNHD_2$ | $oHD_2^+$ | $ppNH_2D_2^+$ | $pD_2$ | 4.2e-11 | 0.00 | 0.0 |
| 1691 | $pNHD_2$ | $pHD_2^+$ | $opNH_2D_2^+$ | $oD_2$ | 2.5e-10 | 0.00 | 0.0 |
| 1692 | $pNHD_2$ | $pHD_2^+$ | $opNH_2D_2^+$ | $pD_2$ | 8.4e-11 | 0.00 | 0.0 |
| 1693 | $pNHD_2$ | $pHD_2^+$ | $ppNH_2D_2^+$ | $oD_2$ | 8.4e-11 | 0.00 | 0.0 |
| 1694 | $pNHD_2$ | $pHD_2^+$ | $ppNH_2D_2^+$ | $pD_2$ | 3.4e-10 | 0.00 | 0.0 |
| 1695 | $oNHD_2$ | $oHD_2^+$ | $mNHD_3^+$ | HD | 3.4e-09 | 0.00 | 0.0 |
| 1696 | $oNHD_2$ | $oHD_2^+$ | $oNHD_3^+$ | HD | 2.7e-09 | 0.00 | 0.0 |
| 1697 | $oNHD_2$ | $pHD_2^+$ | $mNHD_3^+$ | HD | 1.7e-09 | 0.00 | 0.0 |
| 1698 | $oNHD_2$ | $pHD_2^+$ | $pNHD_3^+$ | HD | 3.4e-09 | 0.00 | 0.0 |
| 1699 | $oNHD_2$ | $pHD_2^+$ | $oNHD_3^+$ | HD | 4.0e-09 | 0.00 | 0.0 |
| 1700 | $pNHD_2$ | $oHD_2^+$ | $mNHD_3^+$ | HD | 1.7e-09 | 0.00 | 0.0 |
| 1701 | $pNHD_2$ | $oHD_2^+$ | $pNHD_3^+$ | HD | 3.4e-10 | 0.00 | 0.0 |
| 1702 | $pNHD_2$ | $oHD_2^+$ | $oNHD_3^+$ | HD | 4.0e-09 | 0.00 | 0.0 |

**Table D6** – *continued* (part 24)

| # | Reactants | | Products | | α | β | γ |
|---|---|---|---|---|---|---|---|
| 1703 | pNHD$_2$ | pHD$_2^+$ | pNHD$_3^+$ | HD | 6.7e-10 | 0.00 | 0.0 |
| 1704 | pNHD$_2$ | pHD$_2^+$ | oNHD$_3^+$ | HD | 5.4e-09 | 0.00 | 0.0 |
| 1705 | oNHD$_2$ | mD$_3^+$ | mNHD$_3^+$ | oD$_2$ | 4.6e-09 | 0.00 | 0.0 |
| 1706 | oNHD$_2$ | mD$_3^+$ | mNHD$_3^+$ | pD$_2$ | 9.1e-10 | 0.00 | 0.0 |
| 1707 | oNHD$_2$ | mD$_3^+$ | oNHD$_3^+$ | oD$_2$ | 2.7e-09 | 0.00 | 0.0 |
| 1708 | oNHD$_2$ | mD$_3^+$ | oNHD$_3^+$ | pD$_2$ | 9.1e-10 | 0.00 | 0.0 |
| 1709 | pNHD$_2$ | mD$_3^+$ | mNHD$_3^+$ | oD$_2$ | 1.8e-09 | 0.00 | 0.0 |
| 1710 | pNHD$_2$ | mD$_3^+$ | mNHD$_3^+$ | pD$_2$ | 2.1e-09 | 0.00 | 0.0 |
| 1711 | pNHD$_2$ | mD$_3^+$ | pNHD$_3^+$ | oD$_2$ | 3.0e-10 | 0.00 | 0.0 |
| 1712 | pNHD$_2$ | mD$_3^+$ | oNHD$_3^+$ | oD$_2$ | 4.2e-09 | 0.00 | 0.0 |
| 1713 | pNHD$_2$ | mD$_3^+$ | oNHD$_3^+$ | pD$_2$ | 6.1e-10 | 0.00 | 0.0 |
| 1714 | oNHD$_2$ | pD$_3^+$ | mNHD$_3^+$ | pD$_2$ | 1.5e-09 | 0.00 | 0.0 |
| 1715 | oNHD$_2$ | pD$_3^+$ | pNHD$_3^+$ | oD$_2$ | 1.5e-09 | 0.00 | 0.0 |
| 1716 | oNHD$_2$ | pD$_3^+$ | oNHD$_3^+$ | oD$_2$ | 3.0e-09 | 0.00 | 0.0 |
| 1717 | oNHD$_2$ | pD$_3^+$ | oNHD$_3^+$ | pD$_2$ | 3.0e-09 | 0.00 | 0.0 |
| 1718 | pNHD$_2$ | pD$_3^+$ | pNHD$_3^+$ | pD$_2$ | 1.8e-09 | 0.00 | 0.0 |
| 1719 | pNHD$_2$ | pD$_3^+$ | oNHD$_3^+$ | oD$_2$ | 3.6e-09 | 0.00 | 0.0 |
| 1720 | pNHD$_2$ | pD$_3^+$ | oNHD$_3^+$ | pD$_2$ | 3.6e-09 | 0.00 | 0.0 |
| 1721 | oNHD$_2$ | oD$_3^+$ | mNHD$_3^+$ | oD$_2$ | 1.7e-09 | 0.00 | 0.0 |
| 1722 | oNHD$_2$ | oD$_3^+$ | mNHD$_3^+$ | pD$_2$ | 1.3e-09 | 0.00 | 0.0 |
| 1723 | oNHD$_2$ | oD$_3^+$ | pNHD$_3^+$ | oD$_2$ | 1.9e-10 | 0.00 | 0.0 |
| 1724 | oNHD$_2$ | oD$_3^+$ | pNHD$_3^+$ | pD$_2$ | 1.1e-10 | 0.00 | 0.0 |
| 1725 | oNHD$_2$ | oD$_3^+$ | oNHD$_3^+$ | oD$_2$ | 4.0e-09 | 0.00 | 0.0 |
| 1726 | oNHD$_2$ | oD$_3^+$ | oNHD$_3^+$ | pD$_2$ | 1.7e-09 | 0.00 | 0.0 |
| 1727 | pNHD$_2$ | oD$_3^+$ | mNHD$_3^+$ | oD$_2$ | 1.1e-09 | 0.00 | 0.0 |
| 1728 | pNHD$_2$ | oD$_3^+$ | mNHD$_3^+$ | pD$_2$ | 3.8e-10 | 0.00 | 0.0 |
| 1729 | pNHD$_2$ | oD$_3^+$ | pNHD$_3^+$ | oD$_2$ | 3.8e-10 | 0.00 | 0.0 |
| 1730 | pNHD$_2$ | oD$_3^+$ | pNHD$_3^+$ | pD$_2$ | 2.3e-10 | 0.00 | 0.0 |
| 1731 | pNHD$_2$ | oD$_3^+$ | oNHD$_3^+$ | oD$_2$ | 3.5e-09 | 0.00 | 0.0 |
| 1732 | pNHD$_2$ | oD$_3^+$ | oNHD$_3^+$ | pD$_2$ | 3.5e-09 | 0.00 | 0.0 |
| 1733 | mND$_3$ | H$^+$ | mND$_3^+$ | H | 5.2e-09 | 0.00 | 0.0 |
| 1734 | pND$_3$ | H$^+$ | pND$_3^+$ | H | 5.2e-09 | 0.00 | 0.0 |
| 1735 | oND$_3$ | H$^+$ | oND$_3^+$ | H | 5.2e-09 | 0.00 | 0.0 |
| 1736 | mND$_3$ | oH$_3^+$ | mNHD$_3^+$ | oH$_2$ | 9.1e-09 | 0.00 | 0.0 |
| 1737 | pND$_3$ | oH$_3^+$ | pNHD$_3^+$ | oH$_2$ | 9.1e-09 | 0.00 | 0.0 |
| 1738 | oND$_3$ | oH$_3^+$ | oNHD$_3^+$ | oH$_2$ | 9.1e-09 | 0.00 | 0.0 |
| 1739 | mND$_3$ | pH$_3^+$ | mNHD$_3^+$ | oH$_2$ | 4.6e-09 | 0.00 | 0.0 |
| 1740 | pND$_3$ | pH$_3^+$ | pNHD$_3^+$ | oH$_2$ | 4.6e-09 | 0.00 | 0.0 |
| 1741 | oND$_3$ | pH$_3^+$ | oNHD$_3^+$ | oH$_2$ | 4.6e-09 | 0.00 | 0.0 |
| 1742 | mND$_3$ | pH$_3^+$ | mNHD$_3^+$ | pH$_2$ | 4.6e-09 | 0.00 | 0.0 |
| 1743 | pND$_3$ | pH$_3^+$ | pNHD$_3^+$ | pH$_2$ | 4.6e-09 | 0.00 | 0.0 |
| 1744 | oND$_3$ | pH$_3^+$ | oNHD$_3^+$ | pH$_2$ | 4.6e-09 | 0.00 | 0.0 |
| 1745 | mND$_3$ | oH$_2$D$^+$ | mNHD$_3^+$ | HD | 4.0e-09 | 0.00 | 0.0 |
| 1746 | mND$_3$ | oH$_2$D$^+$ | oNHD$_3^+$ | HD | 2.0e-09 | 0.00 | 0.0 |
| 1747 | pND$_3$ | oH$_2$D$^+$ | pNHD$_3^+$ | HD | 2.0e-09 | 0.00 | 0.0 |
| 1748 | pND$_3$ | oH$_2$D$^+$ | oNHD$_3^+$ | HD | 4.0e-09 | 0.00 | 0.0 |
| 1749 | oND$_3$ | oH$_2$D$^+$ | mNHD$_3^+$ | HD | 1.3e-09 | 0.00 | 0.0 |
| 1750 | oND$_3$ | oH$_2$D$^+$ | pNHD$_3^+$ | HD | 2.5e-10 | 0.00 | 0.0 |
| 1751 | oND$_3$ | oH$_2$D$^+$ | oNHD$_3^+$ | HD | 4.5e-09 | 0.00 | 0.0 |
| 1752 | mND$_3$ | pH$_2$D$^+$ | mNHD$_3^+$ | HD | 4.0e-09 | 0.00 | 0.0 |
| 1753 | mND$_3$ | pH$_2$D$^+$ | oNHD$_3^+$ | HD | 2.0e-09 | 0.00 | 0.0 |
| 1754 | pND$_3$ | pH$_2$D$^+$ | pNHD$_3^+$ | HD | 2.0e-09 | 0.00 | 0.0 |
| 1755 | pND$_3$ | pH$_2$D$^+$ | oNHD$_3^+$ | HD | 4.0e-09 | 0.00 | 0.0 |
| 1756 | oND$_3$ | pH$_2$D$^+$ | mNHD$_3^+$ | HD | 1.3e-09 | 0.00 | 0.0 |
| 1757 | oND$_3$ | pH$_2$D$^+$ | pNHD$_3^+$ | HD | 2.5e-10 | 0.00 | 0.0 |
| 1758 | oND$_3$ | pH$_2$D$^+$ | oNHD$_3^+$ | HD | 4.5e-09 | 0.00 | 0.0 |
| 1759 | mND$_3$ | oH$_2$D$^+$ | lND$_4^+$ | oH$_2$ | 1.5e-09 | 0.00 | 0.0 |
| 1760 | mND$_3$ | oH$_2$D$^+$ | oND$_4^+$ | oH$_2$ | 1.5e-09 | 0.00 | 0.0 |
| 1761 | pND$_3$ | oH$_2$D$^+$ | pND$_4^+$ | oH$_2$ | 3.0e-09 | 0.00 | 0.0 |
| 1762 | oND$_3$ | oH$_2$D$^+$ | mND$_4^+$ | oH$_2$ | 7.6e-10 | 0.00 | 0.0 |
| 1763 | oND$_3$ | oH$_2$D$^+$ | oND$_4^+$ | oH$_2$ | 1.9e-09 | 0.00 | 0.0 |
| 1764 | oND$_3$ | oH$_2$D$^+$ | pND$_4^+$ | oH$_2$ | 3.8e-10 | 0.00 | 0.0 |
| 1765 | mND$_3$ | pH$_2$D$^+$ | lND$_4^+$ | pH$_2$ | 1.5e-09 | 0.00 | 0.0 |
| 1766 | mND$_3$ | pH$_2$D$^+$ | oND$_4^+$ | pH$_2$ | 1.5e-09 | 0.00 | 0.0 |
| 1767 | pND$_3$ | pH$_2$D$^+$ | pND$_4^+$ | pH$_2$ | 3.0e-09 | 0.00 | 0.0 |
| 1768 | oND$_3$ | pH$_2$D$^+$ | mND$_4^+$ | pH$_2$ | 7.6e-10 | 0.00 | 0.0 |
| 1769 | oND$_3$ | pH$_2$D$^+$ | oND$_4^+$ | pH$_2$ | 1.9e-09 | 0.00 | 0.0 |
| 1770 | oND$_3$ | pH$_2$D$^+$ | pND$_4^+$ | pH$_2$ | 3.8e-10 | 0.00 | 0.0 |
| 1771 | mND$_3$ | oHD$_2^+$ | mNHD$_3^+$ | oD$_2$ | 1.5e-09 | 0.00 | 0.0 |
| 1772 | mND$_3$ | oHD$_2^+$ | mNHD$_3^+$ | pD$_2$ | 3.0e-09 | 0.00 | 0.0 |
| 1773 | mND$_3$ | oHD$_2^+$ | oNHD$_3^+$ | oD$_2$ | 9.1e-10 | 0.00 | 0.0 |
| 1774 | mND$_3$ | oHD$_2^+$ | oNHD$_3^+$ | pD$_2$ | 3.0e-10 | 0.00 | 0.0 |
| 1775 | mND$_3$ | pHD$_2^+$ | mNHD$_3^+$ | oD$_2$ | 6.1e-10 | 0.00 | 0.0 |
| 1776 | mND$_3$ | pHD$_2^+$ | mNHD$_3^+$ | pD$_2$ | 7.1e-10 | 0.00 | 0.0 |

**Table D6** – *continued* (part 25)

| # | Reactants | | Products | | $\alpha$ | $\beta$ | $\gamma$ |
|---|---|---|---|---|---|---|---|
| 1777 | $mND_3$ | $pHD_2^+$ | $pNHD_3^+$ | $oD_2$ | 1.0e-10 | 0.00 | 0.0 |
| 1778 | $mND_3$ | $pHD_2^+$ | $oNHD_3^+$ | $oD_2$ | 1.4e-09 | 0.00 | 0.0 |
| 1779 | $mND_3$ | $pHD_2^+$ | $oNHD_3^+$ | $pD_2$ | 2.0e-10 | 0.00 | 0.0 |
| 1780 | $pND_3$ | $oHD_2^+$ | $mNHD_3^+$ | $pD_2$ | 5.1e-10 | 0.00 | 0.0 |
| 1781 | $pND_3$ | $oHD_2^+$ | $pNHD_3^+$ | $oD_2$ | 5.1e-10 | 0.00 | 0.0 |
| 1782 | $pND_3$ | $oHD_2^+$ | $oNHD_3^+$ | $oD_2$ | 1.0e-09 | 0.00 | 0.0 |
| 1783 | $pND_3$ | $oHD_2^+$ | $oNHD_3^+$ | $pD_2$ | 1.0e-09 | 0.00 | 0.0 |
| 1784 | $pND_3$ | $pHD_2^+$ | $pNHD_3^+$ | $pD_2$ | 6.1e-10 | 0.00 | 0.0 |
| 1785 | $pND_3$ | $pHD_2^+$ | $oNHD_3^+$ | $oD_2$ | 1.2e-09 | 0.00 | 0.0 |
| 1786 | $pND_3$ | $pHD_2^+$ | $oNHD_3^+$ | $pD_2$ | 1.2e-09 | 0.00 | 0.0 |
| 1787 | $oND_3$ | $oHD_2^+$ | $mNHD_3^+$ | $oD_2$ | 5.7e-10 | 0.00 | 0.0 |
| 1788 | $oND_3$ | $oHD_2^+$ | $mNHD_3^+$ | $pD_2$ | 4.4e-10 | 0.00 | 0.0 |
| 1789 | $oND_3$ | $oHD_2^+$ | $pNHD_3^+$ | $oD_2$ | 6.3e-11 | 0.00 | 0.0 |
| 1790 | $oND_3$ | $oHD_2^+$ | $pNHD_3^+$ | $pD_2$ | 3.8e-11 | 0.00 | 0.0 |
| 1791 | $oND_3$ | $oHD_2^+$ | $oNHD_3^+$ | $oD_2$ | 1.3e-09 | 0.00 | 0.0 |
| 1792 | $oND_3$ | $oHD_2^+$ | $oNHD_3^+$ | $pD_2$ | 5.8e-10 | 0.00 | 0.0 |
| 1793 | $oND_3$ | $pHD_2^+$ | $mNHD_3^+$ | $oD_2$ | 3.8e-10 | 0.00 | 0.0 |
| 1794 | $oND_3$ | $pHD_2^+$ | $mNHD_3^+$ | $pD_2$ | 1.3e-10 | 0.00 | 0.0 |
| 1795 | $oND_3$ | $pHD_2^+$ | $pNHD_3^+$ | $oD_2$ | 1.3e-10 | 0.00 | 0.0 |
| 1796 | $oND_3$ | $pHD_2^+$ | $pNHD_3^+$ | $pD_2$ | 7.6e-11 | 0.00 | 0.0 |
| 1797 | $oND_3$ | $pHD_2^+$ | $oNHD_3^+$ | $oD_2$ | 1.2e-09 | 0.00 | 0.0 |
| 1798 | $oND_3$ | $pHD_2^+$ | $oNHD_3^+$ | $pD_2$ | 1.2e-09 | 0.00 | 0.0 |
| 1799 | $mND_3$ | $oHD_2^+$ | $lND_4^+$ | HD | 2.7e-09 | 0.00 | 0.0 |
| 1800 | $mND_3$ | $oHD_2^+$ | $mND_4^+$ | HD | 6.1e-10 | 0.00 | 0.0 |
| 1801 | $mND_3$ | $oHD_2^+$ | $oND_4^+$ | HD | 2.7e-09 | 0.00 | 0.0 |
| 1802 | $mND_3$ | $pHD_2^+$ | $lND_4^+$ | HD | 1.2e-09 | 0.00 | 0.0 |
| 1803 | $mND_3$ | $pHD_2^+$ | $oND_4^+$ | HD | 4.2e-09 | 0.00 | 0.0 |
| 1804 | $mND_3$ | $pHD_2^+$ | $pND_4^+$ | HD | 6.1e-10 | 0.00 | 0.0 |
| 1805 | $pND_3$ | $oHD_2^+$ | $oND_4^+$ | HD | 3.0e-09 | 0.00 | 0.0 |
| 1806 | $pND_3$ | $oHD_2^+$ | $pND_4^+$ | HD | 3.0e-09 | 0.00 | 0.0 |
| 1807 | $pND_3$ | $pHD_2^+$ | $mND_4^+$ | HD | 2.4e-09 | 0.00 | 0.0 |
| 1808 | $pND_3$ | $pHD_2^+$ | $pND_4^+$ | HD | 3.6e-09 | 0.00 | 0.0 |
| 1809 | $oND_3$ | $oHD_2^+$ | $lND_4^+$ | HD | 7.6e-10 | 0.00 | 0.0 |
| 1810 | $oND_3$ | $oHD_2^+$ | $mND_4^+$ | HD | 9.1e-10 | 0.00 | 0.0 |
| 1811 | $oND_3$ | $oHD_2^+$ | $oND_4^+$ | HD | 3.8e-09 | 0.00 | 0.0 |
| 1812 | $oND_3$ | $oHD_2^+$ | $pND_4^+$ | HD | 6.1e-10 | 0.00 | 0.0 |
| 1813 | $oND_3$ | $pHD_2^+$ | $mND_4^+$ | HD | 1.8e-09 | 0.00 | 0.0 |
| 1814 | $oND_3$ | $pHD_2^+$ | $oND_4^+$ | HD | 3.0e-09 | 0.00 | 0.0 |
| 1815 | $oND_3$ | $pHD_2^+$ | $pND_4^+$ | HD | 1.2e-09 | 0.00 | 0.0 |
| 1816 | $mND_3$ | $mD_3^+$ | $lND_4^+$ | $oD_2$ | 3.5e-09 | 0.00 | 0.0 |
| 1817 | $mND_3$ | $mD_3^+$ | $lND_4^+$ | $pD_2$ | 6.4e-10 | 0.00 | 0.0 |
| 1818 | $mND_3$ | $mD_3^+$ | $mND_4^+$ | $oD_2$ | 5.5e-10 | 0.00 | 0.0 |
| 1819 | $mND_3$ | $mD_3^+$ | $mND_4^+$ | $pD_2$ | 3.6e-10 | 0.00 | 0.0 |
| 1820 | $mND_3$ | $mD_3^+$ | $oND_4^+$ | $oD_2$ | 3.3e-09 | 0.00 | 0.0 |
| 1821 | $mND_3$ | $mD_3^+$ | $oND_4^+$ | $pD_2$ | 8.2e-10 | 0.00 | 0.0 |
| 1822 | $mND_3$ | $pD_3^+$ | $lND_4^+$ | $pD_2$ | 9.1e-10 | 0.00 | 0.0 |
| 1823 | $mND_3$ | $pD_3^+$ | $oND_4^+$ | $oD_2$ | 2.7e-09 | 0.00 | 0.0 |
| 1824 | $mND_3$ | $pD_3^+$ | $oND_4^+$ | $pD_2$ | 2.7e-09 | 0.00 | 0.0 |
| 1825 | $mND_3$ | $pD_3^+$ | $pND_4^+$ | $oD_2$ | 2.7e-09 | 0.00 | 0.0 |
| 1826 | $mND_3$ | $oD_3^+$ | $lND_4^+$ | $oD_2$ | 1.1e-09 | 0.00 | 0.0 |
| 1827 | $mND_3$ | $oD_3^+$ | $lND_4^+$ | $pD_2$ | 9.1e-10 | 0.00 | 0.0 |
| 1828 | $mND_3$ | $oD_3^+$ | $mND_4^+$ | $oD_2$ | 8.0e-10 | 0.00 | 0.0 |
| 1829 | $mND_3$ | $oD_3^+$ | $mND_4^+$ | $pD_2$ | 1.1e-09 | 0.00 | 0.0 |
| 1830 | $mND_3$ | $oD_3^+$ | $oND_4^+$ | $oD_2$ | 3.9e-09 | 0.00 | 0.0 |
| 1831 | $mND_3$ | $oD_3^+$ | $oND_4^+$ | $pD_2$ | 1.5e-09 | 0.00 | 0.0 |
| 1832 | $mND_3$ | $oD_3^+$ | $pND_4^+$ | $oD_2$ | 5.1e-10 | 0.00 | 0.0 |
| 1833 | $mND_3$ | $oD_3^+$ | $pND_4^+$ | $pD_2$ | 1.7e-10 | 0.00 | 0.0 |
| 1834 | $pND_3$ | $mD_3^+$ | $lND_4^+$ | $pD_2$ | 9.1e-10 | 0.00 | 0.0 |
| 1835 | $pND_3$ | $mD_3^+$ | $oND_4^+$ | $oD_2$ | 2.7e-09 | 0.00 | 0.0 |
| 1836 | $pND_3$ | $mD_3^+$ | $oND_4^+$ | $pD_2$ | 2.7e-09 | 0.00 | 0.0 |
| 1837 | $pND_3$ | $mD_3^+$ | $pND_4^+$ | $oD_2$ | 2.7e-09 | 0.00 | 0.0 |
| 1838 | $pND_3$ | $pD_3^+$ | $mND_4^+$ | $oD_2$ | 3.6e-09 | 0.00 | 0.0 |
| 1839 | $pND_3$ | $pD_3^+$ | $pND_4^+$ | $pD_2$ | 5.5e-09 | 0.00 | 0.0 |
| 1840 | $pND_3$ | $oD_3^+$ | $mND_4^+$ | $oD_2$ | 1.1e-09 | 0.00 | 0.0 |
| 1841 | $pND_3$ | $oD_3^+$ | $mND_4^+$ | $pD_2$ | 1.1e-09 | 0.00 | 0.0 |
| 1842 | $pND_3$ | $oD_3^+$ | $oND_4^+$ | $oD_2$ | 1.7e-09 | 0.00 | 0.0 |
| 1843 | $pND_3$ | $oD_3^+$ | $oND_4^+$ | $pD_2$ | 1.7e-09 | 0.00 | 0.0 |
| 1844 | $pND_3$ | $oD_3^+$ | $pND_4^+$ | $oD_2$ | 1.7e-09 | 0.00 | 0.0 |
| 1845 | $pND_3$ | $oD_3^+$ | $pND_4^+$ | $pD_2$ | 1.7e-09 | 0.00 | 0.0 |
| 1846 | $oND_3$ | $mD_3^+$ | $lND_4^+$ | $oD_2$ | 1.1e-09 | 0.00 | 0.0 |
| 1847 | $oND_3$ | $mD_3^+$ | $lND_4^+$ | $pD_2$ | 9.1e-10 | 0.00 | 0.0 |
| 1848 | $oND_3$ | $mD_3^+$ | $mND_4^+$ | $oD_2$ | 8.0e-10 | 0.00 | 0.0 |
| 1849 | $oND_3$ | $mD_3^+$ | $mND_4^+$ | $pD_2$ | 1.1e-10 | 0.00 | 0.0 |
| 1850 | $oND_3$ | $mD_3^+$ | $oND_4^+$ | $oD_2$ | 3.9e-09 | 0.00 | 0.0 |

**Table D6** – *continued* (part 26)

| # | Reactants | | Products | | | α | β | γ |
|---|---|---|---|---|---|---|---|---|
| 1851 | $oND_3$ | $mD_3^+$ | $oND_4^+$ | $pD_2$ | | 1.5e-09 | 0.00 | 0.0 |
| 1852 | $oND_3$ | $mD_3^+$ | $pND_4^+$ | $oD_2$ | | 5.1e-10 | 0.00 | 0.0 |
| 1853 | $oND_3$ | $mD_3^+$ | $pND_4^+$ | $pD_2$ | | 1.7e-10 | 0.00 | 0.0 |
| 1854 | $oND_3$ | $pD_3^+$ | $mND_4^+$ | $oD_2$ | | 1.1e-09 | 0.00 | 0.0 |
| 1855 | $oND_3$ | $pD_3^+$ | $mND_4^+$ | $pD_2$ | | 1.1e-09 | 0.00 | 0.0 |
| 1856 | $oND_3$ | $pD_3^+$ | $oND_4^+$ | $oD_2$ | | 1.7e-09 | 0.00 | 0.0 |
| 1857 | $oND_3$ | $pD_3^+$ | $oND_4^+$ | $pD_2$ | | 1.7e-09 | 0.00 | 0.0 |
| 1858 | $oND_3$ | $pD_3^+$ | $pND_4^+$ | $oD_2$ | | 1.7e-09 | 0.00 | 0.0 |
| 1859 | $oND_3$ | $pD_3^+$ | $pND_4^+$ | $pD_2$ | | 1.7e-09 | 0.00 | 0.0 |
| 1860 | $oND_3$ | $oD_3^+$ | $lND_4^+$ | $oD_2$ | | 4.3e-10 | 0.00 | 0.0 |
| 1861 | $oND_3$ | $oD_3^+$ | $lND_4^+$ | $pD_2$ | | 1.4e-10 | 0.00 | 0.0 |
| 1862 | $oND_3$ | $oD_3^+$ | $mND_4^+$ | $oD_2$ | | 1.2e-09 | 0.00 | 0.0 |
| 1863 | $oND_3$ | $oD_3^+$ | $mND_4^+$ | $pD_2$ | | 8.5e-10 | 0.00 | 0.0 |
| 1864 | $oND_3$ | $oD_3^+$ | $oND_4^+$ | $oD_2$ | | 3.0e-09 | 0.00 | 0.0 |
| 1865 | $oND_3$ | $oD_3^+$ | $oND_4^+$ | $pD_2$ | | 2.1e-09 | 0.00 | 0.0 |
| 1866 | $oND_3$ | $oD_3^+$ | $pND_4^+$ | $oD_2$ | | 8.5e-10 | 0.00 | 0.0 |
| 1867 | $oND_3$ | $oD_3^+$ | $pND_4^+$ | $pD_2$ | | 5.1e-10 | 0.00 | 0.0 |
| 1868 | $N_2$ | $oH_2D^+$ | $N_2H^+$ | HD | | 8.7e-10 | 0.00 | 0.0 |
| 1869 | $N_2$ | $pH_2D^+$ | $N_2H^+$ | HD | | 8.7e-10 | 0.00 | 0.0 |
| 1870 | $N_2$ | $oH_2D^+$ | $oH_2$ | $N_2D^+$ | | 4.3e-10 | 0.00 | 0.0 |
| 1871 | $N_2$ | $pH_2D^+$ | $pH_2$ | $N_2D^+$ | | 4.3e-10 | 0.00 | 0.0 |
| 1872 | $N_2$ | $oHD_2^+$ | $oD_2$ | $N_2H^+$ | | 4.3e-10 | 0.00 | 0.0 |
| 1873 | $N_2$ | $pHD_2^+$ | $pD_2$ | $N_2H^+$ | | 4.3e-10 | 0.00 | 0.0 |
| 1874 | $N_2$ | $oHD_2^+$ | $N_2D^+$ | HD | | 8.7e-10 | 0.00 | 0.0 |
| 1875 | $N_2$ | $pHD_2^+$ | $N_2D^+$ | HD | | 8.7e-10 | 0.00 | 0.0 |
| 1876 | $N_2$ | $mD_3^+$ | $oD_2$ | $N_2D^+$ | | 1.3e-09 | 0.00 | 0.0 |
| 1877 | $N_2$ | $pD_3^+$ | $pD_2$ | $N_2D^+$ | | 1.3e-09 | 0.00 | 0.0 |
| 1878 | $N_2$ | $oD_3^+$ | $oD_2$ | $N_2D^+$ | | 6.5e-10 | 0.00 | 0.0 |
| 1879 | $N_2$ | $oD_3^+$ | $pD_2$ | $N_2D^+$ | | 6.5e-10 | 0.00 | 0.0 |
| 1880 | $N^+$ | $e^-$ | N | γ | | 3.8e-12 | -0.62 | 0.0 |
| 1881 | $NH^+$ | $e^-$ | N | H | | 2.0e-07 | -0.50 | 0.0 |
| 1882 | $oNH_2^+$ | $e^-$ | NH | H | | 1.2e-07 | -0.50 | 0.0 |
| 1883 | $pNH_2^+$ | $e^-$ | NH | H | | 1.2e-07 | -0.50 | 0.0 |
| 1884 | $oNH_2^+$ | $e^-$ | $oH_2$ | N | | 1.2e-08 | -0.50 | 0.0 |
| 1885 | $pNH_2^+$ | $e^-$ | $pH_2$ | N | | 1.2e-08 | -0.50 | 0.0 |
| 1886 | $oNH_2^+$ | $e^-$ | H | H | N | 1.7e-07 | -0.50 | 0.0 |
| 1887 | $pNH_2^+$ | $e^-$ | H | H | N | 1.7e-07 | -0.50 | 0.0 |
| 1888 | $oNH_3^+$ | $e^-$ | $oNH_2$ | H | | 1.5e-07 | -0.50 | 0.0 |
| 1889 | $pNH_3^+$ | $e^-$ | $oNH_2$ | H | | 7.7e-08 | -0.50 | 0.0 |
| 1890 | $pNH_3^+$ | $e^-$ | $pNH_2$ | H | | 7.7e-08 | -0.50 | 0.0 |
| 1891 | $oNH_3^+$ | $e^-$ | NH | H | H | 1.5e-07 | -0.50 | 0.0 |
| 1892 | $pNH_3^+$ | $e^-$ | NH | H | H | 1.5e-07 | -0.50 | 0.0 |
| 1893 | $mNH_4^+$ | $e^-$ | $oNH_2$ | $oH_2$ | | 1.9e-08 | -0.60 | 0.0 |
| 1894 | $pNH_4^+$ | $e^-$ | $oNH_2$ | $oH_2$ | | 9.4e-09 | -0.60 | 0.0 |
| 1895 | $pNH_4^+$ | $e^-$ | $pNH_2$ | $pH_2$ | | 9.4e-09 | -0.60 | 0.0 |
| 1896 | $oNH_4^+$ | $e^-$ | $oNH_2$ | $oH_2$ | | 6.3e-09 | -0.60 | 0.0 |
| 1897 | $oNH_4^+$ | $e^-$ | $oNH_2$ | $pH_2$ | | 6.3e-09 | -0.60 | 0.0 |
| 1898 | $oNH_4^+$ | $e^-$ | $pNH_2$ | $oH_2$ | | 6.3e-09 | -0.60 | 0.0 |
| 1899 | $mNH_4^+$ | $e^-$ | $oNH_3$ | H | | 8.0e-07 | -0.60 | 0.0 |
| 1900 | $pNH_4^+$ | $e^-$ | $pNH_3$ | H | | 8.0e-07 | -0.60 | 0.0 |
| 1901 | $oNH_4^+$ | $e^-$ | $oNH_3$ | H | | 2.7e-07 | -0.60 | 0.0 |
| 1902 | $oNH_4^+$ | $e^-$ | $pNH_3$ | H | | 5.3e-07 | -0.60 | 0.0 |
| 1903 | $N_2^+$ | $e^-$ | N | N | | 3.6e-08 | -0.42 | 0.0 |
| 1904 | $N_2H^+$ | $e^-$ | $N_2$ | H | | 2.8e-07 | -0.74 | 0.0 |
| 1905 | $N_2H^+$ | $e^-$ | NH | N | | 2.1e-08 | -0.74 | 0.0 |
| 1906 | $ND^+$ | $e^-$ | N | D | | 2.0e-07 | -0.50 | 0.0 |
| 1907 | $NHD^+$ | $e^-$ | N | H | D | 1.5e-07 | -0.50 | 0.0 |
| 1908 | $NHD^+$ | $e^-$ | NH | D | | 7.5e-08 | -0.50 | 0.0 |
| 1909 | $NHD^+$ | $e^-$ | ND | H | | 7.5e-08 | -0.50 | 0.0 |
| 1910 | $oND_2^+$ | $e^-$ | ND | D | | 1.5e-07 | -0.50 | 0.0 |
| 1911 | $pND_2^+$ | $e^-$ | ND | D | | 1.5e-07 | -0.50 | 0.0 |
| 1912 | $oND_2^+$ | $e^-$ | D | D | N | 1.5e-07 | -0.50 | 0.0 |
| 1913 | $pND_2^+$ | $e^-$ | D | D | N | 1.5e-07 | -0.50 | 0.0 |
| 1914 | $oNH_2D^+$ | $e^-$ | $oNH_2$ | D | | 1.0e-07 | -0.50 | 0.0 |
| 1915 | $pNH_2D^+$ | $e^-$ | $pNH_2$ | D | | 1.0e-07 | -0.50 | 0.0 |
| 1916 | $oNH_2D^+$ | $e^-$ | NHD | H | | 2.0e-07 | -0.50 | 0.0 |
| 1917 | $pNH_2D^+$ | $e^-$ | NHD | H | | 2.0e-07 | -0.50 | 0.0 |
| 1918 | $oNHD_2^+$ | $e^-$ | $oND_2$ | H | | 1.0e-07 | -0.50 | 0.0 |
| 1919 | $pNHD_2^+$ | $e^-$ | $pND_2$ | H | | 1.0e-07 | -0.50 | 0.0 |
| 1920 | $oNHD_2^+$ | $e^-$ | NHD | D | | 2.0e-07 | -0.50 | 0.0 |
| 1921 | $pNHD_2^+$ | $e^-$ | NHD | D | | 2.0e-07 | -0.50 | 0.0 |
| 1922 | $mND_3^+$ | $e^-$ | $oND_2$ | D | | 3.0e-07 | -0.50 | 0.0 |
| 1923 | $pND_3^+$ | $e^-$ | $pND_2$ | D | | 3.0e-07 | -0.50 | 0.0 |
| 1924 | $oND_3^+$ | $e^-$ | $oND_2$ | D | | 1.5e-07 | -0.50 | 0.0 |

**Table D6** – *continued* (part 27)

| # | Reactants | | Products | | $\alpha$ | $\beta$ | $\gamma$ |
|---|---|---|---|---|---|---|---|
| 1925 | $oND_3^+$ | $e^-$ | $pND_2$ | D | 1.5e-07 | -0.50 | 0.0 |
| 1926 | $oNH_3D^+$ | $e^-$ | $oNH_2$ | HD | 2.5e-07 | -0.50 | 0.0 |
| 1927 | $pNH_3D^+$ | $e^-$ | $oNH_2$ | HD | 1.3e-07 | -0.50 | 0.0 |
| 1928 | $pNH_3D^+$ | $e^-$ | $pNH_2$ | HD | 1.3e-07 | -0.50 | 0.0 |
| 1929 | $oNH_3D^+$ | $e^-$ | $oH_2$ | NHD | 2.5e-07 | -0.50 | 0.0 |
| 1930 | $pNH_3D^+$ | $e^-$ | $oH_2$ | NHD | 1.3e-07 | -0.50 | 0.0 |
| 1931 | $pNH_3D^+$ | $e^-$ | $pH_2$ | NHD | 1.3e-07 | -0.50 | 0.0 |
| 1932 | $oNH_3D^+$ | $e^-$ | $oNH_3$ | D | 1.9e-07 | -0.50 | 0.0 |
| 1933 | $pNH_3D^+$ | $e^-$ | $pNH_3$ | D | 1.9e-07 | -0.50 | 0.0 |
| 1934 | $oNH_3D^+$ | $e^-$ | $oNH_2D$ | H | 5.7e-07 | -0.50 | 0.0 |
| 1935 | $oNH_3D^+$ | $e^-$ | $oNH_2D$ | H | 2.9e-07 | -0.50 | 0.0 |
| 1936 | $pNH_3D^+$ | $e^-$ | $pNH_2D$ | H | 2.9e-07 | -0.50 | 0.0 |
| 1937 | $ooNH_2D_2^+$ | $e^-$ | $oD_2$ | $oNH_2$ | 8.5e-08 | -0.50 | 0.0 |
| 1938 | $poNH_2D_2^+$ | $e^-$ | $pD_2$ | $oNH_2$ | 8.5e-08 | -0.50 | 0.0 |
| 1939 | $opNH_2D_2^+$ | $e^-$ | $oD_2$ | $pNH_2$ | 8.5e-08 | -0.50 | 0.0 |
| 1940 | $ppNH_2D_2^+$ | $e^-$ | $pD_2$ | $pNH_2$ | 8.5e-08 | -0.50 | 0.0 |
| 1941 | $ooNH_2D_2^+$ | $e^-$ | NHD | HD | 3.4e-07 | -0.50 | 0.0 |
| 1942 | $poNH_2D_2^+$ | $e^-$ | NHD | HD | 3.4e-07 | -0.50 | 0.0 |
| 1943 | $opNH_2D_2^+$ | $e^-$ | NHD | HD | 3.4e-07 | -0.50 | 0.0 |
| 1944 | $ppNH_2D_2^+$ | $e^-$ | NHD | HD | 3.4e-07 | -0.50 | 0.0 |
| 1945 | $ooNH_2D_2^+$ | $e^-$ | $oND_2$ | $oH_2$ | 8.5e-08 | -0.50 | 0.0 |
| 1946 | $poNH_2D_2^+$ | $e^-$ | $pND_2$ | $oH_2$ | 8.5e-08 | -0.50 | 0.0 |
| 1947 | $opNH_2D_2^+$ | $e^-$ | $oND_2$ | $pH_2$ | 8.5e-08 | -0.50 | 0.0 |
| 1948 | $ppNH_2D_2^+$ | $e^-$ | $pND_2$ | $pH_2$ | 8.5e-08 | -0.50 | 0.0 |
| 1949 | $ooNH_2D_2^+$ | $e^-$ | $oNH_2D$ | D | 3.8e-07 | -0.50 | 0.0 |
| 1950 | $poNH_2D_2^+$ | $e^-$ | $oNH_2D$ | D | 3.8e-07 | -0.50 | 0.0 |
| 1951 | $opNH_2D_2^+$ | $e^-$ | $pNH_2D$ | D | 3.8e-07 | -0.50 | 0.0 |
| 1952 | $ppNH_2D_2^+$ | $e^-$ | $pNH_2D$ | D | 3.8e-07 | -0.50 | 0.0 |
| 1953 | $ooNH_2D_2^+$ | $e^-$ | $oNHD_2$ | H | 3.8e-07 | -0.50 | 0.0 |
| 1954 | $poNH_2D_2^+$ | $e^-$ | $pNHD_2$ | H | 3.8e-07 | -0.50 | 0.0 |
| 1955 | $opNH_2D_2^+$ | $e^-$ | $oNHD_2$ | H | 3.8e-07 | -0.50 | 0.0 |
| 1956 | $ppNH_2D_2^+$ | $e^-$ | $pNHD_2$ | H | 3.8e-07 | -0.50 | 0.0 |
| 1957 | $mNHD_3^+$ | $e^-$ | $oD_2$ | NHD | 2.5e-07 | -0.50 | 0.0 |
| 1958 | $pNHD_3^+$ | $e^-$ | $pD_2$ | NHD | 2.5e-07 | -0.50 | 0.0 |
| 1959 | $oNHD_3^+$ | $e^-$ | $oD_2$ | NHD | 1.3e-07 | -0.50 | 0.0 |
| 1960 | $oNHD_3^+$ | $e^-$ | $pD_2$ | NHD | 1.3e-07 | -0.50 | 0.0 |
| 1961 | $mNHD_3^+$ | $e^-$ | $oND_2$ | HD | 2.5e-07 | -0.50 | 0.0 |
| 1962 | $pNHD_3^+$ | $e^-$ | $pND_2$ | HD | 2.5e-07 | -0.50 | 0.0 |
| 1963 | $oNHD_3^+$ | $e^-$ | $oND_2$ | HD | 1.3e-07 | -0.50 | 0.0 |
| 1964 | $oNHD_3^+$ | $e^-$ | $pND_2$ | HD | 1.3e-07 | -0.50 | 0.0 |
| 1965 | $mNHD_3^+$ | $e^-$ | $mND_3$ | H | 1.9e-07 | -0.50 | 0.0 |
| 1966 | $pNHD_3^+$ | $e^-$ | $pND_3$ | H | 1.9e-07 | -0.50 | 0.0 |
| 1967 | $oNHD_3^+$ | $e^-$ | $oND_3$ | H | 1.9e-07 | -0.50 | 0.0 |
| 1968 | $mNHD_3^+$ | $e^-$ | $oNHD_2$ | D | 5.7e-07 | -0.50 | 0.0 |
| 1969 | $pNHD_3^+$ | $e^-$ | $pNHD_2$ | D | 5.7e-07 | -0.50 | 0.0 |
| 1970 | $oNHD_3^+$ | $e^-$ | $oNHD_2$ | D | 2.9e-07 | -0.50 | 0.0 |
| 1971 | $oNHD_3^+$ | $e^-$ | $pNHD_2$ | D | 2.9e-07 | -0.50 | 0.0 |
| 1972 | $lND_4^+$ | $e^-$ | $oND_2$ | $oD_2$ | 5.1e-07 | -0.50 | 0.0 |
| 1973 | $mND_4^+$ | $e^-$ | $oND_2$ | $oD_2$ | 2.5e-07 | -0.50 | 0.0 |
| 1974 | $mND_4^+$ | $e^-$ | $pND_2$ | $pD_2$ | 2.5e-07 | -0.50 | 0.0 |
| 1975 | $oND_4^+$ | $e^-$ | $oND_2$ | $oD_2$ | 1.7e-07 | -0.50 | 0.0 |
| 1976 | $oND_4^+$ | $e^-$ | $oND_2$ | $pD_2$ | 1.7e-07 | -0.50 | 0.0 |
| 1977 | $oND_4^+$ | $e^-$ | $pND_2$ | $oD_2$ | 1.7e-07 | -0.50 | 0.0 |
| 1978 | $pND_4^+$ | $e^-$ | $oND_2$ | $pD_2$ | 1.7e-07 | -0.50 | 0.0 |
| 1979 | $pND_4^+$ | $e^-$ | $pND_2$ | $oD_2$ | 1.7e-07 | -0.50 | 0.0 |
| 1980 | $pND_4^+$ | $e^-$ | $pND_2$ | $pD_2$ | 1.7e-07 | -0.50 | 0.0 |
| 1981 | $lND_4^+$ | $e^-$ | $mND_3$ | D | 7.6e-07 | -0.50 | 0.0 |
| 1982 | $mND_4^+$ | $e^-$ | $oND_3$ | D | 7.6e-07 | -0.50 | 0.0 |
| 1983 | $oND_4^+$ | $e^-$ | $mND_3$ | D | 2.5e-07 | -0.50 | 0.0 |
| 1984 | $oND_4^+$ | $e^-$ | $oND_3$ | D | 5.1e-07 | -0.50 | 0.0 |
| 1985 | $pND_4^+$ | $e^-$ | $pND_3$ | D | 2.5e-07 | -0.50 | 0.0 |
| 1986 | $pND_4^+$ | $e^-$ | $oND_3$ | D | 5.1e-07 | -0.50 | 0.0 |
| 1987 | $N_2D^+$ | $e^-$ | $N_2$ | D | 2.8e-07 | -0.74 | 0.0 |
| 1988 | $N_2D^+$ | $e^-$ | ND | N | 2.1e-08 | -0.74 | 0.0 |
| 1989 | $NH^+$ | $\gamma$ | $N^+$ | H | 2.2e+01 | 0.00 | 0.0 |
| 1990 | $N_2$ | $\gamma$ | N | N | 3.9e+01 | 0.00 | 0.0 |
| 1991 | NH | $\gamma$ | N | H | 3.7e+02 | 0.00 | 0.0 |
| 1992 | NH | $\gamma$ | $NH^+$ | $e^-$ | 7.1e+00 | 0.00 | 0.0 |
| 1993 | $oNH_2$ | $\gamma$ | NH | H | 7.2e+02 | 0.00 | 0.0 |
| 1994 | $pNH_2$ | $\gamma$ | NH | H | 7.2e+02 | 0.00 | 0.0 |
| 1995 | $oNH_2$ | $\gamma$ | $oNH_2^+$ | $e^-$ | 1.4e+02 | 0.00 | 0.0 |
| 1996 | $pNH_2$ | $\gamma$ | $pNH_2^+$ | $e^-$ | 1.4e+02 | 0.00 | 0.0 |
| 1997 | $oNH_3$ | $\gamma$ | $oH_2$ | NH | 1.1e+03 | 0.00 | 0.0 |
| 1998 | $pNH_3$ | $\gamma$ | $oH_2$ | NH | 5.5e+02 | 0.00 | 0.0 |

**Table D6** – *continued* (part 28)

| # | Reactants | | Products | | | $\alpha$ | $\beta$ | $\gamma$ |
|---|---|---|---|---|---|---|---|---|
| 1999 | $pNH_3$ | $\gamma$ | $pH_2$ | NH | | 5.5e+02 | 0.00 | 0.0 |
| 2000 | $oNH_3$ | $\gamma$ | $oNH_3^+$ | $e^-$ | | 2.2e+02 | 0.00 | 0.0 |
| 2001 | $pNH_3$ | $\gamma$ | $pNH_3^+$ | $e^-$ | | 2.2e+02 | 0.00 | 0.0 |
| 2002 | ND | $\gamma$ | N | D | | 3.7e+02 | 0.00 | 0.0 |
| 2003 | ND | $\gamma$ | $ND^+$ | $e^-$ | | 7.1e+00 | 0.00 | 0.0 |
| 2004 | NHD | $\gamma$ | NH | D | | 3.6e+02 | 0.00 | 0.0 |
| 2005 | NHD | $\gamma$ | ND | H | | 3.6e+02 | 0.00 | 0.0 |
| 2006 | NHD | $\gamma$ | $NHD^+$ | $e^-$ | | 1.4e+02 | 0.00 | 0.0 |
| 2007 | $oND_2$ | $\gamma$ | ND | D | | 7.2e+02 | 0.00 | 0.0 |
| 2008 | $pND_2$ | $\gamma$ | ND | D | | 7.2e+02 | 0.00 | 0.0 |
| 2009 | $oND_2$ | $\gamma$ | $oND_2^+$ | $e^-$ | | 1.4e+02 | 0.00 | 0.0 |
| 2010 | $pND_2$ | $\gamma$ | $pND_2^+$ | $e^-$ | | 1.4e+02 | 0.00 | 0.0 |
| 2011 | $oNH_2D$ | $\gamma$ | NH | HD | | 4.5e+02 | 0.00 | 0.0 |
| 2012 | $pNH_2D$ | $\gamma$ | NH | HD | | 4.5e+02 | 0.00 | 0.0 |
| 2013 | $oNH_2D$ | $\gamma$ | $oH_2$ | ND | | 2.2e+02 | 0.00 | 0.0 |
| 2014 | $pNH_2D$ | $\gamma$ | $pH_2$ | ND | | 2.2e+02 | 0.00 | 0.0 |
| 2015 | $oNH_2D$ | $\gamma$ | NHD | H | | 2.9e+02 | 0.00 | 0.0 |
| 2016 | $pNH_2D$ | $\gamma$ | NHD | H | | 2.9e+02 | 0.00 | 0.0 |
| 2017 | $oNH_2D$ | $\gamma$ | $oNH_2$ | D | | 1.4e+02 | 0.00 | 0.0 |
| 2018 | $pNH_2D$ | $\gamma$ | $pNH_2$ | D | | 1.4e+02 | 0.00 | 0.0 |
| 2019 | $oNH_2D$ | $\gamma$ | $oNH_2D^+$ | $e^-$ | | 2.2e+02 | 0.00 | 0.0 |
| 2020 | $pNH_2D$ | $\gamma$ | $pNH_2D^+$ | $e^-$ | | 2.2e+02 | 0.00 | 0.0 |
| 2021 | $oNHD_2$ | $\gamma$ | ND | HD | | 4.5e+02 | 0.00 | 0.0 |
| 2022 | $pNHD_2$ | $\gamma$ | ND | HD | | 4.5e+02 | 0.00 | 0.0 |
| 2023 | $oNHD_2$ | $\gamma$ | $oD_2$ | NH | | 2.2e+02 | 0.00 | 0.0 |
| 2024 | $pNHD_2$ | $\gamma$ | $pD_2$ | NH | | 2.2e+02 | 0.00 | 0.0 |
| 2025 | $oNHD_2$ | $\gamma$ | NHD | D | | 2.9e+02 | 0.00 | 0.0 |
| 2026 | $pNHD_2$ | $\gamma$ | NHD | D | | 2.9e+02 | 0.00 | 0.0 |
| 2027 | $oNHD_2$ | $\gamma$ | $oND_2$ | H | | 1.4e+02 | 0.00 | 0.0 |
| 2028 | $pNHD_2$ | $\gamma$ | $pND_2$ | H | | 1.4e+02 | 0.00 | 0.0 |
| 2029 | $oNHD_2$ | $\gamma$ | $oNHD_2^+$ | $e^-$ | | 2.2e+02 | 0.00 | 0.0 |
| 2030 | $pNHD_2$ | $\gamma$ | $pNHD_2^+$ | $e^-$ | | 2.2e+02 | 0.00 | 0.0 |
| 2031 | $mND_3$ | $\gamma$ | $oD_2$ | ND | | 6.7e+02 | 0.00 | 0.0 |
| 2032 | $pND_3$ | $\gamma$ | $pD_2$ | ND | | 6.7e+02 | 0.00 | 0.0 |
| 2033 | $oND_3$ | $\gamma$ | $oD_2$ | ND | | 3.4e+02 | 0.00 | 0.0 |
| 2034 | $oND_3$ | $\gamma$ | $pD_2$ | ND | | 3.4e+02 | 0.00 | 0.0 |
| 2035 | $mND_3$ | $\gamma$ | $oND_2$ | D | | 4.3e+02 | 0.00 | 0.0 |
| 2036 | $pND_3$ | $\gamma$ | $pND_2$ | D | | 4.3e+02 | 0.00 | 0.0 |
| 2037 | $oND_3$ | $\gamma$ | $oND_2$ | D | | 2.1e+02 | 0.00 | 0.0 |
| 2038 | $oND_3$ | $\gamma$ | $pND_2$ | D | | 2.1e+02 | 0.00 | 0.0 |
| 2039 | $mND_3$ | $\gamma$ | $mND_3^+$ | $e^-$ | | 2.2e+02 | 0.00 | 0.0 |
| 2040 | $pND_3$ | $\gamma$ | $pND_3^+$ | $e^-$ | | 2.2e+02 | 0.00 | 0.0 |
| 2041 | $oND_3$ | $\gamma$ | $oND_3^+$ | $e^-$ | | 2.2e+02 | 0.00 | 0.0 |
| 2042 | $mNH_4^+$ | Gr | $oNH_3$ | H | $Gr^+$ | 3.8e-07 | 0.50 | 0.0 |
| 2043 | $pNH_4^+$ | Gr | $pNH_3$ | H | $Gr^+$ | 3.8e-07 | 0.50 | 0.0 |
| 2044 | $oNH_4^+$ | Gr | $oNH_3$ | H | $Gr^+$ | 1.3e-07 | 0.50 | 0.0 |
| 2045 | $oNH_4^+$ | Gr | $pNH_3$ | H | $Gr^+$ | 2.5e-07 | 0.50 | 0.0 |
| 2046 | $mNH_4^+$ | $Gr^-$ | $oNH_3$ | H | Gr | 3.8e-07 | 0.50 | 0.0 |
| 2047 | $pNH_4^+$ | $Gr^-$ | $pNH_3$ | H | Gr | 3.8e-07 | 0.50 | 0.0 |
| 2048 | $oNH_4^+$ | $Gr^-$ | $oNH_3$ | H | Gr | 1.3e-07 | 0.50 | 0.0 |
| 2049 | $oNH_4^+$ | $Gr^-$ | $pNH_3$ | H | Gr | 2.5e-07 | 0.50 | 0.0 |
| 2050 | $N_2H^+$ | Gr | $Gr^+$ | $N_2$ | H | 1.0e-07 | 0.50 | 0.0 |
| 2051 | $N_2H^+$ | Gr | $Gr^+$ | NH | N | 1.9e-07 | 0.50 | 0.0 |
| 2052 | $N_2H^+$ | $Gr^-$ | Gr | $N_2$ | H | 1.0e-07 | 0.50 | 0.0 |
| 2053 | $N_2H^+$ | $Gr^-$ | Gr | NH | N | 1.9e-07 | 0.50 | 0.0 |
| 2054 | $N_2D^+$ | Gr | $Gr^+$ | $N_2$ | D | 3.0e-07 | 0.50 | 0.0 |
| 2055 | $N_2D^+$ | $Gr^-$ | Gr | $N_2$ | D | 3.0e-07 | 0.50 | 0.0 |
| 2056 | S | $H^+$ | $S^+$ | H | | 1.0e-15 | 0.00 | 0.0 |
| 2057 | S | $oH_3^+$ | $oH_2$ | $SH^+$ | | 2.6e-09 | 0.00 | 0.0 |
| 2058 | S | $pH_3^+$ | $oH_2$ | $SH^+$ | | 1.3e-09 | 0.00 | 0.0 |
| 2059 | S | $pH_3^+$ | $pH_2$ | $SH^+$ | | 1.3e-09 | 0.00 | 0.0 |
| 2060 | SH | H | $oH_2$ | S | | 1.9e-11 | 0.00 | 0.0 |
| 2061 | SH | H | $pH_2$ | S | | 6.3e-12 | 0.00 | 0.0 |
| 2062 | SH | $He^+$ | $S^+$ | H | He | 1.7e-09 | 0.00 | 0.0 |
| 2063 | SH | $H^+$ | $SH^+$ | H | | 1.6e-09 | 0.00 | 0.0 |
| 2064 | SH | $H^+$ | $oH_2$ | $S^+$ | | 1.2e-09 | 0.00 | 0.0 |
| 2065 | SH | $H^+$ | $pH_2$ | $S^+$ | | 4.0e-10 | 0.00 | 0.0 |
| 2066 | SH | $oH_3^+$ | $oH_2S^+$ | $oH_2$ | | 1.4e-09 | 0.00 | 0.0 |
| 2067 | SH | $oH_3^+$ | $oH_2S^+$ | $pH_2$ | | 2.4e-10 | 0.00 | 0.0 |
| 2068 | SH | $oH_3^+$ | $pH_2S^+$ | $oH_2$ | | 2.4e-10 | 0.00 | 0.0 |
| 2069 | SH | $pH_3^+$ | $oH_2S^+$ | $oH_2$ | | 7.1e-10 | 0.00 | 0.0 |
| 2070 | SH | $pH_3^+$ | $oH_2S^+$ | $pH_2$ | | 4.8e-10 | 0.00 | 0.0 |
| 2071 | SH | $pH_3^+$ | $pH_2S^+$ | $oH_2$ | | 4.8e-10 | 0.00 | 0.0 |
| 2072 | SH | $pH_3^+$ | $pH_2S^+$ | $pH_2$ | | 2.4e-10 | 0.00 | 0.0 |

**Table D6** – *continued* (part 29)

| # | Reactants | | Products | | | $\alpha$ | $\beta$ | $\gamma$ |
|---|---|---|---|---|---|---|---|---|
| 2073 | $oH_2S$ | $He^+$ | $oH_2$ | $S^+$ | He | 3.6e-09 | 0.00 | 0.0 |
| 2074 | $pH_2S$ | $He^+$ | $pH_2$ | $S^+$ | He | 3.6e-09 | 0.00 | 0.0 |
| 2075 | $oH_2S$ | $He^+$ | $SH^+$ | H | He | 4.8e-10 | 0.00 | 0.0 |
| 2076 | $pH_2S$ | $He^+$ | $SH^+$ | H | He | 4.8e-10 | 0.00 | 0.0 |
| 2077 | $oH_2S$ | $He^+$ | $oH_2S^+$ | He | | 3.1e-10 | 0.00 | 0.0 |
| 2078 | $pH_2S$ | $He^+$ | $pH_2S^+$ | He | | 3.1e-10 | 0.00 | 0.0 |
| 2079 | $oH_2S$ | $H^+$ | $oH_2S^+$ | H | | 7.6e-09 | 0.00 | 0.0 |
| 2080 | $pH_2S$ | $H^+$ | $pH_2S^+$ | H | | 7.6e-09 | 0.00 | 0.0 |
| 2081 | $oH_2S$ | $oH_3^+$ | $oH_3S^+$ | $oH_2$ | | 2.3e-09 | 0.00 | 0.0 |
| 2082 | $oH_2S$ | $oH_3^+$ | $oH_3S^+$ | $pH_2$ | | 3.1e-10 | 0.00 | 0.0 |
| 2083 | $oH_2S$ | $oH_3^+$ | $pH_3S^+$ | $oH_2$ | | 8.6e-10 | 0.00 | 0.0 |
| 2084 | $oH_2S$ | $oH_3^+$ | $pH_3S^+$ | $pH_2$ | | 2.5e-10 | 0.00 | 0.0 |
| 2085 | $pH_2S$ | $oH_3^+$ | $oH_3S^+$ | $oH_2$ | | 9.3e-10 | 0.00 | 0.0 |
| 2086 | $pH_2S$ | $oH_3^+$ | $oH_3S^+$ | $pH_2$ | | 9.3e-10 | 0.00 | 0.0 |
| 2087 | $pH_2S$ | $oH_3^+$ | $pH_3S^+$ | $oH_2$ | | 1.9e-09 | 0.00 | 0.0 |
| 2088 | $oH_2S$ | $pH_3^+$ | $oH_3S^+$ | $oH_2$ | | 8.6e-10 | 0.00 | 0.0 |
| 2089 | $oH_2S$ | $pH_3^+$ | $oH_3S^+$ | $pH_2$ | | 6.2e-10 | 0.00 | 0.0 |
| 2090 | $oH_2S$ | $pH_3^+$ | $pH_3S^+$ | $oH_2$ | | 1.7e-09 | 0.00 | 0.0 |
| 2091 | $oH_2S$ | $pH_3^+$ | $pH_3S^+$ | $pH_2$ | | 4.9e-10 | 0.00 | 0.0 |
| 2092 | $pH_2S$ | $pH_3^+$ | $oH_3S^+$ | $oH_2$ | | 7.4e-10 | 0.00 | 0.0 |
| 2093 | $pH_2S$ | $pH_3^+$ | $pH_3S^+$ | $oH_2$ | | 1.5e-09 | 0.00 | 0.0 |
| 2094 | $pH_2S$ | $pH_3^+$ | $pH_3S^+$ | $pH_2$ | | 1.5e-09 | 0.00 | 0.0 |
| 2095 | $S^+$ | Fe | $Fe^+$ | S | | 1.8e-10 | 0.00 | 0.0 |
| 2096 | $SH^+$ | H | $oH_2$ | $S^+$ | | 8.2e-11 | 0.00 | 0.0 |
| 2097 | $SH^+$ | H | $pH_2$ | $S^+$ | | 2.7e-11 | 0.00 | 0.0 |
| 2098 | $SH^+$ | $oH_2$ | $oH_3S^+$ | $\gamma$ | | 6.7e-16 | 0.00 | 0.0 |
| 2099 | $SH^+$ | $oH_2$ | $pH_3S^+$ | $\gamma$ | | 3.3e-16 | 0.00 | 0.0 |
| 2100 | $SH^+$ | $pH_2$ | $pH_3S^+$ | $\gamma$ | | 1.0e-15 | 0.00 | 0.0 |
| 2101 | $SH^+$ | S | $S^+$ | SH | | 9.7e-10 | 0.00 | 0.0 |
| 2102 | $SH^+$ | $oH_2S$ | $oH_3S^+$ | S | | 3.3e-10 | 0.00 | 0.0 |
| 2103 | $SH^+$ | $oH_2S$ | $pH_3S^+$ | S | | 1.7e-10 | 0.00 | 0.0 |
| 2104 | $SH^+$ | $pH_2S$ | $pH_3S^+$ | S | | 5.0e-10 | 0.00 | 0.0 |
| 2105 | $SH^+$ | Fe | $Fe^+$ | SH | | 1.6e-09 | 0.00 | 0.0 |
| 2106 | $oH_2S^+$ | H | $oH_2$ | $SH^+$ | | 1.7e-10 | 0.00 | 0.0 |
| 2107 | $oH_2S^+$ | H | $pH_2$ | $SH^+$ | | 3.3e-11 | 0.00 | 0.0 |
| 2108 | $pH_2S^+$ | H | $oH_2$ | $SH^+$ | | 1.0e-10 | 0.00 | 0.0 |
| 2109 | $pH_2S^+$ | H | $pH_2$ | $SH^+$ | | 1.0e-10 | 0.00 | 0.0 |
| 2110 | $oH_2S^+$ | S | $oH_2S$ | $S^+$ | | 1.1e-09 | 0.00 | 0.0 |
| 2111 | $pH_2S^+$ | S | $pH_2S$ | $S^+$ | | 1.1e-09 | 0.00 | 0.0 |
| 2112 | $oH_2S^+$ | SH | $oH_2S$ | $SH^+$ | | 5.0e-10 | 0.00 | 0.0 |
| 2113 | $pH_2S^+$ | SH | $pH_2S$ | $SH^+$ | | 5.0e-10 | 0.00 | 0.0 |
| 2114 | $oH_2S^+$ | Fe | $oH_2S$ | $Fe^+$ | | 1.8e-09 | 0.00 | 0.0 |
| 2115 | $pH_2S^+$ | Fe | $pH_2S$ | $Fe^+$ | | 1.8e-09 | 0.00 | 0.0 |
| 2116 | $oH_3S^+$ | H | $oH_2S^+$ | $oH_2$ | | 4.5e-11 | 0.00 | 0.0 |
| 2117 | $oH_3S^+$ | H | $oH_2S^+$ | $pH_2$ | | 7.5e-12 | 0.00 | 0.0 |
| 2118 | $oH_3S^+$ | H | $pH_2S^+$ | $oH_2$ | | 7.5e-12 | 0.00 | 0.0 |
| 2119 | $pH_3S^+$ | H | $oH_2S^+$ | $oH_2$ | | 2.2e-11 | 0.00 | 0.0 |
| 2120 | $pH_3S^+$ | H | $oH_2S^+$ | $pH_2$ | | 1.5e-11 | 0.00 | 0.0 |
| 2121 | $pH_3S^+$ | H | $pH_2S^+$ | $oH_2$ | | 1.5e-11 | 0.00 | 0.0 |
| 2122 | $pH_3S^+$ | H | $pH_2S^+$ | $pH_2$ | | 7.5e-12 | 0.00 | 0.0 |
| 2123 | S | $oH_2D^+$ | $SH^+$ | HD | | 1.7e-09 | 0.00 | 0.0 |
| 2124 | S | $pH_2D^+$ | $SH^+$ | HD | | 1.7e-09 | 0.00 | 0.0 |
| 2125 | S | $oH_2D^+$ | $oH_2$ | $SD^+$ | | 8.7e-10 | 0.00 | 0.0 |
| 2126 | S | $pH_2D^+$ | $pH_2$ | $SD^+$ | | 8.7e-10 | 0.00 | 0.0 |
| 2127 | S | $oHD_2^+$ | $oD_2$ | $SH^+$ | | 8.7e-10 | 0.00 | 0.0 |
| 2128 | S | $pHD_2^+$ | $pD_2$ | $SH^+$ | | 8.7e-10 | 0.00 | 0.0 |
| 2129 | S | $oHD_2^+$ | $SD^+$ | HD | | 1.7e-09 | 0.00 | 0.0 |
| 2130 | S | $pHD_2^+$ | $SD^+$ | HD | | 1.7e-09 | 0.00 | 0.0 |
| 2131 | S | $mD_3^+$ | $oD_2$ | $SD^+$ | | 2.6e-09 | 0.00 | 0.0 |
| 2132 | S | $pD_3^+$ | $pD_2$ | $SD^+$ | | 2.6e-09 | 0.00 | 0.0 |
| 2133 | S | $oD_3^+$ | $oD_2$ | $SD^+$ | | 1.3e-09 | 0.00 | 0.0 |
| 2134 | S | $oD_3^+$ | $pD_2$ | $SD^+$ | | 1.3e-09 | 0.00 | 0.0 |
| 2135 | SH | $oH_2D^+$ | $oH_2S^+$ | HD | | 1.6e-09 | 0.00 | 0.0 |
| 2136 | SH | $oH_2D^+$ | $pH_2S^+$ | HD | | 3.2e-10 | 0.00 | 0.0 |
| 2137 | SH | $pH_2D^+$ | $oH_2S^+$ | HD | | 9.5e-10 | 0.00 | 0.0 |
| 2138 | SH | $pH_2D^+$ | $pH_2S^+$ | HD | | 9.5e-10 | 0.00 | 0.0 |
| 2139 | SH | $oHD_2^+$ | $oD_2$ | $oH_2S^+$ | | 1.4e-09 | 0.00 | 0.0 |
| 2140 | SH | $pHD_2^+$ | $pD_2$ | $oH_2S^+$ | | 1.4e-09 | 0.00 | 0.0 |
| 2141 | SH | $oHD_2^+$ | $oD_2$ | $pH_2S^+$ | | 4.8e-10 | 0.00 | 0.0 |
| 2142 | SH | $pHD_2^+$ | $pD_2$ | $pH_2S^+$ | | 4.8e-10 | 0.00 | 0.0 |
| 2143 | $oH_2S$ | $oH_2D^+$ | $oH_3S^+$ | HD | | 2.5e-09 | 0.00 | 0.0 |
| 2144 | $oH_2S$ | $oH_2D^+$ | $pH_3S^+$ | HD | | 1.2e-09 | 0.00 | 0.0 |
| 2145 | $oH_2S$ | $pH_2D^+$ | $oH_3S^+$ | HD | | 1.2e-09 | 0.00 | 0.0 |
| 2146 | $oH_2S$ | $pH_2D^+$ | $pH_3S^+$ | HD | | 2.5e-09 | 0.00 | 0.0 |

**Table D6** – *continued* (part 30)

| # | Reactants | | Products | | | $\alpha$ | $\beta$ | $\gamma$ |
|---|---|---|---|---|---|---|---|---|
| 2147 | $pH_2S$ | $oH_2D^+$ | $oH_3S^+$ | HD | | 1.2e-09 | 0.00 | 0.0 |
| 2148 | $pH_2S$ | $oH_2D^+$ | $pH_3S^+$ | HD | | 2.5e-09 | 0.00 | 0.0 |
| 2149 | $pH_2S$ | $pH_2D^+$ | $pH_3S^+$ | HD | | 3.7e-09 | 0.00 | 0.0 |
| 2150 | $oH_2S$ | $oHD_2^+$ | $oD_2$ | $oH_3S^+$ | | 2.5e-09 | 0.00 | 0.0 |
| 2151 | $oH_2S$ | $pHD_2^+$ | $pD_2$ | $oH_3S^+$ | | 2.5e-09 | 0.00 | 0.0 |
| 2152 | $oH_2S$ | $oHD_2^+$ | $oD_2$ | $pH_3S^+$ | | 1.2e-09 | 0.00 | 0.0 |
| 2153 | $oH_2S$ | $pHD_2^+$ | $pD_2$ | $pH_3S^+$ | | 1.2e-09 | 0.00 | 0.0 |
| 2154 | $pH_2S$ | $oHD_2^+$ | $oD_2$ | $pH_3S^+$ | | 3.7e-09 | 0.00 | 0.0 |
| 2155 | $pH_2S$ | $pHD_2^+$ | $pD_2$ | $pH_3S^+$ | | 3.7e-09 | 0.00 | 0.0 |
| 2156 | $S^+$ | Gr | $Gr^+$ | S | | 2.8e-07 | 0.50 | 0.0 |
| 2157 | $S^+$ | $Gr^-$ | Gr | S | | 2.8e-07 | 0.50 | 0.0 |
| 2158 | $oH_3S^+$ | Gr | $oH_2S$ | H | $Gr^+$ | 2.7e-07 | 0.50 | 0.0 |
| 2159 | $pH_3S^+$ | Gr | $oH_2S$ | H | $Gr^+$ | 1.4e-07 | 0.50 | 0.0 |
| 2160 | $pH_3S^+$ | Gr | $pH_2S$ | H | $Gr^+$ | 1.4e-07 | 0.50 | 0.0 |
| 2161 | $oH_3S^+$ | $Gr^-$ | $oH_2S$ | H | Gr | 2.7e-07 | 0.50 | 0.0 |
| 2162 | $pH_3S^+$ | $Gr^-$ | $oH_2S$ | H | Gr | 1.4e-07 | 0.50 | 0.0 |
| 2163 | $pH_3S^+$ | $Gr^-$ | $pH_2S$ | H | Gr | 1.4e-07 | 0.50 | 0.0 |
| 2164 | $S^+$ | $e^-$ | S | $\gamma$ | | 3.9e-12 | -0.63 | 0.0 |
| 2165 | $SH^+$ | $e^-$ | S | H | | 2.0e-07 | -0.50 | 0.0 |
| 2166 | $oH_2S^+$ | $e^-$ | SH | H | | 1.5e-07 | -0.50 | 0.0 |
| 2167 | $pH_2S^+$ | $e^-$ | SH | H | | 1.5e-07 | -0.50 | 0.0 |
| 2168 | $oH_2S^+$ | $e^-$ | H | H | S | 1.5e-07 | -0.50 | 0.0 |
| 2169 | $pH_2S^+$ | $e^-$ | H | H | S | 1.5e-07 | -0.50 | 0.0 |
| 2170 | $oH_2S^+$ | $e^-$ | $oH_2S$ | $\gamma$ | | 1.1e-10 | -0.70 | 0.0 |
| 2171 | $pH_2S^+$ | $e^-$ | $pH_2S$ | $\gamma$ | | 1.1e-10 | -0.70 | 0.0 |
| 2172 | $oH_3S^+$ | $e^-$ | $oH_2S$ | H | | 3.0e-07 | -0.50 | 0.0 |
| 2173 | $pH_3S^+$ | $e^-$ | $oH_2S$ | H | | 1.5e-07 | -0.50 | 0.0 |
| 2174 | $pH_3S^+$ | $e^-$ | $pH_2S$ | H | | 1.5e-07 | -0.50 | 0.0 |
| 2175 | $oH_3S^+$ | $e^-$ | $oH_2$ | SH | | 1.0e-07 | -0.50 | 0.0 |
| 2176 | $pH_3S^+$ | $e^-$ | $oH_2$ | SH | | 5.0e-08 | -0.50 | 0.0 |
| 2177 | $pH_3S^+$ | $e^-$ | $pH_2$ | SH | | 5.0e-08 | -0.50 | 0.0 |
| 2178 | $SD^+$ | $e^-$ | S | D | | 2.0e-07 | -0.50 | 0.0 |
| 2179 | S | $\gamma_2$ | $S^+$ | $e^-$ | | 8.0e+02 | 0.00 | 0.0 |
| 2180 | SH | $\gamma_2$ | S | H | | 1.1e+03 | 0.00 | 0.0 |
| 2181 | SH | $\gamma_2$ | $SH^+$ | $e^-$ | | 3.4e+01 | 0.00 | 0.0 |
| 2182 | $oH_2S$ | $\gamma_2$ | SH | H | | 3.4e+03 | 0.00 | 0.0 |
| 2183 | $pH_2S$ | $\gamma_2$ | SH | H | | 3.4e+03 | 0.00 | 0.0 |
| 2184 | $oH_2S$ | $\gamma_2$ | $oH_2S^+$ | $e^-$ | | 6.2e+02 | 0.00 | 0.0 |
| 2185 | $pH_2S$ | $\gamma_2$ | $pH_2S^+$ | $e^-$ | | 6.2e+02 | 0.00 | 0.0 |
| 2186 | $SH^+$ | $\gamma_2$ | $S^+$ | H | | 4.6e+02 | 0.00 | 0.0 |
| 2187 | O | CH | $HCO^+$ | $e^-$ | | 2.4e-14 | 0.50 | 0.0 |
| 2188 | O | CH | CO | H | | 6.6e-11 | 0.00 | 0.0 |
| 2189 | O | $oCH_2$ | H | H | CO | 1.0e-10 | 0.00 | 0.0 |
| 2190 | O | $pCH_2$ | H | H | CO | 1.0e-10 | 0.00 | 0.0 |
| 2191 | O | $oCH_2$ | $oH_2$ | CO | | 4.0e-11 | 0.00 | 0.0 |
| 2192 | O | $pCH_2$ | $pH_2$ | CO | | 4.0e-11 | 0.00 | 0.0 |
| 2193 | O | $oCH_3$ | $oH_2$ | H | CO | 1.8e-10 | 0.50 | 0.0 |
| 2194 | O | $pCH_3$ | $oH_2$ | H | CO | 9.0e-11 | 0.50 | 0.0 |
| 2195 | O | $pCH_3$ | $pH_2$ | H | CO | 9.0e-11 | 0.50 | 0.0 |
| 2196 | O | $C_2$ | CO | C | | 2.0e-10 | -0.12 | 0.0 |
| 2197 | O | $C_2H$ | CO | CH | | 1.0e-10 | 0.00 | 250.0 |
| 2198 | O | $C_3$ | CO | $C_2$ | | 5.0e-11 | 0.50 | 0.0 |
| 2199 | O | $C_3H$ | $C_2H$ | CO | | 5.0e-11 | 0.50 | 0.0 |
| 2200 | O | $C_3H_2$ | $C_2H_2$ | CO | | 5.0e-11 | 0.50 | 0.0 |
| 2201 | O | NH | NO | H | | 6.6e-11 | 0.00 | 0.0 |
| 2202 | O | $oNH_2$ | NH | OH | | 7.0e-12 | -0.10 | 0.0 |
| 2203 | O | $pNH_2$ | NH | OH | | 7.0e-12 | -0.10 | 0.0 |
| 2204 | O | CN | CO | N | | 5.0e-11 | 0.00 | 0.0 |
| 2205 | O | HNC | CO | NH | | 2.0e-10 | 0.50 | 200.0 |
| 2206 | O | SH | OH | S | | 1.7e-11 | 0.67 | 956.0 |
| 2207 | O | SH | SO | H | | 1.6e-10 | 0.00 | 0.0 |
| 2208 | O | CS | CO | S | | 2.6e-10 | 0.00 | 760.0 |
| 2209 | O | $oCH_3^+$ | $oH_2$ | $HCO^+$ | | 3.1e-10 | 0.00 | 0.0 |
| 2210 | O | $pCH_3^+$ | $oH_2$ | $HCO^+$ | | 1.6e-10 | 0.00 | 0.0 |
| 2211 | O | $pCH_3^+$ | $pH_2$ | $HCO^+$ | | 1.6e-10 | 0.00 | 0.0 |
| 2212 | O | $oCH_3^+$ | $oH_3^+$ | CO | | 1.3e-11 | 0.00 | 0.0 |
| 2213 | O | $pCH_3^+$ | $pH_3^+$ | CO | | 1.3e-11 | 0.00 | 0.0 |
| 2214 | O | $pCH_5^+$ | $oH_3O^+$ | $oCH_2$ | | 2.2e-10 | 0.00 | 0.0 |
| 2215 | O | $oCH_5^+$ | $oH_3O^+$ | $oCH_2$ | | 5.4e-11 | 0.00 | 0.0 |
| 2216 | O | $oCH_5^+$ | $oH_3O^+$ | $pCH_2$ | | 5.4e-11 | 0.00 | 0.0 |
| 2217 | O | $oCH_5^+$ | $pH_3O^+$ | $oCH_2$ | | 1.1e-10 | 0.00 | 0.0 |
| 2218 | O | $mCH_5^+$ | $oH_3O^+$ | $oCH_2$ | | 4.3e-11 | 0.00 | 0.0 |
| 2219 | O | $mCH_5^+$ | $pH_3O^+$ | $oCH_2$ | | 8.6e-11 | 0.00 | 0.0 |
| 2220 | O | $mCH_5^+$ | $pH_3O^+$ | $pCH_2$ | | 8.6e-11 | 0.00 | 0.0 |

**Table D6** – *continued* (part 31)

| # | | Reactants | | Products | | $\alpha$ | $\beta$ | $\gamma$ |
|---|---|---|---|---|---|---|---|---|
| 2221 | O | $HCO_2^+$ | | $HCO^+$ | $O_2$ | 1.0e-09 | 0.00 | 0.0 |
| 2222 | O | $SH^+$ | | $SO^+$ | H | 2.9e-10 | 0.00 | 0.0 |
| 2223 | O | $SH^+$ | | $S^+$ | OH | 2.9e-10 | 0.00 | 0.0 |
| 2224 | O | $oH_2S^+$ | | $SH^+$ | OH | 3.1e-10 | 0.00 | 0.0 |
| 2225 | O | $pH_2S^+$ | | $SH^+$ | OH | 3.1e-10 | 0.00 | 0.0 |
| 2226 | O | $oH_2S^+$ | | $oH_2$ | $SO^+$ | 3.1e-10 | 0.00 | 0.0 |
| 2227 | O | $pH_2S^+$ | | $pH_2$ | $SO^+$ | 3.1e-10 | 0.00 | 0.0 |
| 2228 | O | $HCS^+$ | | $HCO^+$ | S | 1.0e-09 | 0.00 | 0.0 |
| 2229 | C | OH | | CO | H | 3.1e-11 | -0.36 | 0.0 |
| 2230 | C | $O_2$ | | CO | O | 3.3e-11 | 0.50 | 0.0 |
| 2231 | C | NH | | CN | H | 1.2e-10 | 0.00 | 0.0 |
| 2232 | C | $oNH_2$ | | HCN | H | 3.0e-11 | -0.20 | -6.0 |
| 2233 | C | $pNH_2$ | | HCN | H | 3.0e-11 | -0.20 | -6.0 |
| 2234 | C | $oNH_2$ | | HNC | H | 3.0e-11 | -0.20 | -6.0 |
| 2235 | C | $pNH_2$ | | HNC | H | 3.0e-11 | -0.20 | -6.0 |
| 2236 | C | NO | | CN | O | 6.0e-11 | -0.16 | 0.0 |
| 2237 | C | NO | | CO | N | 9.0e-11 | -0.16 | 0.0 |
| 2238 | C | SH | | CS | H | 2.0e-11 | 0.00 | 0.0 |
| 2239 | C | SO | | CO | S | 7.2e-11 | 0.00 | 0.0 |
| 2240 | C | SO | | CS | O | 1.7e-10 | 0.00 | 0.0 |
| 2241 | C | $oH_3O^+$ | | $oH_2$ | $HCO^+$ | 1.0e-11 | 0.00 | 0.0 |
| 2242 | C | $pH_3O^+$ | | $oH_2$ | $HCO^+$ | 5.0e-12 | 0.00 | 0.0 |
| 2243 | C | $pH_3O^+$ | | $pH_2$ | $HCO^+$ | 5.0e-12 | 0.00 | 0.0 |
| 2244 | C | $HCO^+$ | | $CH^+$ | CO | 1.1e-09 | 0.00 | 0.0 |
| 2245 | C | $O_2^+$ | | $CO^+$ | O | 5.2e-11 | 0.00 | 0.0 |
| 2246 | C | $O_2^+$ | | $C^+$ | $O_2$ | 5.2e-11 | 0.00 | 0.0 |
| 2247 | C | $SH^+$ | | $CS^+$ | H | 9.9e-10 | 0.00 | 0.0 |
| 2248 | C | $oH_2S^+$ | | $HCS^+$ | H | 1.0e-09 | 0.00 | 0.0 |
| 2249 | C | $pH_2S^+$ | | $HCS^+$ | H | 1.0e-09 | 0.00 | 0.0 |
| 2250 | C | $oH_2DO^+$ | | $oH_2$ | $DCO^+$ | 1.0e-11 | 0.00 | 0.0 |
| 2251 | C | $pH_2DO^+$ | | $pH_2$ | $DCO^+$ | 1.0e-11 | 0.00 | 0.0 |
| 2252 | C | $oHD_2O^+$ | | $DCO^+$ | HD | 1.0e-11 | 0.00 | 0.0 |
| 2253 | C | $pHD_2O^+$ | | $DCO^+$ | HD | 1.0e-11 | 0.00 | 0.0 |
| 2254 | C | $mD_3O^+$ | | $oD_2$ | $DCO^+$ | 1.0e-11 | 0.00 | 0.0 |
| 2255 | C | $pD_3O^+$ | | $pD_2$ | $DCO^+$ | 1.0e-11 | 0.00 | 0.0 |
| 2256 | C | $oD_3O^+$ | | $oD_2$ | $DCO^+$ | 5.0e-12 | 0.00 | 0.0 |
| 2257 | C | $oD_3O^+$ | | $pD_2$ | $DCO^+$ | 5.0e-12 | 0.00 | 0.0 |
| 2258 | N | CH | | CN | H | 1.4e-10 | 0.41 | 0.0 |
| 2259 | N | CN | | $N_2$ | C | 8.8e-11 | 0.42 | 0.0 |
| 2260 | N | NO | | $N_2$ | O | 7.3e-11 | 0.44 | 12.7 |
| 2261 | N | $oCH_2$ | | HCN | H | 5.0e-11 | 0.17 | 0.0 |
| 2262 | N | $pCH_2$ | | HCN | H | 5.0e-11 | 0.17 | 0.0 |
| 2263 | N | $oCH_2$ | | HNC | H | 3.0e-11 | 0.17 | 0.0 |
| 2264 | N | $pCH_2$ | | HNC | H | 3.0e-11 | 0.17 | 0.0 |
| 2265 | N | $oCH_3$ | | $oH_2$ | HCN | 1.3e-11 | 0.50 | 0.0 |
| 2266 | N | $pCH_3$ | | $oH_2$ | HCN | 6.5e-12 | 0.50 | 0.0 |
| 2267 | N | $pCH_3$ | | $pH_2$ | HCN | 6.5e-12 | 0.50 | 0.0 |
| 2268 | N | OH | | NO | H | 5.0e-11 | 0.00 | 6.0 |
| 2269 | N | $O_2^+$ | | $NO^+$ | O | 7.8e-11 | 0.00 | 0.0 |
| 2270 | N | $oCH_2^+$ | | $HCN^+$ | H | 9.4e-10 | 0.00 | 0.0 |
| 2271 | N | $pCH_2^+$ | | $HCN^+$ | H | 9.4e-10 | 0.00 | 0.0 |
| 2272 | N | $C_2H^+$ | | $C_2N^+$ | H | 8.3e-10 | 0.00 | 0.0 |
| 2273 | N | $oCH_3^+$ | | $oH_2$ | $HCN^+$ | 6.7e-11 | 0.00 | 0.0 |
| 2274 | N | $pCH_3^+$ | | $oH_2$ | $HCN^+$ | 3.4e-11 | 0.00 | 0.0 |
| 2275 | N | $pCH_3^+$ | | $pH_2$ | $HCN^+$ | 3.4e-11 | 0.00 | 0.0 |
| 2276 | N | $oCH_3^+$ | | $HCNH^+$ | H | 6.7e-11 | 0.00 | 0.0 |
| 2277 | N | $pCH_3^+$ | | $HCNH^+$ | H | 6.7e-11 | 0.00 | 0.0 |
| 2278 | N | $C_2H_2^+$ | | $CH^+$ | HCN | 2.5e-11 | 0.00 | 0.0 |
| 2279 | N | SO | | NO | S | 1.7e-11 | 0.50 | 750.0 |
| 2280 | S | CH | | CS | H | 1.1e-12 | 0.00 | 0.0 |
| 2281 | S | OH | | SO | H | 1.0e-10 | 0.00 | 100.0 |
| 2282 | S | $O_2$ | | SO | O | 5.2e-12 | 0.00 | 265.0 |
| 2283 | S | $CH^+$ | | $S^+$ | CH | 4.7e-10 | 0.00 | 0.0 |
| 2284 | S | $CH^+$ | | $SH^+$ | C | 4.7e-10 | 0.00 | 0.0 |
| 2285 | S | $CH^+$ | | $CS^+$ | H | 4.7e-10 | 0.00 | 0.0 |
| 2286 | S | $oCH_3^+$ | | $oH_2$ | $HCS^+$ | 1.4e-09 | 0.00 | 0.0 |
| 2287 | S | $pCH_3^+$ | | $oH_2$ | $HCS^+$ | 7.0e-10 | 0.00 | 0.0 |
| 2288 | S | $pCH_3^+$ | | $pH_2$ | $HCS^+$ | 7.0e-10 | 0.00 | 0.0 |
| 2289 | S | $pCH_5^+$ | | $mCH_4$ | $SH^+$ | 1.3e-09 | 0.00 | 0.0 |
| 2290 | S | $oCH_5^+$ | | $mCH_4$ | $SH^+$ | 3.3e-10 | 0.00 | 0.0 |
| 2291 | S | $oCH_5^+$ | | $oCH_4$ | $SH^+$ | 9.8e-10 | 0.00 | 0.0 |
| 2292 | S | $mCH_5^+$ | | $pCH_4$ | $SH^+$ | 5.2e-10 | 0.00 | 0.0 |
| 2293 | S | $mCH_5^+$ | | $oCH_4$ | $SH^+$ | 7.8e-10 | 0.00 | 0.0 |
| 2294 | S | $HCO^+$ | | $SH^+$ | CO | 3.3e-10 | 0.00 | 0.0 |

**Table D6** – *continued* (part 32)

| # | Reactants | | Products | | | $\alpha$ | $\beta$ | $\gamma$ |
|---|---|---|---|---|---|---|---|---|
| 2295 | S | $O_2^+$ | $SO^+$ | O | | 5.4e-10 | 0.00 | 0.0 |
| 2296 | S | $O_2^+$ | $S^+$ | $O_2$ | | 5.4e-10 | 0.00 | 0.0 |
| 2297 | S | $HNO^+$ | $SH^+$ | NO | | 1.1e-09 | 0.00 | 0.0 |
| 2298 | S | $N_2H^+$ | $SH^+$ | $N_2$ | | 1.1e-09 | 0.00 | 0.0 |
| 2299 | CO | $He^+$ | $C^+$ | O | He | 1.5e-09 | 0.00 | 0.0 |
| 2300 | SO | $He^+$ | $O^+$ | S | He | 8.3e-10 | 0.00 | 0.0 |
| 2301 | SO | $He^+$ | $S^+$ | O | He | 8.3e-10 | 0.00 | 0.0 |
| 2302 | NO | $He^+$ | $N^+$ | O | He | 1.4e-09 | 0.00 | 0.0 |
| 2303 | NO | $He^+$ | $O^+$ | N | He | 2.2e-10 | 0.00 | 0.0 |
| 2304 | CN | $He^+$ | $C^+$ | N | He | 8.8e-10 | 0.00 | 0.0 |
| 2305 | CN | $He^+$ | $N^+$ | C | He | 8.8e-10 | 0.00 | 0.0 |
| 2306 | CS | $He^+$ | $C^+$ | S | He | 1.3e-09 | 0.00 | 0.0 |
| 2307 | CS | $He^+$ | $S^+$ | C | He | 1.3e-09 | 0.00 | 0.0 |
| 2308 | $N_2$ | $He^+$ | $N^+$ | N | He | 7.9e-10 | 0.00 | 0.0 |
| 2309 | $N_2$ | $He^+$ | $N_2^+$ | He | | 4.1e-10 | 0.00 | 0.0 |
| 2310 | HCN | $He^+$ | $CN^+$ | H | He | 1.5e-09 | 0.00 | 0.0 |
| 2311 | HCN | $He^+$ | $CH^+$ | N | He | 6.2e-10 | 0.00 | 0.0 |
| 2312 | HCN | $He^+$ | $C^+$ | NH | He | 7.8e-10 | 0.00 | 0.0 |
| 2313 | HCN | $He^+$ | $N^+$ | CH | He | 2.5e-10 | 0.00 | 0.0 |
| 2314 | HNC | $He^+$ | $CN^+$ | H | He | 1.6e-09 | 0.00 | 0.0 |
| 2315 | HNC | $He^+$ | $C^+$ | NH | He | 1.6e-09 | 0.00 | 0.0 |
| 2316 | $CO_2$ | $He^+$ | $CO^+$ | O | He | 7.7e-10 | 0.00 | 0.0 |
| 2317 | $CO_2$ | $He^+$ | $O^+$ | CO | He | 1.8e-10 | 0.00 | 0.0 |
| 2318 | $CO_2$ | $He^+$ | $C^+$ | $O_2$ | He | 4.0e-11 | 0.00 | 0.0 |
| 2319 | $SO_2$ | $He^+$ | $S^+$ | $O_2$ | He | 8.6e-10 | 0.00 | 0.0 |
| 2320 | $SO_2$ | $He^+$ | $SO^+$ | O | He | 3.4e-09 | 0.00 | 0.0 |
| 2321 | OCS | $He^+$ | $CS^+$ | O | He | 7.6e-10 | 0.00 | 0.0 |
| 2322 | OCS | $He^+$ | $S^+$ | CO | He | 7.6e-10 | 0.00 | 0.0 |
| 2323 | OCS | $He^+$ | $CO^+$ | S | He | 7.6e-10 | 0.00 | 0.0 |
| 2324 | OCS | $He^+$ | $O^+$ | CS | He | 7.6e-11 | 0.00 | 0.0 |
| 2325 | CH | $SH^+$ | $oCH_2^+$ | S | | 4.3e-10 | 0.00 | 0.0 |
| 2326 | CH | $SH^+$ | $pCH_2^+$ | S | | 1.4e-10 | 0.00 | 0.0 |
| 2327 | $oH_2O$ | $pCH_5^+$ | $mCH_4$ | $oH_3O^+$ | | 1.9e-09 | 0.00 | 0.0 |
| 2328 | $oH_2O$ | $pCH_5^+$ | $pCH_4$ | $oH_3O^+$ | | 1.6e-10 | 0.00 | 0.0 |
| 2329 | $oH_2O$ | $pCH_5^+$ | $oCH_4$ | $oH_3O^+$ | | 8.5e-10 | 0.00 | 0.0 |
| 2330 | $oH_2O$ | $pCH_5^+$ | $mCH_4$ | $pH_3O^+$ | | 5.7e-10 | 0.00 | 0.0 |
| 2331 | $oH_2O$ | $pCH_5^+$ | $oCH_4$ | $pH_3O^+$ | | 2.0e-10 | 0.00 | 0.0 |
| 2332 | $pH_2O$ | $pCH_5^+$ | $mCH_4$ | $oH_3O^+$ | | 6.2e-10 | 0.00 | 0.0 |
| 2333 | $pH_2O$ | $pCH_5^+$ | $oCH_4$ | $oH_3O^+$ | | 1.9e-09 | 0.00 | 0.0 |
| 2334 | $pH_2O$ | $pCH_5^+$ | $mCH_4$ | $pH_3O^+$ | | 1.2e-09 | 0.00 | 0.0 |
| 2335 | $oH_2O$ | $oCH_5^+$ | $mCH_4$ | $oH_3O^+$ | | 4.7e-10 | 0.00 | 0.0 |
| 2336 | $oH_2O$ | $oCH_5^+$ | $pCH_4$ | $oH_3O^+$ | | 2.3e-10 | 0.00 | 0.0 |
| 2337 | $oH_2O$ | $oCH_5^+$ | $oCH_4$ | $oH_3O^+$ | | 1.4e-09 | 0.00 | 0.0 |
| 2338 | $oH_2O$ | $oCH_5^+$ | $mCH_4$ | $pH_3O^+$ | | 8.5e-10 | 0.00 | 0.0 |
| 2339 | $oH_2O$ | $oCH_5^+$ | $pCH_4$ | $pH_3O^+$ | | 1.8e-10 | 0.00 | 0.0 |
| 2340 | $oH_2O$ | $oCH_5^+$ | $oCH_4$ | $pH_3O^+$ | | 5.6e-10 | 0.00 | 0.0 |
| 2341 | $pH_2O$ | $oCH_5^+$ | $mCH_4$ | $oH_3O^+$ | | 3.5e-10 | 0.00 | 0.0 |
| 2342 | $pH_2O$ | $oCH_5^+$ | $pCH_4$ | $oH_3O^+$ | | 7.0e-10 | 0.00 | 0.0 |
| 2343 | $pH_2O$ | $oCH_5^+$ | $oCH_4$ | $oH_3O^+$ | | 1.1e-09 | 0.00 | 0.0 |
| 2344 | $pH_2O$ | $oCH_5^+$ | $mCH_4$ | $pH_3O^+$ | | 7.0e-10 | 0.00 | 0.0 |
| 2345 | $pH_2O$ | $oCH_5^+$ | $oCH_4$ | $pH_3O^+$ | | 8.8e-10 | 0.00 | 0.0 |
| 2346 | $oH_2O$ | $mCH_5^+$ | $mCH_4$ | $oH_3O^+$ | | 3.2e-10 | 0.00 | 0.0 |
| 2347 | $oH_2O$ | $mCH_5^+$ | $pCH_4$ | $oH_3O^+$ | | 4.7e-10 | 0.00 | 0.0 |
| 2348 | $oH_2O$ | $mCH_5^+$ | $oCH_4$ | $oH_3O^+$ | | 9.7e-10 | 0.00 | 0.0 |
| 2349 | $oH_2O$ | $mCH_5^+$ | $mCH_4$ | $pH_3O^+$ | | 4.7e-10 | 0.00 | 0.0 |
| 2350 | $oH_2O$ | $mCH_5^+$ | $pCH_4$ | $pH_3O^+$ | | 3.5e-10 | 0.00 | 0.0 |
| 2351 | $oH_2O$ | $mCH_5^+$ | $oCH_4$ | $pH_3O^+$ | | 1.1e-09 | 0.00 | 0.0 |
| 2352 | $pH_2O$ | $mCH_5^+$ | $mCH_4$ | $oH_3O^+$ | | 2.6e-10 | 0.00 | 0.0 |
| 2353 | $pH_2O$ | $mCH_5^+$ | $oCH_4$ | $oH_3O^+$ | | 7.9e-10 | 0.00 | 0.0 |
| 2354 | $pH_2O$ | $mCH_5^+$ | $pCH_4$ | $pH_3O^+$ | | 1.1e-09 | 0.00 | 0.0 |
| 2355 | $pH_2O$ | $mCH_5^+$ | $oCH_4$ | $pH_3O^+$ | | 1.6e-09 | 0.00 | 0.0 |
| 2356 | $oH_2O$ | $C_3H^+$ | $C_2H_2$ | $HCO^+$ | | 2.5e-10 | 0.00 | 0.0 |
| 2357 | $pH_2O$ | $C_3H^+$ | $C_2H_2$ | $HCO^+$ | | 2.5e-10 | 0.00 | 0.0 |
| 2358 | $oH_2O$ | $C_3H^+$ | $C_2H_3^+$ | CO | | 2.0e-10 | 0.00 | 0.0 |
| 2359 | $pH_2O$ | $C_3H^+$ | $C_2H_3^+$ | CO | | 2.0e-10 | 0.00 | 0.0 |
| 2360 | $oH_2O$ | $oNH_3^+$ | $mNH_4^+$ | OH | | 1.5e-10 | 0.00 | 0.0 |
| 2361 | $oH_2O$ | $oNH_3^+$ | $pNH_4^+$ | OH | | 1.7e-11 | 0.00 | 0.0 |
| 2362 | $oH_2O$ | $oNH_3^+$ | $oNH_4^+$ | OH | | 8.7e-11 | 0.00 | 0.0 |
| 2363 | $pH_2O$ | $oNH_3^+$ | $mNH_4^+$ | OH | | 6.3e-11 | 0.00 | 0.0 |
| 2364 | $pH_2O$ | $oNH_3^+$ | $oNH_4^+$ | OH | | 1.9e-10 | 0.00 | 0.0 |
| 2365 | $oH_2O$ | $pNH_3^+$ | $mNH_4^+$ | OH | | 4.2e-11 | 0.00 | 0.0 |
| 2366 | $oH_2O$ | $pNH_3^+$ | $pNH_4^+$ | OH | | 3.3e-11 | 0.00 | 0.0 |
| 2367 | $oH_2O$ | $pNH_3^+$ | $oNH_4^+$ | OH | | 1.7e-10 | 0.00 | 0.0 |
| 2368 | $pH_2O$ | $pNH_3^+$ | $pNH_4^+$ | OH | | 1.0e-10 | 0.00 | 0.0 |

**Table D6** – *continued* (part 33)

| # | Reactants | | Products | | $\alpha$ | $\beta$ | $\gamma$ |
|---|---|---|---|---|---|---|---|
| 2369 | $pH_2O$ | $pNH_3^+$ | $oNH_4^+$ | OH | 1.5e-10 | 0.00 | 0.0 |
| 2370 | $oH_2O$ | $N_2H^+$ | $oH_3O^+$ | $N_2$ | 1.7e-09 | 0.00 | 0.0 |
| 2371 | $oH_2O$ | $N_2H^+$ | $pH_3O^+$ | $N_2$ | 8.7e-10 | 0.00 | 0.0 |
| 2372 | $pH_2O$ | $N_2H^+$ | $pH_3O^+$ | $N_2$ | 2.6e-09 | 0.00 | 0.0 |
| 2373 | $oH_2O$ | $HNO^+$ | $oH_3O^+$ | NO | 1.5e-09 | 0.00 | 0.0 |
| 2374 | $oH_2O$ | $HNO^+$ | $pH_3O^+$ | NO | 7.7e-10 | 0.00 | 0.0 |
| 2375 | $pH_2O$ | $HNO^+$ | $pH_3O^+$ | NO | 2.3e-09 | 0.00 | 0.0 |
| 2376 | $oH_2O$ | $SH^+$ | $oH_3O^+$ | S | 4.2e-10 | 0.00 | 0.0 |
| 2377 | $oH_2O$ | $SH^+$ | $pH_3O^+$ | S | 2.1e-10 | 0.00 | 0.0 |
| 2378 | $pH_2O$ | $SH^+$ | $pH_3O^+$ | S | 6.3e-10 | 0.00 | 0.0 |
| 2379 | $oH_2O$ | $oH_2S^+$ | $oH_3O^+$ | SH | 5.4e-10 | 0.00 | 0.0 |
| 2380 | $oH_2O$ | $oH_2S^+$ | $pH_3O^+$ | SH | 2.7e-10 | 0.00 | 0.0 |
| 2381 | $oH_2O$ | $pH_2S^+$ | $oH_3O^+$ | SH | 2.7e-10 | 0.00 | 0.0 |
| 2382 | $oH_2O$ | $pH_2S^+$ | $pH_3O^+$ | SH | 5.4e-10 | 0.00 | 0.0 |
| 2383 | $pH_2O$ | $oH_2S^+$ | $oH_3O^+$ | SH | 2.7e-10 | 0.00 | 0.0 |
| 2384 | $pH_2O$ | $oH_2S^+$ | $pH_3O^+$ | SH | 5.4e-10 | 0.00 | 0.0 |
| 2385 | $pH_2O$ | $pH_2S^+$ | $pH_3O^+$ | SH | 8.1e-10 | 0.00 | 0.0 |
| 2386 | $oNH_3$ | $SH^+$ | $oNH_3^+$ | SH | 5.3e-10 | 0.00 | 0.0 |
| 2387 | $pNH_3$ | $SH^+$ | $pNH_3^+$ | SH | 5.3e-10 | 0.00 | 0.0 |
| 2388 | $oNH_3$ | $SH^+$ | $mNH_4^+$ | S | 6.1e-10 | 0.00 | 0.0 |
| 2389 | $oNH_3$ | $SH^+$ | $oNH_4^+$ | S | 3.7e-10 | 0.00 | 0.0 |
| 2390 | $pNH_3$ | $SH^+$ | $pNH_4^+$ | S | 2.4e-10 | 0.00 | 0.0 |
| 2391 | $pNH_3$ | $SH^+$ | $oNH_4^+$ | S | 7.3e-10 | 0.00 | 0.0 |
| 2392 | $oNH_3$ | $oH_2S^+$ | $mNH_4^+$ | SH | 7.9e-10 | 0.00 | 0.0 |
| 2393 | $oNH_3$ | $oH_2S^+$ | $pNH_4^+$ | SH | 9.1e-11 | 0.00 | 0.0 |
| 2394 | $oNH_3$ | $oH_2S^+$ | $oNH_4^+$ | SH | 4.8e-10 | 0.00 | 0.0 |
| 2395 | $oNH_3$ | $pH_2S^+$ | $mNH_4^+$ | SH | 3.4e-10 | 0.00 | 0.0 |
| 2396 | $oNH_3$ | $pH_2S^+$ | $oNH_4^+$ | SH | 1.0e-09 | 0.00 | 0.0 |
| 2397 | $pNH_3$ | $oH_2S^+$ | $mNH_4^+$ | SH | 2.3e-10 | 0.00 | 0.0 |
| 2398 | $pNH_3$ | $oH_2S^+$ | $pNH_4^+$ | SH | 1.8e-10 | 0.00 | 0.0 |
| 2399 | $pNH_3$ | $oH_2S^+$ | $oNH_4^+$ | SH | 9.5e-10 | 0.00 | 0.0 |
| 2400 | $pNH_3$ | $pH_2S^+$ | $pNH_4^+$ | SH | 5.4e-10 | 0.00 | 0.0 |
| 2401 | $pNH_3$ | $pH_2S^+$ | $oNH_4^+$ | SH | 8.2e-10 | 0.00 | 0.0 |
| 2402 | $oNH_3$ | $oH_2S^+$ | $oNH_3^+$ | $oH_2S$ | 3.4e-10 | 0.00 | 0.0 |
| 2403 | $oNH_3$ | $pH_2S^+$ | $oNH_3^+$ | $pH_2S$ | 3.4e-10 | 0.00 | 0.0 |
| 2404 | $pNH_3$ | $oH_2S^+$ | $pNH_3^+$ | $oH_2S$ | 3.4e-10 | 0.00 | 0.0 |
| 2405 | $pNH_3$ | $pH_2S^+$ | $pNH_3^+$ | $pH_2S$ | 3.4e-10 | 0.00 | 0.0 |
| 2406 | $oNH_3$ | $oH_3S^+$ | $mNH_4^+$ | $oH_2S$ | 9.9e-10 | 0.00 | 0.0 |
| 2407 | $oNH_3$ | $oH_3S^+$ | $mNH_4^+$ | $pH_2S$ | 1.2e-10 | 0.00 | 0.0 |
| 2408 | $oNH_3$ | $oH_3S^+$ | $pNH_4^+$ | $oH_2S$ | 7.9e-11 | 0.00 | 0.0 |
| 2409 | $oNH_3$ | $oH_3S^+$ | $pNH_4^+$ | $pH_2S$ | 4.7e-11 | 0.00 | 0.0 |
| 2410 | $oNH_3$ | $oH_3S^+$ | $oNH_4^+$ | $oH_2S$ | 5.5e-10 | 0.00 | 0.0 |
| 2411 | $oNH_3$ | $oH_3S^+$ | $oNH_4^+$ | $pH_2S$ | 1.2e-10 | 0.00 | 0.0 |
| 2412 | $oNH_3$ | $pH_3S^+$ | $mNH_4^+$ | $oH_2S$ | 3.2e-10 | 0.00 | 0.0 |
| 2413 | $oNH_3$ | $pH_3S^+$ | $mNH_4^+$ | $pH_2S$ | 2.4e-10 | 0.00 | 0.0 |
| 2414 | $oNH_3$ | $pH_3S^+$ | $pNH_4^+$ | $oH_2S$ | 1.6e-10 | 0.00 | 0.0 |
| 2415 | $oNH_3$ | $pH_3S^+$ | $oNH_4^+$ | $oH_2S$ | 9.5e-10 | 0.00 | 0.0 |
| 2416 | $oNH_3$ | $pH_3S^+$ | $oNH_4^+$ | $pH_2S$ | 2.4e-10 | 0.00 | 0.0 |
| 2417 | $pNH_3$ | $oH_3S^+$ | $mNH_4^+$ | $oH_2S$ | 3.2e-10 | 0.00 | 0.0 |
| 2418 | $pNH_3$ | $oH_3S^+$ | $mNH_4^+$ | $pH_2S$ | 2.4e-10 | 0.00 | 0.0 |
| 2419 | $pNH_3$ | $oH_3S^+$ | $pNH_4^+$ | $oH_2S$ | 1.6e-10 | 0.00 | 0.0 |
| 2420 | $pNH_3$ | $oH_3S^+$ | $oNH_4^+$ | $oH_2S$ | 9.5e-10 | 0.00 | 0.0 |
| 2421 | $pNH_3$ | $oH_3S^+$ | $oNH_4^+$ | $pH_2S$ | 2.4e-10 | 0.00 | 0.0 |
| 2422 | $pNH_3$ | $pH_3S^+$ | $mNH_4^+$ | $oH_2S$ | 1.6e-10 | 0.00 | 0.0 |
| 2423 | $pNH_3$ | $pH_3S^+$ | $pNH_4^+$ | $oH_2S$ | 3.2e-10 | 0.00 | 0.0 |
| 2424 | $pNH_3$ | $pH_3S^+$ | $pNH_4^+$ | $pH_2S$ | 1.9e-10 | 0.00 | 0.0 |
| 2425 | $pNH_3$ | $pH_3S^+$ | $oNH_4^+$ | $oH_2S$ | 7.6e-10 | 0.00 | 0.0 |
| 2426 | $pNH_3$ | $pH_3S^+$ | $oNH_4^+$ | $pH_2S$ | 4.8e-10 | 0.00 | 0.0 |
| 2427 | $oNH_3$ | $SO^+$ | $oNH_3^+$ | SO | 1.3e-09 | 0.00 | 0.0 |
| 2428 | $pNH_3$ | $SO^+$ | $pNH_3^+$ | SO | 1.3e-09 | 0.00 | 0.0 |
| 2429 | $oNH_3$ | $O_2^+$ | $oNH_3^+$ | $O_2$ | 2.0e-09 | 0.00 | 0.0 |
| 2430 | $pNH_3$ | $O_2^+$ | $pNH_3^+$ | $O_2$ | 2.0e-09 | 0.00 | 0.0 |
| 2431 | CO | OH | $CO_2$ | H | 4.4e-13 | -1.15 | 390.0 |
| 2432 | CO | $oH_2^+$ | $HCO^+$ | H | 2.2e-09 | 0.00 | 0.0 |
| 2433 | CO | $pH_2^+$ | $HCO^+$ | H | 2.2e-09 | 0.00 | 0.0 |
| 2434 | CO | $oH_2^+$ | $oH_2$ | $CO^+$ | 6.4e-10 | 0.00 | 0.0 |
| 2435 | CO | $pH_2^+$ | $pH_2$ | $CO^+$ | 6.4e-10 | 0.00 | 0.0 |
| 2436 | CO | $oH_3^+$ | $oH_2$ | $HCO^+$ | 1.7e-09 | 0.00 | 0.0 |
| 2437 | CO | $pH_3^+$ | $oH_2$ | $HCO^+$ | 8.5e-10 | 0.00 | 0.0 |
| 2438 | CO | $pH_3^+$ | $pH_2$ | $HCO^+$ | 8.5e-10 | 0.00 | 0.0 |
| 2439 | CO | $pCH_5^+$ | $mCH_4$ | $HCO^+$ | 9.9e-10 | 0.00 | 0.0 |
| 2440 | CO | $oCH_5^+$ | $mCH_4$ | $HCO^+$ | 2.5e-10 | 0.00 | 0.0 |
| 2441 | CO | $oCH_5^+$ | $oCH_4$ | $HCO^+$ | 7.4e-10 | 0.00 | 0.0 |
| 2442 | CO | $mCH_5^+$ | $pCH_4$ | $HCO^+$ | 4.0e-10 | 0.00 | 0.0 |

**Table D6** – *continued* (part 34)

| # | Reactants | | Products | | $\alpha$ | $\beta$ | $\gamma$ |
|---|---|---|---|---|---|---|---|
| 2443 | CO | $mCH_5^+$ | $oCH_4$ | $HCO^+$ | 5.9e-10 | 0.00 | 0.0 |
| 2444 | CO | $oH_2D^+$ | $HCO^+$ | HD | 1.1e-09 | 0.00 | 0.0 |
| 2445 | CO | $pH_2D^+$ | $HCO^+$ | HD | 1.1e-09 | 0.00 | 0.0 |
| 2446 | CO | $oH_2D^+$ | $oH_2$ | $DCO^+$ | 5.7e-10 | 0.00 | 0.0 |
| 2447 | CO | $pH_2D^+$ | $pH_2$ | $DCO^+$ | 5.7e-10 | 0.00 | 0.0 |
| 2448 | CO | $oHD_2^+$ | $oD_2$ | $HCO^+$ | 5.7e-10 | 0.00 | 0.0 |
| 2449 | CO | $pHD_2^+$ | $pD_2$ | $HCO^+$ | 5.7e-10 | 0.00 | 0.0 |
| 2450 | CO | $oHD_2^+$ | $DCO^+$ | HD | 1.1e-09 | 0.00 | 0.0 |
| 2451 | CO | $pHD_2^+$ | $DCO^+$ | HD | 1.1e-09 | 0.00 | 0.0 |
| 2452 | CO | $mD_3^+$ | $oD_2$ | $DCO^+$ | 1.7e-09 | 0.00 | 0.0 |
| 2453 | CO | $pD_3^+$ | $pD_2$ | $DCO^+$ | 1.7e-09 | 0.00 | 0.0 |
| 2454 | CO | $oD_3^+$ | $oD_2$ | $DCO^+$ | 8.5e-10 | 0.00 | 0.0 |
| 2455 | CO | $oD_3^+$ | $pD_2$ | $DCO^+$ | 8.5e-10 | 0.00 | 0.0 |
| 2456 | SO | OH | $SO_2$ | H | 2.0e-10 | -0.17 | 0.0 |
| 2457 | SO | $H^+$ | $SO^+$ | H | 3.2e-09 | 0.00 | 0.0 |
| 2458 | SO | $oH_3^+$ | $oH_2$ | $HSO^+$ | 1.9e-09 | 0.00 | 0.0 |
| 2459 | SO | $pH_3^+$ | $oH_2$ | $HSO^+$ | 9.5e-10 | 0.00 | 0.0 |
| 2460 | SO | $pH_3^+$ | $pH_2$ | $HSO^+$ | 9.5e-10 | 0.00 | 0.0 |
| 2461 | SO | $CH^+$ | $OH^+$ | CS | 1.0e-09 | 0.00 | 0.0 |
| 2462 | SO | $CH^+$ | $SH^+$ | CO | 1.0e-09 | 0.00 | 0.0 |
| 2463 | SO | $oCH_3^+$ | $oH_2$ | $HOCS^+$ | 9.5e-10 | 0.00 | 0.0 |
| 2464 | SO | $pCH_3^+$ | $oH_2$ | $HOCS^+$ | 4.8e-10 | 0.00 | 0.0 |
| 2465 | SO | $pCH_3^+$ | $pH_2$ | $HOCS^+$ | 4.8e-10 | 0.00 | 0.0 |
| 2466 | SO | $HCO^+$ | $HSO^+$ | CO | 7.5e-10 | 0.00 | 0.0 |
| 2467 | NO | CH | HCN | O | 1.2e-11 | -0.13 | 0.0 |
| 2468 | NO | $H^+$ | $NO^+$ | H | 1.9e-09 | 0.00 | 0.0 |
| 2469 | NO | $oH_3^+$ | $oH_2$ | $HNO^+$ | 1.1e-09 | 0.00 | 0.0 |
| 2470 | NO | $pH_3^+$ | $oH_2$ | $HNO^+$ | 5.5e-10 | 0.00 | 0.0 |
| 2471 | NO | $pH_3^+$ | $pH_2$ | $HNO^+$ | 5.5e-10 | 0.00 | 0.0 |
| 2472 | NO | $HCO_2^+$ | $HNO^+$ | $CO_2$ | 1.0e-10 | 0.00 | 0.0 |
| 2473 | NO | $O_2^+$ | $NO^+$ | $O_2$ | 4.4e-10 | 0.00 | 0.0 |
| 2474 | NO | $SH^+$ | $NO^+$ | SH | 3.3e-10 | 0.00 | 0.0 |
| 2475 | NO | $oH_2S^+$ | $oH_2S$ | $NO^+$ | 3.7e-10 | 0.00 | 0.0 |
| 2476 | NO | $pH_2S^+$ | $pH_2S$ | $NO^+$ | 3.7e-10 | 0.00 | 0.0 |
| 2477 | NO | $oH_2D^+$ | $HNO^+$ | HD | 7.3e-10 | 0.00 | 0.0 |
| 2478 | NO | $pH_2D^+$ | $HNO^+$ | HD | 7.3e-10 | 0.00 | 0.0 |
| 2479 | NO | $oHD_2^+$ | $oD_2$ | $HNO^+$ | 3.7e-10 | 0.00 | 0.0 |
| 2480 | NO | $pHD_2^+$ | $pD_2$ | $HNO^+$ | 3.7e-10 | 0.00 | 0.0 |
| 2481 | CN | $oNH_3$ | $oNH_2$ | HCN | 2.8e-11 | -0.85 | 0.0 |
| 2482 | CN | $pNH_3$ | $oNH_2$ | HCN | 1.4e-11 | -0.85 | 0.0 |
| 2483 | CN | $pNH_3$ | $pNH_2$ | HCN | 1.4e-11 | -0.85 | 0.0 |
| 2484 | CN | $oH_3^+$ | $oH_2$ | $HCN^+$ | 1.0e-09 | 0.00 | 0.0 |
| 2485 | CN | $pH_3^+$ | $oH_2$ | $HCN^+$ | 5.0e-10 | 0.00 | 0.0 |
| 2486 | CN | $pH_3^+$ | $pH_2$ | $HCN^+$ | 5.0e-10 | 0.00 | 0.0 |
| 2487 | CN | $oH_3^+$ | $HCNH^+$ | H | 1.0e-09 | 0.00 | 0.0 |
| 2488 | CN | $pH_3^+$ | $HCNH^+$ | H | 1.0e-09 | 0.00 | 0.0 |
| 2489 | CN | $oH_3O^+$ | $HCNH^+$ | OH | 4.5e-09 | 0.00 | 0.0 |
| 2490 | CN | $pH_3O^+$ | $HCNH^+$ | OH | 4.5e-09 | 0.00 | 0.0 |
| 2491 | CN | $oH_2D^+$ | $HCNH^+$ | D | 1.0e-09 | 0.00 | 0.0 |
| 2492 | CN | $pH_2D^+$ | $HCNH^+$ | D | 1.0e-09 | 0.00 | 0.0 |
| 2493 | CN | $oH_2D^+$ | $HCN^+$ | HD | 1.0e-09 | 0.00 | 0.0 |
| 2494 | CN | $pH_2D^+$ | $HCN^+$ | HD | 1.0e-09 | 0.00 | 0.0 |
| 2495 | CS | OH | OCS | H | 1.7e-10 | 0.00 | 0.0 |
| 2496 | CS | OH | CO | SH | 3.0e-11 | 0.00 | 0.0 |
| 2497 | CS | $H^+$ | $CS^+$ | H | 4.9e-09 | 0.00 | 0.0 |
| 2498 | CS | $oH_3^+$ | $oH_2$ | $HCS^+$ | 2.9e-09 | 0.00 | 0.0 |
| 2499 | CS | $pH_3^+$ | $oH_2$ | $HCS^+$ | 1.4e-09 | 0.00 | 0.0 |
| 2500 | CS | $pH_3^+$ | $pH_2$ | $HCS^+$ | 1.4e-09 | 0.00 | 0.0 |
| 2501 | CS | $oH_2D^+$ | $HCS^+$ | HD | 2.9e-09 | 0.00 | 0.0 |
| 2502 | CS | $pH_2D^+$ | $HCS^+$ | HD | 2.9e-09 | 0.00 | 0.0 |
| 2503 | CS | $oHD_2^+$ | $oD_2$ | $HCS^+$ | 2.9e-09 | 0.00 | 0.0 |
| 2504 | CS | $pHD_2^+$ | $pD_2$ | $HCS^+$ | 2.9e-09 | 0.00 | 0.0 |
| 2505 | SO | $oH_2D^+$ | $HSO^+$ | HD | 1.9e-09 | 0.00 | 0.0 |
| 2506 | SO | $pH_2D^+$ | $HSO^+$ | HD | 1.9e-09 | 0.00 | 0.0 |
| 2507 | SO | $oHD_2^+$ | $oD_2$ | $HSO^+$ | 1.9e-09 | 0.00 | 0.0 |
| 2508 | SO | $pHD_2^+$ | $pD_2$ | $HSO^+$ | 1.9e-09 | 0.00 | 0.0 |
| 2509 | HCN | $H^+$ | $HCN^+$ | H | 1.1e-08 | 0.00 | 0.0 |
| 2510 | HCN | $oH_3^+$ | $oH_2$ | $HCNH^+$ | 9.5e-09 | 0.00 | 0.0 |
| 2511 | HCN | $pH_3^+$ | $oH_2$ | $HCNH^+$ | 4.8e-09 | 0.00 | 0.0 |
| 2512 | HCN | $pH_3^+$ | $pH_2$ | $HCNH^+$ | 4.8e-09 | 0.00 | 0.0 |
| 2513 | HCN | $oH_3O^+$ | $oH_2O$ | $HCNH^+$ | 4.5e-09 | 0.00 | 0.0 |
| 2514 | HCN | $pH_3O^+$ | $oH_2O$ | $HCNH^+$ | 2.2e-09 | 0.00 | 0.0 |
| 2515 | HCN | $pH_3O^+$ | $pH_2O$ | $HCNH^+$ | 2.2e-09 | 0.00 | 0.0 |
| 2516 | HCN | $oH_3S^+$ | $oH_2S$ | $HCNH^+$ | 1.9e-09 | 0.00 | 0.0 |

**Table D6** – *continued* (part 35)

| # | Reactants | | Products | | $\alpha$ | $\beta$ | $\gamma$ |
|---|---|---|---|---|---|---|---|
| 2517 | HCN | $pH_3S^+$ | $oH_2S$ | $HCNH^+$ | 9.5e-10 | 0.00 | 0.0 |
| 2518 | HCN | $pH_3S^+$ | $pH_2S$ | $HCNH^+$ | 9.5e-10 | 0.00 | 0.0 |
| 2519 | HCN | $HCO^+$ | $HCNH^+$ | CO | 3.7e-09 | 0.00 | 0.0 |
| 2520 | HCN | $oH_2D^+$ | $HCNH^+$ | HD | 9.5e-09 | 0.00 | 0.0 |
| 2521 | HCN | $pH_2D^+$ | $HCNH^+$ | HD | 9.5e-09 | 0.00 | 0.0 |
| 2522 | HNC | H | HCN | H | 1.0e-15 | 0.00 | 0.0 |
| 2523 | HNC | $H^+$ | $H^+$ | HCN | 1.0e-09 | 0.00 | 0.0 |
| 2524 | HNC | $oH_3^+$ | $oH_2$ | $HCNH^+$ | 9.5e-09 | 0.00 | 0.0 |
| 2525 | HNC | $pH_3^+$ | $oH_2$ | $HCNH^+$ | 4.8e-09 | 0.00 | 0.0 |
| 2526 | HNC | $pH_3^+$ | $pH_2$ | $HCNH^+$ | 4.8e-09 | 0.00 | 0.0 |
| 2527 | HNC | $oH_2D^+$ | $HCNH^+$ | HD | 9.5e-09 | 0.00 | 0.0 |
| 2528 | HNC | $pH_2D^+$ | $HCNH^+$ | HD | 9.5e-09 | 0.00 | 0.0 |
| 2529 | HNC | $oH_3O^+$ | $oH_2O$ | $HCNH^+$ | 4.5e-09 | 0.00 | 0.0 |
| 2530 | HNC | $pH_3O^+$ | $oH_2O$ | $HCNH^+$ | 2.2e-09 | 0.00 | 0.0 |
| 2531 | HNC | $pH_3O^+$ | $pH_2O$ | $HCNH^+$ | 2.2e-09 | 0.00 | 0.0 |
| 2532 | HNC | $HCO^+$ | $HCNH^+$ | CO | 3.7e-09 | 0.00 | 0.0 |
| 2533 | $CO_2$ | $H^+$ | $HCO^+$ | O | 4.2e-09 | 0.00 | 0.0 |
| 2534 | $CO_2$ | $oH_3^+$ | $oH_2$ | $HCO_2^+$ | 2.0e-09 | 0.00 | 0.0 |
| 2535 | $CO_2$ | $pH_3^+$ | $oH_2$ | $HCO_2^+$ | 1.0e-09 | 0.00 | 0.0 |
| 2536 | $CO_2$ | $pH_3^+$ | $pH_2$ | $HCO_2^+$ | 1.0e-09 | 0.00 | 0.0 |
| 2537 | $CO_2$ | $oH_2D^+$ | $HCO_2^+$ | HD | 1.3e-09 | 0.00 | 0.0 |
| 2538 | $CO_2$ | $pH_2D^+$ | $HCO_2^+$ | HD | 1.3e-09 | 0.00 | 0.0 |
| 2539 | $CO_2$ | $oH_2D^+$ | $oH_2$ | $DCO_2^+$ | 6.7e-10 | 0.00 | 0.0 |
| 2540 | $CO_2$ | $pH_2D^+$ | $pH_2$ | $DCO_2^+$ | 6.7e-10 | 0.00 | 0.0 |
| 2541 | $CO_2$ | $oHD_2^+$ | $oD_2$ | $HCO_2^+$ | 6.7e-10 | 0.00 | 0.0 |
| 2542 | $CO_2$ | $pHD_2^+$ | $pD_2$ | $HCO_2^+$ | 6.7e-10 | 0.00 | 0.0 |
| 2543 | $CO_2$ | $oHD_2^+$ | $DCO_2^+$ | HD | 1.3e-09 | 0.00 | 0.0 |
| 2544 | $CO_2$ | $pHD_2^+$ | $DCO_2^+$ | HD | 1.3e-09 | 0.00 | 0.0 |
| 2545 | $CO_2$ | $mD_3^+$ | $oD_2$ | $DCO_2^+$ | 2.0e-09 | 0.00 | 0.0 |
| 2546 | $CO_2$ | $pD_3^+$ | $pD_2$ | $DCO_2^+$ | 2.0e-09 | 0.00 | 0.0 |
| 2547 | $CO_2$ | $oD_3^+$ | $oD_2$ | $DCO_2^+$ | 1.0e-09 | 0.00 | 0.0 |
| 2548 | $CO_2$ | $oD_3^+$ | $pD_2$ | $DCO_2^+$ | 1.0e-09 | 0.00 | 0.0 |
| 2549 | $CO_2$ | $N_2D^+$ | $DCO_2^+$ | $N_2$ | 1.4e-09 | 0.00 | 0.0 |
| 2550 | $SO_2$ | $oH_3^+$ | $oH_2$ | $HSO_2^+$ | 1.3e-09 | 0.00 | 0.0 |
| 2551 | $SO_2$ | $pH_3^+$ | $oH_2$ | $HSO_2^+$ | 6.5e-10 | 0.00 | 0.0 |
| 2552 | $SO_2$ | $pH_3^+$ | $pH_2$ | $HSO_2^+$ | 6.5e-10 | 0.00 | 0.0 |
| 2553 | $SO_2$ | $oH_2D^+$ | $HSO_2^+$ | HD | 1.3e-09 | 0.00 | 0.0 |
| 2554 | $SO_2$ | $pH_2D^+$ | $HSO_2^+$ | HD | 1.3e-09 | 0.00 | 0.0 |
| 2555 | $SO_2$ | $oHD_2^+$ | $oD_2$ | $HSO_2^+$ | 1.3e-09 | 0.00 | 0.0 |
| 2556 | $SO_2$ | $pHD_2^+$ | $pD_2$ | $HSO_2^+$ | 1.3e-09 | 0.00 | 0.0 |
| 2557 | OCS | $H^+$ | $SH^+$ | CO | 5.9e-09 | 0.00 | 0.0 |
| 2558 | OCS | $oH_3^+$ | $oH_2$ | $HOCS^+$ | 1.9e-09 | 0.00 | 0.0 |
| 2559 | OCS | $pH_3^+$ | $oH_2$ | $HOCS^+$ | 9.5e-10 | 0.00 | 0.0 |
| 2560 | OCS | $pH_3^+$ | $pH_2$ | $HOCS^+$ | 9.5e-10 | 0.00 | 0.0 |
| 2561 | OCS | $HCO^+$ | $HOCS^+$ | CO | 1.1e-09 | 0.00 | 0.0 |
| 2562 | $oH_2S$ | $oH_3O^+$ | $oH_3S^+$ | $oH_2O$ | 1.2e-09 | 0.00 | 0.0 |
| 2563 | $oH_2S$ | $oH_3O^+$ | $oH_3S^+$ | $pH_2O$ | 1.6e-10 | 0.00 | 0.0 |
| 2564 | $oH_2S$ | $oH_3O^+$ | $pH_3S^+$ | $oH_2O$ | 4.4e-10 | 0.00 | 0.0 |
| 2565 | $oH_2S$ | $oH_3O^+$ | $pH_3S^+$ | $pH_2O$ | 1.3e-10 | 0.00 | 0.0 |
| 2566 | $pH_2S$ | $oH_3O^+$ | $oH_3S^+$ | $oH_2O$ | 4.8e-10 | 0.00 | 0.0 |
| 2567 | $pH_2S$ | $oH_3O^+$ | $oH_3S^+$ | $pH_2O$ | 4.8e-10 | 0.00 | 0.0 |
| 2568 | $pH_2S$ | $oH_3O^+$ | $pH_3S^+$ | $oH_2O$ | 9.5e-10 | 0.00 | 0.0 |
| 2569 | $oH_2S$ | $pH_3O^+$ | $oH_3S^+$ | $oH_2O$ | 4.4e-10 | 0.00 | 0.0 |
| 2570 | $oH_2S$ | $pH_3O^+$ | $oH_3S^+$ | $pH_2O$ | 3.2e-10 | 0.00 | 0.0 |
| 2571 | $oH_2S$ | $pH_3O^+$ | $pH_3S^+$ | $oH_2O$ | 8.9e-10 | 0.00 | 0.0 |
| 2572 | $oH_2S$ | $pH_3O^+$ | $pH_3S^+$ | $pH_2O$ | 2.5e-10 | 0.00 | 0.0 |
| 2573 | $pH_2S$ | $pH_3O^+$ | $oH_3S^+$ | $oH_2O$ | 3.8e-10 | 0.00 | 0.0 |
| 2574 | $pH_2S$ | $pH_3O^+$ | $pH_3S^+$ | $oH_2O$ | 7.6e-10 | 0.00 | 0.0 |
| 2575 | $pH_2S$ | $pH_3O^+$ | $pH_3S^+$ | $pH_2O$ | 7.6e-10 | 0.00 | 0.0 |
| 2576 | SH | $HCO^+$ | $oH_2S^+$ | CO | 6.2e-10 | 0.00 | 0.0 |
| 2577 | SH | $HCO^+$ | $pH_2S^+$ | CO | 2.0e-10 | 0.00 | 0.0 |
| 2578 | CS | $HCO^+$ | $HCS^+$ | CO | 1.2e-09 | 0.00 | 0.0 |
| 2579 | $oH_2S$ | $HCO^+$ | $oH_3S^+$ | CO | 1.1e-09 | 0.00 | 0.0 |
| 2580 | $oH_2S$ | $HCO^+$ | $pH_3S^+$ | CO | 5.3e-10 | 0.00 | 0.0 |
| 2581 | $pH_2S$ | $HCO^+$ | $pH_3S^+$ | CO | 1.6e-09 | 0.00 | 0.0 |
| 2582 | $oH_2S$ | $O_2^+$ | $oH_2S^+$ | $O_2$ | 1.4e-09 | 0.00 | 0.0 |
| 2583 | $pH_2S$ | $O_2^+$ | $pH_2S^+$ | $O_2$ | 1.4e-09 | 0.00 | 0.0 |
| 2584 | $oH_2S$ | $oNH_3^+$ | $mNH_4^+$ | SH | 3.5e-10 | 0.00 | 0.0 |
| 2585 | $oH_2S$ | $oNH_3^+$ | $pNH_4^+$ | SH | 4.0e-11 | 0.00 | 0.0 |
| 2586 | $oH_2S$ | $oNH_3^+$ | $oNH_4^+$ | SH | 2.1e-10 | 0.00 | 0.0 |
| 2587 | $pH_2S$ | $oNH_3^+$ | $mNH_4^+$ | SH | 1.5e-10 | 0.00 | 0.0 |
| 2588 | $pH_2S$ | $oNH_3^+$ | $oNH_4^+$ | SH | 4.5e-10 | 0.00 | 0.0 |
| 2589 | $oH_2S$ | $pNH_3^+$ | $mNH_4^+$ | SH | 1.0e-10 | 0.00 | 0.0 |
| 2590 | $oH_2S$ | $pNH_3^+$ | $pNH_4^+$ | SH | 8.0e-11 | 0.00 | 0.0 |

**Table D6** – *continued* (part 36)

| # | Reactants | | Products | | $\alpha$ | $\beta$ | $\gamma$ |
|---|---|---|---|---|---|---|---|
| 2591 | $oH_2S$ | $pNH_3^+$ | $oNH_4^+$ | SH | 4.2e-10 | 0.00 | 0.0 |
| 2592 | $pH_2S$ | $pNH_3^+$ | $pNH_4^+$ | SH | 2.4e-10 | 0.00 | 0.0 |
| 2593 | $pH_2S$ | $pNH_3^+$ | $oNH_4^+$ | SH | 3.6e-10 | 0.00 | 0.0 |
| 2594 | OCS | $oH_2D^+$ | $HOCS^+$ | HD | 1.9e-09 | 0.00 | 0.0 |
| 2595 | OCS | $pH_2D^+$ | $HOCS^+$ | HD | 1.9e-09 | 0.00 | 0.0 |
| 2596 | OCS | $oHD_2^+$ | $oD_2$ | $HOCS^+$ | 1.9e-09 | 0.00 | 0.0 |
| 2597 | OCS | $pHD_2^+$ | $pD_2$ | $HOCS^+$ | 1.9e-09 | 0.00 | 0.0 |
| 2598 | S | $N_2D^+$ | $SD^+$ | $N_2$ | 1.1e-09 | 0.00 | 0.0 |
| 2599 | $oH_2O$ | $N_2D^+$ | $oH_2DO^+$ | $N_2$ | 2.6e-09 | 0.00 | 0.0 |
| 2600 | $pH_2O$ | $N_2D^+$ | $pH_2DO^+$ | $N_2$ | 2.6e-09 | 0.00 | 0.0 |
| 2601 | CO | $N_2D^+$ | $DCO^+$ | $N_2$ | 8.8e-10 | 0.00 | 0.0 |
| 2602 | $C^+$ | S | $S^+$ | C | 5.5e-12 | 0.86 | 681.0 |
| 2603 | $C^+$ | OH | $CO^+$ | H | 8.0e-10 | 0.00 | 0.0 |
| 2604 | $C^+$ | OH | $H^+$ | CO | 8.0e-10 | 0.00 | 0.0 |
| 2605 | $C^+$ | $oH_2O$ | $HCO^+$ | H | 2.4e-09 | -0.63 | 0.0 |
| 2606 | $C^+$ | $pH_2O$ | $HCO^+$ | H | 2.4e-09 | -0.63 | 0.0 |
| 2607 | $C^+$ | $O_2$ | $O^+$ | CO | 5.1e-10 | 0.00 | 0.0 |
| 2608 | $C^+$ | $O_2$ | $CO^+$ | O | 3.1e-10 | 0.00 | 0.0 |
| 2609 | $C^+$ | $CO_2$ | $CO^+$ | CO | 1.1e-09 | 0.00 | 0.0 |
| 2610 | $C^+$ | NH | $CN^+$ | H | 7.8e-10 | 0.00 | 0.0 |
| 2611 | $C^+$ | $oNH_2$ | $HCN^+$ | H | 1.1e-09 | 0.00 | 0.0 |
| 2612 | $C^+$ | $pNH_2$ | $HCN^+$ | H | 1.1e-09 | 0.00 | 0.0 |
| 2613 | $C^+$ | $oNH_3$ | $oNH_3^+$ | C | 5.3e-10 | 0.00 | 0.0 |
| 2614 | $C^+$ | $pNH_3$ | $pNH_3^+$ | C | 5.3e-10 | 0.00 | 0.0 |
| 2615 | $C^+$ | $oNH_3$ | $H_2NC^+$ | H | 7.8e-10 | 0.00 | 0.0 |
| 2616 | $C^+$ | $pNH_3$ | $H_2NC^+$ | H | 7.8e-10 | 0.00 | 0.0 |
| 2617 | $C^+$ | $oNH_3$ | $HCNH^+$ | H | 7.8e-10 | 0.00 | 0.0 |
| 2618 | $C^+$ | $pNH_3$ | $HCNH^+$ | H | 7.8e-10 | 0.00 | 0.0 |
| 2619 | $C^+$ | $oNH_3$ | $oH_2$ | $HCN^+$ | 2.1e-10 | 0.00 | 0.0 |
| 2620 | $C^+$ | $pNH_3$ | $oH_2$ | $HCN^+$ | 1.0e-10 | 0.00 | 0.0 |
| 2621 | $C^+$ | $pNH_3$ | $pH_2$ | $HCN^+$ | 1.0e-10 | 0.00 | 0.0 |
| 2622 | $C^+$ | HCN | $C_2N^+$ | H | 3.4e-09 | 0.00 | 0.0 |
| 2623 | $C^+$ | HNC | $C_2N^+$ | H | 3.4e-09 | 0.00 | 0.0 |
| 2624 | $C^+$ | NO | $NO^+$ | C | 3.4e-09 | 0.00 | 0.0 |
| 2625 | $C^+$ | NO | $N^+$ | CO | 9.0e-10 | 0.00 | 0.0 |
| 2626 | $C^+$ | SH | $CS^+$ | H | 1.1e-09 | 0.00 | 0.0 |
| 2627 | $C^+$ | $oH_2S$ | $HCS^+$ | H | 1.3e-09 | 0.00 | 0.0 |
| 2628 | $C^+$ | $pH_2S$ | $HCS^+$ | H | 1.3e-09 | 0.00 | 0.0 |
| 2629 | $C^+$ | $oH_2S$ | $oH_2S^+$ | C | 4.2e-10 | 0.00 | 0.0 |
| 2630 | $C^+$ | $pH_2S$ | $pH_2S^+$ | C | 4.2e-10 | 0.00 | 0.0 |
| 2631 | $C^+$ | SO | $S^+$ | CO | 2.6e-10 | 0.00 | 0.0 |
| 2632 | $C^+$ | SO | $CS^+$ | O | 2.6e-10 | 0.00 | 0.0 |
| 2633 | $C^+$ | SO | $SO^+$ | C | 2.6e-10 | 0.00 | 0.0 |
| 2634 | $C^+$ | SO | $CO^+$ | S | 2.6e-10 | 0.00 | 0.0 |
| 2635 | $C^+$ | $SO_2$ | $SO^+$ | CO | 2.3e-09 | 0.00 | 0.0 |
| 2636 | $C^+$ | CS | $CS^+$ | C | 1.6e-09 | 0.00 | 700.0 |
| 2637 | $C^+$ | OCS | $CS^+$ | CO | 1.6e-09 | 0.00 | 0.0 |
| 2638 | $C^+$ | OD | $CO^+$ | D | 8.0e-10 | 0.00 | 0.0 |
| 2639 | $C^+$ | OD | $D^+$ | CO | 8.0e-10 | 0.00 | 0.0 |
| 2640 | $C^+$ | HDO | $DCO^+$ | H | 1.2e-09 | -0.63 | 0.0 |
| 2641 | $C^+$ | HDO | $HCO^+$ | D | 1.2e-09 | -0.63 | 0.0 |
| 2642 | $C^+$ | $oD_2O$ | $DCO^+$ | D | 2.4e-09 | -0.63 | 0.0 |
| 2643 | $C^+$ | $pD_2O$ | $DCO^+$ | D | 2.4e-09 | -0.63 | 0.0 |
| 2644 | $N^+$ | $O_2$ | $O_2^+$ | N | 2.8e-10 | 0.00 | 0.0 |
| 2645 | $N^+$ | $O_2$ | $NO^+$ | O | 2.4e-10 | 0.00 | 0.0 |
| 2646 | $N^+$ | $O_2$ | $O^+$ | NO | 3.3e-11 | 0.00 | 0.0 |
| 2647 | $N^+$ | CO | $CO^+$ | N | 8.3e-10 | 0.00 | 0.0 |
| 2648 | $N^+$ | CO | $NO^+$ | C | 1.5e-10 | 0.00 | 0.0 |
| 2649 | $N^+$ | NO | $NO^+$ | N | 4.5e-10 | 0.00 | 0.0 |
| 2650 | $N^+$ | NO | $N_2^+$ | O | 7.9e-11 | 0.00 | 0.0 |
| 2651 | $S^+$ | CH | $CS^+$ | H | 6.2e-10 | 0.00 | 0.0 |
| 2652 | $S^+$ | $oCH_2$ | $HCS^+$ | H | 1.0e-11 | 0.00 | 0.0 |
| 2653 | $S^+$ | $pCH_2$ | $HCS^+$ | H | 1.0e-11 | 0.00 | 0.0 |
| 2654 | $S^+$ | OH | $SO^+$ | H | 6.1e-10 | 0.00 | 0.0 |
| 2655 | $S^+$ | NO | $NO^+$ | S | 3.2e-10 | 0.00 | 0.0 |
| 2656 | $S^+$ | $oNH_3$ | $oNH_3^+$ | S | 1.6e-09 | 0.00 | 0.0 |
| 2657 | $S^+$ | $pNH_3$ | $pNH_3^+$ | S | 1.6e-09 | 0.00 | 0.0 |
| 2658 | $S^+$ | $O_2$ | $SO^+$ | O | 2.3e-11 | 0.00 | 0.0 |
| 2659 | $S^+$ | $oNH_2D$ | $oNH_2D^+$ | S | 1.6e-09 | 0.00 | 0.0 |
| 2660 | $S^+$ | $pNH_2D$ | $pNH_2D^+$ | S | 1.6e-09 | 0.00 | 0.0 |
| 2661 | $S^+$ | $oNHD_2$ | $oNHD_2^+$ | S | 1.6e-09 | 0.00 | 0.0 |
| 2662 | $S^+$ | $pNHD_2$ | $pNHD_2^+$ | S | 1.6e-09 | 0.00 | 0.0 |
| 2663 | $S^+$ | $mND_3$ | $mND_3^+$ | S | 1.6e-09 | 0.00 | 0.0 |
| 2664 | $S^+$ | $pND_3$ | $pND_3^+$ | S | 1.6e-09 | 0.00 | 0.0 |

**Table D6** – *continued* (part 37)

| # | Reactants | | Products | | | $\alpha$ | $\beta$ | $\gamma$ |
|---|---|---|---|---|---|---|---|---|
| 2665 | S$^+$ | oND$_3$ | oND$_3^+$ | S | | 1.6e-09 | 0.00 | 0.0 |
| 2666 | CO$^+$ | oH$_2$ | HCO$^+$ | H | | 1.3e-09 | 0.00 | 0.0 |
| 2667 | CO$^+$ | pH$_2$ | HCO$^+$ | H | | 1.3e-09 | 0.00 | 0.0 |
| 2668 | CO$^+$ | H | H$^+$ | CO | | 7.5e-10 | 0.00 | 0.0 |
| 2669 | HCO$^+$ | CH | oCH$_2^+$ | CO | | 4.7e-10 | 0.00 | 0.0 |
| 2670 | HCO$^+$ | CH | pCH$_2^+$ | CO | | 1.6e-10 | 0.00 | 0.0 |
| 2671 | HCO$^+$ | oCH$_2$ | oCH$_3^+$ | CO | | 5.7e-10 | 0.00 | 0.0 |
| 2672 | HCO$^+$ | oCH$_2$ | pCH$_3^+$ | CO | | 2.9e-10 | 0.00 | 0.0 |
| 2673 | HCO$^+$ | pCH$_2$ | pCH$_3^+$ | CO | | 8.6e-10 | 0.00 | 0.0 |
| 2674 | HCO$^+$ | oH$_2$O | oH$_3$O$^+$ | CO | | 1.7e-09 | 0.00 | 0.0 |
| 2675 | HCO$^+$ | oH$_2$O | pH$_3$O$^+$ | CO | | 8.3e-10 | 0.00 | 0.0 |
| 2676 | HCO$^+$ | pH$_2$O | pH$_3$O$^+$ | CO | | 2.5e-09 | 0.00 | 0.0 |
| 2677 | HCO$^+$ | OH | HCO$_2^+$ | H | | 1.0e-09 | 0.00 | 0.0 |
| 2678 | HCO$^+$ | C$_2$H | C$_2$H$_2^+$ | CO | | 7.8e-10 | 0.00 | 0.0 |
| 2679 | HCO$^+$ | C$_2$H$_2$ | C$_2$H$_3^+$ | CO | | 1.4e-09 | 0.00 | 0.0 |
| 2680 | HCO$^+$ | C$_3$H | C$_3$H$_2^+$ | CO | | 1.4e-09 | 0.00 | 0.0 |
| 2681 | HCO$^+$ | C$_3$H$_2$ | C$_3$H$_3^+$ | CO | | 1.4e-09 | 0.00 | 0.0 |
| 2682 | HCO$^+$ | NH | oNH$_2^+$ | CO | | 4.8e-10 | 0.00 | 0.0 |
| 2683 | HCO$^+$ | NH | pNH$_2^+$ | CO | | 1.6e-10 | 0.00 | 0.0 |
| 2684 | HCO$^+$ | oNH$_2$ | oNH$_3^+$ | CO | | 5.9e-10 | 0.00 | 0.0 |
| 2685 | HCO$^+$ | oNH$_2$ | pNH$_3^+$ | CO | | 3.0e-10 | 0.00 | 0.0 |
| 2686 | HCO$^+$ | pNH$_2$ | pNH$_3^+$ | CO | | 8.9e-10 | 0.00 | 0.0 |
| 2687 | HCO$^+$ | oNH$_3$ | mNH$_4^+$ | CO | | 1.2e-09 | 0.00 | 0.0 |
| 2688 | HCO$^+$ | oNH$_3$ | oNH$_4^+$ | CO | | 7.1e-10 | 0.00 | 0.0 |
| 2689 | HCO$^+$ | pNH$_3$ | pNH$_4^+$ | CO | | 4.8e-10 | 0.00 | 0.0 |
| 2690 | HCO$^+$ | pNH$_3$ | oNH$_4^+$ | CO | | 1.4e-09 | 0.00 | 0.0 |
| 2691 | HCO$^+$ | Fe | Fe$^+$ | CO | H | 1.9e-09 | 0.00 | 0.0 |
| 2692 | HCO$_2^+$ | CO | HCO$^+$ | CO$_2$ | | 1.0e-09 | 0.00 | 0.0 |
| 2693 | HCO$_2^+$ | mCH$_4$ | pCH$_5^+$ | CO$_2$ | | 4.7e-10 | 0.00 | 0.0 |
| 2694 | HCO$_2^+$ | mCH$_4$ | oCH$_5^+$ | CO$_2$ | | 3.1e-10 | 0.00 | 0.0 |
| 2695 | HCO$_2^+$ | pCH$_4$ | mCH$_5^+$ | CO$_2$ | | 7.8e-10 | 0.00 | 0.0 |
| 2696 | HCO$_2^+$ | oCH$_4$ | oCH$_5^+$ | CO$_2$ | | 5.2e-10 | 0.00 | 0.0 |
| 2697 | HCO$_2^+$ | oCH$_4$ | mCH$_5^+$ | CO$_2$ | | 2.6e-10 | 0.00 | 0.0 |
| 2698 | oH$_3$O$^+$ | CH | oCH$_2^+$ | oH$_2$O | | 5.1e-10 | 0.00 | 0.0 |
| 2699 | oH$_3$O$^+$ | CH | oCH$_2^+$ | pH$_2$O | | 8.5e-11 | 0.00 | 0.0 |
| 2700 | oH$_3$O$^+$ | CH | pCH$_2^+$ | oH$_2$O | | 8.5e-11 | 0.00 | 0.0 |
| 2701 | pH$_3$O$^+$ | CH | oCH$_2^+$ | oH$_2$O | | 2.5e-10 | 0.00 | 0.0 |
| 2702 | pH$_3$O$^+$ | CH | oCH$_2^+$ | pH$_2$O | | 1.7e-10 | 0.00 | 0.0 |
| 2703 | pH$_3$O$^+$ | CH | pCH$_2^+$ | oH$_2$O | | 1.7e-10 | 0.00 | 0.0 |
| 2704 | pH$_3$O$^+$ | CH | pCH$_2^+$ | pH$_2$O | | 8.5e-11 | 0.00 | 0.0 |
| 2705 | oH$_3$O$^+$ | oCH$_2$ | oCH$_3^+$ | oH$_2$O | | 5.8e-10 | 0.00 | 0.0 |
| 2706 | oH$_3$O$^+$ | oCH$_2$ | oCH$_3^+$ | pH$_2$O | | 7.8e-11 | 0.00 | 0.0 |
| 2707 | oH$_3$O$^+$ | oCH$_2$ | pCH$_3^+$ | oH$_2$O | | 2.2e-10 | 0.00 | 0.0 |
| 2708 | oH$_3$O$^+$ | oCH$_2$ | pCH$_3^+$ | pH$_2$O | | 6.3e-11 | 0.00 | 0.0 |
| 2709 | oH$_3$O$^+$ | pCH$_2$ | oCH$_3^+$ | oH$_2$O | | 2.4e-10 | 0.00 | 0.0 |
| 2710 | oH$_3$O$^+$ | pCH$_2$ | oCH$_3^+$ | pH$_2$O | | 2.4e-10 | 0.00 | 0.0 |
| 2711 | oH$_3$O$^+$ | pCH$_2$ | pCH$_3^+$ | oH$_2$O | | 4.7e-10 | 0.00 | 0.0 |
| 2712 | pH$_3$O$^+$ | oCH$_2$ | oCH$_3^+$ | oH$_2$O | | 2.2e-10 | 0.00 | 0.0 |
| 2713 | pH$_3$O$^+$ | oCH$_2$ | oCH$_3^+$ | pH$_2$O | | 1.6e-10 | 0.00 | 0.0 |
| 2714 | pH$_3$O$^+$ | oCH$_2$ | pCH$_3^+$ | oH$_2$O | | 4.4e-10 | 0.00 | 0.0 |
| 2715 | pH$_3$O$^+$ | oCH$_2$ | pCH$_3^+$ | pH$_2$O | | 1.3e-10 | 0.00 | 0.0 |
| 2716 | pH$_3$O$^+$ | pCH$_2$ | oCH$_3^+$ | oH$_2$O | | 1.9e-10 | 0.00 | 0.0 |
| 2717 | pH$_3$O$^+$ | pCH$_2$ | pCH$_3^+$ | oH$_2$O | | 3.8e-10 | 0.00 | 0.0 |
| 2718 | pH$_3$O$^+$ | pCH$_2$ | pCH$_3^+$ | pH$_2$O | | 3.8e-10 | 0.00 | 0.0 |
| 2719 | oH$_3$O$^+$ | oNH$_3$ | mNH$_4^+$ | oH$_2$O | | 1.1e-09 | 0.00 | 0.0 |
| 2720 | oH$_3$O$^+$ | oNH$_3$ | mNH$_4^+$ | pH$_2$O | | 1.4e-10 | 0.00 | 0.0 |
| 2721 | oH$_3$O$^+$ | oNH$_3$ | pNH$_4^+$ | oH$_2$O | | 9.2e-11 | 0.00 | 0.0 |
| 2722 | oH$_3$O$^+$ | oNH$_3$ | pNH$_4^+$ | pH$_2$O | | 5.5e-11 | 0.00 | 0.0 |
| 2723 | oH$_3$O$^+$ | oNH$_3$ | oNH$_4^+$ | oH$_2$O | | 6.3e-10 | 0.00 | 0.0 |
| 2724 | oH$_3$O$^+$ | oNH$_3$ | oNH$_4^+$ | pH$_2$O | | 1.4e-10 | 0.00 | 0.0 |
| 2725 | oH$_3$O$^+$ | pNH$_3$ | mNH$_4^+$ | oH$_2$O | | 3.7e-10 | 0.00 | 0.0 |
| 2726 | oH$_3$O$^+$ | pNH$_3$ | mNH$_4^+$ | pH$_2$O | | 2.7e-10 | 0.00 | 0.0 |
| 2727 | oH$_3$O$^+$ | pNH$_3$ | pNH$_4^+$ | oH$_2$O | | 1.8e-10 | 0.00 | 0.0 |
| 2728 | oH$_3$O$^+$ | pNH$_3$ | oNH$_4^+$ | oH$_2$O | | 1.1e-09 | 0.00 | 0.0 |
| 2729 | oH$_3$O$^+$ | pNH$_3$ | oNH$_4^+$ | pH$_2$O | | 2.7e-10 | 0.00 | 0.0 |
| 2730 | pH$_3$O$^+$ | oNH$_3$ | mNH$_4^+$ | oH$_2$O | | 3.7e-10 | 0.00 | 0.0 |
| 2731 | pH$_3$O$^+$ | oNH$_3$ | mNH$_4^+$ | pH$_2$O | | 2.7e-10 | 0.00 | 0.0 |
| 2732 | pH$_3$O$^+$ | oNH$_3$ | pNH$_4^+$ | oH$_2$O | | 1.8e-10 | 0.00 | 0.0 |
| 2733 | pH$_3$O$^+$ | oNH$_3$ | oNH$_4^+$ | oH$_2$O | | 1.1e-09 | 0.00 | 0.0 |
| 2734 | pH$_3$O$^+$ | oNH$_3$ | oNH$_4^+$ | pH$_2$O | | 2.7e-10 | 0.00 | 0.0 |
| 2735 | pH$_3$O$^+$ | pNH$_3$ | mNH$_4^+$ | oH$_2$O | | 1.8e-10 | 0.00 | 0.0 |
| 2736 | pH$_3$O$^+$ | pNH$_3$ | pNH$_4^+$ | oH$_2$O | | 3.7e-10 | 0.00 | 0.0 |
| 2737 | pH$_3$O$^+$ | pNH$_3$ | pNH$_4^+$ | pH$_2$O | | 2.2e-10 | 0.00 | 0.0 |
| 2738 | pH$_3$O$^+$ | pNH$_3$ | oNH$_4^+$ | oH$_2$O | | 8.8e-10 | 0.00 | 0.0 |

**Table D6** – *continued* (part 38)

| # | Reactants | | Products | | | $\alpha$ | $\beta$ | $\gamma$ |
|---|---|---|---|---|---|---|---|---|
| 2739 | $pH_3O^+$ | $pNH_3$ | $oNH_4^+$ | $pH_2O$ | | 5.5e-10 | 0.00 | 0.0 |
| 2740 | $CN^+$ | $oH_2$ | $HCN^+$ | H | | 1.0e-09 | 0.00 | 0.0 |
| 2741 | $CN^+$ | $pH_2$ | $HCN^+$ | H | | 1.0e-09 | 0.00 | 0.0 |
| 2742 | $HCN^+$ | $oH_2$ | $HCNH^+$ | H | | 9.8e-10 | 0.00 | 0.0 |
| 2743 | $HCN^+$ | $pH_2$ | $HCNH^+$ | H | | 9.8e-10 | 0.00 | 0.0 |
| 2744 | $HCNH^+$ | CH | $oCH_2^+$ | HCN | | 2.4e-10 | 0.00 | 0.0 |
| 2745 | $HCNH^+$ | CH | $pCH_2^+$ | HCN | | 7.9e-11 | 0.00 | 0.0 |
| 2746 | $HCNH^+$ | CH | $oCH_2^+$ | HNC | | 2.4e-10 | 0.00 | 0.0 |
| 2747 | $HCNH^+$ | CH | $pCH_2^+$ | HNC | | 7.9e-11 | 0.00 | 0.0 |
| 2748 | $HCNH^+$ | $oCH_2$ | $oCH_3^+$ | HCN | | 2.9e-10 | 0.00 | 0.0 |
| 2749 | $HCNH^+$ | $oCH_2$ | $pCH_3^+$ | HCN | | 1.4e-10 | 0.00 | 0.0 |
| 2750 | $HCNH^+$ | $pCH_2$ | $pCH_3^+$ | HCN | | 4.3e-10 | 0.00 | 0.0 |
| 2751 | $HCNH^+$ | $oCH_2$ | $oCH_3^+$ | HNC | | 2.9e-10 | 0.00 | 0.0 |
| 2752 | $HCNH^+$ | $oCH_2$ | $pCH_3^+$ | HNC | | 1.4e-10 | 0.00 | 0.0 |
| 2753 | $HCNH^+$ | $pCH_2$ | $pCH_3^+$ | HNC | | 4.3e-10 | 0.00 | 0.0 |
| 2754 | $HCNH^+$ | $oNH_2$ | $oNH_3^+$ | HCN | | 3.0e-10 | 0.00 | 0.0 |
| 2755 | $HCNH^+$ | $oNH_2$ | $pNH_3^+$ | HCN | | 1.5e-10 | 0.00 | 0.0 |
| 2756 | $HCNH^+$ | $pNH_2$ | $pNH_3^+$ | HCN | | 4.5e-10 | 0.00 | 0.0 |
| 2757 | $HCNH^+$ | $oNH_2$ | $oNH_3^+$ | HNC | | 3.0e-10 | 0.00 | 0.0 |
| 2758 | $HCNH^+$ | $oNH_2$ | $pNH_3^+$ | HNC | | 1.5e-10 | 0.00 | 0.0 |
| 2759 | $HCNH^+$ | $pNH_2$ | $pNH_3^+$ | HNC | | 4.5e-10 | 0.00 | 0.0 |
| 2760 | $HCNH^+$ | $oNH_3$ | $mNH_4^+$ | HCN | | 6.9e-10 | 0.00 | 0.0 |
| 2761 | $HCNH^+$ | $oNH_3$ | $oNH_4^+$ | HCN | | 4.1e-10 | 0.00 | 0.0 |
| 2762 | $HCNH^+$ | $pNH_3$ | $pNH_4^+$ | HCN | | 2.7e-10 | 0.00 | 0.0 |
| 2763 | $HCNH^+$ | $pNH_3$ | $oNH_4^+$ | HCN | | 8.3e-10 | 0.00 | 0.0 |
| 2764 | $HCNH^+$ | $oNH_3$ | $mNH_4^+$ | HNC | | 6.9e-10 | 0.00 | 0.0 |
| 2765 | $HCNH^+$ | $oNH_3$ | $oNH_4^+$ | HNC | | 4.1e-10 | 0.00 | 0.0 |
| 2766 | $HCNH^+$ | $pNH_3$ | $pNH_4^+$ | HNC | | 2.7e-10 | 0.00 | 0.0 |
| 2767 | $HCNH^+$ | $pNH_3$ | $oNH_4^+$ | HNC | | 8.3e-10 | 0.00 | 0.0 |
| 2768 | $HCNH^+$ | $oH_2S$ | $oH_3S^+$ | HCN | | 1.1e-10 | 0.00 | 0.0 |
| 2769 | $HCNH^+$ | $oH_2S$ | $pH_3S^+$ | HCN | | 5.7e-11 | 0.00 | 0.0 |
| 2770 | $HCNH^+$ | $pH_2S$ | $pH_3S^+$ | HCN | | 1.7e-10 | 0.00 | 0.0 |
| 2771 | $HCNH^+$ | $oH_2S$ | $oH_3S^+$ | HNC | | 1.1e-10 | 0.00 | 0.0 |
| 2772 | $HCNH^+$ | $oH_2S$ | $pH_3S^+$ | HNC | | 5.7e-11 | 0.00 | 0.0 |
| 2773 | $HCNH^+$ | $pH_2S$ | $pH_3S^+$ | HNC | | 1.7e-10 | 0.00 | 0.0 |
| 2774 | $N_2H^+$ | CO | $HCO^+$ | $N_2$ | | 8.8e-10 | 0.00 | 0.0 |
| 2775 | $N_2H^+$ | $CO_2$ | $HCO_2^+$ | $N_2$ | | 1.4e-09 | 0.00 | 0.0 |
| 2776 | $N_2H^+$ | NO | $HNO^+$ | $N_2$ | | 3.4e-10 | 0.00 | 0.0 |
| 2777 | $C_2N^+$ | $oNH_3$ | $C_2H_2$ | $N_2H^+$ | | 1.9e-09 | 0.00 | 0.0 |
| 2778 | $C_2N^+$ | $pNH_3$ | $C_2H_2$ | $N_2H^+$ | | 1.9e-09 | 0.00 | 0.0 |
| 2779 | $NO^+$ | Fe | $Fe^+$ | NO | | 1.0e-09 | 0.00 | 0.0 |
| 2780 | $HNO^+$ | C | $CH^+$ | NO | | 1.0e-09 | 0.00 | 0.0 |
| 2781 | $HNO^+$ | CO | $HCO^+$ | NO | | 1.0e-09 | 0.00 | 0.0 |
| 2782 | $HNO^+$ | $CO_2$ | $HCO_2^+$ | NO | | 1.0e-10 | 0.00 | 0.0 |
| 2783 | $HNO^+$ | OH | $oH_2O^+$ | NO | | 4.6e-10 | 0.00 | 0.0 |
| 2784 | $HNO^+$ | OH | $pH_2O^+$ | NO | | 1.6e-10 | 0.00 | 0.0 |
| 2785 | $SO^+$ | Fe | $Fe^+$ | SO | | 1.6e-09 | 0.00 | 0.0 |
| 2786 | $CS^+$ | $oH_2$ | $HCS^+$ | H | | 4.8e-10 | 0.00 | 0.0 |
| 2787 | $CS^+$ | $pH_2$ | $HCS^+$ | H | | 4.8e-10 | 0.00 | 0.0 |
| 2788 | $HCO^+$ | Gr | $Gr^+$ | CO | H | 3.0e-07 | 0.50 | 0.0 |
| 2789 | $HCS^+$ | Gr | $Gr^+$ | CS | H | 2.4e-07 | 0.50 | 0.0 |
| 2790 | $HCO^+$ | $Gr^-$ | Gr | CO | H | 3.0e-07 | 0.50 | 0.0 |
| 2791 | $HCS^+$ | $Gr^-$ | Gr | CS | H | 2.4e-07 | 0.50 | 0.0 |
| 2792 | $CO^+$ | $e^-$ | C | O | | 1.0e-07 | -0.46 | 0.0 |
| 2793 | $HCO^+$ | $e^-$ | CO | H | | 2.4e-07 | -0.69 | 0.0 |
| 2794 | $HCO_2^+$ | $e^-$ | $CO_2$ | H | | 2.2e-07 | -0.50 | 0.0 |
| 2795 | $HCO_2^+$ | $e^-$ | CO | OH | | 1.2e-07 | -0.50 | 0.0 |
| 2796 | $CN^+$ | $e^-$ | C | N | | 1.8e-07 | -0.50 | 0.0 |
| 2797 | $C_2N^+$ | $e^-$ | $C_2$ | N | | 1.0e-07 | -0.50 | 0.0 |
| 2798 | $C_2N^+$ | $e^-$ | CN | C | | 2.0e-07 | -0.50 | 0.0 |
| 2799 | $HCN^+$ | $e^-$ | CN | H | | 1.5e-07 | -0.50 | 0.0 |
| 2800 | $HCN^+$ | $e^-$ | CH | N | | 1.5e-07 | -0.50 | 0.0 |
| 2801 | $HCNH^+$ | $e^-$ | HCN | H | | 9.6e-08 | -0.65 | 0.0 |
| 2802 | $HCNH^+$ | $e^-$ | HNC | H | | 9.6e-08 | -0.65 | 0.0 |
| 2803 | $HCNH^+$ | $e^-$ | CN | H | H | 9.1e-08 | -0.65 | 0.0 |
| 2804 | $H_2NC^+$ | $e^-$ | HNC | H | | 1.8e-07 | -0.50 | 0.0 |
| 2805 | $H_2NC^+$ | $e^-$ | CN | H | H | 1.8e-08 | -0.50 | 0.0 |
| 2806 | $NO^+$ | $e^-$ | N | O | | 4.3e-07 | -0.37 | 0.0 |
| 2807 | $HNO^+$ | $e^-$ | NO | H | | 3.0e-07 | -0.50 | 0.0 |
| 2808 | $CS^+$ | $e^-$ | C | S | | 2.0e-07 | -0.50 | 0.0 |
| 2809 | $HCS^+$ | $e^-$ | CS | H | | 7.0e-07 | -0.50 | 0.0 |
| 2810 | $SO^+$ | $e^-$ | S | O | | 2.0e-07 | -0.50 | 0.0 |
| 2811 | $HSO^+$ | $e^-$ | SO | H | | 2.0e-07 | -0.50 | 0.0 |
| 2812 | $HSO_2^+$ | $e^-$ | SO | H | O | 1.0e-07 | -0.50 | 0.0 |

**Table D6** – *continued* (part 39)

| # | Reactants | | Products | | | $\alpha$ | $\beta$ | $\gamma$ |
|---|---|---|---|---|---|---|---|---|
| 2813 | $HSO_2^+$ | $e^-$ | SO | OH | | 1.0e-07 | -0.50 | 0.0 |
| 2814 | $HOCS^+$ | $e^-$ | OH | CS | | 2.0e-07 | -0.50 | 0.0 |
| 2815 | $HOCS^+$ | $e^-$ | OCS | H | | 2.0e-07 | -0.50 | 0.0 |
| 2816 | $Fe^+$ | $e^-$ | Fe | $\gamma$ | | 3.7e-12 | -0.65 | 0.0 |
| 2817 | $DCO^+$ | $e^-$ | CO | D | | 2.4e-07 | -0.69 | 0.0 |
| 2818 | $DCO_2^+$ | $e^-$ | $CO_2$ | D | | 2.2e-07 | -0.50 | 0.0 |
| 2819 | $DCO_2^+$ | $e^-$ | CO | OD | | 1.2e-07 | -0.50 | 0.0 |
| 2820 | $CO_2$ | $\gamma_2$ | CO | O | | 6.0e+02 | 0.00 | 0.0 |
| 2821 | CO | $\gamma_2$ | C | O | | 4.6e+01 | 0.00 | 0.0 |
| 2822 | CO | $\gamma_2$ | $CO^+$ | $e^-$ | | 1.4e+01 | 0.00 | 0.0 |
| 2823 | $HCO^+$ | $\gamma_2$ | $CO^+$ | H | | 3.3e+00 | 0.00 | 0.0 |
| 2824 | $CO^+$ | $\gamma_2$ | $C^+$ | O | | 7.7e+01 | 0.00 | 0.0 |
| 2825 | CN | $\gamma_2$ | C | N | | 4.5e+02 | 0.00 | 0.0 |
| 2826 | CN | $\gamma_2$ | $CN^+$ | $e^-$ | | 8.3e+00 | 0.00 | 0.0 |
| 2827 | HCN | $\gamma_2$ | CN | H | | 2.0e+03 | 0.00 | 0.0 |
| 2828 | HCN | $\gamma_2$ | $HCN^+$ | $e^-$ | | 1.4e+00 | 0.00 | 0.0 |
| 2829 | HNC | $\gamma_2$ | CN | H | | 2.0e+03 | 0.00 | 0.0 |
| 2830 | NO | $\gamma_2$ | N | O | | 3.0e+02 | 0.00 | 0.0 |
| 2831 | NO | $\gamma_2$ | $NO^+$ | $e^-$ | | 2.4e+02 | 0.00 | 0.0 |
| 2832 | SO | $\gamma_2$ | S | O | | 5.5e+03 | 0.00 | 0.0 |
| 2833 | SO | $\gamma_2$ | $SO^+$ | $e^-$ | | 4.5e+02 | 0.00 | 0.0 |
| 2834 | CS | $\gamma_2$ | S | C | | 1.9e+03 | 0.00 | 0.0 |
| 2835 | CS | $\gamma_2$ | $CS^+$ | $e^-$ | | 2.0e+01 | 0.00 | 0.0 |
| 2836 | OCS | $\gamma_2$ | CO | S | | 5.2e+03 | 0.00 | 0.0 |
| 2837 | $SO_2$ | $\gamma_2$ | SO | O | | 2.7e+03 | 0.00 | 0.0 |
| 2838 | N | $C_2H_2^+$ | $CH^+$ | HNC | | 2.5e-11 | 0.00 | 2600.0 |
| 2839 | HNC | $pCH_5^+$ | $C_2H_3^+$ | $oNH_3$ | | 8.3e-10 | 0.00 | 0.0 |
| 2840 | HNC | $pCH_5^+$ | $C_2H_3^+$ | $pNH_3$ | | 1.7e-10 | 0.00 | 0.0 |
| 2841 | HNC | $oCH_5^+$ | $C_2H_3^+$ | $oNH_3$ | | 5.0e-10 | 0.00 | 0.0 |
| 2842 | HNC | $oCH_5^+$ | $C_2H_3^+$ | $pNH_3$ | | 5.0e-10 | 0.00 | 0.0 |
| 2843 | HNC | $mCH_5^+$ | $C_2H_3^+$ | $oNH_3$ | | 3.0e-10 | 0.00 | 0.0 |
| 2844 | HNC | $mCH_5^+$ | $C_2H_3^+$ | $pNH_3$ | | 7.0e-10 | 0.00 | 0.0 |
| 2845 | $oH_3O^+$ | $C_3H$ | $C_3H_2^+$ | $oH_2O$ | | 1.7e-09 | 0.00 | 0.0 |
| 2846 | $oH_3O^+$ | $C_3H$ | $C_3H_2^+$ | $pH_2O$ | | 2.5e-10 | 0.00 | 0.0 |
| 2847 | $pH_3O^+$ | $C_3H$ | $C_3H_2^+$ | $oH_2O$ | | 1.3e-09 | 0.00 | 0.0 |
| 2848 | $pH_3O^+$ | $C_3H$ | $C_3H_2^+$ | $pH_2O$ | | 7.5e-10 | 0.00 | 0.0 |
| 2849 | $C_2H_3^+$ | $e^-$ | $oH_2$ | $C_2H$ | | 1.0e-07 | -0.50 | 0.0 |
| 2850 | $C_2H_3^+$ | $e^-$ | $oCH_2$ | CH | | 1.0e-07 | -0.50 | 0.0 |
| 2851 | $C_2H_3^+$ | $e^-$ | $pCH_2$ | CH | | 6.8e-08 | -0.50 | 0.0 |
| 2852 | $C_2H_2$ | $H^+$ | $oH_2$ | $C_2H^+$ | | 1.3e-09 | 0.00 | 0.0 |
| 2853 | $C_2H_2$ | $H^+$ | $pH_2$ | $C_2H^+$ | | 6.7e-10 | 0.00 | 0.0 |
| 2854 | $C_3H_2$ | $H^+$ | $oH_2$ | $C_3H^+$ | | 1.3e-09 | 0.00 | 0.0 |
| 2855 | $C_3H_2$ | $H^+$ | $pH_2$ | $C_3H^+$ | | 6.7e-10 | 0.00 | 0.0 |
| 2856 | $C_2H_2$ | $He^+$ | $pH_2$ | $C_2^+$ | He | 1.6e-09 | 0.00 | 0.0 |
| 2857 | $C_3H_2$ | $He^+$ | $pH_2$ | $C_3^+$ | He | 1.0e-09 | 0.00 | 0.0 |
| 2858 | $oH_3^+$ | $C_2H_2$ | $C_2H_3^+$ | $oH_2$ | | 2.3e-09 | 0.00 | 0.0 |
| 2859 | $oH_3^+$ | $C_2H_2$ | $C_2H_3^+$ | $pH_2$ | | 5.1e-10 | 0.00 | 0.0 |
| 2860 | $pH_3^+$ | $C_2H_2$ | $C_2H_3^+$ | $oH_2$ | | 1.9e-09 | 0.00 | 0.0 |
| 2861 | $pH_3^+$ | $C_2H_2$ | $C_2H_3^+$ | $pH_2$ | | 1.0e-09 | 0.00 | 0.0 |
| 2862 | $oH_3^+$ | $C_3H_2$ | $C_3H_3^+$ | $oH_2$ | | 1.6e-09 | 0.00 | 0.0 |
| 2863 | $oH_3^+$ | $C_3H_2$ | $C_3H_3^+$ | $pH_2$ | | 4.0e-10 | 0.00 | 0.0 |
| 2864 | $pH_3^+$ | $C_3H_2$ | $C_3H_3^+$ | $oH_2$ | | 1.3e-09 | 0.00 | 0.0 |
| 2865 | $pH_3^+$ | $C_3H_2$ | $C_3H_3^+$ | $pH_2$ | | 7.0e-10 | 0.00 | 0.0 |
| 2866 | $oH_2D^+$ | $C_2H_2$ | $C_2H_3^+$ | HD | | 2.9e-09 | 0.00 | 0.0 |
| 2867 | $pH_2D^+$ | $C_2H_2$ | $C_2H_3^+$ | HD | | 2.9e-09 | 0.00 | 0.0 |
| 2868 | $oH_2D^+$ | $C_3H_2$ | $C_3H_3^+$ | HD | | 2.0e-09 | 0.00 | 0.0 |
| 2869 | $pH_2D^+$ | $C_3H_2$ | $C_3H_3^+$ | HD | | 2.0e-09 | 0.00 | 0.0 |
| 2870 | $C_2H_2$ | $pHD_2^+$ | $C_2H_3^+$ | $pD_2$ | | 2.9e-09 | 0.00 | 0.0 |
| 2871 | $C_3H_2$ | $pHD_2^+$ | $C_3H_3^+$ | $pD_2$ | | 2.0e-09 | 0.00 | 0.0 |
| 2872 | $C_2H_2^+$ | $oH_2O$ | $oH_3O^+$ | $C_2H$ | | 1.1e-10 | 0.00 | 0.0 |
| 2873 | $C_2H_2^+$ | $oH_2O$ | $pH_3O^+$ | $C_2H$ | | 1.1e-10 | 0.00 | 0.0 |
| 2874 | $C_2H_2^+$ | $pH_2O$ | $oH_3O^+$ | $C_2H$ | | 7.3e-11 | 0.00 | 0.0 |
| 2875 | $C_2H_2^+$ | $pH_2O$ | $pH_3O^+$ | $C_2H$ | | 1.8e-10 | 0.00 | 0.0 |
| 2876 | $C_2H_2^+$ | N | $oH_2$ | $C_2N^+$ | | 2.2e-10 | 0.00 | 0.0 |
| 2877 | $C_2H_2^+$ | N | $pH_2$ | $C_2N^+$ | | 2.2e-10 | 0.00 | 0.0 |
| 2878 | $H_2CO$ | $\gamma_2$ | $oH_2$ | CO | | 1.1e+04 | 0.00 | 0.0 |
| 2879 | $H_2CO$ | $\gamma_2$ | $pH_2$ | CO | | 1.1e+04 | 0.00 | 0.0 |
| 2880 | $oH_3O^+$ | $C_3H_2$ | $C_3H_3^+$ | $oH_2O$ | | 2.4e-09 | 0.00 | 0.0 |
| 2881 | $oH_3O^+$ | $C_3H_2$ | $C_3H_3^+$ | $pH_2O$ | | 6.0e-10 | 0.00 | 0.0 |
| 2882 | $pH_3O^+$ | $C_3H_2$ | $C_3H_3^+$ | $oH_2O$ | | 2.0e-09 | 0.00 | 0.0 |
| 2883 | $pH_3O^+$ | $C_3H_2$ | $C_3H_3^+$ | $pH_2O$ | | 1.1e-09 | 0.00 | 0.0 |
| 2884 | $C_2H_3^+$ | $oH_2O$ | $oH_3O^+$ | $C_2H_2$ | | 6.1e-10 | 0.00 | 0.0 |
| 2885 | $C_2H_3^+$ | $oH_2O$ | $pH_3O^+$ | $C_2H_2$ | | 5.0e-10 | 0.00 | 0.0 |
| 2886 | $C_2H_3^+$ | $pH_2O$ | $oH_3O^+$ | $C_2H_2$ | | 3.9e-10 | 0.00 | 0.0 |

**Table D6** – *continued* (part 40)

| # | Reactants | | Products | | $\alpha$ | $\beta$ | $\gamma$ |
|---|---|---|---|---|---|---|---|
| 2887 | $C_2H_3^+$ | $pH_2O$ | $pH_3O^+$ | $C_2H_2$ | 7.2e-10 | 0.00 | 0.0 |
| 2888 | $oH_3^+$ | HD | $pH_3^+$ | HD | 7.7e-11 | 0.44 | -4.8 |
| 2889 | $oH_3^+$ | HD | $pH_2D^+$ | $oH_2$ | 1.6e-10 | -0.02 | -0.4 |
| 2890 | $oH_3^+$ | HD | $oH_2D^+$ | $pH_2$ | 1.5e-10 | -0.16 | 1.1 |
| 2891 | $oH_3^+$ | HD | $oH_2D^+$ | $oH_2$ | 1.1e-09 | 0.01 | 0.3 |
| 2892 | $pH_3^+$ | HD | $oH_3^+$ | HD | 1.2e-10 | 0.33 | 29.2 |
| 2893 | $pH_3^+$ | HD | $pH_2D^+$ | $pH_2$ | 1.1e-10 | -0.41 | 2.9 |
| 2894 | $pH_3^+$ | HD | $pH_2D^+$ | $oH_2$ | 2.5e-10 | -0.27 | 3.3 |
| 2895 | $pH_3^+$ | HD | $oH_2D^+$ | $pH_2$ | 2.8e-10 | -0.32 | 1.9 |
| 2896 | $pH_3^+$ | HD | $oH_2D^+$ | $oH_2$ | 1.2e-09 | 0.30 | 22.8 |
| 2897 | $oH_3^+$ | $oH_2$ | $pH_3^+$ | $pH_2$ | 1.3e-10 | 0.08 | -0.7 |
| 2898 | $oH_3^+$ | $oH_2$ | $oH_3^+$ | $pH_2$ | 9.7e-11 | 0.00 | -0.2 |
| 2899 | $oH_3^+$ | $pH_2$ | $pH_3^+$ | $oH_2$ | 3.5e-10 | -0.90 | 154.2 |
| 2900 | $oH_3^+$ | $pH_2$ | $oH_3^+$ | $oH_2$ | 5.0e-10 | -0.42 | 180.4 |
| 2901 | $pH_3^+$ | $oH_2$ | $pH_3^+$ | $oH_2$ | 4.1e-10 | 0.02 | -0.5 |
| 2902 | $pH_3^+$ | $oH_2$ | $pH_3^+$ | $pH_2$ | 1.9e-10 | -0.18 | 1.1 |
| 2903 | $pH_3^+$ | $oH_2$ | $oH_3^+$ | $pH_2$ | 1.7e-10 | -0.28 | 1.7 |
| 2904 | $pH_3^+$ | $oH_2$ | $oH_3^+$ | $oH_2$ | 6.7e-10 | -0.07 | 33.3 |
| 2905 | $pH_3^+$ | $pH_2$ | $pH_3^+$ | $oH_2$ | 1.0e-09 | -0.57 | 180.4 |
| 2906 | $pH_3^+$ | $pH_2$ | $oH_3^+$ | $oH_2$ | 9.2e-10 | -0.54 | 216.9 |
| 2907 | $oH_3^+$ | $oD_2$ | $oH_2D^+$ | HD | 1.2e-09 | 0.34 | -0.8 |
| 2908 | $oH_3^+$ | $oD_2$ | $oHD_2^+$ | $oH_2$ | 6.2e-10 | -0.22 | 1.2 |
| 2909 | $oH_3^+$ | $pD_2$ | $oH_2D^+$ | HD | 9.1e-10 | 0.05 | -0.4 |
| 2910 | $oH_3^+$ | $pD_2$ | $pHD_2^+$ | $oH_2$ | 6.5e-10 | -0.06 | 0.5 |
| 2911 | $pH_3^+$ | $oD_2$ | $pH_2D^+$ | HD | 5.3e-10 | 0.24 | -1.6 |
| 2912 | $pH_3^+$ | $oD_2$ | $oH_2D^+$ | HD | 5.8e-10 | 0.38 | -3.7 |
| 2913 | $pH_3^+$ | $oD_2$ | $oHD_2^+$ | $pH_2$ | 2.6e-10 | -0.27 | 2.2 |
| 2914 | $pH_3^+$ | $oD_2$ | $oHD_2^+$ | $oH_2$ | 4.0e-10 | -0.13 | 1.3 |
| 2915 | $pH_3^+$ | $pD_2$ | $pH_2D^+$ | HD | 4.0e-10 | 0.06 | -1.1 |
| 2916 | $pH_3^+$ | $pD_2$ | $oH_2D^+$ | HD | 4.7e-10 | -0.03 | 0.2 |
| 2917 | $pH_3^+$ | $pD_2$ | $pHD_2^+$ | $pH_2$ | 2.6e-10 | -0.06 | 0.5 |
| 2918 | $pH_3^+$ | $pD_2$ | $pHD_2^+$ | $oH_2$ | 4.1e-10 | 0.00 | 0.6 |
| 2919 | $oH_2D^+$ | $oH_2$ | $pH_3^+$ | HD | 7.9e-11 | 0.27 | -4.0 |
| 2920 | $oH_2D^+$ | $oH_2$ | $oH_3^+$ | HD | 1.3e-10 | -0.12 | 7.4 |
| 2921 | $oH_2D^+$ | $oH_2$ | $pH_2D^+$ | $pH_2$ | 7.6e-11 | -0.00 | -1.3 |
| 2922 | $oH_2D^+$ | $oH_2$ | $pH_2D^+$ | $oH_2$ | 1.5e-10 | -0.04 | -0.7 |
| 2923 | $oH_2D^+$ | $oH_2$ | $oH_2D^+$ | $pH_2$ | 1.4e-10 | -0.16 | 0.6 |
| 2924 | $oH_2D^+$ | $pH_2$ | $pH_3^+$ | HD | 9.4e-11 | -0.79 | 154.6 |
| 2925 | $oH_2D^+$ | $pH_2$ | $oH_3^+$ | HD | 1.1e-10 | -0.52 | 184.4 |
| 2926 | $oH_2D^+$ | $pH_2$ | $pH_2D^+$ | $oH_2$ | 8.2e-10 | -0.04 | 82.2 |
| 2927 | $oH_2D^+$ | $pH_2$ | $oH_2D^+$ | $oH_2$ | 9.3e-10 | -0.41 | 177.1 |
| 2928 | $pH_2D^+$ | $oH_2$ | $pH_3^+$ | HD | 9.0e-11 | -0.69 | 68.2 |
| 2929 | $pH_2D^+$ | $oH_2$ | $oH_3^+$ | HD | 8.7e-11 | -0.58 | 99.6 |
| 2930 | $pH_2D^+$ | $oH_2$ | $oH_2D^+$ | $pH_2$ | 4.9e-10 | -0.40 | 3.8 |
| 2931 | $pH_2D^+$ | $oH_2$ | $oH_2D^+$ | $oH_2$ | 7.5e-10 | -0.43 | 91.3 |
| 2932 | $pH_2D^+$ | $pH_2$ | $pH_3^+$ | HD | 2.7e-10 | -1.08 | 245.4 |
| 2933 | $pH_2D^+$ | $pH_2$ | $oH_2D^+$ | $oH_2$ | 3.1e-09 | -0.55 | 267.1 |
| 2934 | $oH_2D^+$ | HD | $pH_3^+$ | $pD_2$ | 4.8e-12 | -0.61 | 155.0 |
| 2935 | $oH_2D^+$ | HD | $pH_3^+$ | $oD_2$ | 4.7e-12 | -0.10 | 65.3 |
| 2936 | $oH_2D^+$ | HD | $oH_3^+$ | $pD_2$ | 1.5e-11 | -0.61 | 188.4 |
| 2937 | $oH_2D^+$ | HD | $oH_3^+$ | $oD_2$ | 1.1e-11 | -0.49 | 106.9 |
| 2938 | $oH_2D^+$ | HD | $pH_2D^+$ | HD | 1.7e-10 | 0.31 | -3.5 |
| 2939 | $oH_2D^+$ | HD | $pHD_2^+$ | $pH_2$ | 3.2e-11 | -0.30 | 3.2 |
| 2940 | $oH_2D^+$ | HD | $pHD_2^+$ | $oH_2$ | 2.0e-10 | -0.16 | 0.8 |
| 2941 | $oH_2D^+$ | HD | $oHD_2^+$ | $pH_2$ | 6.0e-11 | -0.39 | 3.7 |
| 2942 | $oH_2D^+$ | HD | $oHD_2^+$ | $oH_2$ | 4.7e-10 | -0.04 | -0.3 |
| 2943 | $pH_2D^+$ | HD | $pH_3^+$ | $pD_2$ | 2.5e-11 | -0.49 | 218.2 |
| 2944 | $pH_2D^+$ | HD | $pH_3^+$ | $oD_2$ | 2.2e-11 | -0.40 | 140.6 |
| 2945 | $pH_2D^+$ | HD | $oH_2D^+$ | HD | 8.4e-10 | -0.11 | 88.8 |
| 2946 | $pH_2D^+$ | HD | $pHD_2^+$ | $pH_2$ | 8.6e-11 | -0.65 | 5.4 |
| 2947 | $pH_2D^+$ | HD | $pHD_2^+$ | $oH_2$ | 1.0e-10 | -0.76 | 53.4 |
| 2948 | $pH_2D^+$ | HD | $oHD_2^+$ | $pH_2$ | 2.3e-10 | -0.48 | 4.9 |
| 2949 | $pH_2D^+$ | HD | $oHD_2^+$ | $oH_2$ | 4.7e-10 | 0.35 | -3.5 |
| 2950 | $oH_2D^+$ | $oD_2$ | $oH_2D^+$ | $pD_2$ | 1.1e-10 | 0.27 | 83.4 |
| 2951 | $oH_2D^+$ | $oD_2$ | $pHD_2^+$ | HD | 2.7e-10 | -0.02 | -0.1 |
| 2952 | $oH_2D^+$ | $oD_2$ | $oHD_2^+$ | HD | 9.3e-11 | 0.02 | 0.3 |
| 2953 | $oH_2D^+$ | $oD_2$ | $oD_3^+$ | $oH_2$ | 6.0e-11 | -0.10 | -0.2 |
| 2954 | $oH_2D^+$ | $oD_2$ | $mD_3^+$ | $oH_2$ | 7.2e-11 | -0.32 | 1.2 |
| 2955 | $oH_2D^+$ | $pD_2$ | $oH_2D^+$ | $oD_2$ | 8.1e-11 | 0.30 | -2.5 |
| 2956 | $oH_2D^+$ | $pD_2$ | $pHD_2^+$ | HD | 5.8e-10 | -0.05 | 0.7 |
| 2957 | $oH_2D^+$ | $pD_2$ | $oHD_2^+$ | HD | 5.1e-10 | -0.01 | -0.4 |
| 2958 | $oH_2D^+$ | $pD_2$ | $pD_3^+$ | $oH_2$ | 1.4e-11 | -0.03 | -0.2 |
| 2959 | $oH_2D^+$ | $pD_2$ | $oD_3^+$ | $oH_2$ | 1.1e-10 | -0.10 | 1.4 |
| 2960 | $pH_2D^+$ | $oD_2$ | $pH_2D^+$ | $pD_2$ | 8.5e-11 | 0.28 | 73.5 |

**Table D6** – *continued* (part 41)

| # | Reactants | | Products | | $\alpha$ | $\beta$ | $\gamma$ |
|---|---|---|---|---|---|---|---|
| 2961 | $pH_2D^+$ | $oD_2$ | $pHD_2^+$ | HD | 2.7e-10 | 0.02 | -0.8 |
| 2962 | $pH_2D^+$ | $oD_2$ | $oHD_2^+$ | HD | 1.2e-09 | 0.21 | -1.2 |
| 2963 | $pH_2D^+$ | $oD_2$ | $oD_3^+$ | $pH_2$ | 4.6e-11 | -0.52 | 5.0 |
| 2964 | $pH_2D^+$ | $oD_2$ | $mD_3^+$ | $pH_2$ | 3.8e-11 | -0.74 | 5.8 |
| 2965 | $pH_2D^+$ | $pD_2$ | $pH_2D^+$ | $oD_2$ | 3.5e-11 | -0.11 | -0.4 |
| 2966 | $pH_2D^+$ | $pD_2$ | $pHD_2^+$ | HD | 6.9e-10 | 0.01 | -0.2 |
| 2967 | $pH_2D^+$ | $pD_2$ | $oHD_2^+$ | HD | 5.6e-10 | 0.01 | 0.7 |
| 2968 | $pH_2D^+$ | $pD_2$ | $pD_3^+$ | $pH_2$ | 5.6e-12 | -0.45 | 3.4 |
| 2969 | $pH_2D^+$ | $pD_2$ | $oD_3^+$ | $pH_2$ | 5.7e-11 | -0.37 | 2.2 |
| 2970 | $oHD_2^+$ | $oH_2$ | $pH_3^+$ | $oD_2$ | 7.1e-12 | -1.09 | 182.4 |
| 2971 | $oHD_2^+$ | $oH_2$ | $oH_3^+$ | $oD_2$ | 1.9e-11 | -1.24 | 215.7 |
| 2972 | $oHD_2^+$ | $oH_2$ | $pH_2D^+$ | HD | 2.6e-10 | 0.44 | 12.3 |
| 2973 | $oHD_2^+$ | $oH_2$ | $oH_2D^+$ | HD | 1.2e-09 | -0.42 | 109.4 |
| 2974 | $oHD_2^+$ | $oH_2$ | $oHD_2^+$ | $pH_2$ | 1.4e-10 | -0.47 | 4.2 |
| 2975 | $oHD_2^+$ | $pH_2$ | $pH_3^+$ | $oD_2$ | 4.0e-11 | -1.29 | 357.6 |
| 2976 | $oHD_2^+$ | $pH_2$ | $pH_2D^+$ | HD | 8.1e-10 | -0.64 | 196.4 |
| 2977 | $oHD_2^+$ | $pH_2$ | $oH_2D^+$ | HD | 1.2e-09 | -0.92 | 288.2 |
| 2978 | $oHD_2^+$ | $pH_2$ | $oHD_2^+$ | $oH_2$ | 9.2e-10 | -0.68 | 178.8 |
| 2979 | $pHD_2^+$ | $oH_2$ | $pH_3^+$ | $pD_2$ | 9.8e-12 | -0.87 | 220.4 |
| 2980 | $pHD_2^+$ | $oH_2$ | $oH_3^+$ | $pD_2$ | 3.5e-11 | -0.82 | 249.9 |
| 2981 | $pHD_2^+$ | $oH_2$ | $pH_2D^+$ | HD | 1.9e-10 | 0.37 | -1.4 |
| 2982 | $pHD_2^+$ | $oH_2$ | $oH_2D^+$ | HD | 8.1e-10 | -0.06 | 53.5 |
| 2983 | $pHD_2^+$ | $oH_2$ | $pHD_2^+$ | $pH_2$ | 1.3e-10 | -0.31 | 2.1 |
| 2984 | $pHD_2^+$ | $pH_2$ | $pH_3^+$ | $pD_2$ | 9.3e-11 | -0.68 | 388.7 |
| 2985 | $pHD_2^+$ | $pH_2$ | $pH_2D^+$ | HD | 3.3e-10 | -0.55 | 144.3 |
| 2986 | $pHD_2^+$ | $pH_2$ | $oH_2D^+$ | HD | 7.0e-10 | -0.59 | 236.4 |
| 2987 | $pHD_2^+$ | $pH_2$ | $pHD_2^+$ | $oH_2$ | 6.5e-10 | -0.73 | 182.4 |
| 2988 | $oHD_2^+$ | HD | $pH_2D^+$ | $pD_2$ | 4.3e-11 | -0.49 | 200.3 |
| 2989 | $oHD_2^+$ | HD | $pH_2D^+$ | $oD_2$ | 6.8e-11 | -0.27 | 113.0 |
| 2990 | $oHD_2^+$ | HD | $oH_2D^+$ | $pD_2$ | 2.2e-10 | -0.85 | 290.0 |
| 2991 | $oHD_2^+$ | HD | $oH_2D^+$ | $oD_2$ | 2.7e-10 | -0.82 | 205.4 |
| 2992 | $oHD_2^+$ | HD | $pHD_2^+$ | HD | 6.7e-10 | 0.03 | 53.0 |
| 2993 | $oHD_2^+$ | HD | $oD_3^+$ | $pH_2$ | 1.3e-11 | -0.93 | 8.6 |
| 2994 | $oHD_2^+$ | HD | $oD_3^+$ | $oH_2$ | 4.7e-11 | -0.77 | 7.0 |
| 2995 | $oHD_2^+$ | HD | $mD_3^+$ | $pH_2$ | 2.0e-11 | -0.84 | 8.1 |
| 2996 | $oHD_2^+$ | HD | $mD_3^+$ | $oH_2$ | 5.9e-11 | -0.52 | 4.7 |
| 2997 | $pHD_2^+$ | HD | $pH_2D^+$ | $pD_2$ | 7.3e-11 | -0.11 | 143.7 |
| 2998 | $pHD_2^+$ | HD | $pH_2D^+$ | $oD_2$ | 2.5e-11 | 0.05 | 56.0 |
| 2999 | $pHD_2^+$ | HD | $oH_2D^+$ | $pD_2$ | 2.9e-10 | -0.64 | 238.6 |
| 3000 | $pHD_2^+$ | HD | $oH_2D^+$ | $oD_2$ | 1.0e-10 | -0.49 | 149.5 |
| 3001 | $pHD_2^+$ | HD | $oHD_2^+$ | HD | 9.0e-10 | 0.40 | -1.7 |
| 3002 | $pHD_2^+$ | HD | $pD_3^+$ | $pH_2$ | 3.6e-12 | -0.58 | 4.6 |
| 3003 | $pHD_2^+$ | HD | $pD_3^+$ | $oH_2$ | 1.2e-11 | -0.49 | 3.6 |
| 3004 | $pHD_2^+$ | HD | $oD_3^+$ | $pH_2$ | 3.2e-11 | -0.51 | 3.9 |
| 3005 | $pHD_2^+$ | HD | $oD_3^+$ | $oH_2$ | 1.2e-10 | -0.48 | 5.1 |
| 3006 | $oHD_2^+$ | $oD_2$ | $pHD_2^+$ | $pD_2$ | 1.8e-10 | -0.36 | 143.7 |
| 3007 | $oHD_2^+$ | $oD_2$ | $pHD_2^+$ | $oD_2$ | 1.6e-10 | -0.05 | 55.6 |
| 3008 | $oHD_2^+$ | $oD_2$ | $oHD_2^+$ | $pD_2$ | 2.5e-10 | 0.17 | 83.8 |
| 3009 | $oHD_2^+$ | $oD_2$ | $oD_3^+$ | HD | 3.3e-10 | -0.24 | 2.7 |
| 3010 | $oHD_2^+$ | $oD_2$ | $mD_3^+$ | HD | 4.7e-10 | -0.09 | 0.9 |
| 3011 | $oHD_2^+$ | $pD_2$ | $pHD_2^+$ | $pD_2$ | 1.0e-10 | 0.07 | 54.2 |
| 3012 | $oHD_2^+$ | $pD_2$ | $pHD_2^+$ | $oD_2$ | 1.3e-10 | -0.14 | 0.5 |
| 3013 | $oHD_2^+$ | $pD_2$ | $oHD_2^+$ | $oD_2$ | 1.8e-10 | 0.22 | -2.6 |
| 3014 | $oHD_2^+$ | $pD_2$ | $pD_3^+$ | HD | 2.3e-11 | -0.45 | 3.2 |
| 3015 | $oHD_2^+$ | $pD_2$ | $oD_3^+$ | HD | 4.9e-10 | -0.13 | 2.3 |
| 3016 | $oHD_2^+$ | $pD_2$ | $mD_3^+$ | HD | 1.9e-10 | -0.15 | 0.6 |
| 3017 | $pHD_2^+$ | $oD_2$ | $pHD_2^+$ | $pD_2$ | 7.8e-11 | -0.01 | 86.3 |
| 3018 | $pHD_2^+$ | $oD_2$ | $oHD_2^+$ | $pD_2$ | 2.6e-12 | 0.20 | 31.9 |
| 3019 | $pHD_2^+$ | $oD_2$ | $oHD_2^+$ | $oD_2$ | 2.1e-10 | 0.32 | 0.8 |
| 3020 | $pHD_2^+$ | $oD_2$ | $pD_3^+$ | HD | 4.1e-11 | -0.26 | 2.7 |
| 3021 | $pHD_2^+$ | $oD_2$ | $oD_3^+$ | HD | 5.0e-10 | -0.16 | 0.8 |
| 3022 | $pHD_2^+$ | $oD_2$ | $mD_3^+$ | HD | 2.2e-10 | -0.12 | 1.4 |
| 3023 | $pHD_2^+$ | $pD_2$ | $pHD_2^+$ | $oD_2$ | 5.7e-11 | 0.04 | -0.3 |
| 3024 | $pHD_2^+$ | $pD_2$ | $oHD_2^+$ | $pD_2$ | 1.4e-10 | 0.44 | -0.7 |
| 3025 | $pHD_2^+$ | $pD_2$ | $oHD_2^+$ | $oD_2$ | 2.1e-10 | 0.17 | -0.4 |
| 3026 | $pHD_2^+$ | $pD_2$ | $pD_3^+$ | HD | 7.0e-11 | -0.08 | -0.5 |
| 3027 | $pHD_2^+$ | $pD_2$ | $oD_3^+$ | HD | 5.7e-10 | -0.15 | 1.3 |
| 3028 | $oD_3^+$ | $oH_2$ | $oH_2D^+$ | $pD_2$ | 2.3e-10 | -1.01 | 324.3 |
| 3029 | $oD_3^+$ | $oH_2$ | $oH_2D^+$ | $oD_2$ | 1.3e-10 | -0.67 | 227.0 |
| 3030 | $oD_3^+$ | $oH_2$ | $pHD_2^+$ | HD | 1.1e-09 | -0.21 | 71.6 |
| 3031 | $oD_3^+$ | $oH_2$ | $oHD_2^+$ | HD | 5.8e-10 | -0.13 | 18.7 |
| 3032 | $mD_3^+$ | $oH_2$ | $oH_2D^+$ | $oD_2$ | 8.8e-10 | -0.52 | 270.7 |
| 3033 | $mD_3^+$ | $oH_2$ | $oHD_2^+$ | HD | 3.7e-09 | 0.41 | 60.2 |
| 3034 | $mD_3^+$ | $pH_2$ | $pH_2D^+$ | $oD_2$ | 8.1e-10 | -0.57 | 355.0 |

*Chemical models of collapsing prestellar sources* 75**Table D6** – *continued* (part 42)

| # | Reactants | | Products | | | $\alpha$ | $\beta$ | $\gamma$ |
|---|---|---|---|---|---|---|---|---|
| 3035 | $mD_3^+$ | $pH_2$ | $oHD_2^+$ | HD | | 9.6e-09 | -0.01 | 236.9 |
| 3036 | $pD_3^+$ | $oH_2$ | $oH_2D^+$ | $pD_2$ | | 6.1e-10 | -0.55 | 304.6 |
| 3037 | $pD_3^+$ | $oH_2$ | $pHD_2^+$ | HD | | 2.3e-09 | 0.17 | 50.1 |
| 3038 | $oD_3^+$ | $pH_2$ | $pH_2D^+$ | $pD_2$ | | 3.1e-10 | -0.67 | 400.1 |
| 3039 | $oD_3^+$ | $pH_2$ | $pH_2D^+$ | $oD_2$ | | 1.5e-10 | -0.85 | 315.3 |
| 3040 | $oD_3^+$ | $pH_2$ | $pHD_2^+$ | HD | | 1.3e-09 | -0.75 | 253.9 |
| 3041 | $oD_3^+$ | $pH_2$ | $oHD_2^+$ | HD | | 9.2e-10 | -0.59 | 197.8 |
| 3042 | $pD_3^+$ | $pH_2$ | $pH_2D^+$ | $pD_2$ | | 8.2e-10 | -0.18 | 378.0 |
| 3043 | $pD_3^+$ | $pH_2$ | $pHD_2^+$ | HD | | 3.5e-09 | -0.34 | 233.0 |
| 3044 | $oD_3^+$ | HD | $pHD_2^+$ | $pD_2$ | | 4.2e-10 | -0.85 | 259.0 |
| 3045 | $oD_3^+$ | HD | $pHD_2^+$ | $oD_2$ | | 3.8e-10 | -0.56 | 166.4 |
| 3046 | $oD_3^+$ | HD | $oHD_2^+$ | $pD_2$ | | 7.3e-10 | -0.18 | 199.3 |
| 3047 | $oD_3^+$ | HD | $oHD_2^+$ | $oD_2$ | | 3.6e-10 | -0.24 | 114.1 |
| 3048 | $oD_3^+$ | HD | $pD_3^+$ | HD | | 4.5e-11 | -0.33 | 19.2 |
| 3049 | $oD_3^+$ | HD | $mD_3^+$ | HD | | 2.1e-10 | -0.07 | 0.4 |
| 3050 | $mD_3^+$ | HD | $pHD_2^+$ | $oD_2$ | | 7.5e-10 | -0.32 | 213.9 |
| 3051 | $mD_3^+$ | HD | $oHD_2^+$ | $pD_2$ | | 2.0e-09 | 0.31 | 238.1 |
| 3052 | $mD_3^+$ | HD | $oHD_2^+$ | $oD_2$ | | 2.9e-09 | 0.30 | 151.8 |
| 3053 | $mD_3^+$ | HD | $oD_3^+$ | HD | | 1.1e-09 | 0.25 | 44.1 |
| 3054 | $pD_3^+$ | HD | $pHD_2^+$ | $pD_2$ | | 1.6e-09 | -0.13 | 232.6 |
| 3055 | $pD_3^+$ | HD | $pHD_2^+$ | $oD_2$ | | 6.9e-10 | -0.21 | 146.1 |
| 3056 | $pD_3^+$ | HD | $oHD_2^+$ | $pD_2$ | | 8.9e-10 | 0.05 | 176.9 |
| 3057 | $pD_3^+$ | HD | $oD_3^+$ | HD | | 9.5e-10 | 0.08 | -0.7 |
| 3058 | $oD_3^+$ | $oD_2$ | $pD_3^+$ | $pD_2$ | | 2.5e-11 | -0.60 | 110.1 |
| 3059 | $oD_3^+$ | $oD_2$ | $pD_3^+$ | $oD_2$ | | 2.0e-11 | -0.57 | 20.8 |
| 3060 | $oD_3^+$ | $oD_2$ | $oD_3^+$ | $pD_2$ | | 4.9e-10 | -0.26 | 88.4 |
| 3061 | $oD_3^+$ | $oD_2$ | $mD_3^+$ | $pD_2$ | | 2.2e-10 | -0.23 | 41.9 |
| 3062 | $oD_3^+$ | $oD_2$ | $mD_3^+$ | $oD_2$ | | 1.3e-10 | -0.33 | 2.4 |
| 3063 | $oD_3^+$ | $pD_2$ | $pD_3^+$ | $pD_2$ | | 3.2e-11 | -0.24 | 18.1 |
| 3064 | $oD_3^+$ | $pD_2$ | $pD_3^+$ | $oD_2$ | | 4.4e-11 | -0.00 | -0.3 |
| 3065 | $oD_3^+$ | $pD_2$ | $oD_3^+$ | $oD_2$ | | 3.5e-10 | -0.21 | 1.9 |
| 3066 | $oD_3^+$ | $pD_2$ | $mD_3^+$ | $pD_2$ | | 3.7e-11 | -0.06 | 0.1 |
| 3067 | $oD_3^+$ | $pD_2$ | $mD_3^+$ | $oD_2$ | | 6.2e-10 | 0.99 | -8.3 |
| 3068 | $mD_3^+$ | $oD_2$ | $oD_3^+$ | $pD_2$ | | 3.4e-09 | 1.08 | 126.4 |
| 3069 | $mD_3^+$ | $oD_2$ | $oD_3^+$ | $oD_2$ | | 6.5e-10 | -0.02 | 46.2 |
| 3070 | $mD_3^+$ | $oD_2$ | $mD_3^+$ | $pD_2$ | | 2.9e-10 | 0.11 | 84.5 |
| 3071 | $mD_3^+$ | $pD_2$ | $pD_3^+$ | $oD_2$ | | 3.2e-11 | -0.42 | 3.5 |
| 3072 | $mD_3^+$ | $pD_2$ | $oD_3^+$ | $pD_2$ | | 1.9e-10 | 0.24 | 43.9 |
| 3073 | $mD_3^+$ | $pD_2$ | $oD_3^+$ | $oD_2$ | | 8.1e-10 | 0.11 | -0.5 |
| 3074 | $mD_3^+$ | $pD_2$ | $mD_3^+$ | $oD_2$ | | 2.2e-10 | 0.18 | -2.6 |
| 3075 | $pD_3^+$ | $oD_2$ | $oD_3^+$ | $pD_2$ | | 1.3e-09 | 0.35 | 66.1 |
| 3076 | $pD_3^+$ | $oD_2$ | $oD_3^+$ | $oD_2$ | | 4.1e-10 | -0.17 | 0.9 |
| 3077 | $pD_3^+$ | $oD_2$ | $mD_3^+$ | $pD_2$ | | 1.8e-10 | -0.36 | 25.9 |
| 3078 | $pD_3^+$ | $pD_2$ | $oD_3^+$ | $pD_2$ | | 6.6e-10 | 0.15 | -1.8 |
| 3079 | $pD_3^+$ | $pD_2$ | $oD_3^+$ | $oD_2$ | | 4.7e-10 | -0.00 | 0.6 |
| 3080 | $pH_3^+$ | $e^-$ | $pH_2$ | H | | 9.2e-09 | -0.73 | 1.0 |
| 3081 | $pH_3^+$ | $e^-$ | $oH_2$ | H | | 9.2e-09 | -0.73 | 1.0 |
| 3082 | $pH_3^+$ | $e^-$ | H | H | H | 3.6e-08 | -0.73 | 1.0 |
| 3083 | $oH_3^+$ | $e^-$ | $oH_2$ | H | | 2.5e-08 | 0.16 | -1.0 |
| 3084 | $oH_3^+$ | $e^-$ | H | H | H | 4.9e-08 | 0.16 | -1.0 |
| 3085 | $pH_2D^+$ | $e^-$ | H | H | D | 5.6e-07 | 0.44 | -2.8 |
| 3086 | $pH_2D^+$ | $e^-$ | HD | H | | 1.3e-07 | 0.44 | -2.8 |
| 3087 | $pH_2D^+$ | $e^-$ | $pH_2$ | D | | 5.2e-09 | 0.44 | -2.8 |
| 3088 | $oH_2D^+$ | $e^-$ | H | H | D | 1.5e-07 | -0.00 | -3.5 |
| 3089 | $oH_2D^+$ | $e^-$ | HD | H | | 3.7e-08 | -0.00 | -3.5 |
| 3090 | $oH_2D^+$ | $e^-$ | $oH_2$ | D | | 1.4e-08 | -0.00 | -3.5 |
| 3091 | $pHD_2^+$ | $e^-$ | D | D | H | 1.7e-07 | 0.69 | -9.4 |
| 3092 | $pHD_2^+$ | $e^-$ | HD | D | | 2.2e-08 | 0.69 | -9.4 |
| 3093 | $pHD_2^+$ | $e^-$ | $pD_2$ | H | | 2.0e-08 | 0.69 | -9.4 |
| 3094 | $oHD_2^+$ | $e^-$ | D | D | H | 9.4e-08 | 0.66 | -12.5 |
| 3095 | $oHD_2^+$ | $e^-$ | HD | D | | 1.2e-08 | 0.66 | -12.5 |
| 3096 | $oHD_2^+$ | $e^-$ | $oD_2$ | H | | 1.1e-08 | 0.66 | -12.5 |
| 3097 | $pD_2^+$ | $e^-$ | D | D | | 2.3e-08 | -0.69 | 0.0 |
| 3098 | $oD_2^+$ | $e^-$ | D | D | | 2.3e-08 | -0.69 | 0.0 |
| 3099 | $pD_3^+$ | $e^-$ | D | D | D | 5.8e-08 | -0.60 | 9.2 |
| 3100 | $pD_3^+$ | $e^-$ | $pD_2$ | D | | 1.9e-08 | -0.60 | 9.2 |
| 3101 | $mD_3^+$ | $e^-$ | D | D | D | 3.7e-08 | -0.49 | -2.8 |
| 3102 | $mD_3^+$ | $e^-$ | $oD_2$ | D | | 1.2e-08 | -0.49 | -2.8 |
| 3103 | $oD_3^+$ | $e^-$ | D | D | D | 7.4e-08 | -0.77 | 15.0 |
| 3104 | $oD_3^+$ | $e^-$ | $pD_2$ | D | | 1.2e-08 | -0.77 | 15.0 |
| 3105 | $oD_3^+$ | $e^-$ | $oD_2$ | D | | 1.2e-08 | -0.77 | 15.0 |
| 3106 | $oH_2$ | $H^+$ | $pH_2$ | $H^+$ | | 1.8e-10 | 0.13 | -0.0 |
| 3107 | $pH_2$ | $H^+$ | $oH_2$ | $H^+$ | | 1.6e-09 | 0.13 | 170.5 |
| 3108 | $N^+$ | $pH_2$ | $NH^+$ | H | | 8.4e-10 | 0.00 | 168.5 |

MNRAS **000**, 1–?? (2017)

**Table D6** – *continued* (part 43)

| # | Reactants | | Products | | | | $\alpha$ | $\beta$ | $\gamma$ |
|---|---|---|---|---|---|---|---|---|---|
| 3109 | $N^+$ | $oH_2$ | $NH^+$ | H | | | 4.2e-10 | -0.15 | 44.1 |
| 3110 | $oH_2$ | $HCO^+$ | $pH_2$ | $HCO^+$ | | | 1.3e-10 | 0.00 | 0.0 |
| 3111 | $pH_2$ | $HCO^+$ | $oH_2$ | $HCO^+$ | | | 1.1e-09 | 0.00 | 170.7 |
| 3112 | HD | $H^+$ | $D^+$ | $pH_2$ | | | 4.4e-10 | 0.00 | 458.0 |
| 3113 | HD | $H^+$ | $D^+$ | $oH_2$ | | | 7.1e-10 | 0.00 | 628.7 |
| 3114 | $oD_2$ | $D^+$ | $D^+$ | $pD_2$ | | | 6.0e-10 | 0.00 | 86.0 |
| 3115 | $pD_2$ | $D^+$ | $D^+$ | $oD_2$ | | | 4.0e-10 | 0.00 | 0.0 |
| 3116 | $oD_2$ | $DCO^+$ | $pD_2$ | $DCO^+$ | | | 3.7e-10 | 0.00 | 86.0 |
| 3117 | $pD_2$ | $DCO^+$ | $oD_2$ | $DCO^+$ | | | 2.5e-10 | 0.00 | 0.0 |
| 3118 | $C_2H_2^+$ | $pH_2$ | $C_2H_3^+$ | H | | | 5.0e-10 | 0.00 | 800.0 |
| 3119 | $C_2H_2^+$ | $oH_2$ | $C_2H_3^+$ | H | | | 5.0e-10 | 0.00 | 629.3 |
| 3120 | $HD^+$ | H | $pH_2^+$ | D | | | 1.0e-09 | 0.00 | 285.0 |
| 3121 | $pH_2D^+$ | H | $pH_3^+$ | D | | | 1.0e-09 | 0.00 | 644.2 |
| 3122 | $oH_2D^+$ | H | $pH_3^+$ | D | | | 1.0e-09 | 0.00 | 644.2 |
| 3123 | $pHD_2^+$ | H | $pH_2D^+$ | D | | | 1.0e-09 | 0.00 | 599.6 |
| 3124 | $oHD_2^+$ | H | $pH_2D^+$ | D | | | 1.0e-09 | 0.00 | 599.6 |
| 3125 | $pD_3^+$ | H | $pHD_2^+$ | D | | | 1.0e-09 | 0.00 | 646.2 |
| 3126 | $mD_3^+$ | H | $pHD_2^+$ | D | | | 1.0e-09 | 0.00 | 646.2 |
| 3127 | $oD_3^+$ | H | $pHD_2^+$ | D | | | 1.0e-09 | 0.00 | 646.2 |
| 3128 | $pNH_3$ | OH | $pNH_2$ | $pH_2O$ | | | 3.5e-12 | 0.00 | 925.0 |
| 3129 | $oNH_3$ | OH | $pNH_2$ | $pH_2O$ | | | 3.5e-12 | 0.00 | 925.0 |
| 3130 | $pH_2S$ | OH | SH | $pH_2O$ | | | 6.1e-12 | 0.00 | 80.0 |
| 3131 | $oH_2S$ | OH | SH | $pH_2O$ | | | 6.1e-12 | 0.00 | 80.0 |
| 3132 | $oCH_4$ | $He^+$ | $CH^+$ | $oH_2$ | H | He | 1.7e-10 | 0.00 | 0.0 |
| 3133 | $oCH_4$ | $He^+$ | $CH^+$ | $pH_2$ | H | He | 8.5e-11 | 0.00 | 0.0 |
| 3134 | $mCH_4$ | $He^+$ | $CH^+$ | $oH_2$ | H | He | 2.6e-10 | 0.00 | 0.0 |
| 3135 | $pCH_4$ | $He^+$ | $CH^+$ | $oH_2$ | H | He | 1.3e-10 | 0.00 | 0.0 |
| 3136 | $pCH_4$ | $He^+$ | $CH^+$ | $pH_2$ | H | He | 1.3e-10 | 0.00 | 0.0 |
| 3137 | $oCH_4^+$ | $e^-$ | $oCH_2$ | H | H | | 2.0e-07 | -0.60 | 0.0 |
| 3138 | $oCH_4^+$ | $e^-$ | $pCH_2$ | H | H | | 1.0e-07 | -0.60 | 0.0 |
| 3139 | $mCH_4^+$ | $e^-$ | $oCH_2$ | H | H | | 3.0e-07 | -0.60 | 0.0 |
| 3140 | $pCH_4^+$ | $e^-$ | $oCH_2$ | H | H | | 1.5e-07 | -0.60 | 0.0 |
| 3141 | $pCH_4^+$ | $e^-$ | $pCH_2$ | H | H | | 1.5e-07 | -0.60 | 0.0 |
| 3142 | $oCH_5^+$ | $e^-$ | CH | $oH_2$ | $oH_2$ | | 4.4e-08 | -0.30 | 0.0 |
| 3143 | $oCH_5^+$ | $e^-$ | CH | $oH_2$ | $pH_2$ | | 4.4e-08 | -0.30 | 0.0 |
| 3144 | $mCH_5^+$ | $e^-$ | CH | $oH_2$ | $oH_2$ | | 3.5e-08 | -0.30 | 0.0 |
| 3145 | $mCH_5^+$ | $e^-$ | CH | $pH_2$ | $oH_2$ | | 3.5e-08 | -0.30 | 0.0 |
| 3146 | $mCH_5^+$ | $e^-$ | CH | $pH_2$ | $pH_2$ | | 1.8e-08 | -0.30 | 0.0 |
| 3147 | $pCH_5^+$ | $e^-$ | CH | $oH_2$ | $oH_2$ | | 8.7e-08 | -0.30 | 0.0 |
| 3148 | $oCH_5^+$ | $e^-$ | $oCH_2$ | $oH_2$ | H | | 4.4e-08 | -0.30 | 0.0 |
| 3149 | $oCH_5^+$ | $e^-$ | $oCH_2$ | $pH_2$ | H | | 2.2e-08 | -0.30 | 0.0 |
| 3150 | $oCH_5^+$ | $e^-$ | $pCH_2$ | $oH_2$ | H | | 2.2e-08 | -0.30 | 0.0 |
| 3151 | $mCH_5^+$ | $e^-$ | $oCH_2$ | $oH_2$ | H | | 3.5e-08 | -0.30 | 0.0 |
| 3152 | $mCH_5^+$ | $e^-$ | $oCH_2$ | $pH_2$ | H | | 1.8e-08 | -0.30 | 0.0 |
| 3153 | $mCH_5^+$ | $e^-$ | $pCH_2$ | $oH_2$ | H | | 1.8e-08 | -0.30 | 0.0 |
| 3154 | $mCH_5^+$ | $e^-$ | $pCH_2$ | $pH_2$ | H | | 1.8e-08 | -0.30 | 0.0 |
| 3155 | $pCH_5^+$ | $e^-$ | $oCH_2$ | $oH_2$ | H | | 8.7e-08 | -0.30 | 0.0 |
| 3156 | $oNH_4^+$ | $e^-$ | $oNH_2$ | H | H | | 8.1e-08 | -0.60 | 0.0 |
| 3157 | $oNH_4^+$ | $e^-$ | $pNH_2$ | H | H | | 4.1e-08 | -0.60 | 0.0 |
| 3158 | $mNH_4^+$ | $e^-$ | $oNH_2$ | H | H | | 1.2e-07 | -0.60 | 0.0 |
| 3159 | $pNH_4^+$ | $e^-$ | $oNH_2$ | H | H | | 6.1e-08 | -0.60 | 0.0 |
| 3160 | $pNH_4^+$ | $e^-$ | $pNH_2$ | H | H | | 6.1e-08 | -0.60 | 0.0 |
| 3161 | $CH_3OH$ | $\gamma_2$ | $pCH_3$ | OH | | | 3.0e+03 | 0.00 | 0.0 |
| 3162 | Gr | $\gamma_2$ | $Gr^+$ | $e^-$ | | | 6.3e+07 | 0.00 | 0.0 |
| 3163 | $Gr^-$ | $\gamma_2$ | Gr | $e^-$ | | | 4.2e+08 | 0.00 | 0.0 |
| 3164 | Gr | $e^-$ | $Gr^-$ | $\gamma$ | | | 6.9e-05 | 0.50 | 0.0 |
| 3165 | $Gr^+$ | $e^-$ | Gr | $\gamma$ | | | 6.9e-05 | 0.50 | 0.0 |
| 3166 | Fe | $\gamma_2$ | $Fe^+$ | $e^-$ | | | 4.8e+02 | 0.00 | 0.0 |

**Table D7.** Species-to-species rates for the inter-conversion reactions of $H_3^+$ with HD, $H_2$, and $D_2$. The rates have been fitted, in the 5 to 50 K temperature range, by a modified Arrhenius function of the form $k(T) = \alpha (T/300)^\beta \exp(-\gamma/T)$ in $cm^3\,s^{-1}$. Elastic channels are not reported. Rates at 10 K are also listed.

| # | Reactants | | Products | | $\alpha$ $cm^3\,s^{-1}$ | $\beta$ | $\gamma$ K | $k(10)$ $cm^3\,s^{-1}$ |
|---|---|---|---|---|---|---|---|---|
| 1 | $oH_3^+$ | HD | $pH_3^+$ | HD | 7.7e-11 | 0.44 | -4.8 | 2.8e-11 |
| 2 | $oH_3^+$ | HD | $pH_2D^+$ | $oH_2$ | 1.6e-10 | -0.02 | -0.4 | 1.8e-10 |
| 3 | $oH_3^+$ | HD | $oH_2D^+$ | $pH_2$ | 1.5e-10 | -0.16 | 1.1 | 2.3e-10 |
| 4 | $oH_3^+$ | HD | $oH_2D^+$ | $oH_2$ | 1.1e-09 | 0.01 | 0.3 | 1.0e-09 |
| 5 | $pH_3^+$ | HD | $oH_3^+$ | HD | 1.2e-10 | 0.33 | 29.2 | 2.1e-12 |
| 6 | $pH_3^+$ | HD | $pH_2D^+$ | $pH_2$ | 1.1e-10 | -0.41 | 2.9 | 3.3e-10 |
| 7 | $pH_3^+$ | HD | $pH_2D^+$ | $oH_2$ | 2.5e-10 | -0.27 | 3.3 | 4.5e-10 |
| 8 | $pH_3^+$ | HD | $oH_2D^+$ | $pH_2$ | 2.8e-10 | -0.32 | 1.9 | 6.9e-10 |
| 9 | $pH_3^+$ | HD | $oH_2D^+$ | $oH_2$ | 1.2e-09 | 0.30 | 22.8 | 4.4e-11 |
| 10 | $oH_3^+$ | $oH_2$ | $pH_3^+$ | $pH_2$ | 1.3e-10 | 0.08 | -0.7 | 1.1e-10 |
| 11 | $oH_3^+$ | $oH_2$ | $oH_3^+$ | $pH_2$ | 9.7e-11 | 0.00 | -0.2 | 9.9e-11 |
| 12 | $oH_3^+$ | $pH_2$ | $pH_3^+$ | $oH_2$ | 3.5e-10 | -0.90 | 154.2 | 1.5e-15 |
| 13 | $oH_3^+$ | $pH_2$ | $oH_3^+$ | $oH_2$ | 5.0e-10 | -0.42 | 180.4 | 3.1e-17 |
| 14 | $oH_3^+$ | $oH_2$ | $pH_3^+$ | $oH_2$ | 4.1e-10 | 0.02 | -0.5 | 4.0e-10 |
| 15 | $pH_3^+$ | $oH_2$ | $pH_3^+$ | $pH_2$ | 1.9e-10 | -0.18 | 1.1 | 3.1e-10 |
| 16 | $pH_3^+$ | $oH_2$ | $oH_3^+$ | $pH_2$ | 1.7e-10 | -0.28 | 1.7 | 3.7e-10 |
| 17 | $pH_3^+$ | $oH_2$ | $oH_3^+$ | $oH_2$ | 6.7e-10 | -0.07 | 33.3 | 3.0e-11 |
| 18 | $pH_3^+$ | $pH_2$ | $pH_3^+$ | $oH_2$ | 1.0e-09 | -0.57 | 180.4 | 1.0e-16 |
| 19 | $pH_3^+$ | $pH_2$ | $oH_3^+$ | $oH_2$ | 9.2e-10 | -0.54 | 216.9 | 2.2e-18 |
| 20 | $oH_3^+$ | $oD_2$ | $oH_2D^+$ | HD | 1.2e-09 | 0.34 | -0.8 | 4.1e-10 |
| 21 | $oH_3^+$ | $oD_2$ | $oHD_2^+$ | $oH_2$ | 6.2e-10 | -0.22 | 1.2 | 1.2e-09 |
| 22 | $oH_3^+$ | $pD_2$ | $oH_2D^+$ | HD | 9.1e-10 | 0.05 | -0.4 | 8.0e-10 |
| 23 | $oH_3^+$ | $pD_2$ | $pHD_2^+$ | $oH_2$ | 6.5e-10 | -0.06 | 0.5 | 7.6e-10 |
| 24 | $pH_3^+$ | $oD_2$ | $pH_2D^+$ | HD | 5.3e-10 | 0.24 | -1.6 | 2.7e-10 |
| 25 | $pH_3^+$ | $oD_2$ | $oH_2D^+$ | HD | 5.8e-10 | 0.38 | -3.7 | 2.3e-10 |
| 26 | $pH_3^+$ | $oD_2$ | $oHD_2^+$ | $pH_2$ | 2.6e-10 | -0.27 | 2.2 | 5.2e-10 |
| 27 | $pH_3^+$ | $oD_2$ | $oHD_2^+$ | $oH_2$ | 4.0e-10 | -0.13 | 1.3 | 5.5e-10 |
| 28 | $pH_3^+$ | $pD_2$ | $pH_2D^+$ | HD | 4.0e-10 | 0.06 | -1.1 | 3.6e-10 |
| 29 | $pH_3^+$ | $pD_2$ | $oH_2D^+$ | HD | 4.7e-10 | -0.03 | 0.2 | 5.1e-10 |
| 30 | $pH_3^+$ | $pD_2$ | $pHD_2^+$ | $pH_2$ | 2.6e-10 | -0.06 | 0.5 | 3.0e-10 |
| 31 | $pH_3^+$ | $pD_2$ | $pHD_2^+$ | $oH_2$ | 4.1e-10 | 0.00 | 0.6 | 3.9e-10 |

**Table D8.** Same as Table D7 for reactions involving $H_2D^+$.

| # | Reactants | | Products | | $\alpha$ cm$^3$ s$^{-1}$ | $\beta$ | $\gamma$ K | $k(10)$ cm$^3$ s$^{-1}$ |
|---|---|---|---|---|---|---|---|---|
| 1 | $oH_2D^+$ | $oH_2$ | $pH_3^+$ | HD | 7.9e-11 | 0.27 | -4.0 | 4.7e-11 |
| 2 | $oH_2D^+$ | $oH_2$ | $oH_3^+$ | HD | 1.3e-10 | -0.12 | 7.4 | 9.3e-11 |
| 3 | $oH_2D^+$ | $oH_2$ | $pH_2D^+$ | $pH_2$ | 7.6e-11 | -0.00 | -1.3 | 8.7e-11 |
| 4 | $oH_2D^+$ | $oH_2$ | $pH_2D^+$ | $oH_2$ | 1.5e-10 | -0.04 | -0.7 | 1.8e-10 |
| 5 | $oH_2D^+$ | $oH_2$ | $oH_2D^+$ | $pH_2$ | 1.4e-10 | -0.16 | 0.6 | 2.3e-10 |
| 6 | $oH_2D^+$ | $pH_2$ | $pH_3^+$ | HD | 9.4e-11 | -0.79 | 154.6 | 2.7e-16 |
| 7 | $oH_2D^+$ | $pH_2$ | $oH_3^+$ | HD | 1.1e-10 | -0.52 | 184.4 | 6.3e-18 |
| 8 | $oH_2D^+$ | $pH_2$ | $pH_2D^+$ | $oH_2$ | 8.2e-10 | -0.04 | 82.2 | 2.5e-13 |
| 9 | $oH_2D^+$ | $pH_2$ | $oH_2D^+$ | $oH_2$ | 9.3e-10 | -0.41 | 177.1 | 7.6e-17 |
| 10 | $pH_2D^+$ | $oH_2$ | $pH_3^+$ | HD | 9.0e-11 | -0.69 | 68.2 | 1.0e-12 |
| 11 | $pH_2D^+$ | $oH_2$ | $oH_3^+$ | HD | 8.7e-11 | -0.58 | 99.6 | 3.0e-14 |
| 12 | $pH_2D^+$ | $oH_2$ | $oH_2D^+$ | $pH_2$ | 4.9e-10 | -0.40 | 3.8 | 1.3e-09 |
| 13 | $pH_2D^+$ | $oH_2$ | $oH_2D^+$ | $oH_2$ | 7.5e-10 | -0.43 | 91.3 | 3.5e-13 |
| 14 | $pH_2D^+$ | $pH_2$ | $pH_3^+$ | HD | 2.7e-10 | -1.08 | 245.4 | 2.3e-19 |
| 15 | $pH_2D^+$ | $pH_2$ | $oH_2D^+$ | $oH_2$ | 3.1e-09 | -0.55 | 267.1 | 5.1e-20 |
| 16 | $oH_2D^+$ | HD | $pH_3^+$ | $pD_2$ | 4.8e-12 | -0.61 | 155.0 | 7.1e-18 |
| 17 | $oH_2D^+$ | HD | $pH_3^+$ | $oD_2$ | 4.7e-12 | -0.10 | 65.3 | 9.6e-15 |
| 18 | $oH_2D^+$ | HD | $oH_3^+$ | $pD_2$ | 1.5e-11 | -0.61 | 188.4 | 7.9e-19 |
| 19 | $oH_2D^+$ | HD | $oH_3^+$ | $oD_2$ | 1.1e-11 | -0.49 | 106.9 | 1.3e-15 |
| 20 | $oH_2D^+$ | HD | $pH_2D^+$ | HD | 1.7e-10 | 0.31 | -3.5 | 8.4e-11 |
| 21 | $oH_2D^+$ | HD | $pHD_2^+$ | $pH_2$ | 3.2e-11 | -0.30 | 3.2 | 6.4e-11 |
| 22 | $oH_2D^+$ | HD | $pHD_2^+$ | $oH_2$ | 2.0e-10 | -0.16 | 0.8 | 3.2e-10 |
| 23 | $oH_2D^+$ | HD | $oHD_2^+$ | $pH_2$ | 6.0e-11 | -0.39 | 3.7 | 1.6e-10 |
| 24 | $oH_2D^+$ | HD | $oHD_2^+$ | $oH_2$ | 4.7e-10 | -0.04 | -0.3 | 5.5e-10 |
| 25 | $pH_2D^+$ | HD | $pH_3^+$ | $pD_2$ | 2.5e-11 | -0.49 | 218.2 | 4.4e-20 |
| 26 | $pH_2D^+$ | HD | $pH_3^+$ | $oD_2$ | 2.2e-11 | -0.40 | 140.6 | 6.7e-17 |
| 27 | $pH_2D^+$ | HD | $oH_2D^+$ | HD | 8.4e-10 | -0.11 | 88.8 | 1.7e-13 |
| 28 | $pH_2D^+$ | HD | $pHD_2^+$ | $pH_2$ | 8.6e-11 | -0.65 | 5.4 | 4.6e-10 |
| 29 | $pH_2D^+$ | HD | $pHD_2^+$ | $oH_2$ | 1.0e-10 | -0.76 | 53.4 | 6.4e-12 |
| 30 | $pH_2D^+$ | HD | $oHD_2^+$ | $pH_2$ | 2.3e-10 | -0.48 | 4.9 | 7.2e-10 |
| 31 | $pH_2D^+$ | HD | $oHD_2^+$ | $oH_2$ | 4.7e-10 | 0.35 | -3.5 | 2.0e-10 |
| 32 | $oH_2D^+$ | $oD_2$ | $oH_2D^+$ | $pD_2$ | 1.1e-10 | 0.27 | 83.4 | 1.0e-14 |
| 33 | $oH_2D^+$ | $oD_2$ | $pHD_2^+$ | HD | 2.7e-10 | -0.02 | -0.1 | 2.9e-10 |
| 34 | $oH_2D^+$ | $oD_2$ | $oHD_2^+$ | HD | 9.3e-10 | 0.02 | 0.3 | 8.4e-10 |
| 35 | $oH_2D^+$ | $oD_2$ | $oD_3^+$ | $oH_2$ | 6.0e-11 | -0.10 | -0.2 | 8.6e-11 |
| 36 | $oH_2D^+$ | $oD_2$ | $mD_3^+$ | $oH_2$ | 7.2e-11 | -0.32 | 1.2 | 1.9e-10 |
| 37 | $oH_2D^+$ | $pD_2$ | $oH_2D^+$ | $oD_2$ | 8.1e-11 | 0.30 | -2.5 | 3.7e-11 |
| 38 | $oH_2D^+$ | $pD_2$ | $pHD_2^+$ | HD | 5.8e-10 | -0.05 | 0.7 | 6.4e-10 |
| 39 | $oH_2D^+$ | $pD_2$ | $oHD_2^+$ | HD | 5.1e-10 | -0.01 | -0.4 | 5.5e-10 |
| 40 | $oH_2D^+$ | $pD_2$ | $pD_3^+$ | $oH_2$ | 1.4e-11 | -0.03 | -0.2 | 1.6e-11 |
| 41 | $oH_2D^+$ | $pD_2$ | $oD_3^+$ | $oH_2$ | 1.1e-10 | -0.10 | 1.4 | 1.3e-10 |
| 42 | $pH_2D^+$ | $oD_2$ | $pH_2D^+$ | $pD_2$ | 8.5e-11 | 0.28 | 73.5 | 2.1e-14 |
| 43 | $pH_2D^+$ | $oD_2$ | $pHD_2^+$ | HD | 2.7e-10 | 0.02 | -0.8 | 2.7e-10 |
| 44 | $pH_2D^+$ | $oD_2$ | $oHD_2^+$ | HD | 1.2e-09 | 0.21 | -1.2 | 6.6e-10 |
| 45 | $pH_2D^+$ | $oD_2$ | $oD_3^+$ | $pH_2$ | 4.6e-11 | -0.52 | 5.0 | 1.6e-10 |
| 46 | $pH_2D^+$ | $oD_2$ | $mD_3^+$ | $pH_2$ | 3.8e-11 | -0.74 | 5.8 | 2.6e-10 |
| 47 | $pH_2D^+$ | $pD_2$ | $pH_2D^+$ | $oD_2$ | 3.5e-11 | -0.11 | -0.4 | 5.3e-11 |
| 48 | $pH_2D^+$ | $pD_2$ | $pHD_2^+$ | HD | 6.9e-10 | 0.01 | -0.2 | 6.8e-10 |
| 49 | $pH_2D^+$ | $pD_2$ | $oHD_2^+$ | HD | 5.6e-10 | 0.01 | 0.7 | 5.0e-10 |
| 50 | $pH_2D^+$ | $pD_2$ | $pD_3^+$ | $pH_2$ | 5.6e-12 | -0.45 | 3.4 | 1.8e-11 |
| 51 | $pH_2D^+$ | $pD_2$ | $oD_3^+$ | $pH_2$ | 5.7e-11 | -0.37 | 2.2 | 1.6e-10 |

**Table D9.** Same as Table D7 for reactions involving $D_2H^+$.

| # | Reactants | | Products | | $\alpha$ cm$^3$ s$^{-1}$ | $\beta$ | $\gamma$ K | $k(10)$ cm$^3$ s$^{-1}$ |
|---|---|---|---|---|---|---|---|---|
| 1  | oHD$_2^+$ | oH$_2$ | pH$_3^+$   | oD$_2$ | 7.1e-12 | -1.09 | 182.4 | 3.5e-18 |
| 2  | oHD$_2^+$ | oH$_2$ | oH$_3^+$   | oD$_2$ | 1.9e-11 | -1.24 | 215.7 | 5.5e-19 |
| 3  | oHD$_2^+$ | oH$_2$ | pH$_2$D$^+$ | HD    | 2.6e-10 | 0.44  | 12.3  | 1.7e-11 |
| 4  | oHD$_2^+$ | oH$_2$ | oH$_2$D$^+$ | HD    | 1.2e-09 | -0.42 | 109.4 | 8.9e-14 |
| 5  | oHD$_2^+$ | oH$_2$ | oHD$_2^+$  | pH$_2$ | 1.4e-10 | -0.47 | 4.2   | 4.5e-10 |
| 6  | oHD$_2^+$ | pH$_2$ | pH$_3^+$   | oD$_2$ | 4.0e-11 | -1.29 | 357.6 | 9.5e-25 |
| 7  | oHD$_2^+$ | pH$_2$ | pH$_2$D$^+$ | HD    | 8.1e-10 | -0.64 | 196.4 | 2.1e-17 |
| 8  | oHD$_2^+$ | pH$_2$ | oH$_2$D$^+$ | HD    | 1.2e-09 | -0.92 | 288.2 | 8.4e-21 |
| 9  | oHD$_2^+$ | pH$_2$ | oHD$_2^+$  | oH$_2$ | 9.2e-10 | -0.68 | 178.8 | 1.6e-16 |
| 10 | pHD$_2^+$ | oH$_2$ | pH$_3^+$   | pD$_2$ | 9.8e-12 | -0.87 | 220.4 | 5.1e-20 |
| 11 | pHD$_2^+$ | oH$_2$ | oH$_3^+$   | pD$_2$ | 3.5e-11 | -0.82 | 249.9 | 8.0e-21 |
| 12 | pHD$_2^+$ | oH$_2$ | pH$_2$D$^+$ | HD    | 1.9e-10 | 0.37  | -1.4  | 6.2e-11 |
| 13 | pHD$_2^+$ | oH$_2$ | oH$_2$D$^+$ | HD    | 8.1e-10 | -0.06 | 53.5  | 4.7e-12 |
| 14 | pHD$_2^+$ | oH$_2$ | pHD$_2^+$  | pH$_2$ | 1.3e-10 | -0.31 | 2.1   | 3.0e-10 |
| 15 | pHD$_2^+$ | pH$_2$ | pH$_3^+$   | pD$_2$ | 9.3e-11 | -0.68 | 388.7 | 1.2e-26 |
| 16 | pHD$_2^+$ | pH$_2$ | pH$_2$D$^+$ | HD    | 3.3e-10 | -0.55 | 144.3 | 1.2e-15 |
| 17 | pHD$_2^+$ | pH$_2$ | oH$_2$D$^+$ | HD    | 7.0e-10 | -0.59 | 236.4 | 2.8e-19 |
| 18 | pHD$_2^+$ | pH$_2$ | pHD$_2^+$  | oH$_2$ | 6.5e-10 | -0.73 | 182.4 | 9.3e-17 |
| 19 | oHD$_2^+$ | HD     | pH$_2$D$^+$ | pD$_2$ | 4.3e-11 | -0.49 | 200.3 | 4.6e-19 |
| 20 | oHD$_2^+$ | HD     | pH$_2$D$^+$ | oD$_2$ | 6.8e-11 | -0.27 | 113.0 | 2.1e-15 |
| 21 | oHD$_2^+$ | HD     | oH$_2$D$^+$ | pD$_2$ | 2.2e-10 | -0.85 | 290.0 | 1.0e-21 |
| 22 | oHD$_2^+$ | HD     | oH$_2$D$^+$ | oD$_2$ | 2.7e-10 | -0.82 | 205.4 | 5.3e-18 |
| 23 | oHD$_2^+$ | HD     | pHD$_2^+$  | HD    | 6.7e-10 | 0.03  | 53.0  | 3.0e-12 |
| 24 | oHD$_2^+$ | HD     | oD$_3^+$   | pH$_2$ | 1.3e-11 | -0.93 | 8.6   | 1.3e-10 |
| 25 | oHD$_2^+$ | HD     | oD$_3^+$   | oH$_2$ | 4.7e-11 | -0.77 | 7.0   | 3.2e-10 |
| 26 | oHD$_2^+$ | HD     | mD$_3^+$   | pH$_2$ | 2.0e-11 | -0.84 | 8.1   | 1.5e-10 |
| 27 | oHD$_2^+$ | HD     | mD$_3^+$   | oH$_2$ | 5.9e-11 | -0.52 | 4.7   | 2.2e-10 |
| 28 | pHD$_2^+$ | HD     | pH$_2$D$^+$ | pD$_2$ | 7.3e-11 | -0.11 | 143.7 | 6.1e-17 |
| 29 | pHD$_2^+$ | HD     | pH$_2$D$^+$ | oD$_2$ | 2.5e-11 | 0.05  | 56.0  | 7.8e-14 |
| 30 | pHD$_2^+$ | HD     | oH$_2$D$^+$ | pD$_2$ | 2.9e-10 | -0.64 | 238.6 | 1.1e-19 |
| 31 | pHD$_2^+$ | HD     | oH$_2$D$^+$ | oD$_2$ | 1.0e-10 | -0.49 | 149.5 | 1.7e-16 |
| 32 | pHD$_2^+$ | HD     | oHD$_2^+$  | HD    | 9.0e-10 | 0.40  | -1.7  | 2.7e-10 |
| 33 | pHD$_2^+$ | HD     | pD$_3^+$   | pH$_2$ | 3.6e-12 | -0.58 | 4.6   | 1.6e-11 |
| 34 | pHD$_2^+$ | HD     | pD$_3^+$   | oH$_2$ | 1.2e-11 | -0.49 | 3.6   | 4.4e-11 |
| 35 | pHD$_2^+$ | HD     | oD$_3^+$   | pH$_2$ | 3.2e-11 | -0.51 | 3.9   | 1.2e-10 |
| 36 | pHD$_2^+$ | HD     | oD$_3^+$   | oH$_2$ | 1.2e-10 | -0.48 | 5.1   | 3.7e-10 |
| 37 | oHD$_2^+$ | oD$_2$ | pHD$_2^+$  | pD$_2$ | 1.8e-10 | -0.36 | 143.7 | 3.5e-16 |
| 38 | oHD$_2^+$ | oD$_2$ | pHD$_2^+$  | oD$_2$ | 1.6e-10 | -0.05 | 55.6  | 7.3e-13 |
| 39 | oHD$_2^+$ | oD$_2$ | oHD$_2^+$  | pD$_2$ | 2.5e-10 | 0.17  | 83.8  | 3.2e-14 |
| 40 | oHD$_2^+$ | oD$_2$ | oD$_3^+$   | HD    | 3.3e-10 | -0.24 | 2.7   | 5.7e-10 |
| 41 | oHD$_2^+$ | oD$_2$ | mD$_3^+$   | HD    | 4.7e-10 | -0.09 | 0.9   | 5.8e-10 |
| 42 | oHD$_2^+$ | pD$_2$ | pHD$_2^+$  | pD$_2$ | 1.0e-10 | 0.07  | 54.2  | 3.5e-13 |
| 43 | oHD$_2^+$ | pD$_2$ | pHD$_2^+$  | oD$_2$ | 1.3e-10 | -0.14 | 0.5   | 2.0e-10 |
| 44 | oHD$_2^+$ | pD$_2$ | oHD$_2^+$  | oD$_2$ | 1.8e-10 | 0.22  | -2.6  | 1.1e-10 |
| 45 | oHD$_2^+$ | pD$_2$ | pD$_3^+$   | HD    | 2.3e-11 | -0.45 | 3.2   | 7.7e-11 |
| 46 | oHD$_2^+$ | pD$_2$ | oD$_3^+$   | HD    | 4.9e-10 | -0.13 | 2.3   | 6.1e-10 |
| 47 | oHD$_2^+$ | pD$_2$ | mD$_3^+$   | HD    | 1.9e-10 | -0.15 | 0.6   | 3.0e-10 |
| 48 | pHD$_2^+$ | oD$_2$ | pHD$_2^+$  | pD$_2$ | 7.8e-11 | -0.01 | 86.3  | 1.4e-14 |
| 49 | pHD$_2^+$ | oD$_2$ | oHD$_2^+$  | pD$_2$ | 2.6e-10 | 0.20  | 31.9  | 5.4e-12 |
| 50 | pHD$_2^+$ | oD$_2$ | oHD$_2^+$  | oD$_2$ | 2.1e-10 | 0.32  | 0.8   | 6.5e-11 |
| 51 | pHD$_2^+$ | oD$_2$ | pD$_3^+$   | HD    | 4.1e-11 | -0.26 | 2.7   | 7.6e-11 |
| 52 | pHD$_2^+$ | oD$_2$ | oD$_3^+$   | HD    | 5.0e-10 | -0.16 | 0.8   | 8.0e-10 |
| 53 | pHD$_2^+$ | oD$_2$ | mD$_3^+$   | HD    | 2.2e-10 | -0.12 | 1.4   | 2.9e-10 |
| 54 | pHD$_2^+$ | pD$_2$ | pHD$_2^+$  | oD$_2$ | 5.7e-11 | 0.04  | -0.3  | 5.1e-11 |
| 55 | pHD$_2^+$ | pD$_2$ | oHD$_2^+$  | pD$_2$ | 1.4e-10 | 0.44  | -0.7  | 3.4e-11 |
| 56 | pHD$_2^+$ | pD$_2$ | oHD$_2^+$  | oD$_2$ | 2.1e-10 | 0.17  | -0.4  | 1.2e-10 |
| 57 | pHD$_2^+$ | pD$_2$ | pD$_3^+$   | HD    | 7.0e-11 | -0.08 | -0.5  | 9.7e-11 |
| 58 | pHD$_2^+$ | pD$_2$ | oD$_3^+$   | HD    | 5.7e-10 | -0.15 | 1.3   | 8.3e-10 |

**Table D10.** Same as Table D7 for reactions involving $D_3^+$.

| # | Reactants | | Products | | $\alpha$ cm$^3$ s$^{-1}$ | $\beta$ | $\gamma$ K | $k(10)$ cm$^3$ s$^{-1}$ |
|---|---|---|---|---|---|---|---|---|
| 1  | oD$_3^+$ | oH$_2$ | oH$_2$D$^+$ | pD$_2$ | 2.3e-10 | -1.01 | 324.3 | 5.9e-23 |
| 2  | oD$_3^+$ | oH$_2$ | oH$_2$D$^+$ | oD$_2$ | 1.3e-10 | -0.67 | 227.0 | 1.8e-19 |
| 3  | oD$_3^+$ | oH$_2$ | pHD$_2^+$ | HD | 1.1e-09 | -0.21 | 71.6 | 1.7e-12 |
| 4  | oD$_3^+$ | oH$_2$ | oHD$_2^+$ | HD | 5.8e-10 | -0.13 | 18.7 | 1.4e-10 |
| 5  | mD$_3^+$ | oH$_2$ | oH$_2$D$^+$ | oD$_2$ | 8.8e-10 | -0.52 | 270.7 | 9.0e-21 |
| 6  | mD$_3^+$ | oH$_2$ | oHD$_2^+$ | HD | 3.7e-09 | 0.41 | 60.2 | 2.2e-12 |
| 7  | mD$_3^+$ | pH$_2$ | pH$_2$D$^+$ | oD$_2$ | 8.1e-10 | -0.57 | 355.0 | 2.2e-24 |
| 8  | mD$_3^+$ | pH$_2$ | oHD$_2^+$ | HD | 9.6e-09 | -0.01 | 236.9 | 5.1e-19 |
| 9  | pD$_3^+$ | oH$_2$ | oH$_2$D$^+$ | pD$_2$ | 6.1e-10 | -0.55 | 304.6 | 2.3e-22 |
| 10 | pD$_3^+$ | oH$_2$ | pHD$_2^+$ | HD | 2.3e-09 | 0.17 | 50.1 | 8.6e-12 |
| 11 | oD$_3^+$ | pH$_2$ | pH$_2$D$^+$ | pD$_2$ | 3.1e-10 | -0.67 | 400.1 | 1.3e-26 |
| 12 | oD$_3^+$ | pH$_2$ | pH$_2$D$^+$ | oD$_2$ | 1.5e-10 | -0.85 | 315.3 | 5.5e-23 |
| 13 | oD$_3^+$ | pH$_2$ | pHD$_2^+$ | HD | 1.3e-09 | -0.75 | 253.9 | 1.6e-19 |
| 14 | oD$_3^+$ | pH$_2$ | oHD$_2^+$ | HD | 9.2e-10 | -0.59 | 197.8 | 1.8e-17 |
| 15 | pD$_3^+$ | pH$_2$ | pH$_2$D$^+$ | pD$_2$ | 8.2e-10 | -0.18 | 378.0 | 5.8e-26 |
| 16 | pD$_3^+$ | pH$_2$ | pHD$_2^+$ | HD | 3.5e-09 | -0.34 | 233.0 | 8.5e-19 |
| 17 | oD$_3^+$ | HD | pHD$_2^+$ | pD$_2$ | 4.2e-10 | -0.85 | 259.0 | 4.3e-20 |
| 18 | oD$_3^+$ | HD | pHD$_2^+$ | oD$_2$ | 3.8e-10 | -0.56 | 166.4 | 1.5e-16 |
| 19 | oD$_3^+$ | HD | oHD$_2^+$ | pD$_2$ | 7.3e-10 | -0.18 | 199.3 | 3.0e-18 |
| 20 | oD$_3^+$ | HD | oHD$_2^+$ | oD$_2$ | 3.6e-10 | -0.24 | 114.1 | 9.0e-15 |
| 21 | oD$_3^+$ | HD | pD$_3^+$ | HD | 4.5e-11 | -0.33 | 19.2 | 2.0e-11 |
| 22 | oD$_3^+$ | HD | mD$_3^+$ | HD | 2.1e-10 | -0.07 | 0.4 | 2.6e-10 |
| 23 | mD$_3^+$ | HD | pHD$_2^+$ | oD$_2$ | 7.5e-10 | -0.32 | 213.9 | 1.1e-18 |
| 24 | mD$_3^+$ | HD | oHD$_2^+$ | pD$_2$ | 2.0e-09 | 0.31 | 238.1 | 3.2e-20 |
| 25 | mD$_3^+$ | HD | oHD$_2^+$ | oD$_2$ | 2.9e-09 | 0.30 | 151.8 | 2.7e-16 |
| 26 | mD$_3^+$ | HD | oD$_3^+$ | HD | 1.1e-09 | 0.25 | 44.1 | 5.7e-12 |
| 27 | pD$_3^+$ | HD | pHD$_2^+$ | pD$_2$ | 1.6e-09 | -0.13 | 232.6 | 2.0e-19 |
| 28 | pD$_3^+$ | HD | pHD$_2^+$ | oD$_2$ | 6.9e-10 | -0.21 | 146.1 | 6.4e-16 |
| 29 | pD$_3^+$ | HD | oHD$_2^+$ | pD$_2$ | 8.9e-10 | 0.05 | 176.9 | 1.6e-17 |
| 30 | pD$_3^+$ | HD | oD$_3^+$ | HD | 9.5e-10 | 0.08 | -0.7 | 7.8e-10 |
| 31 | oD$_3^+$ | oD$_2$ | pD$_3^+$ | pD$_2$ | 2.5e-11 | -0.60 | 110.1 | 3.2e-15 |
| 32 | oD$_3^+$ | oD$_2$ | pD$_3^+$ | oD$_2$ | 2.0e-11 | -0.57 | 20.8 | 1.7e-11 |
| 33 | oD$_3^+$ | oD$_2$ | oD$_3^+$ | pD$_2$ | 4.9e-10 | -0.26 | 88.4 | 1.7e-13 |
| 34 | oD$_3^+$ | oD$_2$ | mD$_3^+$ | pD$_2$ | 2.2e-10 | -0.23 | 41.9 | 7.3e-12 |
| 35 | oD$_3^+$ | oD$_2$ | mD$_3^+$ | oD$_2$ | 1.3e-10 | -0.33 | 2.4 | 3.1e-10 |
| 36 | oD$_3^+$ | pD$_2$ | pD$_3^+$ | pD$_2$ | 3.2e-11 | -0.24 | 18.1 | 1.2e-11 |
| 37 | oD$_3^+$ | pD$_2$ | pD$_3^+$ | oD$_2$ | 4.4e-11 | -0.00 | -0.3 | 4.5e-11 |
| 38 | oD$_3^+$ | pD$_2$ | oD$_3^+$ | oD$_2$ | 3.5e-10 | -0.21 | 1.9 | 5.9e-10 |
| 39 | oD$_3^+$ | pD$_2$ | mD$_3^+$ | pD$_2$ | 3.7e-11 | -0.06 | 0.1 | 4.5e-11 |
| 40 | oD$_3^+$ | pD$_2$ | mD$_3^+$ | oD$_2$ | 6.2e-10 | 0.99 | -8.3 | 4.9e-11 |
| 41 | mD$_3^+$ | oD$_2$ | oD$_3^+$ | pD$_2$ | 3.4e-09 | 1.08 | 126.4 | 2.8e-16 |
| 42 | mD$_3^+$ | oD$_2$ | oD$_3^+$ | oD$_2$ | 6.5e-10 | -0.02 | 46.2 | 6.9e-12 |
| 43 | mD$_3^+$ | oD$_2$ | mD$_3^+$ | pD$_2$ | 2.9e-10 | 0.11 | 84.5 | 4.3e-14 |
| 44 | mD$_3^+$ | pD$_2$ | pD$_3^+$ | oD$_2$ | 3.2e-11 | -0.42 | 3.5 | 9.4e-11 |
| 45 | mD$_3^+$ | pD$_2$ | oD$_3^+$ | pD$_2$ | 1.9e-10 | 0.24 | 43.9 | 1.0e-12 |
| 46 | mD$_3^+$ | pD$_2$ | oD$_3^+$ | oD$_2$ | 8.1e-10 | 0.11 | -0.5 | 5.9e-10 |
| 47 | mD$_3^+$ | pD$_2$ | mD$_3^+$ | oD$_2$ | 2.2e-10 | 0.18 | -2.6 | 1.5e-10 |
| 48 | pD$_3^+$ | oD$_2$ | oD$_3^+$ | pD$_2$ | 1.3e-09 | 0.35 | 66.1 | 5.3e-13 |
| 49 | pD$_3^+$ | oD$_2$ | oD$_3^+$ | oD$_2$ | 4.1e-10 | -0.17 | 0.9 | 6.7e-10 |
| 50 | pD$_3^+$ | oD$_2$ | mD$_3^+$ | pD$_2$ | 1.8e-10 | -0.36 | 25.9 | 4.6e-11 |
| 51 | pD$_3^+$ | pD$_2$ | oD$_3^+$ | pD$_2$ | 6.6e-10 | 0.15 | -1.8 | 4.7e-10 |
| 52 | pD$_3^+$ | pD$_2$ | oD$_3^+$ | oD$_2$ | 4.7e-10 | -0.00 | 0.6 | 4.4e-10 |